\documentclass[11pt,a4paper, twoside, openright]{book}
\pdfoutput=1

\usepackage{hyperref}
\usepackage{bm}
\usepackage{amsmath,amssymb}
\hypersetup{colorlinks=true, citecolor=green, filecolor=black, linkcolor=blue, urlcolor=blue }
\usepackage{cite}
\usepackage{graphicx}
\usepackage[utf8]{inputenc}

\newcommand\ie{\textit{i.e.}\ }
\newcommand\eg{\textit{e.g.}\ }

\newcommand{\derarg}{\left(\frac{D^2}{M^2}\right)}
\newcommand{\derbarg}{\left(\frac{\nabla^2}{M^2}\right)}
\newcommand{\w}{\underset{\rm weak}{=}}

\newcommand{\dbar}{\mathchar'26\mkern-11mu\mathrm{d}}
\newcommand{\delbar}{\mathchar'26\mkern-11mu\mathrm{\delta}}
\newcommand{\cutarg}{(-\nabla^2)}
\newcommand{\pcutarg}{\left( \frac{p^2}{\Lambda^2} \right)}
\newcommand{\KoL}{\dot{\Delta}}
\newcommand{\tr}{{\rm tr}}

\newcommand{\Sv}{\mathcal{S}}
\newcommand{\Dm}{\mathcal{D}}
\newcommand{\cL}{\mathcal{L}}
\newcommand{\cO}{\mathcal{O}}
\newcommand{\ca}{\mathsf{a}}

\numberwithin{figure}{section}
\numberwithin{equation}{section}
\numberwithin{table}{section}

\begin{document}
\frontmatter

\thispagestyle{empty}

\begin{center}
\mbox{}


{\Huge University of Southampton}	\\
\vspace{0.5cm}
{\huge Faculty of Physical Sciences and Engineering}
{\Large Department of Physics and Astronomy}
\vfill

\vspace{2cm}

\textbf{\huge{Diffeomorphism-invariant averaging in quantum gravity and cosmology}}\\

\vspace{4cm}

{\huge
Anthony William Henry Preston
}
 \vspace{3cm}

{\Large Submitted for the degree of}\\
\vspace{0.5cm}
\begin{Large}\MakeUppercase{Doctor of Philosophy} \end{Large}\\

\vspace{1cm}
\vfill

{\Large
September 2016
}
\vfill
\end{center}


\newpage
\thispagestyle{empty}
\mbox{}


\newpage
\thispagestyle{empty}

\begin{center}

\MakeUppercase{University of Southampton}
\vfill
\underline{\MakeUppercase{{\Large Abstract}}}
\vfill
\MakeUppercase{Faculty of Physical Sciences and Engineering}

Department of Physics and Astronomy
\vfill
\underline{Doctor of Philosophy}
\vfill
\MakeUppercase{{\Large Diffeomorphism-invariant averaging in quantum gravity and cosmology}}
\vfill
Anthony William Henry Preston
\vfill 
\end{center}

\noindent
This thesis concerns research undertaken in two related topics concerning high-energy gravitational physics. The first is the construction of a manifestly diffeomorphism-invariant Exact Renormalization Group (ERG). This is a procedure that constructs effective theories of gravity by integrating out high-energy modes down to an ultraviolet cutoff scale without gauge-fixing. The manifest diffeomorphism invariance enables us to construct a fully background-independent formulation. This thesis will explore both the fixed-background and background-independent forms of the manifestly diffeomorphism-invariant ERG. The second topic is cosmological backreaction, which concerns the effect of averaging over high-frequency metric perturbations to the gravitational field equations describing the universe at large scales. This has been much studied the context of the unmodified form of General Relativity, but has been much less studied in the context of higher-derivative effective theories obtained by integrating out the high-energy modes of some more fundamental (quantum) theory of gravity. The effective stress-energy tensor for backreaction can be used directly as a diffeomorphism-invariant effective stress-energy tensor for gravitational waves without specifying the background metric.

This thesis will construct the manifestly diffeomorphism-invariant ERG and compute the effective action at the classical level in two different schemes. We will then turn to cosmological backreaction in higher-derivative gravity, deriving the general form of the effective stress-energy tensor due to inhomogeneity for local diffeomorphism-invariant effective theories of gravity. This an exciting research direction, as it begins the construction of a quantum theory of gravity as well as investigating possible implications for cosmology.
\vfill

\newpage
\thispagestyle{empty}

\tableofcontents

\listoffigures

\newpage
\thispagestyle{empty}

\newpage

\thispagestyle{empty}
\noindent
\textbf{{\LARGE Declaration of Authorship}}
\vspace{1cm}

\noindent
I, Anthony Preston, declare that this thesis, entitled ``Diffeomorphism-invariant averaging in quantum gravity and cosmology'', and the work presented in it, are my own and have been generated by me as the result of my own original research.

\mbox{}

\noindent
I confirm that:
\begin{enumerate}
\item This work was done wholly or mainly while in candidature for a research degree at this University;
\item No part of this thesis has previously been submitted for a degree or any other qualification at this University or any other institution;
\item Where I have consulted the published work of others, this is always clearly attributed;
\item Where I have quoted from the work of others, the source is always given. With the exception of such quotations, this thesis is entirely my own work;
\item I have acknowledged all main sources of help;
\item Chapters \ref{INTRO}, \ref{AWHPreview1} and \ref{AWHPreview2} are reviews, written by me, describing the previously existing literature. Chapters \ref{ChapterMDIERG} and \ref{ChapterBack1} report on original research conducted in collaboration with my supervisor, Tim Morris \cite{Preston:2014tua,Morris:2016nda}. Chapter \ref{ChapterBack2} reports on original research I conducted alone \cite{Preston:2016sip}. Chapter \ref{Summary} is a summary of the whole thesis.
\end{enumerate}

\mbox{}

Signed:	........................ \hspace{4cm} Date: ..........................

\vfill

\newpage
\thispagestyle{empty}
\noindent
\textbf{{\LARGE Acknowledgements}}
\vspace{1cm}

First of all, I would like to thank my supervisor, Tim Morris, for guiding me on this path and entertaining my wilder speculations while also teaching me how to formulate this research rigorously. I thank the University of Southampton for funding my PhD with a Mayflower Scholarship. I thank my parents and my brother for patiently supporting and encouraging me throughout my long education.

During my time in Southampton, I have met many brilliant people who have made PhD life colourful and rewarding. In particular, I would like to thank my fellow PhD students in the Southampton High Energy Physics Group for exchange of ideas and friendship. Working clockwise through the weird and wonderful world of my office, the amazing 46/4007, I thank Thomas Neder, the Big Friendly German, for lively discussions on physics and the meaning of life; I thank Hamzaan Bridle for regularly surprising us with edible treats and unexplained machinery; I thank Marc Scott for his artistry in decorating the office and for lending a sympathetic ear throughout the highs and lows of the past four years; I thank the visiting students, David Garner, Diana Rojas and Rafael Delgado for the completely different ways they each enriched our office culture and mythology; I thank Andrew Meadowcroft for his surgical wit and unexpected pranks; I thank Tadeusz Janowski for outstanding contributions to 4007's physics knowledge, creative chaos and reliquary; I thank Alex Titterton for an endless stream of puns; I thank Jeff Choi for his helpfulness and stoicism; I thank Simon King for keeping us all on our toes; I thank Matt Spraggs for his entertainingly heated monologues; I thank Ronnie Rodgers for his remarkable enthusiasm for formal physics; I thank Maien Binjonaid for being the office's wise elder statesman; I thank Azaria Coupe for injecting some optimism amid the madness; I thank Edwin Lizarazo for keeping the peace during the frequent office dramas; I thank Andrew Lawson for sound programming advice.

Beyond the walls of 4007, I thank Jason Hammett for helping me to get settled into the office and maintaining a cheerful disposition amid the slings and arrows of outrageous leptoquarks. I thank the double-doctor Marc Thomas for inspirational conversations. I thank the Asymptotic Safety students, J\"urgen Dietz and Zo\"e Slade, for interesting discussions and good humour in the face of my surreal tomfoolery. Many other people in Southampton and beyond have had a positive impact on my time here, and I thank you all for helping me along this adventure.

\newpage 
\thispagestyle{empty}
\mainmatter

\chapter{Introduction}\label{INTRO}

The layout of this introduction is as follows. Section \ref{IntroGentle} will give a broad discussion of how varying scales of length are studied in different physical contexts. Section \ref{IntroStatMech} will discuss how problems with many degrees of freedom are tackled in statistical mechanics. This is of interest because there is a strong overlap between the methods of statistical mechanics and the methods of Quantum Field Theory (QFT), which are relevant to this thesis. In particular, Renormalization Group (RG) methods will be discussed in Subsections \ref{IntroKadanoff} and \ref{IntroFlow}, while a supersymmetric method for studying disordered systems will be discussed in \ref{IntroPSSUSY}. Section \ref{IntroGR} provides a summary of General Relativity as our current best theory of gravity. Section \ref{LCDM} is an overview of cosmology, discussing its history and present status.
In a addition to this introduction, there will be a short review of the manifestly gauge-invariant Exact RG (ERG) in Chapter \ref{AWHPreview1} and a short review of backreaction in Einstein gravity in Chapter \ref{AWHPreview2} to prepare the ground before discussing original research presented in Chapters \ref{ChapterMDIERG}, \ref{ChapterBack1} and \ref{ChapterBack2}, which report on work published in \cite{Preston:2014tua,Morris:2016nda,Preston:2016sip}. Chapter \ref{ChapterMDIERG} concerns the manifestly diffeomorphism-invariant ERG \cite{Morris:2016nda}, Chapter \ref{ChapterBack1} covers backreaction in the context of a simple $R+R^2/6M^2$ theory of gravity \cite{Preston:2014tua}, and finally \ref{ChapterBack2} delves into backreaction in general local diffeomorphism-invariant higher-derivative gravity expansions \cite{Preston:2016sip}. Chapter \ref{Summary} is a summary of the whole thesis.

\section{Physics at many scales of length}\label{IntroGentle}

When addressing any physics problem, one always has particular scales of distance and time in mind.
Usually, an approximation is made where interactions over some range of physical scales are explicitly considered and others are either neglected or averaged over.
For a mundane example, consider fluid flow down a domestic water pipe. To calculate this, one would usually ignore the microscopic physics of individual water molecules. These approximations are necessary because it is not feasible to gather all theoretically possible information on every microscopic degree of freedom, nor is it feasible to perform computations at such a level. Such an approximation is acceptable because the difference between the approximation and reality is small enough to be neglected. There are a variety of ways to express such choices of approximation. In the example of fluid mechanics, practitioners in that field favour the use of certain named dimensionless numbers \cite{citeulike:6043967}. To decide whether it is important to use the microscopic understanding of a fluid as composed of particles, or if it is valid to simply use the continuum approximation of fluid mechanics, the Knudson number of the system is consulted \cite{ANDP:ANDP19093330106}. The Knudson number is defined as
\begin{equation}
 {\rm Kn} = \frac{\lambda}{L},
\end{equation}
where $\lambda$ is the mean free path of the fluid particles and $L$ is the length scale one wishes to study. When the Knudson number is small, the fluid may be safely approximated by continuous fields. Other aspects of fluid physics at different scales of length are captured by the famous ``Reynolds number'' \cite{1851TCaPS...9....8S,10.2307/109431}, which is the ratio of inertial to viscous forces:
\begin{equation}
 {\rm Re} = \frac{UL}{\nu},
\end{equation}
where $U$ is the characteristic speed of the fluid and $\nu$ is the kinematic viscosity. When the Reynolds number is small, the fluid flow is laminar, dominated by viscous forces and easily calculable. Conversely, a large Reynolds number corresponds to a turbulent system that is can exhibit chaotic behaviour and is typically full of complicated features such as eddies. The dependence on $L$ tells us that very small fluid systems are usually dominated by viscous forces, whereas larger systems can be dominated by inertial forces and thus enter a turbulent flow regime.

The challenge of studying systems with many degrees of freedom also appears in solid-state physics. A unit commonly favoured by chemists for the number of particles (such as atoms or molecules) in a sample is the ``mole'' \cite{0026-1394-8-1-006}, where 1 mole $\approx 6\times 10^{23}$ particles. An ``everyday'' solid object might contain many moles of atoms. Each of the atoms can independently oscillate in three dimensions of space and carry its own spin state. In order to be able to understand the macroscopic behaviours of such as system, it is impractical to attempt to directly simulate all of these coupled degrees of freedom, rather one needs to apply an averaging scheme. The method that is usually applied in statistical mechanics and indeed high-energy physics is the Renormalization Group (RG) \cite{Wilson:1971bg,Wilson:1971dh,Wilson:1973jj,Wilson:1974mb,Fisher:1974uq,Fisher:1998kv}.

Born out of nuclear physics in the mid-twentieth century, high-energy particle physics is the study of particle interactions at extremely short length scales, usually by means of collisions with a large centre-of-mass energy. For a rough quantitative guide, energies at or below $\mathcal{O}$(10 eV) are typical in the study of atomic physics, while energies at or below $\mathcal{O}$(10 MeV) are typical in nuclear physics. Typically, any energies above about 1 GeV (which is roughly the mass of a proton) belong to particle physics. 

Particle physicists tend to have a very energy-centric view of physics, adopting a system of ``natural units'' such that all physical scales can be described in terms of an energy scale. Since Lorentz invariance requires that the dimensions of space and time are related in a single ``spacetime'' geometry, it is natural to adopt a system of units in which separations in all dimensions of space and time have the same units, \ie to set $c=1$. Similarly, since quantum mechanics directly relates the energy of a particle to the frequency of a corresponding wave via Planck's constant (and thus position and momentum representations are related via Fourier transformations), it becomes natural to set $\hbar = 1$. If we also wish to study the physics of a large number of interacting particles, it becomes convenient to use the same units for temperature as for energy, \ie to set $k_{B}=1$. Thus we have a system of natural units where we can express all physical quantities using only a single unit, \ie by specifying just one scale of length to refer to. The current energy frontier from direct experiment is set by the Large Hadron Collider (LHC), which, at the time of writing, is performing proton-proton collisions with a centre of mass energy of 13 TeV. 

In July 2012, the ATLAS and CMS experiments at the LHC found conclusive evidence for the Higgs boson \cite{Aad:2012tfa,Chatrchyan:2012xdj}, which is the final particle required to complete the Standard Model of particle physics. Presently, the results of the various LHC experiments are in remarkably good agreement with the Standard Model. There is, however, direct experimental evidence for particle physics Beyond the Standard Model (BSM) from neutrino oscillations, which imply that neutrinos possess a mass, in direct contradiction to the Standard Model \cite{Fukuda:1998mi,Ahmad:2002jz,Eguchi:2002dm}. There is also indirect evidence for new high-energy physics from cosmology, as will be discussed in Section \ref{LCDM}. The most starkly obvious omission in the Standard Model is gravity, which is the core subject of this thesis.

\section{Statistical mechanics and the Renormalization Group}\label{IntroStatMech}

\subsection{The Ising model}

Statistical mechanics is the established theoretical framework for studying complex systems in physics \cite{Yeom92}. There are strong links between the techniques employed in statistical mechanics and those employed in QFT \cite{citeulike:12927284}, especially concerning RG methods, which are of special interest to this thesis.
A commonly used toy model for exploring calculational methods in statistical mechanics (and by extension other areas, such as particle physics) is the Ising model \cite{Lenz1920,Ising:1925em}. The Ising model envisages a $D$-dimensional (in space) lattice of spins with nearest-neighbour interactions in a canonical ensemble. The most commonly considered case is the spin-1/2 Ising model, whose Hamiltonian, $\mathcal{H}$, can be written as
\begin{equation}
 \mathcal{H} = -J\sum_{\left<ij\right>}s_i s_j - h\sum_i s_i.
\end{equation}
where $J$ is the exchange energy, $s_i$ is a classical spin variable that can take values of $\pm 1$, $h$ is an external magnetic field and the notation $\left<ij\right>$ specifies that the sum is over pairs of nearest neighbours. A positive value for $J$ energetically favours parallel alignment of the spins (ferromagnetism), whereas a negative value favours antiparallel alignment (antiferromagnetism). A zero value for $J$ simply favours alignment with $h$ (paramagnetism).

When the temperature is taken to zero, the system drops into its ground state of either perfectly parallel or antiparallel alignment of the spins (unless both $J$ and $h$ are zero). As the temperature is raised, the system gains entropy as some spins move out of the energetically favoured alignment with their neighbours and ultimately take random values as the temperature tends to infinity. To extract the values for the various functions of state in statistical mechanics, a partition function is calculated. The partition function, which (here) is a function of temperature $T$ and magnetic field $h$, is
\begin{equation}\label{TraditionalPartition}
 \mathcal{Z}(T,h) = \sum_{n} e^{-\beta E_n},
\end{equation}
where $n$ is the set of labels for microstates of the system and $\beta$ is defined as $1/k_B T$.
From the partition function, it is easy to calculate the Helmholtz free energy, $F$:
\begin{equation}
 F = -k_B T {\rm ln} \mathcal{Z}.
\end{equation}
With this information, all the other functions of state can be calculated from their definitions and the fundamental relation of thermodynamics for this system:
\begin{equation}
 {\rm d}U = T{\rm d}S -M{\rm d}h,
\end{equation}
where $U$ is the internal energy, $S$ is the entropy and $M$ is the magnetisation. The exact partition function for the Ising model is easily derived in one spatial dimension (a polymer-like spin-chain). The two-dimensional model was solved exactly by Onsager in 1944 \cite{Onsager:1943jn}. The three-dimensional Ising model is much more difficult and there is no known exact solution. This motivated Wilson's development of the RG as a method to compute scaling behaviours for complex systems without requiring an exact solution. For modern studies, see for example \cite{ElShowk:2012ht,El-Showk:2014dwa}.
 
\subsection{Phase transitions}\label{IntroPhaseTransitions}

A strong motivation for the development of RG techniques is the study of second-order phase transitions. A phase transition is a sharp change in the macroscopic behaviour of a substance, as indicated by a singularity in the Helmholtz free energy, or one of its derivatives. A first order phase transition is when there is a singularity in one of the first derivatives. A second order transition is when there is a singularity in the second derivatives: this is otherwise known as a continuous or critical phase transition. To understand how the physics changes close to a phase transition, it is helpful to look at the correlation between pairs of spins as a function of distance, \ie the 2-point correlation function. A common convention for this is the mean product of the deviation from the average spin value of pairs of spins, \ie
\begin{equation}
 C^{(2)}\left(\vec{r}_i,\vec{r}_j\right) = \left<\left(s_i-\left<s_i\right>\right)\left(s_j-\left<s_j\right>\right)\right>,
\end{equation}
where $\vec{r}_i$ is the position vector for the $i$th spin. For a translationally-invariant system, this can be expressed as a function of a single position vector:
\begin{equation}
 C^{(2)}\left(\vec{r}_i - \vec{r}_j\right) = \left<s_i s_j\right>-\left<s\right>^2.
\end{equation}
An alternative convention is simply to take the mean product of pairs of spins, \ie to leave out the last term. 

The 2-point correlation gives us information about the size of structures in the system.
The value of the 2-point correlation above is usually suppressed at large distances by an exponential decay, thus pairs of spins that are separated by a large distance usually have a negligible correlation. More precisely, the correlation function decays as
\begin{equation}
 C^{(2)}\left(\vec{r}\right) \sim r^{-k}{\rm exp}(-r/\xi),
\end{equation}
where $k$ is a dimensionless number called the ``critical exponent'' and $\xi$ has dimensions of length and is called the ``correlation length''. As one approaches a second order phase transition, $\xi\to\infty$. When this happens, the correlation function simply follows a power law as the system becomes scale-invariant, \ie the system appears unchanged under rescaling all distances by some factor. In this regime, structures can be found at all scales of length. The value of a parameter (such as temperature) where this transition happens is called its ``critical'' value.
The classical prediction for the critical exponent is $k=D-2$, where $D$ is the dimension of the system. This is not what is typically observed in nature. The measurable critical exponent is often written as $k=D-2+\eta$, where $\eta$ is the anomalous dimension for the system that appears due to interactions. In QFT, $\eta=0$ corresponds to theories of zero-mass non-interacting particles. Prior to the development of RG for statistical mechanics, there was no framework for calculating $\eta$. This made it especially mysterious that there exist ``universality classes'', which are sets of values for critical exponents that are especially common, appearing in systems that are seemingly unrelated.

\subsection{Kadanoff blocking and RG transformations}\label{IntroKadanoff}

Wilson's RG is theoretically underpinned by Kadanoff blocking \cite{Kadanoff:1966wm,Wilson:1971bg}, which is a transformation used to reduce the number of degrees of freedom in a system, thus allowing one to study the macroscopic properties of a system with a large number of microscopic degrees of freedom in a way that is computationally tractable. This is feasible to do because, for most physical systems, most degrees of freedom in the system are individually unimportant in describing its macroscopic properties. In non-local systems, however, where interactions between individual microscopic degrees of freedom can take place over very large length scales without suppression, one could not apply an averaging scheme like this without losing important information.

Kadanoff blocking was developed for studying scaling properties of the Ising model as it approaches its critical temperature \cite{Kadanoff:1966wm}. It works by grouping together the microscopic degrees of freedom (\ie the spins) into a regular set of simply-connected, tessellated blocks, each retaining only a single degree of freedom. The lattice of spins is then rescaled to form a new lattice such that the physical separation between two lattice points, $a$, rescales according to
\begin{equation}\label{IntroRGscale}
 b^D := \left(\frac{a_{\rm after}}{a_{\rm before}}\right)^D = \frac{N_{\rm before}}{N_{\rm after}}, 
\end{equation}
where $b$ is the scale factor of the transformation and $N$ is the number of degrees of freedom. The ``before'' and ``after'' refer to values before and after blocking respectively.
A simple example of this is to consider the two-dimensional Ising model and a blocking scheme in which the spins are grouped into three-by-three blocks. The blocking scheme can be chosen such that the macroscopic ``blocked'' spins take their values from a modal average of the microscopic spins in the block. This is one example of a Kadanoff blocking scheme: one could potentially devise infinitely many others for a given system. The application of the blocking procedure followed by subsequent rescaling of the lattice is also known as an RG transformation. 

Because RG transformations reduce the number of degrees of freedom, they are not invertible and it is technically incorrect to describe them as forming a symmetry group. It is formally better to refer to them as a ``semi-group''. For most particle physicists, this is an academic point because ``renormalizable'' theories possess a self-similarity at different scales of length such that RG flow applies to a set of couplings that transforms. Since the transformation of the couplings is described by calculable $\beta$-functions, one can use the value of a coupling at one scale to know its value at all scales. This does suppose that the original theory is complete, not missing out on some suppressed physics that only becomes apparent at short length scales.

Typically, the Hamiltonian of the system is changed after the RG transformation, but a necessary condition for consistency is that the partition function remains invariant. This ensures that the RG transformation does not change the macroscopic observables of the system, which can all be derived from the partition function \eg via calculating the Helmholtz free energy from the partition function and then using standard thermodynamic relations to derive the others. While the macroscopic observables remain unchanged, the blocking does rescale the microscopic degrees of freedom. 
Since the lattice spacings are rescaled by the RG transformation, as given in (\ref{IntroRGscale}), parameters measured in lattice units, \ie by setting $a=1$, are also rescaled according to their dimension in lattice units. For example, momentum scales as $\vec{p}_{\rm after} = b\vec{p}_{\rm before}$, spatial displacements rescale as $\vec{r}_{\rm after} = b^{-1}\vec{r}_{\rm before}$ and lengths, such as correlation lengths, also rescale as $\xi_{\rm after} = b^{-1}\xi_{\rm before}$.
It is convenient to define a ``reduced'' Hamiltonian $\bar{\mathcal{H}}:=\beta\mathcal{H}$ and reduced free energy $\bar{F}:=\beta F$. Since the free energy $F$ is extensive, and the blocking reduces the number of spins by a factor of $b^D$, the reduced free energy per spin $\bar{f}$ rescales as
\begin{equation}
 \bar{f}_{\rm after} = b^D \bar{f}_{\rm before}.
\end{equation}

A fixed-point of the RG transformation is where the Hamiltonian remains unchanged after blocking. In this case, the correlation length also remains unchanged. This can only be reconciled with an RG step that has reduced the degrees of freedom (and hence rescaled the correlation length) if the correlation length is infinite or zero. If the correlation length is zero, the system is trivially scale-invariant in the sense that it has no structure present at all. The correlation length becomes infinite at criticality, \ie at a second-order phase transition. In the latter scenario, there are structures present at all scales of length. A physical example of this is the critical point of water. This occurs at a temperature of 647 K and a pressure for 218 atm, where gas and liquid structures can be found across a wide (formally infinite) range of length scales, causing a scattering of light across the full range of wavelengths, which in turn makes the gas/liquid mixture take a white appearance. Other substances have critical points at different temperatures and pressures.

\subsection{RG flow}\label{IntroFlow}

Consider an RG transformation with scale factor $b$, such that a reduced Hamiltonian $\bar{\mathcal{H}}$ transforms to ${\bf R}_b\bar{\mathcal{H}}$, where ${\bf R}_b$ is the operator that effects the transformation. A fixed-point reduced Hamiltonian $\bar{\mathcal{H}}^*$ is invariant under RG transformations. Of more interest is a system that is close to a fixed point, but perturbed slightly away from it. Consider a system described by a reduced Hamiltonian of the form
\begin{equation}
 \bar{\mathcal{H}} = \bar{\mathcal{H}}^* + \sum_i g_i \mathcal{O}_i,
\end{equation}
where $\mathcal{O}_i$ are a basis of operators that perturb the Hamiltonian away from the fixed point and $g_i$ are conjugate fields for those operators. The RG transformation on the Hamiltonian is then
\begin{equation}
 {\bf R}_b\left[\bar{\mathcal{H}}^* + \sum_i g_i \mathcal{O}_i\right] = \bar{\mathcal{H}}^* + \sum_i g_i {\bf L}_b\mathcal{O}_i,
\end{equation}
where ${\bf L}_b$ is a linear operator. For the set of eigenoperators $\mathcal{O}_k$ of ${\bf L}_b$,
\begin{equation}
 {\bf L}_b\mathcal{O}_k = b^{\lambda_k}\mathcal{O}_k.
\end{equation}
This effectively rescales the conjugate fields as
\begin{equation}
 g_i \to g_i b^{\lambda_k}.  
\end{equation}
The basis of eigenoperators gives the set of trajectories for the RG flow of the system. If $\lambda_k$ is positive, the operator is called ``relevant''. Relevant operators become more important as the system becomes coarse-grained by RG transformations, thus they are important for describing the system at large scales. Successive RG transformations drive the system further from the fixed point along these relevant directions. If $\lambda_k$ is negative, the operator is called ``irrelevant''. Along these irrelevant directions, RG transformations take the Hamiltonian back towards the fixed point, thus they are unimportant for describing the system at large scales. If $\lambda_k$ is zero, the operator is called ``marginal''. The values of $\lambda_k$ determine the scaling relations of the system near criticality. Universality classes exist because RG flow describes transformations in the space of Hamiltonians and many different Hamiltonians can be found by RG transformations in the vicinity of a given fixed point. 

\subsection{Statistical field theory}

Consider a continuous theory parametrized by a continuous scalar field $\phi$ (known as the order parameter) such that the Hamiltonian is a functional of the order parameter. The order parameter is a generic field that can take different physical meanings in different systems. For example, the order parameter might be the magnetization of a system, or a density perturbation. The generalization of the partition function (\ref{TraditionalPartition}) replaces the sum over microstates with an integral over field configurations, known as the path integral \cite{Feynman:1948ur}:
\begin{equation}\label{StatFieldPartition}
 \mathcal{Z} = \int \mathcal{D} \phi\ e^{-\bar{\mathcal{H}}[\phi]}.
\end{equation}
A frequently used form of $\bar{\mathcal{H}}$ is one where symmetries are assumed under spatial translations, rotations and $\phi\to-\phi$ transformations:
\begin{equation}\label{sampleHamiltonian}
 \bar{\mathcal{H}} = \int d^D x\left(-\frac{1}{2}\phi(x)\left(\partial_t^2+\partial_x^2\right) \phi(x) + \frac{1}{2}m\phi(x)^2 + \frac{g}{4!}\phi(x)^4+\cdots\right),
\end{equation}
where $\partial^2_t$ and $\partial_x^2$ are squared partial derivative operators with respect to time and space coordinates respectively (there are usually several spatial dimensions which are not distinguished here).
The $\phi\to-\phi$ symmetry may be broken after a phase transition, such as dropping the temperature below a critical value, or applying an external magnetic field above a critical value. As was touched on in Subsection \ref{IntroPhaseTransitions}, one can extract physical observables from correlation functions formed from the mean values of products of fields:
\begin{equation}
 \left<\phi(x_1)\phi(x_2)\cdots\phi(x_n)\right> = \frac{1}{\mathcal{Z}}\int \mathcal{D}\phi\ \phi(x_1)\phi(x_2)\cdots\phi(x_n)e^{-\bar{\mathcal{H}}[\phi]}.
\end{equation}
Note that the right hand side is an integral over field configurations weighted by the probability of each configuration. These can be derived using a generalized form of 
(\ref{StatFieldPartition}) with an external source field $h(x)$:
\begin{equation}\label{StatFieldPartSource}
 \mathcal{Z}[h] = \int \mathcal{D} \phi\ {\rm exp}\left(-\bar{\mathcal{H}}[\phi]+\int dt\ d^D x\ h(x)\phi(x)\right).
\end{equation}
The correlation functions can be calculated from successive functional derivatives of (\ref{StatFieldPartSource}) by
\begin{equation}
 \left<\phi(x_1)\phi(x_2)\cdots\phi(x_n)\right> = \left.\frac{1}{\mathcal{Z}[0]}\frac{\delta}{\delta h(x_1)}\frac{\delta}{\delta h(x_2)}\cdots\frac{\delta}{\delta h(x_n)}\mathcal{Z}[h]\right|_{h=0}.
\end{equation}
Since the partition function is invariant under RG transformations, the correlation functions derived in this way are also invariant, so RG transformations do not alter the macroscopic observables of the system. It is often more convenient, especially in relativistic theories, to use a Lagrangian formulation. The Lagrangian $L$ is related to the Hamiltonian $\mathcal{H}$ via a Legendre transform:
\begin{equation}
 \mathcal{H} = \sum_i \dot{q}_i\frac{\partial L}{\partial \dot{q}_i}-L,
\end{equation}
where $q_i$ are ``generalized coordinates''. For the Hamiltonian in (\ref{sampleHamiltonian}), we have just one field $q=\phi$, so we have
\begin{equation}\label{sampleLagrangian}
 \bar{L} = \int d^D x\left(-\frac{1}{2}\phi(x)\left(\partial_t^2-\partial_x^2\right) \phi(x) - \frac{1}{2}m\phi(x)^2 - \frac{g}{4!}\phi(x)^4-\cdots\right).
\end{equation}
Incidentally, a Wick rotation, as one would perform in order to transform a Lorentz-invariant theory from Minkowski to Euclidean spacetime, $t\to it$, would also transform (\ref{sampleLagrangian}) to minus the form seen in (\ref{sampleHamiltonian}) (noting that the integrals shown here do not carry a time dimension, as they would in a full QFT, which would be the correct context for such a transformation). This also gives an insight into how the path integral in QFT is related to the partition function in statistical mechanics. For an action $S = \int dt L$, the path integral in QFT for a field $\phi$ and source $J$ is
\begin{equation}
 \mathcal{Z}[J] = \int \mathcal{D}\phi\ {\rm exp}\left(iS - i\int dt\ d^Dx\ J\phi\right).
\end{equation}
After performing a Wick rotation to Euclidean space such that we now have the Euclidean action, $S_E=\int dt L_E$, the path integral reads as
\begin{equation}
 \mathcal{Z}[J] = \int \mathcal{D}\phi\ {\rm exp}\left(-S_E + \int dt\ d^Dx\ J\phi\right),
\end{equation}
where the Euclidean Lagrangian $L_E$, for the choice of $L$ in (\ref{sampleLagrangian}), is now the same as the form in (\ref{sampleHamiltonian}). More generally, $S_E$ is always positive definite, giving the advantage that the Boltzmann factor is always exponentially damped as the field or its derivatives become large. It is natural then to use the terms ``path integral'' and ``partition function'' interchangeably in QFT.

\subsection{Parisi-Sourlas supersymmetry}\label{IntroPSSUSY}

A supersymmetry is a symmetry under transformations of bosonic fields to fermionic fields and vice versa \cite{Ramond:1971gb,Neveu:1971rx,Gervais:1971ji,Wess:1974tw,Delbourgo:1974jg,Salam:1974yz,Salam:1974pp,Fayet:1976cr}. It is most famously of interest to the particle physics phenomenology community as a possible means to extend the Standard Model \cite{Haber:1984rc}. However, for this thesis, supersymmetry is of interest purely for formal reasons that have no phenomenological implications in themselves. In this subsection, let us consider the application of supersymmetry as a formal device in statistical field theory, as used in understanding spin glasses, or disordered systems more generally. This is the supersymmetry formalism developed by Parisi and Sourlas in 1979 \cite{Parisi:1979ka}. Ordinary glasses are solids where the spatial distribution of the atoms is fixed but disorderly, unlike a crystal, where the atoms are arranged according to some symmetry property of the spatial distribution. Spin glasses are solids where, instead of the spatial positions, it is the magnetic spin states that are disorderly, \ie there is no correlation between nearby spins \cite{Binder:1986zz}. In this subsection, all energies are reduced energies, which will be left implicit for notational convenience.

Such a system can be modelled using a field theory where the Helmholtz free energy is perturbed by a magnetic source, $h$, whose value at each position coordinate, $x$, is a Gaussian-distributed random variable, \ie there is a zero correlation length. The free energy, $F$, is a functional of $h(x)$:
\begin{equation}
 F[h] = {\rm ln}\int \mathcal{D}\phi \ {\rm exp}\left(-\int d^D x \left[\mathcal{L}(x)+h(x)\phi(x)\right]\right),
\end{equation}
where $\phi(x)$ is the order parameter, which is a scalar field, $\mathcal{L}(x)$ is the Lagrangian density and $D$ is the dimension of the system. The Lagrangian can be written as
\begin{equation}
 \mathcal{L}(x) = -\frac{1}{2}\phi(x)\partial^2 \phi(x) + V\left(\phi(x)\right),
\end{equation}
where $V(\phi)$ is the potential for the scalar field. A mathematical curiosity of spin systems that are acted on by random external magnetic fields is that their macroscopic observables are the same as those of equivalent spin systems with no external fields and two fewer dimensions \cite{Imry:1975zz,Aharony:1976jx,0022-3719-10-9-007}. For example, a three-dimensional spin system acted upon by a random external source would have the same functions of state as the corresponding one-dimensional system with no external source. This motivated the supersymmetric formalism that demonstrated why this is. To see this, they started by writing down the form of the 2-point function:
\begin{equation}\label{PSav2p}
 \left<\phi(x)\phi(0)\right> \sim \int \mathcal{D} h \ \phi_h(x)\phi_h(0)\ {\rm exp}\left(-\frac{1}{2}\int d^D y\ h^2(y)\right),
\end{equation}
where $\phi_h$ is the solution to the classical field equation for $\phi$,
\begin{equation}
 -\partial^2 \phi + V'(\phi) + h = 0.
\end{equation}
The first step is to rewrite (\ref{PSav2p}) as
\begin{eqnarray}
 \left<\phi(x)\phi(0)\right> & \sim & \int\mathcal{D}\phi\ \mathcal{D}h\ \phi(x)\phi(0)\delta\left(-\partial^2 \phi + V'(\phi) + h\right){\rm det}\left[-\partial^2 + V''(\phi)\right] \nonumber \\ && \times\ {\rm exp}\left[-\frac{1}{2}\int d^D y h^2(y)\right],
\end{eqnarray}
which can be the seen to be the same as (\ref{PSav2p}) by performing the functional integral over $\phi$. The next step is to reparametrize the fields into a more convenient form:
\begin{equation}
 \left<\phi(x)\phi(0)\right> \sim \int \mathcal{D}\phi\ \mathcal{D}\omega\ \mathcal{D}\psi\ \mathcal{D}\bar{\psi}\ {\rm exp}\left[-\int d^D y\mathcal{L}_R(y)\right]\phi(x)\phi(0),
\end{equation}
where $\psi$ is an anticommuting scalar field and $\mathcal{L}_R$ is the averaged Lagrangian:
\begin{equation}\label{PSavLag}
 \mathcal{L}_R = -\frac{1}{2}\omega^2 + \omega \left[-\partial^2\phi + V'(\phi)\right] + \bar{\psi}\left[-\partial^2+V''(\phi)\right]\psi.
\end{equation}
This averaged Lagrangian is now supersymmetric under the following transformations:
\begin{eqnarray}\label{PSsusytran}
 \delta \phi & = & -\bar{a}\epsilon_\mu x_\mu \psi,\nonumber\\
 \delta \omega & = & 2\bar{a}\epsilon_\mu\partial_\mu\psi,\nonumber\\
 \delta\psi & = & 0, \nonumber\\
 \delta \bar{\psi} & = & \bar{a}\left(\epsilon_\mu x_\mu \omega + 2\epsilon_\mu\partial_\mu\phi\right),
\end{eqnarray}
where $\bar{a}$ is an infinitesimal anticommuting number, $\epsilon_\mu$ is an arbitrary vector and the repeated Greek indices indicate a summation with the identity as the metric. This supersymmetry can be made more manifest by re-expressing the Lagrangian, which is the integral over $D$ dimensions of the Lagrangian density (\ref{PSavLag}), in terms of a superfield. The superfield, $\Phi$ is a function of the ``bosonic'' coordinates, $x$, and fermionic coordinates, $\theta$. It can be written as
\begin{equation}
 \Phi(x,\theta) = \phi(x) + \bar{\theta}\psi(x) + \bar{\psi}(x)\theta + \theta\bar{\theta}\omega(x).
\end{equation}
The manifestly supersymmetric action is then
\begin{equation}
 L = \int d^D x\ d\bar{\theta}\ d\theta\left(-\frac{1}{2}\Phi\left[\partial^2 + \frac{\partial^2}{\partial \bar{\theta}\partial\theta}\right]\Phi+V(\Phi)\right).
\end{equation}
The integration over the fermionic coordinates effectively selects the term proportional to $\theta\bar{\theta}$. In the chosen normalization, this gives
\begin{equation}
 \int d\bar{\theta}\ d\theta\ \Phi = -\frac{1}{\pi}\omega(x).
\end{equation}
For some function, $f$,
\begin{equation}
 \int d^D x\ d\bar{\theta}\ d\theta\ f\left(x^2 + \bar{\theta}\theta\right) = -\frac{1}{\pi}\int d^D x\ \frac{d}{dx^2}f(x^2).
\end{equation}
Integrating isotropically, using the standard result for a $D$-dimensional spherical integral, this can be written as
\begin{equation}\label{PShsi}
 -\frac{1}{\pi}\frac{2\pi^{D/2}}{\Gamma(D/2)}\int dx\ x^{D-1}\frac{d}{dx^2}f(x^2).
\end{equation}
Using the chain rule, this rearranges to
\begin{equation}
 -\frac{2\pi^{D/2-1}}{\Gamma(D/2)}\int dx\ \frac{x^{D-2}}{2}\frac{d}{dx}f(x^2).
\end{equation}
Integrating by parts gives
\begin{equation}
 2\pi^{D/2-1}\frac{D/2-1}{\Gamma(D/2)}\int dx\ x^{D-3}f(x^2) = \frac{2\pi^{D/2-1}}{\Gamma(D/2-1)}\int dx x^{D-3}f(x^2).
\end{equation}
From this follows the result that indicates the effective reduction in dimension by two:
\begin{equation}
  \int d^D x\ d\bar{\theta}\ d\theta\ f\left(x^2 + \bar{\theta}\theta\right) = \int d^{D-2}x\ f(x^2).
\end{equation}
The supersymmetry transformations given in (\ref{PSsusytran}) are effectively rotations in superspace such that the interval $x^2+\bar{\theta}\theta$ remains invariant. Parisi and Sourlas gave this argument to demonstrate that the superspace with $D$ bosonic coordinates was equivalent to a ($D-2$)-dimensional pure bosonic space. From this, it then follows that the macroscopic observables for the system forced with a random external field are the same as for an isolated system with two fewer dimensions, as had previously been seen by explicitly computed examples. In Section \ref{MGIERGSUSY}, I will discuss how this is related to an elegant regularization scheme for the manifestly gauge-invariant ERG that is a special form of Pauli-Villars regularization. The success of this method for gauge theories motivates its adaptation to the manifestly diffeomorphism-invariant ERG in the form of supermanifold regularization, which we will return to in Section \ref{MDIERGSUSY}.

\section{Overview of General Relativity}\label{IntroGR}
This section gives an overview of Einstein's general theory of relativity \cite{Einstein:1911vc,Einstein:1914bw,Einstein:1914bx,Einstein:1915by,Einstein:1915bz,Einstein:1915ca,Einstein:1916vd}, starting from its foundations and paying particular attention to those aspects that are relevant for this thesis. This is the subject of many textbooks, this overview has been particularly influenced by \cite{Waldbook,Rindlerbook,GRHob,ellis2012relativistic}.
\subsection{Conceptual background}

Prior to General Relativity (GR), the accepted theory of gravity came from Newton \cite{principia}, who described gravity as a force $F$, with a coupling constant (Newton's constant) $G$, exerted by any body of mass $M_1$ on any other body of mass $M_2$, separated by a distance $r$, such that
\begin{equation}\label{IntroNewton}
 \vec{F} = -\frac{GM_1M_2}{r^2}\hat{r}.
\end{equation}
This fits well with the overall formalism of Newton's Classical Mechanics, for which the geometry follows from the Galilean transformations between the reference frames of different observers:
\begin{eqnarray}\label{IntroGalTran}
 t' & = & t,\nonumber\\
 x' & = & x-v_xt,\nonumber\\
 y' & = & y-v_yt,\nonumber\\
 z' & = & z-v_zt,
\end{eqnarray}
where $(t,x,y,z)$ and $(t',x',y',z')$ are the spacetime coordinates of two different observers with relative velocity $(v_x,v_y,v_z)$. In this geometry, there exist two separate invariant intervals, which are the Euclidean separation in the three spatial dimensions and the time separation by itself. Einstein's motivation for constructing his Special Theory of Relativity was that Maxwell's equations for electromagnetism give us wave equations for electric and magnetic fields in the vacuum with propagation speed equal to the measured speed of light \cite{Einstein:1905ve}. The implication of this is that the speed of electromagnetic waves is independent of the reference frame, which is directly in contradiction with the Galilean transformations given in (\ref{IntroGalTran}). Instead, this observation implies an invariant interval of the form\footnote{This is subject to a sign convention, this thesis uses the mostly positive sign convention throughout.}
\begin{equation}
 ds^2 = -c^2 dt^2 + dx^2 + dy^2 + dz^2,
\end{equation}
where $c$ is the speed of light. Instead of (\ref{IntroGalTran}), the spacetime coordinates of different observers are related using the famous Lorentz transformations. This reasoning underpins Special Relativity, which is a cornerstone of modern physics. However, it is not compatible with Newton's theory of gravity, since Newton's theory of gravity is instantly interacting at all distances, rather than satisfying the causal structure of Special Relativity. Na\"{i}vely, one might think that one could simply take Maxwell's equations for electromagnetism and replace ``charge'' with ``mass'' to obtain a candidate gravity theory. This is not a viable option because mass-energy density Lorentz transforms differently to charge density, so a new construction is required. 

A clue that indicates the correct solution is the apparent coincidence that gravitational charge and inertial mass are well-demonstrated by experiment to be be in fixed proportion. This is currently known to be true to a precision of roughly one part in $10^{13}$ \cite{Muller:2012sea,Wagner:2012ui,Williams:2012nc}. As already noted by Newton, this implies that the acceleration due to gravity of a body is independent of its mass. Einstein's insight, the Equivalence Principle, states that a non-rotating laboratory freely falling in a uniform gravitational field will find the outcome of any internal experiment to be consistent with special relativity, \ie the gravitational field has no effect on those experiments \cite{Einstein:1911vc}. Instead of requiring a uniform gravitational field, one could alternatively require that the laboratory is small enough not to be able to measure the variation in field strength. This led Einstein to the realization that gravity can be expressed as an effect of the spacetime geometry itself.

\subsection{Geometry of curved spacetime}\label{IntroGeometry}

In this section, we will be concerned with the geometry of intrinsically curved spacetime. This is to be contrasted with geometries that are merely extrinsically curved, such as a cylinder, on which the invariant interval between two nearby spacetime points takes the same form as a Euclidean interval on a flat surface. Although not intrinsically curved, the extrinsic curvature of the cylinder is intrinsically detectable by the topology, \ie the periodic boundary condition on one of the directions. More precisely, we will be concerned with (pseudo-)Riemannian geometry. The ``pseudo-'' refers to geometries where the metric is not in Euclidean signature. Like Special Relativity, GR uses metrics with a Lorentzian signature, which is what we will consider from here on unless otherwise stated. The ``pseudo-'' prefix will be dropped from here on. Riemannian geometries are characterized by invariant intervals of the form
\begin{equation}\label{Riemannmetric}
 ds^2 = g_{\mu\nu}(x)dx^\mu dx^\nu,
\end{equation}
where $g_{\mu\nu}$ is the metric, and $x^\mu$ are a set of spacetime coordinates. Lorentz indices are always denoted by Greek letters in this thesis\footnote{There exists a convention where Lorentz indices belonging to a particular coordinate scheme (usually Greek letters) are distinguished from abstract indices (usually Latin letters). Since this thesis has a strong focus on diffeomorphism-invariant formulations, there is no demand for this distinction here.}. The metric in GR is a 2-component covariant tensor field that is symmetric under exchange of Lorentz index labels. The inverse metric, given by $g^{\mu\nu}(x)$ has the property that $g^{\mu\nu}g_{\mu\nu}=D$, where $D$ is the dimension of the spacetime. To see that the metric should be symmetric under exchange of index labels, note that any 2-component covariant tensor $T_{\mu\nu}$ can be split into a symmetric part and an antisymmetric part:
\begin{equation}\label{splitmetricsymmetry}
 T_{\mu\nu}(x) = T_{(\mu\nu)}(x)+T_{[\mu\nu]}(x),
\end{equation}
where round brackets indicate symmetrization and square brackets indicate antisymmetrization. More precisely, for some tensor $T_{\mu\cdots\nu}$, 
\begin{equation}
T_{(\mu|\cdots|\nu)}=\frac{1}{2}\left(T_{\mu\cdots\nu}+T_{\nu\cdots\mu}\right) 
\end{equation}
and
\begin{equation} 
T_{[\mu|\cdots|\nu]}=\frac{1}{2}\left(T_{\mu\cdots\nu}-T_{\nu\cdots\mu}\right). 
\end{equation}
Applying (\ref{splitmetricsymmetry}) to the metric tensor in (\ref{Riemannmetric}), it is easy to see that an antisymmetric component of the metric would not contribute to the invariant interval if the coordinates are bosonic, \ie commutative: $dx^{\mu}dx^{\nu}=dx^\nu dx^\mu$. Thus we can build the metric exclusively out of bosonic coordinates.

Now equipped with a metric, we can calculate invariant lengths, areas, volumes and higher dimensional regions of the manifold. The invariant length, $L$, of a given path from a point $A$ to a point $B$ on the manifold is given by
\begin{equation}
 L = \left|\int^{B}_{A}dx^\mu dx^\nu\ g_{\mu\nu}(x)\right|^{1/2}. 
\end{equation}
As with flat geometry, a 2-dimensional area element $dA$ comes from a multiplication of two orthogonal length elements, $dL_1$ and $dL_2$:
\begin{equation}
 dA = \sqrt{g_{11}g_{22}}\ dx^1 dx^2,
\end{equation}
where the index labels 1 and 2 correspond to the directions of length elements $dL_1$ and $dL_2$ respectively.
This idea is straightforward to apply to higher-dimension regions of the manifold, such that an $N$-dimensional volume element can be written as
\begin{equation}
 dV = \sqrt{g_{11}\cdots g_{NN}}\ dx^1\cdots dx^N.
\end{equation}
Finally, an integral over all spacetime dimensions on an $D$-dimensional manifold goes as
\begin{equation}
 dV = \sqrt{-{\rm det}(g)}\ d^Dx.
\end{equation}
where the minus sign originates from the time-time component in Lorentzian signature.
An important example of this is in the action, which is the $D$-dimensional spacetime integral of the Lagrangian density in $D$ spacetime dimensions:
\begin{equation}
 S = \int d^{D}x\ \sqrt{-g}\mathcal{L},
\end{equation}
where $g$ is used as shorthand notation for det($g$).

To perform tensor calculus on a manifold, it is necessary to evaluate a difference in the value of a tensor at different points on the manifold. On a curved spacetime, the tensor lies in different tangent spaces at different points, so it is not possible in general to sum instances of the tensor evaluated at different points in the usual way. To construct the metric-compatible covariant derivative at a point $P$, instances of tensors at neighbouring points are projected onto the tangent space at $P$ via an affine connection, $\Gamma^{\lambda}_{\ \rho\sigma}$\footnote{This is sometimes also referred to as the Christoffel symbol of the second kind. The Christoffel symbol of the first kind has the first index lowered by the metric.}. More precisely, the covariant derivative of a tensor field, $T_{\alpha_1\cdots\alpha_m}^{\ \ \ \ \ \ \ \beta_1\cdots\beta_n}$ is given by
\begin{eqnarray}
 \nabla_\mu T_{\alpha_1\cdots\alpha_m}^{\ \ \ \ \ \ \ \beta_1\cdots\beta_n} & = & \partial_\mu T_{\alpha_1\cdots\alpha_m}^{\ \ \ \ \ \ \ \beta_1\cdots\beta_n} - \sum_{i=1}^{m}\Gamma^{\lambda}_{\ \ \mu\alpha_i}T_{\alpha_1\cdots\lambda\cdots\alpha_m}^{\ \ \ \ \ \ \ \ \ \ \ \beta_1\cdots\beta_n} \nonumber \\ &&  + \sum_{i=1}^{n}\Gamma^{\beta_i}_{\ \ \mu\lambda}T_{\alpha_1\cdots\alpha_m}^{\ \ \ \ \ \ \ \beta_1\cdots\lambda\cdots\beta_n}.
\end{eqnarray}
In the case where the covariant derivative is metric-compatible,
\begin{equation}
 \nabla_\mu g_{\alpha\beta} = 0.
\end{equation}
The connection used throughout this thesis will be the torsionless metric connection, otherwise known as the Levi-Civita connection,
\begin{equation}\label{LeviCivitaConn}
 \Gamma^{\alpha}_{\ \beta\gamma} = \frac{1}{2}g^{\alpha\lambda}\left(\partial_\beta g_{\gamma\lambda}+\partial_\gamma g_{\beta\lambda}-\partial_\lambda g_{\beta\gamma}\right),
\end{equation}
which is symmetric under exchange of its lower two Lorentz indices $\beta$ and $\gamma$.
The Levi-Civita connection does not transform as a tensor under diffeomorphisms (coordinate transformations), although variations of it (\ie $\delta \Gamma^\alpha_{\beta\gamma}$) do transform as tensors. 
An alternative to this is Einstein-Cartan theory \cite{Cartan1922,Cartan1923,Cartan1924,Cartan1925,Kibble:1961ba,Sciama:1964wt,Hehl:1976kj,Shapiro:2001rz}, where the assumption of torsionlessness is relaxed, such that the torsion tensor, which is formed from an antisymmetrization of the lower indices of the connection, is non-zero: $\Gamma^{\alpha}_{\ \beta\gamma}-\Gamma^{\alpha}_{\ \gamma\beta}\ne 0$. We will not consider this possibility any further. 

Spacetime curvature is represented in GR via the Riemann tensor $R^{\alpha}_{\ \beta\gamma\delta}$, which is related to the commutator of covariant derivative operators, for example acting on a vector $v_\alpha$:
\begin{equation}\label{EinsteinFieldStrengthAnalogy}
 \left[\nabla_{\mu},\nabla_{\nu}\right]v_{\alpha} = R^{\lambda}_{\ \alpha\nu\mu}v_{\lambda}.
\end{equation}
More generally, the Riemann tensor is defined (according to a sign convention) as
\begin{equation}\label{RiemannTensorDef}
 R^{\alpha}_{\ \beta\gamma\delta} := 2\partial_{[\gamma}\Gamma^{\alpha}_{\ \delta]\beta} + 2\Gamma^{\alpha}_{\ \lambda[\gamma}\Gamma^{\lambda}_{\ \delta]\beta}.
\end{equation}
A contraction of this gives the Ricci tensor: $R_{\mu\nu} := R^{\alpha}_{\ \mu\alpha\nu}$ (again according to a sign convention). These two choices of sign convention and the chosen metric sign convention are collectively called the Landau-Lifshitz spacelike sign convention (+,+,+) and it will be used throughout this thesis (except when the metric is rotated into Euclidean signature to study local QFT). Further to that, the Ricci scalar is a contraction of the Ricci tensor such that $R :=g^{\alpha\beta}R_{\alpha\beta}$.

The Riemann tensor $R_{\alpha\beta\gamma\delta}$ is antisymmetric under exchange of the index labels $\alpha$ and $\beta$. It is also antisymmetric under exchange of the labels of the other pair of indices $\gamma$ and $\delta$. It is, however, symmetric under exchange of labels of the two pairs, \ie $\alpha\beta \leftrightarrow \gamma\delta$. From these properties follows the first Bianchi identity (the cyclic identity),
\begin{equation}
  R_{\alpha\beta\gamma\delta} + R_{\alpha\gamma\delta\beta} + R_{\alpha\delta\beta\gamma} = 0.
\end{equation}
It will frequently also be useful in this thesis to refer to the second Bianchi identity,
 \begin{equation}\label{SecondBianchi}
  \nabla_{\lambda}R_{\alpha\beta\gamma\delta} + \nabla_{\gamma}R_{\alpha\beta\delta\lambda} + \nabla_{\delta}R_{\alpha\beta\lambda\gamma} = 0.
 \end{equation}
An especially powerful specialization of the second Bianchi identity is
\begin{equation}\label{SpecialSecondBianchi}
 g^{\alpha\beta}\nabla_{\alpha}R_{\beta\gamma} = \frac{1}{2}\nabla_{\gamma}R.
\end{equation}
It is common to refer to two Bianchi identities as the ``cyclic'' and ``Bianchi'' identities respectively, \ie the ``second'' is often left implicit when referring to the latter.
\subsection{Derivation of field equations from an action}

As with other areas of physics, GR can be constructed via a principle of least action. The action for 4-dimensional GR is given by
\begin{equation}\label{EinsteinGravAct}
 S = \int d^4 x\ \sqrt{-g}\frac{1}{16\pi G}\left(R - 2\Lambda\right) + S_{\rm matter},
\end{equation}
where $G$ is Newton's constant, as also seen in (\ref{IntroNewton}), $\Lambda$ is the cosmological constant and $S_{\rm matter}$ is the action for the matter content of the theory, which it is sufficient to leave general for this thesis. It is conventional to introduce $\kappa := 8\pi G$ for notational convenience. This is the most general form of the action that is local, diffeomorphism-invariant and expanded only to second-order in derivatives. The diffeomorphism invariance (coordinate independence), discussed further in Section \ref{IntroDiffSym}, is an important issue that will be returned to frequently in this thesis. The constraint that the action is expanded only to second order is usually imposed to avoid Ostrogradsky instabilities \cite{Ostrogradski1850}. An Ostrogradsky instability is when the Hamiltonian becomes unbounded from below due to the appearance of negative kinetic terms. Ostrogradsky instabilities translate into QFT as a loss of unitarity of the $S$-matrix due to the appearance of negative norm states \cite{Woodard:2015zca}. An instability arises if the Lagrangian cannot be expressed (\eg using integration by parts in the action) as consisting only of fields that have been differentiated with respect to time no more than once. Typically, this means theories whose field equations possess third order or higher time derivatives have instabilities. An exception to this is where the Lagrangian is purely a function of the Ricci scalar \cite{Woodard:2006nt,Faraoni:2008mf}. This case is called $f(R)$ gravity, it escapes the Ostrogradsky instability by being reparametrizable to a scalar-tensor theory that is only at second order in derivatives, as we will see with equation (\ref{Legendrefr}). Effective field theories can be constructed with higher-order derivatives provided that they are only applied to energy scales below their cutoff mass scale \cite{Hawking:2001yt}. Higher-derivative theories of gravity are of interest because they are renormalizable \cite{Stelle:1976gc}, although unitarity is lost at a given truncation of a higher-derivative effective theory.

To derive Einstein's field equations, one takes the functional derivative of (\ref{EinsteinGravAct}). It is helpful to note that, since $g^{\alpha\beta}g_{\beta\gamma} = \delta^{\alpha}_{\ \gamma}$,
\begin{equation}\label{raiselowerminus}
 \delta\left(g^{\alpha\beta}g_{\beta\gamma}\right) = g_{\beta\gamma}\ \delta g^{\alpha\beta} + g^{\alpha\beta}\delta g_{\beta\gamma} = 0.
\end{equation}
This tells us that
\begin{equation}
 \delta g^{\alpha\beta} = -g^{\alpha\gamma}g^{\beta\delta}\ \delta g_{\gamma\delta}.
\end{equation}
From which it follows that
\begin{equation}
 \delta\sqrt{-g} = -\frac{1}{2}\sqrt{-g}g_{\mu\nu}\ \delta g^{\mu\nu}.
\end{equation}
It is also helpful to note that, since $R=g^{\alpha\beta}R_{\alpha\beta}$,
\begin{equation}
 \delta R = R_{\alpha\beta}\ \delta g^{\alpha\beta} + g^{\alpha\beta}\ \delta R_{\alpha\beta},
\end{equation}
where
\begin{equation}
 \delta R_{\alpha\beta} = \nabla_\gamma \ \delta\Gamma^{\gamma}_{\ \alpha\beta} - \nabla_{(\alpha|}\ \delta\Gamma^{\gamma}_{\ |\beta)\gamma}.
\end{equation}
Note that $\delta\Gamma^{\alpha}_{\ \beta\gamma}$, being the difference of two connections, transforms as a tensor under diffeomorphisms. It can be written as
\begin{equation}
 \delta \Gamma^{\alpha}_{\ \beta\gamma} = \frac{1}{2}g^{\alpha\delta}\left(2\nabla_{(\beta}\ \delta g_{\gamma)\delta} - \nabla_\delta\ \delta g_{\beta\gamma}\right).
\end{equation}
The stress-energy tensor (also known as the energy-momentum tensor) is defined as
\begin{equation}\label{IntroDefStress}
 T_{\mu\nu} := -\frac{2}{\sqrt{-g}}\frac{\delta S_{\rm matter}}{\delta g^{\mu\nu}}.
\end{equation}
Finally, these ingredients lead to the following form for Einstein's field equation:
\begin{equation}\label{EinsteinGravFieldEq}
 R_{\mu\nu}-\frac{1}{2}g_{\mu\nu}R + \Lambda g_{\mu\nu}= \kappa T_{\mu\nu},
\end{equation}
where a factor of $\sqrt{-g}$ has been divided out, which can be done without loss of generality because $\sqrt{-g}$ is always non-zero. The left hand side is often written in condensed notation by defining the ``Einstein tensor'' $\mathcal{G}_{\mu\nu}$ to be
\begin{equation}\label{EinsteinTensor}
 \mathcal{G}_{\mu\nu} := R_{\mu\nu}-\frac{1}{2}g_{\mu\nu}R.
\end{equation}
The trace of the field equation is
\begin{equation}\label{EinsteinFieldEqTrace}
 -R + 4\Lambda = \kappa T,
\end{equation}
which can be substituted into (\ref{EinsteinGravFieldEq}) to obtain the ``Ricci form'' of the field equation:
\begin{equation}\label{EinsteinFieldEqRicci}
 R_{\mu\nu} - \Lambda g_{\mu\nu} = \kappa\left(T_{\mu\nu} -\frac{1}{2}g_{\mu\nu}T\right).
\end{equation}

Using the (second) Bianchi identity given in (\ref{SpecialSecondBianchi}), it is clear that
\begin{equation}
 \nabla_\mu\left(R^{\mu\nu}-\frac{1}{2}g^{\mu\nu}R\right) = 0.
\end{equation}
From this, we can also see the generalization of energy/momentum conservation to curved spacetime:
\begin{equation}\label{EnergyConservation}
 \nabla_\mu T^{\mu\nu} = 0.
\end{equation}
Similar ingredients can be used to derive the field equations in more general theories of gravity, as will be returned to later in the thesis.
\subsection{The stress-energy tensor}\label{IntroStressEnergy}

The stress-energy tensor, introduced in (\ref{IntroDefStress}), contains information about the matter content of the spacetime. The field equation (\ref{EinsteinGravFieldEq}) tells us that the presence of matter introduces curvature into the spacetime purely as a consequence of the calculus of variations applied to the geometry of spacetime. For an observer moving with a 4-velocity $u_\alpha$ relative to the matter, such that there exists a projection tensor, $g_{\perp \alpha\beta} := g_{\alpha\beta} + u_\alpha u_\beta$, that maps onto the 3-dimensional tangent plane orthogonal to $u_\alpha$, the stress-energy tensor can be written as
\begin{equation}\label{IntroGenStress}
 T_{\mu\nu} = \rho u_{\mu}u_{\nu} + 2q_{(\mu} u_{\nu)} + pg_{\perp \mu\nu} + \pi_{\mu\nu},
\end{equation}
where $\rho$ is the relativistic energy density, $p$ is the relativistic pressure, $q_\alpha$ is the relativistic momentum density and $\pi_{\alpha\beta}$ is the relativistic trace-free anisotropic stress tensor. 

To apply these ideas to useful physics, one usually imposes some further physical constraints on (\ref{IntroGenStress}). Such constraints are called ``energy conditions''. The choice of energy condition for a model defines the assumptions about the nature of matter in that model. The rest of this subsection concerns some popular choices of energy condition.

The null energy condition is usually assumed to be true in realistic applications of GR. It states that, for any future-pointing null vector $n_{\alpha}$\footnote{A null vector $n^\alpha$ is one where $g_{\alpha\beta}n^\alpha n^\beta=0$.},
\begin{equation}
 T_{\alpha\beta}n^\alpha n^\beta \ge 0.
\end{equation}
Exceptions to this energy condition are very rarely considered and are quite exotic.
The weak energy condition states that, for any timelike vector field $t_\alpha$\footnote{A timelike vector field $t^\alpha$ is one where $g_{\alpha\beta}t^\alpha t^\beta < 0$ in the mostly positive metric sign convention.},
\begin{equation}\label{WeakEnergy}
 T_{\alpha\beta}t^\alpha t^\beta \ge 0.
\end{equation}
Given this energy condition, the relativistic energy density is non-negative for all observers. Exceptions to the weak energy condition are also rarely considered in realistic physics. There also exists the dominant energy condition where, in addition to the weak energy condition, it is asserted that, for a causal vector field (\ie timelike or null) $c_\alpha$, the vector field defined by $-c_\alpha T^{\alpha\beta}$ is causal and future-pointing, which is also usually considered to be a fair assumption for sensible physics. This energy condition prevents superluminal propagation of energy.
The strong energy condition states that, for any timelike vector field $t_\alpha$,
\begin{equation}
 \left(T_{\alpha\beta}-\frac{1}{2}g_{\alpha\beta}T\right)t^\alpha t^\beta \ge 0.
\end{equation}
Unlike the other energy conditions mentioned, this energy condition is frequently violated in realistic cosmological models, including all inflationary models.

If one supposes that the stress-energy tensor is describing a perfect fluid, \ie one with no shear stress or heat flux, then we can specialize (\ref{IntroDefStress}) to
\begin{equation}\label{IntroPerfectFluid}
 T_{\alpha\beta} = (\rho + p)u_\alpha u_\beta + pg_{\alpha\beta}.
\end{equation}
The trace is given by
\begin{equation}\label{IntroPerfectFluidTrace}
 T = -\rho + 3p.
\end{equation}
Perfect fluids are defined by their equation of state:
\begin{equation}\label{FluidEqState}
 p = w\rho,
\end{equation}
where $w$ is a parameter that is different for different fluids. For non-relativistic ``dust'', $w=0$. For radiation in $D$-dimensional spacetime, $w=1/(D-1)$. For vacuum energy, $T_{\mu\nu}$ is simply a constant (proportional to the cosmological constant) multiplied by the metric, \ie $w=-1$.

\subsection{Metric perturbations}\label{IntroMetricPert}

There are many known exact solutions to Einstein's field equations, but the real universe has a complicated arrangement of matter that yields a complicated form for the metric. Even for the simple second-order action given in (\ref{EinsteinGravAct}), the field equations (\ref{EinsteinGravFieldEq}) are highly non-linear and it is unreasonable to suppose that the real universe would follow one of the known exact solutions. Instead, it is common to suppose that the metric is close to, but not precisely equal to, one of the known exact solutions and then define a metric perturbation as the difference between the actual metric and that simpler ``background metric'':
\begin{equation}\label{MetricPertDef}
 h_{\mu\nu} := g_{\mu\nu} - g^{(0)}_{\mu\nu},
\end{equation}
where $h_{\mu\nu}$ is the metric perturbation and $g^{(0)}_{\mu\nu}$ is the background metric. Defining a background has the benefit that it is then possible to describe the propagation of gravitational waves on that background: the gravitational waves are identified with the perturbations. Indeed, one can switch between position and momentum representations of the metric perturbation via Fourier transformations:
\begin{equation}\label{PertFourier}
h_{\mu\nu}(x) = \int \dbar p \,{\rm e}^{-ip\cdot x} h_{\mu\nu}(p)\,,
\end{equation}
where a convenient shorthand notation has been used such that
\begin{equation}\label{dbarpshort}
\dbar p := \frac{d^{D}p}{(2\pi)^{D}}.
\end{equation}
This is especially of interest when constructing a quantum gravity description in which the metric perturbation is identified with the graviton field and the Fourier modes are identified with momenta propagating through the field. This is analogous to how Fourier modes in the electromagnetic field can be identified with frequencies of electromagnetic waves classically and also with momenta of photons quantum mechanically. These Fourier transformations require that points on the background manifold can be assigned a one-to-one mapping onto $\mathbb{R}^D$, so the procedure runs into difficulty if the spacetime is not simply connected. Since $g_{\mu\nu}g^{\mu\nu}=D$, the inverse metric expands as
\begin{equation}\label{InverseMetricExp}
 g^{\mu\nu}(x) = g^{(0)\mu\nu} - h^{\mu\nu}(x) + h^{\mu}_{\ \rho}(x)h^{\nu\rho}(x) + \cdots\,.
\end{equation}

Since this fixed-background formalism requires one to define two metrics, $g_{\mu\nu}$ and $g_{\mu\nu}^{(0)}$, it is also necessary to define two covariant derivative operators, one that is compatible with each metric. If $g_{\mu\nu}^{(0)}$ is the flat metric, then the associated covariant derivative is simply the partial derivative. This choice is especially convenient for exploring quantum gravity because partial derivative operators commute with one another, with the result that, after transforming into a momentum representation, the momentum operators also commute with one another. This provides a natural (and very conventional) route into the studying the RG flow of gravity, since the Kadanoff blocking integrates out the high-energy Fourier modes from the theory \cite{Wilson:1971bg}. As will be seen, however, the manifestly diffeomorphism-invariant ERG gives us the opportunity to choose either a formalism that perturbs around a fixed background, as discussed here, or a formalism that maintains strict background independence, with consistent results either way \cite{Morris:2016nda}.    

\subsection{The diffeomorphism symmetry}\label{IntroDiffSym}

GR is a theory of spacetime that does not physically depend on the choice of coordinates used to label the points on spacetime. Any coordinate-dependent results are therefore artifacts of the methods used to calculate them, rather than being physical features of the theory. For this reason, there is motivation to construct formalisms that do not impose coordinates. The symmetry under changing the coordinate scheme is called the diffeomorphism symmetry and it is of central importance to this thesis, which concerns the construction of averaging schemes that maintain diffeomorphism invariance. Consider a general transformation on a set of coordinates $x$ of the form
\begin{equation}
 x'^\mu = x^\mu - \xi^\mu(x),
\end{equation}
 where $\xi$ is some infinitesimal displacement of the $x$ coordinates that depends on $x$, and $x'$ is the resulting set of coordinates. Given some choice of covariant derivative $D_\alpha$, the metric transforms under diffeomorphisms like
\begin{equation}
 g'_{\mu\nu}(x') = \frac{\partial x^\rho}{\partial x'^\mu}\frac{\partial x^\sigma}{\partial x'^\nu}g_{\rho\sigma}(x) = g_{\mu\nu}(x) + 2g_{\lambda(\mu}D_{\nu)}\xi^\lambda + \xi\cdot Dg_{\mu\nu}.
\end{equation}
If $D_\alpha$ is metric-compatible, then the last term disappears. An example where $D_\alpha$ is not metric-compatible is where one chooses $D_\alpha$ to be a partial derivative despite having a curved metric, as will be convenient when constructing the fixed-background form of the ERG for gravity in Section \ref{MDIERGfixed}. Relating this to Subsection \ref{IntroMetricPert}, the metric perturbation given in (\ref{MetricPertDef}) transforms as
\begin{equation}\label{dpertintro}
 \delta h_{\mu\nu} = \delta g_{\mu\nu} = 2g_{\lambda(\mu}D_{\nu)}\xi^\lambda + \xi\cdot D g_{\mu\nu}.
\end{equation}
Again, the second term disappears if $D_\alpha$ is metric-compatible. The entire gauge freedom has been taken in by the perturbation such that $\delta g_{\mu\nu}=\delta h_{\mu\nu}$.
More precisely, the diffeomorphism transformation of any tensor is given by its Lie derivative, defined by
\begin{equation}
  \mathsterling_{\xi}T^{\alpha_1\cdots\alpha_m}_{\ \ \ \ \ \ \ \ \beta_1\cdots\beta_n} = \lim_{s\to0}\left(\frac{ \phi^{*}_{s}T^{\alpha_1\cdots\alpha_m}_{\ \ \ \ \ \ \ \ \beta_1\cdots\beta_n} - T^{\alpha_1\cdots\alpha_m}_{\ \ \ \ \ \ \ \ \beta_1\cdots\beta_n}}{s}\right),
\end{equation}
where $\phi^{*}_{s}$ is chart, parametrized by $s$, that maps one coordinate system onto another such that $\phi^{*}_0$ is the identity.
The Lie derivative of a general tensor field is given by
\begin{eqnarray}\label{LieGeneralTens}
 \mathsterling_{\xi}T^{\alpha_1\cdots\alpha_m}_{\ \ \ \ \ \ \ \ \beta_1\cdots\beta_n} & = & \xi\cdot DT^{\alpha_1\cdots\alpha_m}_{\ \ \ \ \ \ \ \ \beta_1\cdots\beta_n} + \sum^n_{i=1}T^{\alpha_1\cdots\alpha_m}_{\ \ \ \ \ \ \ \ \beta_1\cdots\lambda\cdots\beta_n}D_{\beta_i}\xi^\lambda \nonumber \\ &&
 - \sum^n_{i=1}T^{\alpha_1\cdots\lambda\cdots\alpha_m}_{\ \ \ \ \ \ \ \ \ \ \ \ \beta_1\cdots\beta_n}D_{\lambda}\xi^{\alpha_i}. 
\end{eqnarray}
The tensor considered above is a function of only one set of coordinates. Ordinarily, this is a very reasonable specialization, since a manifold only requires one set of coordinates to parametrize its all of its points. However, it will be useful in Chapter \ref{ChapterMDIERG} to construct a diffeomorphism-covariant bitensor that relates two spacetime points, thus giving it a modified diffeomorphism transformation that will be discussed in Section \ref{MDIERGWard}.

\section{Observational status of modern cosmology}\label{LCDM}

\subsection{Evidence for GR}

In 1916, Einstein proposed three tests for GR \cite{Einstein:1916vd}. First of all, an analysis performed as long ago as 1859 by Urbain Le Verrier had already shown that Newtonian gravity failed to give the correct prediction for the precession of the perihelion, which is the point of closest approach to the sun, of the planet Mercury \cite{LeVerrier1859}. Attempts to resolve this by the introduction of an elusive planet ``Vulcan'' had been unsuccessful. Einstein was able to use GR to calculate the precession and show that it agrees with the observations \cite{Einstein:1915bz}.

Secondly, Einstein proposed that the distortion of spacetime due to the sun would deflect the paths of light rays from stars whose angular distance from the sun as seen from the Earth is small. This would be seen on Earth as a shift in the relative positions of these stars in the sky. Newtonian gravity would also predict this effect, as noted by Henry Cavendish in 1784 and Johann Georg von Soldner in 1801 \cite{Soldner1804,WillC1988}, since acceleration is independent of mass, but the Newtonian deflection would be double the prediction from GR. This measurement raises a challenge, since the light from the sun is of much greater intensity than the light from the stars. However, a total eclipse of the sun in 1919 gave a team led by Arthur Eddington the perfect opportunity to precisely measure the relative positions of stars at small angular distance from the sun \cite{Dyson:1920cwa}. The result was an agreement with GR and a disagreement with Newtonian gravity. 

The third of the original tests comes from the gravitational redshift of light. Since there is a time dilation effect induced by gravitational fields, light becomes redshifted as it moves away from a massive body. This was verified precisely by the Pound-Rebka experiment, first reported in 1959 \cite{Pound.3.439,Pound:1960zz,Pound:1964zz,Pound:1965zz}, which uses a principle similar to the M\"{o}ssbauer effect observed in nuclear physics \cite{Mossbauer1958}. The M\"{o}ssbauer effect is where the energy of a gamma ray emitted from a nucleus in an excited state is slightly less than the energy of the transition because of the momentum taken by the nuclear recoil. Because of this, for the gamma ray to be reabsorbed by a similar nucleus, there must be a relative motion between the source and the target to compensate for the recoil. In the case of the Pound-Rebka experiment, a 14 keV gamma source consisting of iron-57 was placed 22.5m vertically above a target of the same isotope as the source. A scintillation counter  was placed below the target to detect unabsorbed gamma rays. The source was moved relative to the target using the vibrations of a loudspeaker of known frequency to produce a Doppler redshift to counter the gravitational blueshift due to the difference in height. Since the unstable atoms were in a solid lattice, the ordinary M\"{o}ssbauer effect was greatly reduced, allowing the gravitational blueshift to dominate. 

The gravitational time-dilation effect responsible for this has been tested in many other ways, such as the Hafele-Keating experiment \cite{Hafele166,Hafele168}, which placed atomic clocks onto commercial aircraft flying around the world to measure the effect due to both the motion (special relativity) and the altitude (general relativity) relative to stationary clocks on the Earth's surface. A more modern experiment has measured the time-dilation effect due to gravity between two clocks vertically separated by only 33 cm \cite{Chou:2010zz}. Modern Global Positioning satellites are also required to use predictions from GR to account for the time dilation effect due to motion and gravity, otherwise their accuracy would suffer from a consistent drift \cite{lrr-2003-1}.

Despite the overwhelming evidence in favour of the current understanding of gravity, researchers continue to find new tests, see for example \cite{Bertotti:2003rm,Fomalont:2003pd,Hoyle:2004cw}. Deviations from GR are most likely to occur in a strong-field regime, \ie where the spacetime curvature is large. For this reason, the relatively exotic astrophysical case of binary pulsars is extremely popular, see for example \cite{Stairs:2003eg,Lyne:2004cj,Kramer:2006nb,Demorest:2010bx,Weisberg:2010zz,Freire:2012mg,Antoniadis:2013pzd}. Pulsars are highly magnetized, rapidly rotating neutron stars that are strong sources of radio waves, emitted as beams from two poles. The pulses measured when the beams align with the Earth provide extremely precise measures of the pulsar's rotation, a regular variation in the pulse rate of a pulsar indicates that it is orbiting about the centre of mass of a binary system. A system of two neutron stars is a suitably strong field regime to do precision tests of GR. Prior to the direct observation of gravitational waves by LIGO in 2015, reported in 2016 \cite{Abbott:2016blz}, the original observational evidence for gravitational waves came from the orbital decay of the Hulse-Taylor binary \cite{Hulse:1974eb,Weisberg:1981bh,Taylor:1982zz}, which is a binary system of a pulsar and another neutron star. The decay of the orbit is driven by release of gravitational waves, as predicted by GR. All currently existing tests of GR agree with Einstein.

\subsection{The big bang theory}\label{IntroBBT}

When Einstein first formulated GR, it was not known that the universe is expanding. Indeed, Einstein himself tried to construct his cosmology with a static universe in mind. This was already in tension with observation via Olber's paradox. Olber's paradox notes that if we live in an infinite universe filled with stars, then every direction one looks in should eventually end with a star, therefore the night sky should be as bright as the sun. The paradox is not resolved by adding dark obscuring objects, because they would eventually heat up and glow brightly too. The paradox would be solved, however, if the universe only has a finite age and light from distant stars has not been able to reach us. 

The first direct evidence for the origins of our universe came from what has been called Hubble's law: that the apparent recessional velocity of galaxies (as measured by their redshift) that are between about 10 Mpc and $\mathcal{O}$(100 Mpc) away is approximately proportional to their distance away (as measured using the cosmological distance ladder). This effect was predicted by Georges Lema\^{i}tre in 1927 \cite{1927ASSB...47...49L}, although Alexander Friedmann had already suggested expanding cosmologies from GR in 1922 \cite{Friedman1922}. The observational discovery is attributed to Edwin Hubble \cite{Hubble:1929ig}. However, Vesto Slipher, who measured the first radial velocity of a galaxy (Andromeda) in 1912 \cite{1913LowOB...2...56S}, had already observed this pattern by 1917 \cite{1915PA.....23...21S,1917PAPhS..56..403S}. Unfortunately, he did not see at that time the full significance of that observation. It is considered unreasonable to believe that the Earth is specially placed at the centre of the universe (the Copernican Principle), so this is interpreted as a homogeneous, isotropic expansion of the universe. Homogeneity and isotropy will be further discussed in Section \ref{IntroHomoIso}. At shorter distances, galaxies are gravitationally bound, indeed the Andromeda galaxy is predicted to merge with the Milky Way in just under six billion years' time \cite{vanderMarel:2012xq}. At larger distances, the expansion rate of the universe has been shown to be accelerating, this is a mysterious effect called ``dark energy'' that will be further discussed in Section \ref{IntroDarkEnergy}. 

The observation that the universe is expanding, together with the strong theoretical motivation from GR that the universe should not be static, implies that the universe has evolved to its present state from a much denser state in the past. Classically, Einstein's field equations can be used to extrapolate back to a finite time when the observable universe would have been confined to a single point of space \cite{Ellis:1968vy,Hawking:1969sw}. This is what is referred to as the ``big bang''. The aforementioned point of infinite density (and therefore infinite spacetime curvature) is referred to as the big bang singularity. In the absence of a quantum theory of gravity, we do not know what really happened at the very beginning of the universe, or even if the universe that we know emerged from some previously existing universe. For this reason, one can refer to the ``big bang'' as a classical notion that makes for an excellent model of the universe at all times that we have observational evidence for. However, early universe cosmology (especially pre-inflation) is still an open subject.

Historically, some cosmologists advocated ``steady state'' models in which expansion was reconciled with an eternal universe via continuous creation \cite{Bondi:1948qk,Hoyle:1948zz}. Strong evidence against the steady state came from the observation that bright radio sources are not evenly distributed in the universe: they only appear at large distances away, implying that the universe has evolved over time \cite{Ryle01041961}. Conclusive evidence came from the chance discovery of the Comic Microwave Background (CMB) by Penzias and Wilson in 1965 \cite{Penzias:1965wn,Dicke:1965zz}. The CMB had already been predicted by Alpher and Herman in 1948, in their study of cosmic nucleosynthesis, which was also influenced by Gamow \cite{PhysRev.74.1737,Alpher:1948ve,Gamow:1946eb}. The CMB is a source of blackbody radiation that is almost uniform across the whole sky with a temperature of 2.726 K \cite{0004-637X-707-2-916} that has variations approximately at the level of one part in $10^5$ \cite{Smoot:1992td}. The CMB was created when the universe transitioned from an opaque plasma of electrons and nuclei to a transparent gas of atoms. This occurred because the expansion of space also causes the wavelengths of photons to expand, cooling the universe. When the universe was hot, electrons and nuclei were unable to form atoms because of the abundance of high-energy photons that would be able to reionize those atoms. When the universe cooled below about 3000 K, the density of such photons dropped sufficiently low that atoms were able to form. This event happened approximately 380,000 years after the big bang \cite{Bennett:2003bz}. Initially, the light, for the first time able to propagate over large distances, was emitted as a blackbody spectrum of the same temperature, but the expansion of space has since caused the wavelengths to be stretched, cooling the spectrum down to the presently observed value. Since the plasma that preceded the formation of atoms absorbed any photons propagating through it, there are no photons pre-dating this time to observe. For this reason, the CMB is also referred to as the surface of last scattering. This is considered to be conclusive piece of evidence that the universe is not in a steady state, since an expanding universe will continue to redshift this light, eventually rendering it unobservable. Since the CMB is the oldest light in the universe, it is extensively studied for hints about the conditions of the early universe, particularly concerning inflation, which is discussed in Subsection \ref{IntroInflation}.

Another success of the big bang theory is in predicting the abundance of light elements and their isotopes in the universe \cite{Yang:1983gn,Olive:1999ij,Cyburt:2015mya}. Heavier elements are created exclusively by stellar nucleosynthesis, however the universe underwent a period of cosmological nucleosynthesis at a much earlier time (approximately between 10 seconds and 20 minutes after the big bang) that fused hydrogen into other light elements and isotopes, especially helium-4 with smaller amounts of deuterium and helium-3, very small amounts of lithium-7 and also the unstable isotopes tritium and beryllium-7, which would then undergo beta decay to helium-3 and lithium-7 respectively. Big bang nucleosynthesis, being the dominant source of such isotopes in the universe, is able to account for the relative abundance of light elements in the modern universe remarkably well, especially the deuterium abundance, however there is a famous deficit in lithium-7 abundance, which might be due to unknown systematics from astrophysics, or might point to new underlying physics \cite{Fields:2011zzb}.

\subsection{Homogeneity and isotropy}\label{IntroHomoIso}

The most general metric in GR that is homogeneous and isotropic is the Friedmann-Lema\^{i}tre-Robertson-Walker (FLRW) metric, which can be expressed via an invariant interval given in spherical polar coordinates $(t,r,\theta,\phi)$ as
\begin{equation}\label{FLRWmetric}
 ds^2 = -dt^2 + a(t)^2\left(\frac{dr^2}{1-kr^2}+r^2\left(d\theta^2 + {\rm sin}^2\theta d\phi^2\right)\right),
\end{equation}
where $a(t)$ is a time-dependent scale factor and $k$ is an overall curvature constant for the spacetime.
The universe is observed to be homogeneous and isotropic at very large distance scale, from which it follows that the metric is well-approximated by an FLRW metric when considering averages over sufficiently large scales. This is the Cosmological Principle, which is an extension to the Copernican Principle that we do not occupy a specially preferred part of the universe.

We now turn to understanding the dynamics of such a universe. Consider the Ricci form of the field equation, given in (\ref{EinsteinFieldEqRicci}) and contract the indices with two instances of the 4-velocity of the observer $u_\alpha$:
\begin{equation}\label{IntroRaychaud1}
 R_{\mu\nu}u^\mu u^\nu - g_{\mu\nu}u^\mu u^\nu \Lambda = \kappa\left(T_{\mu\nu} -\frac{1}{2}g_{\mu\nu}T\right)u^\mu u^\nu .
\end{equation}
Since this chosen universe is homogeneous and isotropic, we can use the stress-energy tensor for a perfect fluid given in (\ref{IntroPerfectFluid}) and its trace in (\ref{IntroPerfectFluidTrace}) to get
\begin{eqnarray}
 T_{\mu\nu}u^\mu u^\nu & = & \rho, \\
 g_{\mu\nu}u^\mu u^\nu T & = & \rho - 3p, \\
 8\pi G\left(T_{\mu\nu} - \frac{1}{2}g_{\mu\nu}T\right)u^\mu u^\nu & =&  4\pi G(\rho + 3p).
\end{eqnarray}
This allows us to rewrite (\ref{IntroRaychaud1}) as
\begin{equation}
 R_{\mu\nu}u^\mu u^\nu = 4\pi G(\rho + 3p) - \Lambda.
\end{equation}
Using (\ref{FLRWmetric}) and (\ref{LeviCivitaConn}), the non-zero connection coefficients for the FLRW metric in spherical polar coordinates can be written out as
\begin{eqnarray}\label{FLRWconnections}
 \Gamma^t_{\ rr} & = & \frac{\dot{a}a}{1-kr^2}, \\
 \Gamma^t_{\ \theta\theta} & = & \dot{a}ar^2, \\
 \Gamma^t_{\ \phi\phi} & = & \dot{a}ar^2 {\rm sin}^2\theta, \\
 \Gamma^r_{\ rt} = \Gamma^\theta_{\ \theta t} = \Gamma^\phi_{\ \phi t}& = & \dot{a}/a, \\
 \Gamma^r_{\ rr} & = & \frac{kr}{1-kr^2}, \\
 \Gamma^r_{\ \theta\theta} & = & r\left(kr^2-1\right), \\
 \Gamma^r_{\ \phi\phi} & = & r\left(kr^2-1\right){\rm sin}^2\theta, \\
 \Gamma^\theta_{\ r\theta} = \Gamma^\phi_{\ r\phi} & = & \frac{1}{r}, \\
 \Gamma^\theta_{\ \phi\phi} & = & -{\rm sin }\theta\ {\rm cos }\theta, \\
 \Gamma^\phi_{\ \theta\phi} & = & 1/{\rm tan }\theta,
\end{eqnarray}
where the over-dot indicates differentiation with respect to time and the symmetry under exchange of the lower Lorentz indices has been left implicit.
Using these coefficients and the definition of the Ricci tensor via the Riemann tensor in (\ref{RiemannTensorDef}), one gets
\begin{eqnarray}
 R_{tt} & = & -3\frac{\ddot{a}}{a}, \\
 R_{rr} & = & \frac{\ddot{a}a+2\dot{a}^2+2k}{1-kr^2}, \\
 R_{\theta\theta} & = & \ddot{a}ar^2+2\dot{a}^2r^2 + 2kr^2, \\
 R_{\phi\phi} & = & \left(\ddot{a}a + 2\dot{a}^2+2k\right)r^2{\rm sin}^2\theta, \\
 R & = & 6\frac{\ddot{a}}{a}+6\frac{\dot{a}^2}{a^2}+6\frac{k}{a^2}.
\end{eqnarray}
Thus we have the Raychaudhuri equation for a homogeneous isotropic universe to be
\begin{equation}\label{FLRWRaychaud}
 3\frac{\ddot{a}}{a} = -4\pi G(\rho + 3p) + \Lambda.
\end{equation}
The energy conservation equation, which is the time component of (\ref{EnergyConservation}), is found to be
\begin{equation}\label{FLRWEnCons}
 \dot{\rho} + 3\frac{\dot{a}}{a}(\rho + p) = 0.
\end{equation}
Using these expressions for the Ricci tensor and the Ricci scalar, the time-time component of the field equation (\ref{EinsteinGravFieldEq}) can be rearranged into the Friedmann equation:
\begin{equation}\label{FLRWFried}
 \frac{\dot{a}^2}{a^2} = \kappa\frac{\rho}{3} + \frac{\Lambda}{3} - \frac{k}{a^2}.
\end{equation}
Equations (\ref{FLRWRaychaud},\ref{FLRWEnCons},\ref{FLRWFried}) are all related via the time derivative of (\ref{FLRWFried}) for $\dot{a}\neq 0$. Sometimes (\ref{FLRWRaychaud}) is also referred to as a Friedmann equation.

Using the Friedmann equation (\ref{FLRWFried}) for inspiration, it is conventional to introduce the following notation:
\begin{eqnarray}
 H & := & \dot{a}/a, \\
 \Omega & := & \frac{\kappa \rho}{3H^2}, \\
 \Omega_\Lambda & := & \frac{\Lambda}{3H^2}, \\
 \Omega_k & := & -\frac{k}{a^2H^2}, \\
 q & := & -\frac{\ddot{a}}{aH^2},
\end{eqnarray}
where $H$ is the Hubble parameter, $\Omega$ is the dimensionless density parameter for matter, $\Omega_\Lambda$ is the dimensionless density parameter for the cosmological constant, $\Omega_k$ is the dimensionless density parameter for curvature and $q$ is the cosmological deceleration parameter. The choice of sign for $q$ to make a ``deceleration'' parameter rather than ``acceleration'' has its historical root in the assumption that the acceleration is given by (\ref{FLRWRaychaud}) with $\Lambda=0$ and positive $\rho$ and $p$, implying a universe whose expansion decelerates as gravity pulls it back together. We now know that this is not the case for the recent universe, as will be discussed in Subsection \ref{IntroDarkEnergy}. Sometimes $\Omega$ is split further into $\Omega_m$ for non-relativistic matter and $\Omega_r$ for radiation. The Friedmann equation relates the density parameters to be
\begin{equation}
 \Omega_{\rm total} := \Omega + \Omega_\Lambda = 1 - \Omega_k.
\end{equation}
The ``critical'' density is the case of $\Omega_{\rm total}=1$, which results in a universe with a ``flat'' geometry. The critical density $\rho_c$ required to make this happen can be written as
\begin{equation}\label{Critdensity}
 \rho_c = \frac{3}{\kappa}H^2.
\end{equation}
The other two cases are ``spherical'' geometry where $k>1$ and $\Omega_{\rm total}>1$, and ``hyperbolic'' geometry where $k<1$ and $\Omega_{\rm total}<1$. These density parameters are functions of time. The measured values for these parameters in the present universe are approximately $\Omega_m = 0.32$, $\Omega_\Lambda = 0.68$ and $\Omega_r$ is only of order $10^{-4}$. The Raychaudhuri (\ref{FLRWRaychaud}) equation can be expressed for a single perfect fluid of the form in (\ref{FluidEqState}) plus cosmological constant as
\begin{equation}
 q = \frac{1}{2}\Omega\left(1+3w\right)-\Omega_{\Lambda}.
\end{equation}

It is possible to construct cosmological models that are homogeneous but not isotropic, such as Bianchi universes, or isotropic but not homogeneous (if we occupy a special location at the centre), such as a Tolman-Lema\^{i}tre-Bondi universe \cite{Lemaitre:1933gd,1934PNAS...20..169T,Bondi:1947fta}. The latter suggestion seems very unreasonable {\it a priori} as a cosmological model, although it is perfectly sensible for describing smaller structures like collapsing dust clouds. Astronomical observations constrain both of these suggestions. At smaller distance scales, it is obvious that the universe is extremely inhomogeneous, since we see sharp contrasts in matter density all around us in everyday life. The ratio of densities of water to air is about 784, making the surface of the Earth a place where a very sharp jump in density occurs. However, this is very small compared with other ratios one could take. The mean density of the Earth is about 5514 kg/m$^3$, whereas the average energy density of the universe (including the dark matter and dark energy) is of order $10^{-26}$ kg/m$^3$. Contrast both of these with the density of an atomic nucleus, which is about 2$\times 10^{17}$ kg/m$^3$.

When the assertion is made that the universe is well-approximated as homogeneous and isotropic, this only applies to averages at very large scales, \ie the average density over a suitably large spherical volume of space is independent of where the centre is placed. The observational question is then: how large an averaging volume must this be? The two main methods to constrain this are observations of the CMB \cite{Ade:2015hxq} (which strictly only measure anisotropy) and galaxy surveys, such as the Sloan Digital Sky Survey (SDSS) \cite{Tegmark:2003ud,Eisenstein:2005su}, or the WiggleZ Dark Energy Survey \cite{Blake:2011rj,Blake:2011en}. There is a small amount of disagreement between studies as to precisely at what scale the universe transitions to homogeneity, but there is wide agreement that the universe transitions to homogeneity above approximately 100$h^{-1}$ Mpc and remains homogeneous at all larger scales \cite{Hogg:2004vw,Sarkar:2009iga,Yadav:2005vv,Scrimgeour:2012wt,Pandey:2015htc,Pandey:2015xea,Laurent:2016eqo}.

\subsection{The dark side of the universe}\label{IntroDarkEnergy}

Observations of the CMB reveal that the value for $\Omega_k$ in (\ref{FLRWmetric}) is less than 0.005, consistent with zero \cite{Hinshaw:2012aka,Ade:2013zuv,Ade:2015xua}. However, the successful predictions for the relative abundance of light elements, in agreement with direct astronomical observations, depend on the density of baryonic matter in the universe being only $\sim5$\% of the critical density required to achieve a flat FLRW metric \cite{Walker:1991ap}. The rest of the energy density of the universe must come from some new ``dark'' physics that is thus far not understood. The standard ``concordance'' model is referred to as $\Lambda$CDM after its main components, which are the cosmological constant and cold dark matter, which we will now discuss. 

The first component of this new physics is called ``dark matter''. Dark matter was originally suggested for use in two contexts. One was observations of the velocities of stars orbiting galaxies implied the existence of extra unobserved mass in the galaxy forming a halo stretching beyond the bright disk. The velocity plotted as a function of radial distance from the centre makes the famous ``rotation curves'' for galaxies that have been extensively studied over many decades \cite{1939LicOB..19...41B,1940ApJ....91..273O,1959BAN....14..323V,1970ApJ...159..379R,1978ApJ...225L.107R,Rubin:1980zd,Persic:1995ru}. The other original context was in estimating the masses of galaxy clusters \cite{Zwicky:1933gu}. This dark matter is observed to be six times more abundant by mass than ordinary matter. Simulations, backed up by observations, indicate that this dark matter is important for the formation of structure in the universe at scales ranging between the size of galaxies through to the scale of transition to homogeneity \cite{Davis:1985rj,White:1991mr,Liddle:1993fq,Springel:2005nw}. Moreover, to produce the correct sizes of structures, this dark matter is thought to be non-relativistic, \ie ``cold'' dark matter (CDM). Highly relativistic ``hot'' dark matter is ruled out observationally. For this reason, neutrinos can be ruled out as dark matter candidates. Another reason why neutrinos are ruled out is that the upper bound on neutrino masses combined with the observed number density of neutrinos in the universe makes clear that neutrinos make up a negligible fraction of the mass density of the universe \cite{Ade:2015hxq}. Gravitational microlensing provides an upper bound on the masses of individual dark matter particles \cite{Paczynski:1985jf,Alcock:1995dm}. For this reason, compact stellar objects can be ruled out as sources of dark matter. At the time of writing, dark matter has never been confirmed to have been detected interacting non-gravitationally with Standard Model particles \cite{Aprile:2012nq,Akerib:2015rjg}. 

Typically, viable dark matter candidates are required to have no electric charge or strong nuclear interaction (otherwise they would have been observed already). However, some interaction with the Standard Model is expected in order to get the correct relic abundance of dark matter in the present universe, given that inflation, discussed in Subsection \ref{IntroInflation}, is expected to have washed-out any pre-existing particles. Dark matter cannot be accounted for in the Standard Model, although many popular speculative BSM candidates exist \cite{Jungman:1995df,ArkaniHamed:1998nn,Bertone:2004pz}. 

It has also be proposed that, since all the observational evidence for dark matter comes from its gravitational effects, it is possible that dark matter can be accounted for as a quirk of gravity rather than as consisting of particles. One popular candidate for such a quirk of gravity is Modified Newtonian Dynamics \cite{Milgrom:1983ca}, which can be formulated relativistically in Tensor-Vector-Scalar models of modified gravity \cite{Bekenstein:2004ne}. One of many problems with this proposal is that it suffers from a need to fit different parameter values to different galaxies to extract the correct rotation curves \cite{1987AJ.....93..816K}. More seriously, modified gravity is unable to explain why collisions of galaxy clusters, such as the famous ``Bullet Cluster'' \cite{Clowe:2006eq} have a concentration of observable matter in the middle of the collision (indicated by X-ray observations), but larger amounts of extra matter indicated by gravitational lensing either side of the collision. The dark matter-based explanation is that, while ordinary matter is stopped in the middle by its relatively strong electromagnetic interaction, dark matter can pass through the collision unimpeded due to its very small interaction cross-section, even with other dark matter particles.

Of more direct interest to this thesis is the remaining $\sim 68$\% of the current energy density of the universe, which is referred to as dark energy.
Einstein constructed GR and began applying it to cosmology before Hubble's observations of the expanding universe, believing that the universe should be static, existing eternally in more or less its present state \cite{1917SPAW.......142E}. As already discussed, GR naturally predicts that the scale factor of the universe should change as space either expands or contracts. The presence of matter in the universe would be expected to cause a negative acceleration in the scale factor that would cause an initially stationary universe to collapse. In order to counter this effect, Einstein proposed the cosmological constant to act as a repulsive force that would balance the gravitational effect of the matter density and maintain a constant scale factor. This proposal is flawed because it requires there to exist an exact cancellation, which would be disturbed if any energy is transferred between non-relativistic matter and radiation, which does indeed happen in the real universe. 

It was discovered at the end of the twentieth century that the acceleration of the scale factor of the universe is not negative, as one would expect from a universe consisting purely of ordinary matter, but rather the expansion rate of the universe is accelerating \cite{Riess:1998cb,Perlmutter:1998np}. This late-time acceleration came greatly to the surprise of most astronomers and cosmologists, resurrecting a small $\sim10^{-12}$ GeV positive vacuum energy as an explanation for late-time acceleration. The effect of the cosmological constant kicks in at late times because, unlike for other cosmological fluids, the energy density for the cosmological constant does not dilute as the universe expands. Matter density dilutes as $a^{-3}$ because the volume containing that matter scales as $a^3$. Radiation density scales as $a^{-4}$ because, in addition to the volume increasing, the wavelength stretches by a factor of $a$, causing the energy per photon to scale as $a^{-1}$. The cosmological constant, on the other hand, remains constant, so it dominates as $a$ becomes large. There is, however, the ``why now?'' problem of why the transition from a matter-dominated universe to an accelerating universe is happening at the time when there are human astronomers to witness it \cite{Zlatev:1998tr,Garriga:1999hu,Velten:2014nra}. The ``why now?'' problem is a curiosity that may or may not have any interesting explanation.

The cosmological constant exists in GR as a free parameter that can be determined by experiment. Once one includes the action for the Standard Model of particle physics into (\ref{EinsteinGravAct}) as $S_{\rm matter}$, one would expect the Standard Model vacuum to introduce contributions that the diffeomorphism symmetry would require to take the form $\int d^4 x\sqrt{-g} \times \left<{\rm operators}\right>$, \ie the Standard Model vacuum offers its own contributions to the cosmological constant. These contributions would be expected to come from every scale of length available in the Standard Model. Even in the absence of any mixing with the Einstein-Hilbert action or new BSM physics, the electroweak sector of particle physics would be expected to provide contributions of energy scale $\mathcal{O}(10^2)$ GeV. If the physical cosmological constant took such a large value, the induced acceleration effect would have made the formation of structure, and indeed life, in the universe impossible. This is the infamous ``cosmological constant problem'' \cite{Weinberg:1988cp}. It would seem that some extreme fine-tuning is required to cancel all quantum contributions to the energy scale of the cosmological constant from the Planck scale, determined by Newton's constant, at $10^{19}$ GeV down to the observed value of $10^{-12}$ GeV. The effective stress-energy tensor due to a cosmological constant is
\begin{equation}
 T_{\mu\nu}^{\Lambda} = \frac{\Lambda}{8\pi G}g_{\mu\nu},
\end{equation}
which is a mass-dimension 4 tensor, which is why the fine-tuning is often said to be at a level of one part in $10^{124}$. This is an extreme example of fine-tuning that dwarfs the famous Hierarchy problem for the Higgs boson mass \cite{Gildener:1976ai}, which arguably forms the biggest motivation for probing energies in the vicinity of the electroweak scale using the Large Hadron Collider and proposed future experiments.

The cosmological constant problem is considered to be a much harder problem than the hierarchy problem. One reason, ironically, is that the energy scale of $10^{-12}$ GeV is easily accessible to experiment. In fact, the length scale is $\mathcal{O}(0.1)$mm, which is easy to see with the naked eye. This energy scale is far too well explored to speculate freely about undiscovered BSM particles and interactions at that scale. So, unlike with the Hierarchy problem, it seems very difficult solve the cosmological constant problem by inserting more particles into the Standard Model \eg as SUSY is applied to the hierarchy problem \cite{Veltman:1980mj}. Despite this, the mystery of dark energy has prompted many to speculate about alternatives to the cosmological constant, popular among which are modified gravity theories that provide late-time acceleration. These alternatives do not provide solutions to the cosmological constant problem, rather they suppose that the cosmological constant problem is solved somehow (with the solution being that the actual cosmological constant is zero) and then seek to explain the observational effects of the cosmological constant by some other means. 

Distinguishing between the actual cosmological constant and these alternative theories is the subject of many telescope-based projects in observational cosmology \cite{Riess:2006fw,Kowalski:2008ez,Amanullah:2010vv,Blake:2011en,Beutler:2011hx,Amendola:2012ys,Betoule:2014frx}. A popular method is to perform a survey of the peculiar velocities of galaxies (\ie the velocity contribution not due to Hubble's law) to look for deviations from what is expected in $\Lambda$CDM. The motivation for supposing a zero cosmological constant is that the aforementioned lack of new physics at the apparent cosmological constant scale might be a hint that actually some deeper principle requires that the cosmological constant should disappear entirely, rather than merely being carefully cancelled in some delicate way down to such a low scale. On the other hand, in the absence of such a principle, a value of zero for the cosmological constant is actually the extreme limit of fine-tuning to one part in infinity. In that sense the discovery of a small, positive cosmological constant might actually be an improvement on the fine-tuning problem by a factor of infinity. Nevertheless, until some compelling argument for why we have a $10^{-12}$ GeV cosmological constant is established, it is important to consider all possibilities for explaining dark energy. In the latter half of this thesis, we will explore cosmological backreaction as one of those alternatives and evaluate its viability as an explanation of the late-time accelerating expansion of the universe \cite{Preston:2014tua,Preston:2016sip}.

\subsection{Inflation}\label{IntroInflation}

Inflation is a proposed phase of the early universe in which there is exponential growth in the scale factor \cite{Guth:1980zm,Starobinsky:1980te,Linde:1981mu,Linde:1983gd,Ratra:1987rm}. The universe at this time is described by a de Sitter metric, which is a special case of the FLRW metric (\ref{FLRWmetric}). For a universe dominated by a positive cosmological constant, we have
\begin{equation}
 a = A\ {\rm exp}\left(\sqrt{\Lambda/3}t\right)+B\ {\rm exp}\left(-\sqrt{\Lambda/3}t\right),
\end{equation}
where $A$ and $B$ are constants that are constrained via the Friedmann equation (\ref{FLRWFried}) to be $4AB\Lambda = 3K$. We then have exponentially expanding solutions of the form
\begin{eqnarray}
 {\rm for}\ k>0,\ a & = & \sqrt{3k/\Lambda}\ {\rm cosh}\sqrt{\Lambda/3}t,\\
 {\rm for}\ k=0,\ a & = & A\ {\rm exp}\sqrt{\Lambda/3}t,\\
 {\rm for}\ k<0,\ a & = & \sqrt{-3k/\Lambda}\ {\rm sinh}\sqrt{\Lambda/3}t.
\end{eqnarray}
Inflation is similar to a cosmological constant-dominated universe except that, instead of the vacuum energy being constant, it is sourced from a slowly changing potential that ultimately drops to zero at the exit to inflation. The simplest models involve a scalar field $\phi$ and potential $V(\phi)$ whose action is given by
\begin{equation}
 S_{\phi} = -\int d^4 x\ \sqrt{-g}\left(\frac{1}{2}\nabla_\alpha \phi \nabla^\alpha \phi + V(\phi)\right).
\end{equation}
The equation of motion for $\phi$ is the Klein-Gordon equation:
\begin{equation}
 \nabla^2 \phi - V'(\phi) = 0,
\end{equation}
where $V'=dV/d\phi$. Using the FLRW metric (\ref{FLRWmetric}) and connection coefficients (\ref{FLRWconnections}), this can be rewritten as
\begin{equation}\label{ExpandedKleinGordon}
 \ddot{\phi} + 3H\dot{\phi} + V'(\phi) = 0,
\end{equation}
where the second term is analogous to a friction term.
The stress-energy tensor for $\phi$ is
\begin{equation}
 T_{\mu\nu}^\phi = \nabla_\mu\phi\nabla_\nu\phi - \frac{1}{2}g_{\mu\nu}\left(\nabla_\alpha\phi\nabla^\alpha\phi + 2V(\phi)\right).
\end{equation}
In the case where $\nabla_\alpha\phi$ is small, the stress-energy tensor is dominated by $V(\phi)$ as an effective cosmological constant $\Lambda_{\rm inf}=\kappa V(\phi)$. The equation of state (\ref{FluidEqState}) for this fluid is
\begin{equation}
 w = \frac{p}{\rho} = \frac{-V(\phi)+\dot{\phi}^2/2}{V(\phi)+\dot{\phi}^2/2},
\end{equation}
which is close to the cosmological constant value of $w=-1$ for small $\dot{\phi}$. If $\ddot{\phi}$ is small then, from the Klein-Gordon equation (\ref{ExpandedKleinGordon}), we have
\begin{equation}\label{Infldotphi}
 \dot{\phi} \approx -\frac{V'}{3H}.
\end{equation}
If $\dot{\phi}$ is small, then, from the Friedmann equation (\ref{FLRWFried}), we have
\begin{equation}\label{InflFried}
 H^2 \approx \kappa\frac{V}{3}.
\end{equation}
To quantify these requirements, the slow-roll parameters $\epsilon$ and $\eta$ are invoked:
\begin{eqnarray}
 \epsilon & := & \kappa\frac{\dot{\phi}^2}{2H^2},\\
 \eta & := & -\frac{1}{H}\frac{\ddot{\phi}}{\dot{\phi}}.
\end{eqnarray}
The conditions for slow-roll inflation are that $\epsilon<<1$ and $|\eta|<<1$. There also exists a different convention for the slow-roll parameters, which is
\begin{eqnarray}
 \epsilon_V & := &  \frac{1}{2\kappa}\left(\frac{V'}{V}\right)^2, \\
 \eta_V & := & \frac{1}{\kappa}\frac{V''}{V}.
\end{eqnarray}
It can be seen from (\ref{Infldotphi}) and (\ref{InflFried}) that $\epsilon \approx \epsilon_V$ in the slow-roll regime, however $\eta_V\approx \eta+\epsilon$.

Three motivations for believing that inflation took place are the ``flatness problem'', the ``horizon problem'' and the ``magnetic monopole problem''. Let us begin with the flatness problem. At very early times, the universe is dominated by radiation, \ie $w=1/3$. Later it becomes dominated by matter with $w=0$. The cosmological constant only becomes important at late times. The Friedmann equation (\ref{FLRWFried}) can be rewritten in terms of the critical density (\ref{Critdensity}) in the case of negligible cosmological constant as
\begin{equation}\label{curvaturegrowshrink}
 \left(\rho-\rho_c\right)a^2 =  \left(1-1/\Omega_{\rm total}\right)\rho a^2= \frac{3k}{\kappa}.
\end{equation}
The right hand side is a constant by inspection. The radiation component of the density $\rho_r\sim a^{-4}$ and the non-relativistic matter component $\rho_m\sim a^{-3}$, so $\rho a^2$ decreases as $a$ becomes large. This implies that $|1-1/\Omega_{\rm total}|$ increases to balance it, implying that $|\Omega_k|$ grows as the universe expands. As noted at the start of Subsection \ref{IntroDarkEnergy}, the value of the curvature parameter is observationally indistinguishable from zero at the time of writing, which implies that it was very finely tuned to be close to zero in the early universe. Indeed, the value at the time of cosmic nucleosynthesis would be $10^{-9}$ of the present value, assuming matter domination. Inflation provides a mechanism to drastically reduce the value of $|\Omega_k|$ at early times via accelerating expansion. One way to see this is that since $\rho$ is a constant for a cosmological constant or for a slow-rolling scalar field, the argument using (\ref{curvaturegrowshrink}) is reversed and $|1-1/\Omega_{\rm total}|$ decreases as $a$ grows. Another way to view the flatness problem is via the Hubble radius $1/H$, which is the distance beyond which we are causally disconnected from due to the expansion rate of the universe. In comoving coordinates, the Hubble radius is $1/aH=1/\dot{a}$. In an accelerating universe, regions of space we are currently in causal contact with can be moved out of causal contact with us. The curvature density parameter is proportional to the squared comoving Hubble radius: $|\Omega_k|=|k|/(aH)^2$. The original flatness problem is that if $\dot{a}$ is decreasing, $\Omega_k$ is increasing, but the inflationary resolution is that an early phase of increasing $\dot{a}$ can decrease the value of $\Omega_k$ enough so that it stays small until the present day. An intuitive understanding is that the accelerating expansion phase pushed neighbouring regions of the pre-inflationary universe out of causal contact with each other, forming the later universe out of only a region of the early universe that was so small that the effect of curvature over that region was negligible.

The horizon problem concerns the apparent uniformity of the CMB. As noted in Subsection \ref{IntroBBT}, the CMB is a blackbody spectrum of an almost uniform 2.7 K temperature in all parts of the sky, having only order $10^{-5}$ deviations from the mean temperature. Put another way, the CMB from all directions in the sky appears to be very close to being in thermal equilibrium. In a decelerating universe, opposite ends of the CMB should not have been in causal contact when the CMB was produced, since the light from these opposing sides has only just been able to reach the Earth-based observers in the middle, and yet they are apparently in thermal equilibrium. According to inflation, the answer is that they were previously in causal contact, but the accelerating expansion of the universe during inflation isolated these different regions from one another, only to come back into causal contact later. Inflation takes a small region of the early universe and makes it very large, so that all the information in the sky corresponds to a small region of the early universe that was isotropic and in thermal equilibrium. The currently observable anisotropies of the CMB would have been created after the end of inflation. Although inflation resolves the fine-tuning in the horizon and flatness problems, it has its own fine-tuning of initial conditions for the scalar field. The derivations above depend on the initial velocity of the scalar field being small, the field being close to homogeneous and the initial value of the field being far enough away from the position of the minimum of the potential.

The magnetic monopole problem is that magnetic monopoles of the form $\vec{\nabla}\cdot\vec{B} \ne 0$ are not observed to exist in our universe\footnote{In condensed matter physics, anologous ``magnetic monopoles'' \eg of the form $\vec{\nabla}\cdot\vec{H} \ne 0$ do exist \cite{Castelnovo:2007qi,Morris:2010ma}, but this consistent with conventional electromagnetism.}. Although magnetic monopoles are forbidden in the Standard Model, they are a necessary part of proposed Grand Unified Theories (GUTs) \cite{'tHooft:1974qc,Polyakov:1974ek}, which offers an explanation of why charge is quantized in elementary particles \cite{Dirac:1948um}. Indeed, such theories typically predict that magnetic monopoles should be produced in large number densities in the high temperatures of the early universe. Worse than that, they would be stable, leaving behind a large number density in the present universe. GUTs are extensions to the Standard Model of particle physics in which the Standard Model gauge couplings unify at some large energy scale (usually $\sim10^{16}$ GeV). Inflation can explain the lack of magnetic monopoles as the accelerating expansion rate causes particles that are initially nearby to be pushed out of causal contact with each other, creating causally isolated regions that are almost totally empty. A downside to this motivation is the complete lack of experimental evidence that these GUTs are actually realized in nature.

As inflation comes to an end, the universe is almost completely empty and the temperature is very close to zero. As the inflaton field decays, it ``reheats'' the universe and repopulates it with particles. Since the mechanism for inflation is currently unknown and, if it exists at all, lies far beyond the Standard Model, how this process works in detail is also not known.

The relevance of inflation to this thesis is that one of the best fits to current CMB observations, such as from WMAP \cite{Hinshaw:2012aka} and Planck \cite{Planck:2013jfk,Ade:2015lrj}, comes from one of the oldest models of inflation that is related to low-energy effective theories of gravity. The Starobinsky model of inflation \cite{Davies:1977ze,Starobinsky:1980te} assumes a gravitational action of the form
\begin{equation}\label{Starobaction}
 S_{\rm grav} =  \int d^4x\ \sqrt{-g}\mathcal{L}_{\rm grav} = \int d^4 x\ \sqrt{-g}\frac{1}{2\kappa}\left(R+\frac{R^2}{6M^2}\right),
\end{equation}
where $M$ is a mass scale called the ``scalaron'' mass. A sensible physical interpretation of the higher-order curvature term is that it comes from integrating out high-energy modes of a more fundamental gravity theory by means of an RG flow down to some scale $\Lambda\sim M$ where $\Lambda$ in this case is a large UV cutoff scale and {\it not} the cosmological constant. One would expect this to be a truncation of a larger gravitational action that contains an infinite series of terms. A possible extension to $\mathcal{L}_{\rm grav}$ is to set it to a Taylor-expandable function of the Ricci scalar:
\begin{equation}
 2\kappa\mathcal{L}_{\rm grav} = R+\frac{R^2}{6M^2}+{\rm const}\times \frac{R^3}{M^4} + \cdots
\end{equation}
This would, however, also be expected to be a truncation of a more complete gravitational action. Non-local (\ie non-Taylor-expandable) $f(R)$ theories are sometimes also considered for late-universe cosmology, although these would not be compatible with Kadanoff blocking, upon which RG relies. Care is required when constructing such models to ensure that other terms in the higher-derivative expansion do not hinder the inflationary effect of the $R^2$ term.

To see that this scalaron mass corresponds to a scalar mode in the gravity theory, one can perform a Legendre transformation:
\begin{equation}\label{Legendrefr}
 f(R) =  \phi R - V(\phi),
\end{equation}
\ie $f'(R)=\phi$ and $V'(\phi)=R$, where $\phi$ is a scalar field and $V(\phi)$ is a potential. Including a cosmological constant for the sake of generality, we have an action of the form
\begin{equation}\label{ScalTensJordAct}
 S = \int d^4 x\ \sqrt{-g}\left(\frac{1}{2\kappa}\left[\phi R-V(\phi)-2\Lambda\right]+\mathcal{L}_{\rm matter}\right).
\end{equation}
This form is referred to as the Jordan Frame. In the action for Starobinsky inflation (\ref{Starobaction}), we have
\begin{equation}\label{ScalTensStarPot}
 V(\phi) = \frac{3M^2}{2}\left(\phi-1\right)^2.
\end{equation}
Since the minimum of $V(\phi)$ is at $\phi_0=1$, the universe at late times returns to pure Einstein gravity (\ref{EinsteinGravAct}). There also exists an ``Einstein'' frame obtained from a conformal rescaling of the metric. Together with a field redefinition of $\phi$, this recasts the action to look like Einstein gravity together with a scalar field with canonically normalized kinetic term and a potential with an exponential form.

The observable effects of inflation are in the CMB. In addition to making the CMB close to isotropic, the inflation also generates small anisotropies in the CMB by taking quantum fluctuations of the inflaton field and expanding them to make scalar perturbations to the CMB temperature. Gravitational radiation produced at the end of inflation would also create tensor perturbations. The latter has, at the time of writing, not been observed in the CMB, which provides constraints on viable inflation models and thus favours $R^2$ inflation or other models based on an exponential scalar potential. The first of two commonly considered observables is the scalar spectral index $n_s$ (or ``primordial tilt''), defined via the power spectrum of the Fourier modes $k$ of scalar fluctuations:
\begin{equation}
 \mathcal{P}(k) \sim k^{n_s-1},
\end{equation}
where a value of $n_s=1$ implies a scale-invariant spectrum. The power spectrum itself is related to the 2-point function of a perturbation $\psi$ via
\begin{equation}
 \left<\psi(0)\psi(x)\right> = \int d{\rm ln}k\ \mathcal{P}(k).
\end{equation}
The observed value is between about 0.95 and 0.98, implying a slight bias towards lower-frequency fluctuations. The second is the ratio of tensor to scalar perturbations $r$, which is currently constrained to be less than about 0.1. Relating $r$ to the slow-roll parameters, one finds that $r\approx 16\epsilon$.

Taking this whole introduction together, we are motivated to study the physics of higher-derivative theories of gravity at different scales of length and their possible cosmological implications. The next chapter of this thesis will review manifestly gauge-invariant ERG methods in preparation for developing the manifestly diffeomorphism-invariant ERG \cite{Morris:2016nda} in Chapter \ref{ChapterMDIERG}. This will give us a stronger theoretical basis for studying higher-derivative gravity expansions. The subject of backreaction, which is the non-linear effect of metric perturbations on cosmological expansion, will be reviewed in Einstein gravity in Chapter \ref{AWHPreview2} in preparation for discussing the original research on backreaction in higher-derivative gravity theories in Chapters \ref{ChapterBack1} and \ref{ChapterBack2}. 


\chapter{Review of the manifestly gauge-invariant ERG}\label{AWHPreview1}

\section{Kadanoff blocking continuous fields}\label{MGIERGKadanoff}

Quantum field theory, unifying quantum mechanics and special relativity, concerns continuous fields, each possessing an infinite number of degrees of freedom, out of which particles can be created and destroyed as excitations of the field. Interaction terms in the action receive quantum corrections in perturbation theory from loops, around which a momentum propagates. This momentum is integrated over to evaluate the total contribution. Since the spacetime is continuous at all scales, the momentum can range from zero to infinity. The lack of an upper bound on the momentum gives rise to ultraviolet divergences (where the result of integrating over a loop momentum is infinite). Ultraviolet divergences must be removed by means of a regularization scheme in order to extract physical results. The regularization scheme is a formal device that does not imply any meaningful physics in itself. Any observables must be independent of the regularization scheme. 

One conceptually simple regularization is to model the spacetime as a lattice of points. The distance between nearest neighbour points is then the smallest length scale explicitly considered. This is the ``ultraviolet cutoff'', which corresponds to a maximum momentum via a Fourier transform. In practice, computations performed on a finite lattice are also infrared regulated, possessing a minimum momentum mode corresponding to an ``infrared cutoff''. The complete physics that explicitly considers all scales of length would emerge in the ``continuum limit'' where the lattice spacing is tended towards zero, corresponding to the limit where the maximum momentum considered is tended towards infinity. In the case of a finite lattice spacing, length scales shorter than the lattice spacing have been averaged out, thus the lattice cannot be used to study the details of physics at such scales. This is related to the idea of an ``effective field theory'', which is valid for computing physics at length scales above the ultraviolet cutoff length. We will not discuss lattice computations any further, since this thesis concerns exact constructions.

In this section, we wish to apply the powerful technique of Kadanoff blocking discussed in Section \ref{IntroKadanoff} to continuous fields \cite{Wilson:1971bg,Wilson:1971dh}, thus creating an exact RG\cite{Wilson:1973jj}. We will also wish to construct the blocking procedure in a momentum representation, since this is of most use for QFT. Instead of group together blocks of spins, we will average over the high-energy Fourier modes to construct an effective action describing low-energy physics. The averaging is effected by an integral over field configurations where a smooth infrared cutoff function is used to restrict which field configurations are integrated out. The result is that Fourier modes above some cutoff scale $\Lambda$ are removed \cite{Wegner:1972ih,Wilson:1973jj,Wilson:1974mb}. As with Kadanoff blocking over spins, this reduction in the degrees of freedom is not invertible, since details of the modes integrated out have been removed. RG flow is then described by lowering the scale of $\Lambda$, \ie integrating out more modes. The objective is then to construct a flow equation that can be used for computing effective actions and, via a loopwise expansion, a calculation of the $\beta$-functions for running couplings.

As in statistical mechanics, it is necessary that the theory is local, \ie that interactions between points separated by a large distance are suppressed so that the blocking does not lose macroscopically important information. As mentioned above, QFT is relativistic, so one would na\"{i}vely use the Minkowski metric to describe the $D$-dimensional spacetime geometry, in which there is 1 dimension of time and $D-1$ dimensions of space. However, this creates a tension with the requirement for locality in Kadanoff blocking. Consider light-like separations as an extreme example, \ie where $ds^2=0$. Phrased in position representation, such separations can have an infinite coordinate separation, so it is impossible to frame-independently distinguish long-distance and short-distance separations via an invariant interval, thus we are unable to construct a blocking scheme that groups together nearby spacetime points. In momentum representation, we see that arbitrarily large energies and momenta can be found in a 4-momentum that has $p^2=0$. This prevents us from integrating out all of the high-energy modes. This can be remedied by performing a Wick rotation ($t\to it$) into a Euclidean metric signature and then performing the blocking.

\begin{figure}[ht]
\begin{center}
\includegraphics[height=8cm]{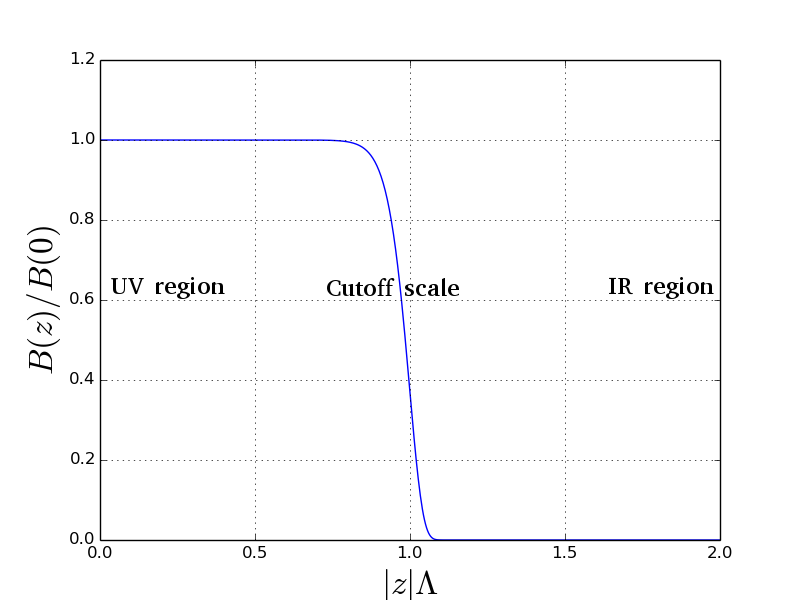}
\end{center}
\caption{Example of a smooth infrared cutoff function in position representation}
\label{Icuttoff}
\end{figure}

Consider a scalar field whose microscopic degrees of freedom are given by the ``bare'' field $\varphi_0$ and whose macroscopic degrees of freedom after blocking are given by the ``renormalized'' field $\varphi$. The blocking functional $b$ is defined via the effective Boltzmann factor:
\begin{equation}\label{Kblocking}
 e^{-S[\varphi]}=\int\Dm\varphi_{0} \ \delta\left[\varphi-b\left[\varphi_{0}\right]\right]e^{-S_{\rm bare}\left[\varphi_{0}\right]}.
\end{equation}
The delta function replaces the bare field with the renormalized field, which is constructed using the blocking functional, in which the high-energy modes are integrated out to give an average. Thus we obtain an effective Boltzmann factor with only the macroscopic degrees of freedom. As with Kadanoff blocking on a lattice of spins, the form of the blocking can be chosen freely. A simple, linear, example of a blocking functional is given by
\begin{equation}
 b[\varphi_{0}](x) = \int_{y} B(x-y)\varphi_{0}(y),
\end{equation}
where $B(x-y)$ is some kernel that contains the infrared cutoff function and a shortened notation has been used such that
\begin{equation}
 \int_x := \int d^D x.
\end{equation}
Figure \ref{Icuttoff} illustrates an example profile for the smooth kernel $B(z)$. In this example, we have $B(z) = B(0)e^{-(z\Lambda)^{2N}}$ where $N$ is a large enough integer to make a suitably rapid transition (for this figure $N=12$ has been selected). We have $B(z)\to0$ for large $|z|\Lambda$, transitioning rapidly at the length scale $1/\Lambda$.

As with statistical mechanics, we require that the partition function remains invariant under Kadanoff blocking, ensuring that macroscopic observables (derivable from correlation functions) also remain invariant. To see that this is true, consider performing a functional integral of (\ref{Kblocking}) with respect to the renormalized field $\varphi$. This trivially integrates out the delta function, leaving the usual form of the partition function for the bare field $\varphi_0$:
\begin{equation}\label{partinv}
 \mathcal{Z} = \int\Dm\varphi\ e^{-S[\varphi]} = \int\Dm\varphi_{0}\ e^{-S_{\rm bare}[\varphi_{0}]}.
\end{equation}
The first steps towards deriving a flow equation is to differentiate the effective Boltzmann factor with respect to RG time, \ie ln$(\Lambda)$:
\begin{equation}\label{BoltzFlow}
 \Lambda\frac{\partial}{\partial\Lambda}e^{-S[\varphi]}=-\int_{x} \frac{\delta}{\delta \varphi(x)}\int\Dm\varphi_{0}\ \delta\left[\varphi-b\left[\varphi_{0}\right]\right]\Lambda\frac{\partial b[\varphi_{0}](x)}{\partial\Lambda}e^{-S_{\rm bare}\left[\varphi_{0}\right]}.
\end{equation}
We then integrate out the bare field $\varphi_0$ to obtain a general expression in terms of $\Psi$, which is a function of $x$ and functional of $\varphi$ that can be loosely interpreted as the rate of change of the blocking functional with respect to RG time:
\begin{equation}\label{generalERG}
 \Lambda \frac{\partial}{\partial \Lambda}e^{-S[\varphi]}=\int_{x} \frac{\delta}{\delta \varphi(x)}\left(\Psi(x)e^{-S[\varphi]}\right)\,,
\end{equation}
Performing the differentiation and dividing out the Boltzmann factor, we obtain a general flow equation for an action with a single scalar field:
\begin{equation} 
\label{generalERG2}
\Lambda \frac{\partial}{\partial \Lambda} S =  \int_{x} \Psi(x)\frac{\delta S}{\delta \varphi(x)}  - \int_{x} \frac{\delta \Psi(x)}{\delta \varphi(x)}\,.
\end{equation}
Since there are an infinite number of possible blocking functionals that could be chosen, there are an infinite number of flow equations that could be constructed \cite{Latorre:2000qc}. They would, however, all be expressible in the form given in (\ref{generalERG2}). We can once again see that the partition function remains invariant under RG flow by seeing that the right hand side of (\ref{generalERG}) is in the form of a total derivative of $\varphi$. The rate of change of the partition function is given by the functional integral of (\ref{generalERG}) with respect to $\varphi$, which is zero, given a suitably damped Boltzmann factor at large $\varphi$, which is reasonably expected in Euclidean signature. 

Concerning notation used in the rest of this chapter, it is convenient in future to represent differentiation with respect to RG time with an over-dot such that, for some function $f(\Lambda)$, we have $\dot{f}(\Lambda) = \Lambda\frac{\partial}{\partial \Lambda}f(\Lambda)$.
We will follow the notation conventions seen in \cite{Morris:1998da,Arnone:2006ie} for a momentum kernel $W$ in the flow equation:
\begin{equation}\label{kernelquick}
 f\cdot W\cdot g := \int_{x} f(x) 
 W\left(-\partial^2\right)g(x),
\end{equation}
where, in position representation, $W$ generally contains a dimensionful factor that is a function of $-\partial^2$ and a dimensionless factor that is a function of $-\partial^2/\Lambda^2$. In momentum representation, the $-\partial^2$ is replaced with $p^2$. The momentum kernel usually used in the flow equation is related to the effective propagator of the field at a fixed point. Indeed, for massless scalar fields, it is identically the same as the effective propagator.

\section{Classical ERG for massless scalar fields}\label{ReviewScalar}

Let us now consider a particular specialization of (\ref{generalERG2}), which is the Polchinski form of the flow equation \cite{Polchinski:1983gv}. In the Polchinski form, the rate of change of blocking functional is written as
\begin{equation}\label{Polchange}
 \Psi(x) = \frac{1}{2}\int_{y} \dot{\Delta}(x,y)\frac{\delta \Sigma}{\delta \varphi(y)},
\end{equation}
where $\Sigma = S -2\hat{S}$ is an action that we will return to shortly and $\Delta$ is precisely the effective propagator. In momentum representation, we have $\Delta = c\left(p^2/\Lambda^2\right)/p^2$, where $c\left(p^2/\Lambda^2\right)$ is an ultraviolet momentum cutoff function used to regulate the propagator, which is otherwise just the usual form given by $1/p^2$. Also note that
\begin{equation}
 \dot{\Delta} = -\frac{2}{\Lambda^2}c'\pcutarg,
\end{equation}
where $c'(p^2/\Lambda^2)$ is the derivative of $c(p^2/\Lambda^2)$ with respect to its argument. If $c$ is a local expansion, then so is $\dot{\Delta}$.
The ``seed action'' $\hat{S}$ is a fixed-point action whose only scale is $\Lambda$. In the Polchinski form of the flow equation, $\hat{S}$ is conveniently chosen to be the regularized kinetic term:
\begin{equation}
\label{scalarSeed}
 \hat{S} = \frac{1}{2} \partial_{\mu}\varphi\cdot c^{-1}\cdot\partial_{\mu}\varphi,
\end{equation}
where we use the notation in (\ref{kernelquick}) and the Greek Lorentz indices are implicitly summed over with a Euclidean metric. Because we have freedom in our choice of blocking scheme, we also have freedom in our choice of seed action. We do, however, still have to insist that both $\hat{S}$ and $\dot{\Delta}$ take the form of local derivative expansions to all orders to ensure the validity of Kadanoff blocking. However, we can freely add 3-point and higher terms to $\hat{S}$ without introducing any changes to the physics.

The seed action in (\ref{scalarSeed}) is identical to the regularized kinetic term for the effective action $S$. Canonical normalization of the kinetic term and the propagator requires that $c(0)=1$. When we solve the classical part of the flow equation at the 2-point level, we will see that the factor of $1/2$ in (\ref{Polchange}) is also necessary to ensure that the kinetic term in $S$ is canonically normalized, and this factor of $1/2$ is similarly required in both the gauge and gravity generalizations of the Polchinski flow equation. Applying the Polchinski form for the rate of change of blocking functional in (\ref{Polchange}) to the general ERG flow equation for a single scalar field in (\ref{generalERG2}), we get the Polchinski flow equation, expressed using the notation in (\ref{kernelquick}):
\begin{equation}\label{scalarSdot}
 \dot{S} = \frac{1}{2} \frac{\delta S}{\delta\varphi}\cdot\dot{\Delta}\cdot\frac{\delta \Sigma}{\delta\varphi} - \frac{1}{2} \frac{\delta}{\delta \varphi}\cdot\dot{\Delta}\cdot\frac{\delta \Sigma}{\delta\varphi}.
\end{equation}
Note that since both $\dot{\Delta}$ and $\hat{S}$ are local, then if $S$ is local prior to RG transformations, it remains local after the RG transformation. Now consider the action (in momentum representation) as a Taylor expansion in the scalar field:
\begin{equation}\label{Scalarnpointexp}
 S = \int \dbar p\ \frac{1}{2}S^{\varphi\varphi}(p,-p)\varphi(p)\varphi(-p) + \int \dbar p\ \dbar q\ \frac{1}{3!}S^{\varphi\varphi\varphi}(p,q,-p-q)\varphi(p)\varphi(q)\varphi(-p-q) + \cdots,
\end{equation}
 where $S^{\varphi\varphi}(p,q)$ is the 2-point function of the action, $S^{\varphi\varphi\varphi}(p,q,r)$ is the 3-point function and so on. These $n$-point functions can be obtained by  performing $n$ functional derivatives on the action and evaluating at $\varphi=0$. Similarly, the flow equation for the $n$-point part of the action can be evaluated by performing the same procedure to equation (\ref{scalarSdot}). Taking the seed action in (\ref{scalarSeed}), we can write this in momentum representation as
\begin{equation}
 \hat{S} = \int \dbar p\ \frac{1}{2}\frac{p^2}{c}\varphi(p)\varphi(-p),
\end{equation}
so $\hat{S}^{\varphi\varphi}(p,-p) = c^{-1}p^2$. The propagator is then the inverse of this: $\Delta = c/p^2$. 

Figure \ref{fig:npointflow} below, illustrates the structure of the flow equation diagrammatically in the momentum representation. 
\begin{figure}[ht]
\begin{center}
\includegraphics[height=1.8cm]{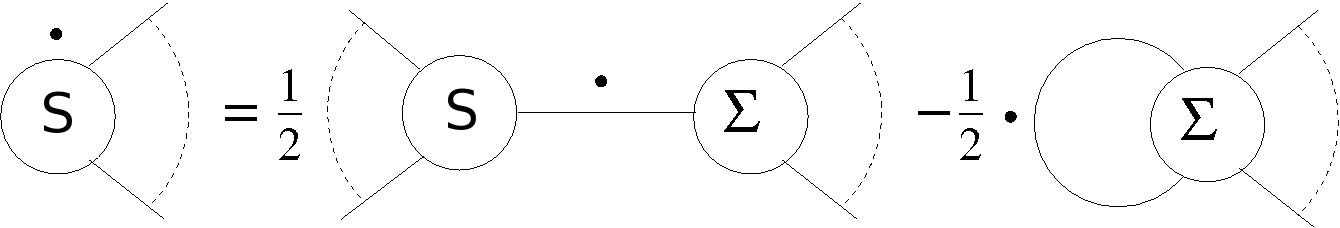}
\end{center}
\caption{General flow equation for an action with a single scalar field at some $n$-point level}
\label{fig:npointflow}
\end{figure}
In Figure \ref{fig:npointflow}, adapted from \cite{Arnone:2006ie} and also appearing in \cite{Morris:2016nda}, large circles with $n$ solid lines attached represent $n$-point functions for an action. The labels inside the circles indicate which action, $S$ or $\Sigma$, the $n$-point function is taken from. The small black dots have the same meaning as over-dots in the equations, \ie they represent differentiation with respect to RG time. An internal line, which is where both ends connect to an $n$-point function, represents an effective propagator given by the flow equation kernel, which carries momentum from one end to the other (although the sign notation convention for this thesis is that momenta always flow out of the actions and the kernel). External lines, which are lines where one end connects to an $n$-point function and one end is free, represent the momenta flowing out of the $n$-point function. An external line on the left-hand side of the flow equation must correspond to an external line on the right hand side and vice versa, carrying the same momentum on both sides.

This diagrammatic illustration helps with the intuitive understanding of the flow equation, as expanded out in $n$-point functions. Immediately, we can see that the second term on the right hand side of the flow equation has no tree-level component, because the internal propagator has both ends connecting to the same $n$-point function, forming a loop with an unconstrained momentum that is integrated over. The loopwise expansion of the action effectively counts powers of $\hbar$, with higher powers of $\hbar$ corresponding to higher loop orders, as will be discussed in Section \ref{MGIERGloopwise}. The $\hbar\to0$ limit corresponds to the classical level, which is the tree-level. Na\"{i}vely, $\hbar\to0$ is nonsense, given that we choose a system of natural units in which $\hbar=1$, however this limit represents a special case that is not our ultimate goal, but it is nevertheless instructive. Setting $\hbar=1$ is safe to do because rescaling $\hbar$ to anything over than 0 or $\infty$ is simply a rescaling of units with no physical implications. An alternative limit that reproduces the classical level is where interaction couplings are tended to zero, \ie the weak coupling limit. In a weak-coupling limit, higher loop orders, which are also higher order in the coupling, become significantly less important than lower loop orders. Given a suitably weak coupling, the theory is well-approximated by classical computations. This is of relevance to motivating the study of classical RG flows for gravity in Chapter \ref{ChapterMDIERG}, because gravity is extremely weakly-coupled at currently observable length scales. However, as we move into regimes with stronger coupling, the quantum corrections from higher loop orders become more important, ultimately requiring one to develop non-perturbative descriptions for physics in strongly-coupled regimes.

To see how the classical flow equation can be solved exactly at the 2-point level, let us consider a single-component scalar field theory with a symmetry under $\varphi\to-\varphi$. Because of the $\varphi\to-\varphi$ symmetry, this is a massless scalar field theory, so quantum corrections to the action begin at the 4-point level. As noted already, the classical level of the flow equation only uses the first term on the right hand side of (\ref{scalarSdot}), so we can neglect the second term at this stage. The $\varphi\to-\varphi$ symmetry imposes that there exist no 1-point functions. The seed action $\hat{S}$ has been required by construction to take the same 2-point function as the effective action $S$, giving us that $\Sigma^{\varphi\varphi}=-S^{\varphi\varphi}$. This tells us that the 2-point level of the flow equation is expressed only in terms of the 2-point function for the action. To see this, note that expanding either $S$ or $\Sigma$ in Figure \ref{fig:npointflow} to the 3-point or higher level would result in there existing 3 or more external lines. The exception to this is if either $S$ or $\Sigma$ were allowed to possess 1-point functions, which they do not under $\varphi\to-\varphi$ symmetry.

With this in mind, we find that the classical flow equation (\ref{scalarSdot}) can be written at the 2-point level as
\begin{equation}
\label{scalar-2pt-flow}
 \dot{S}^{\varphi\varphi} = -S^{\varphi\varphi}\dot{\Delta}S^{\varphi\varphi},
\end{equation}
which is solved by $\Delta = \left(S^{\varphi\varphi}\right)^{-1}$, as required for $\Delta$ to be the effective propagator. Higher-point modifications to $\hat{S}$ do not introduce any new observable physics, even at the quantum level, because such modifications to $\hat{S}$ cause nothing more than reparametrizations of the renormalized field.

Using the exact solution for the 2-point function in $S$ at the classical level, one can proceed iteratively to exactly solve all of the classical $n$-point functions using the flow equation and knowledge of the $(n-1)$-point functions and lower. One can then use the loopwise expansion to exactly solve the flow equation up to the desired number of loops, again working iteratively from the classical solution. An exact solution to the complete flow equation for $S$ would be the exact form of the effective action. A fixed-point action is characterised by having only a single scale $\Lambda$ such that all dimensionful parameters in the action are expressed as a dimensionless number multiplied by some power of $\Lambda$. This is the field theory equivalent to expressing dimensionful quantities in terms of lattice units in a renormalized Ising model. A fixed point exists which is an exact solution to the flow equation of the closed form $S=\hat{S}$, which is a non-interacting theory. For such a theory, the Boltzmann factor is simply a Gaussian. One can then perturb away from this fixed point by inserting extra operators (\ie interaction terms). There is, however, a ``triviality'' problem that such interactions in scalar field theory are either irrelevant or marginally irrelevant, so the RG flow falls back to the fixed point.

\section{Gauge-invariant ERG for Yang-Mills theories}\label{YangMills}

This thesis will be concerned with constructing an ERG for gravity in Chapter \ref{ChapterMDIERG}. As discussed in the Introduction's Subsection \ref{IntroDiffSym}, gravity theories are required to be symmetric under diffeomorphisms, which are infinitesimal coordinate transformations. This is very strongly analogous to the gauge symmetries associated with massless vector bosons that communicate forces such as electromagnetism or strong nuclear interactions. With that in mind, this section will be concerned with partially reviewing the wealth of existing knowledge on performing exact RG calculations in gauge theories without fixing a gauge. This manifestly gauge-invariant ERG provides much of the mathematical machinery required to construct an ERG for gravity that does not gauge-fix the diffeomorphism-invariant theory: such an ERG would then be called the manifestly diffeomorphism-invariant ERG \cite{Morris:2016nda}.

Let us begin with a brief technical history of the manifestly gauge-invariant ERG.
Manifest gauge invariance was first incorporated in the ERG in the context of a pure U(1) (Abelian) gauge theory \cite{Morris:1995he} and, following \cite{Morris:1998kz} it was generalized to pure non-Abelian SU($N$) theories \cite{Morris:1999px,Morris:2000fs,Morris:2000jj} and then developed to include fermions for Quantum Electrodynamics \cite{Arnone:2005vd,Rosten:2008zp} and Quantum Chromodynamics \cite{Morris:2006in}. Its regularization structure, which will be summarized in Section \ref{MGIERGSUSY} was further developed in \cite{Arnone:2000bv,Arnone:2000qd,Arnone:2001iy}. Further development of the method has demonstrated the ability to extract results that are independent of the regularization scheme \cite{Arnone:2002yh,Arnone:2002qi,Arnone:2003pa,Arnone:2002cs,Morris:2005tv,Rosten:2005ka} and to handle general group structures \cite{Arnone:2005fb}. Scheme-independence at all loop orders has been demonstrated in \cite{Rosten:2005ep,Rosten:2006tk} and general expressions for the expectation values of gauge-invariant operators were developed in \cite{Rosten:2006qx,Rosten:2006pd}. Short reviews can be found in \cite{Arnone:2002fa,Rosten:2004aw,Rosten:2011ty}. For a much longer review, see \cite{Rosten:2010vm}. Manifestly gauge-invariant ERG has also been the subject of other PhD theses \cite{Gatti:2002kc,Rosten:2005qs}.

For a much gentler introduction, let us consider a theory in which there is a gauge field with 4-potential $A_\mu$. Relating this to classical electromagnetism in four spacetime dimensions, it is common to express the electromagnetic 4-potential as an electric scalar potential $\varphi$ and a magnetic 3-potential ${\bf A}$ via $A^\mu = (\varphi,{\bf A})$. The familiar electric and magnetic field strengths, $\bf{E}$ and ${\bf B}$ are related to the scalar and 3-vector potentials via ${\bf E}  =  -{\rm grad}(\varphi) -\dot{{\bf A}}$ and ${\bf B}  =  {\rm curl}({\bf A})$. However, it is more convenient for us to express these using covariant notation as components of a field-strength tensor $F_{\mu\nu}$ for electromagnetism (\ie Abelian gauge theory):
\begin{equation}\label{fieldstrengthtensor}
 F_{\mu\nu} = 2\partial_{[\mu}A_{\nu]}.
\end{equation}
The ${\bf E}$ field is related to $F_{\mu\nu}$ via $F^{0i}=E^i$ where $i$ is an index label between 1 and 3 and $F^{ij} = \epsilon^{ijk}B_k$ where $\epsilon^{ijk}$ is the Levi-Civita symbol, representing an antisymmetric tensor. More generally, the field-strength tensor can also be expressed as $F_{\mu\nu}:=i\left[D_\mu,D_\nu\right]$, where $D_\mu$ is the covariant derivative for this gauge theory, defined by
\begin{equation}
\label{gaugeCovariantDerivative}
 D_\mu := \partial_\mu -iA_\mu.
\end{equation}
In non-Abelian gauge theories, such as Quantum Chromodynamics (QCD), $A_\mu$ can be expanded out colour-wise to
\begin{equation}
 A_\mu(x) = A^a_\mu(x)T^a,
\end{equation}
where $a$ is the colour index corresponding to a group generator $T^a$ and repeated colour indices indicate an implicit summation over colours. Usually, the gauge field is defined such that the second term on the right hand side of (\ref{gaugeCovariantDerivative}) is $-igA_\mu$, explicitly containing a coupling $g$ as a separate factor, but here the field is defined to identically be the gauge connection, which is helpful for constructing the manifestly gauge-invariant formalism, as will become apparent soon. The covariant derivative for gauge theories is strongly analogous to the covariant derivative for curved spacetime and the field-strength tensor is similarly analogous to the Riemann tensor, as one can see by consulting equation (\ref{EinsteinFieldStrengthAnalogy}). Similarly to how the effective kinetic term for a scalar field theory was constructed in (\ref{scalarSeed}), the effective action for a pure gauge theory can be written as
\begin{equation}\label{rescaleact}
 S[A](g) = \frac{1}{4g^2}{\rm tr}\int_{x} F_{\mu\nu}\,c^{-1}\!\left(-\frac{D^2}{\Lambda^2}\right)\!F_{\mu\nu} + \mathcal{O}(A^3) + \cdots
\end{equation}
Note that the rescaling of the field with the coupling has enabled us to express the action with the coupling as an overall factor, this will become convenient when performing the loopwise expansion in Section \ref{MGIERGloopwise}.
Analogous to the diffeomorphism transformation is the gauge transformation:
\begin{equation}\label{usualgaugetransform}
 \delta A_\mu = [D_\mu,\omega(x)] = \partial_\mu\omega(x) -i\left[A_\mu(x),\omega(x)\right],
\end{equation}
where $\Omega$ is a scalar field effecting the equivalent of a coordinate transformation for the 4-potential. The action in (\ref{rescaleact}) is invariant under gauge transformations. In commonly-used QFT methods that depend on gauge-fixing, fields such as $A_\mu$ rescale under renormalization from their ``bare'' values to ``renormalized'' values. Consider a covariant derivative defined as $ D_\mu := \partial_\mu -igA_\mu$, the renormalized field $A^R_\mu$ is related to the bare field $A^B_{\mu}$ by
\begin{equation}
 A^R_\mu :=Z^{-1/2}A_\mu^B,
\end{equation}
where $Z$ is a wavefunction renormalization factor that would ordinarily appear as an overall factor in the renormalized kinetic term $F_{\mu\nu}F^{\mu\nu}$ such that the ``counterterm'' to the bare action has an overall factor of $Z-1$. Similarly, the coupling $g$ would also rescale from its bare value $g_B$ as $g := Z^{1/2} g_B$. However, wavefunction renormalization is not compatible with manifest gauge invariance because the gauge transformation of the renormalized field $A^R_\mu$ would be
\begin{equation}
 \delta A^{R}_\mu = Z^{-1/2}\partial_\mu \omega-i[A^{R}_{\mu},\omega],
\end{equation}
which differs from the original form of the gauge transformation in (\ref{usualgaugetransform}) by the factor of $Z^{-1/2}$ in the first term. Methods that depend on gauge-fixing introduce Fadeev-Popov ghost fields to remove the longitudinal mode introduced to the field by the gauge-fixing and also restore of modified form of the gauge invariance called BRST symmetry, in which the ghost fields, which are fermionic scalars, also transform under the symmetry \cite{Faddeev:1967fc,Becchi:1974xu,Becchi:1974md,Becchi:1975nq}. The gauge-fixed action is then invariant under the complete set of BRST transformations. Since we are concerned with manifestly gauge-invariant constructions, we will insist that $Z=1$ at all times. Rescaling the field with $g$ to give the form of the covariant derivative in (\ref{gaugeCovariantDerivative}) and moving the coupling outside the action in (\ref{rescaleact}) has enabled this advantageous property. Furthermore, the loopwise expansion of the action also proceeds in a similar way:
\begin{equation}\label{rescaleactseries}
 S = \frac{1}{g^2}S_0 + S_1 + g^2 S_2 + \cdots
\end{equation}
where $S_n$ is the (coupling-independent) $n$th loop contribution to the action with the coupling rescaled outside the action as seen above. Thus we see explicitly how the loopwise expansion of the action corresponds to a series expansion in increasingly high powers of the coupling, with the classical limit being obtained where $g\to0$. The RG running of the coupling is given by the $\beta$-function, which is the derivative of the coupling with respect to RG time:
\begin{equation}
 \beta := \Lambda \frac{\partial g}{\partial \Lambda}.
\end{equation}
Just as the action possesses a loopwise expansion in which higher loop orders correspond to higher coupling powers, so too does the $\beta$-function:
\begin{equation}\label{rescalebeta}
 \beta := \Lambda\partial_{\Lambda}g = \beta_1 g^3 + \beta_2 g^5 + \cdots
\end{equation}
The first contribution to the $\beta$-function is at the 1-loop level, so the coupling does not run in the classical limit. For this reason, most particle physicists view RG flow as a quantum property of the field theory that originates in the loops. However, as we have already seen, the effective action already has a non-zero derivative with respect to RG time at the classical level.

\begin{figure}[ht]
\begin{center}
\includegraphics[height=1.8cm]{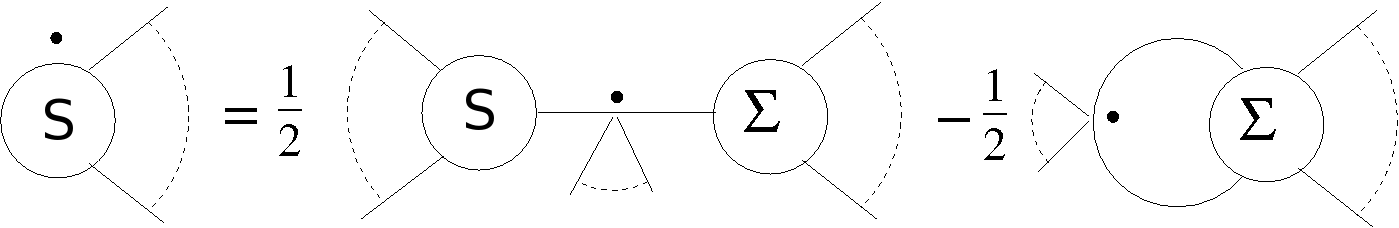}
\end{center}
\caption{Gauge-invariant $n$-point flow equation with a covariantized kernel}
\label{fig:Anpoint}
\end{figure}

The generalization of (\ref{scalarSdot}) to gauge theories is simply
\begin{equation}\label{gaugeflow}
 \dot{S} = \frac{1}{2}\frac{\delta S}{\delta A_\mu}\cdot \{\dot{\Delta}\} \cdot \frac{\delta \Sigma_g}{\delta A_\mu} - \frac{1}{2} \frac{\delta}{\delta A_\mu}\cdot\{\dot{\Delta}\}\cdot\frac{\delta \Sigma_g}{\delta A_\mu},
\end{equation}
where the repeated Lorentz index again has an implicit summation and the shorthand notation in (\ref{kernelquick}) has been used. As with the scalar case, the factors of 1/2 are necessary to ensure that the kinetic term is canonically normalized. Na\"{i}vely, the presence of a momentum kernel $\dot{\Delta}$ would break the gauge invariance because of its derivative expansion. In order to maintain gauge invariance, it is necessary to covariantize the kernel. Once again, we are at liberty to choose from an infinite number of possible schemes for doing this. In this thesis, we will covariantize kernels by replacing the partial derivatives with covariant derivatives. In the flow equation (\ref{gaugeflow}), the covariantization of the kernel is denoted by the braces either side of $\dot{\Delta}$. The $g$ subscript applied to $\Sigma$ denotes that the rescaling of $A_\mu$ with $g$ has led to a coupling rescaling in $\Sigma$ to ensure that $\dot{S}$ has the same order-by-order expansion in $g$ as $S$ does:
\begin{equation} 
\label{Sigma-g}
\Sigma_g =  g^2 S - 2\hat{S}\,.
\end{equation}
As with the scalar field theory, we can write the gauge theory action as a series expansion in $n$-point functions in the spirit of equation (\ref{Scalarnpointexp}):
\begin{eqnarray}\label{Gaugenpointexp}
 S & = & \frac{1}{2}\int \dbar p\ S^{AA}_{\alpha\beta}(p,-p)A_{\alpha}(p)A_{\beta}(-p) + \nonumber \\ && \frac{1}{3!}\int \dbar p\ \dbar q\ S^{AAA}_{\alpha\beta\gamma}(p,q,-p-q)A_{\alpha}(p)A_{\beta}(q)A_{\gamma}(-p-q) + \cdots
\end{eqnarray}
where $S^{AA}_{\alpha\beta}(p,-p)$ is the 2-point function, $S^{AAA}_{\alpha\beta\gamma}$ is the 3-point function and so on.
Once again, we can represent the $n$-point expansion of the flow equation diagrammatically, as seen in Figure \ref{fig:Anpoint}, adapted from \cite{Arnone:2006ie} and also appearing in \cite{Morris:2016nda}.
The $n$-point expansion of the flow equation in gauge theories is very similar to the $n-$point expansion in scalar field theories except that the covariantized kernel also has its own $n$-point expansion. To see why, note that replacing the partial derivatives with covariant derivatives has introduced an expansion in the field $A_\mu$ into the kernel itself, independently of the actions it connects to at each end. Unlike the actions, however, the $n$-point expansion of the kernel begins at the 0-point level and has a complete set of non-zero $n$-point functions for all non-negative integer values of $n$. 

The regularized 2-point function can be written as
\begin{equation}
\label{gauge-2point}
 S^{AA}_{\mu\nu} = (\delta_{\mu\nu}p^2 -p_{\mu}p_{\nu})\,c^{-1}\!\left(\frac{p^2}{\Lambda^2}\right).
\end{equation}
Unlike with scalar fields, the 2-point function for the gauge field $S^{AA}_{\mu\nu}(p,-p)$ is transverse: $p_\mu S^{AA}_{\mu\nu}=0$. The transverse form of the 2-point function is imposed by gauge invariance.
This 2-point function is non-invertible and therefore we cannot construct $\Delta$ to invert the 2-point function in the usual way: $S^{AA}_{\mu\nu}(p,-p)\KoL \neq \delta_{\mu\nu}$. Gauge-fixing introduces a longitudinal mode, allowing the 2-point function to be inverted. The longitudinal mode is then cancelled away again by the BRST ghost fields. However, in this formalism, we do not need to invert the 2-point function. Instead, we define $\KoL$ to map the 2-point function onto the transverse projector:
\begin{equation}\label{gaugeeffectivepropagator}
 \Delta S^{AA}_{\mu\nu} = \delta_{\mu\nu}-p_{\mu}p_{\nu}/p^2.
\end{equation}
Although this is not inverting the effective 2-point function, it is still convenient to refer to $\KoL$ as the effective propagator. 

Let us turn to solving the gauge-invariant 2-point flow equation at the classical level. The seed action should match the effective action at the 2-point level but otherwise be chosen freely without affecting the physics. We will take the simplest case in which the seed action is nothing but the regularized kinetic term:
\begin{equation}
 \hat{S}[A] = \frac{1}{4}\,{\rm tr}\!\int_{x} F_{\mu\nu}\,c^{-1}\!\left(-\frac{D^2}{\Lambda^2}\right)\!F_{\mu\nu}\,.
\end{equation}
 As with the scalar case, the classical flow equation only uses the first term on the right hand side of the flow equation (\ref{gaugeflow}). Lorentz invariance requires that the effective action (\ref{rescaleact}) of a pure gauge theory, expressed as an $n$-point expansion like (\ref{Gaugenpointexp}) begins at the 2-point level, as can be seen by the need to contract Lorentz indices to form the Lagrangian out of Lorentz scalars. The lack of a 1-point function means that, as with scalars possessing the $\varphi\to-\varphi$ symmetry, the only contributions to the right hand side of the flow equation at the 2-point level are found by taking both $S$ and $\Sigma_g$ at the 2-point level and the kernel at the 0-point level. The flow equation then reads as
\begin{equation}
 \dot{S}^{AA}_{\mu\nu} = -S^{AA}_{\mu\alpha}\dot{\Delta}S^{AA}_{\alpha\nu}.
\end{equation}
This equation is solved by (\ref{gauge-2point}) and (\ref{gaugeeffectivepropagator}), as required.

\section{Loopwise expansion}\label{MGIERGloopwise}

Although computing the RG flow at the classical level has its uses, which will be explored in this thesis in the context of gravity, ultimately the objective is to also understand field theories at the quantum level. Perturbatively, this involves expanding the flow equation loopwise and computing $\beta$-functions. Let us first consider the scalar field theory from Section \ref{ReviewScalar}. Consider a theory where there is a coupling $\lambda$ at the 4-point level. Once again, the field is rescaled by the coupling: $\tilde\varphi=\varphi/\sqrt{\lambda}$, and the action is rescaled by the coupling to keep the kinetic term canonically normalized: $\tilde S = S/\lambda$. However, it is convenient to drop the tildes from the notation as we proceed. The action is a function of $\lambda$ and a functional of $\varphi$ with a loopwise expansion akin to that in (\ref{rescaleactseries}):
\begin{equation}\label{scalarloopaction}
 S\left[\varphi;\lambda\right] = S_0 + \lambda S_1 + \lambda^2 S_2 + \cdots.
\end{equation}
The seed action also possesses a similar loopwise expansion where it is convenient to impose that $S^{\varphi\varphi}(p,q)=\hat{S}^{\varphi\varphi}(p,q)$. The $\beta$-function, defined by $\beta(\Lambda):=\Lambda\partial_\Lambda\lambda$ also has a loopwise expansion akin to that in (\ref{rescalebeta}):
\begin{equation}
 \beta(\Lambda) = \beta_1\lambda^2 +\beta_2\lambda^3 + \cdots.
\end{equation}
Because scalar field theories do not possess a gauge symmetry that would be violated by wavefunction renormalization, we have in general that $Z(\Lambda)\neq1$ where $\varphi_B=Z^{1/2}\varphi_R$. As a result, there is also an anomalous dimension $\gamma(\Lambda) := \Lambda\partial_\Lambda Z$, with an expansion $\gamma(\Lambda) = \gamma_1\lambda + \gamma_2\lambda^2 + \cdots$. Differentiating (\ref{scalarloopaction}) with respect to RG time and applying the Polchinski flow equation in (\ref{scalarSdot}) gives us the loopwise expansion of the flow equation. At the classical level, there is the familiar form:
\begin{equation}\label{scalarloop0}
 \dot{S}_0 = \frac{1}{2}\frac{\delta S_0}{\delta \varphi}\cdot\KoL\cdot\frac{\delta \Sigma_0}{\delta \phi}.
\end{equation}
Expanding to one loop, we have to include the second term of the flow equation as well as the 1-loop parts of the $\beta$-function and the anomalous dimension:
\begin{eqnarray}\label{scalarloop1}
 \dot{S}_1 + \beta_1S_0 + \frac{\gamma_1}{2}\varphi\cdot\frac{\delta S_0}{\delta\varphi} & = & \frac{\delta S_1}{\delta\varphi}\cdot\KoL\cdot\frac{\delta S_0}{\delta\varphi} - \frac{\delta S_1}{\delta\varphi}\cdot\KoL\cdot\frac{\delta \hat{S}_0}{\delta\varphi} -\nonumber \\ &&
 \frac{\delta S_0}{\delta\varphi}\cdot\KoL\cdot\frac{\delta \hat{S}_1}{\delta\varphi} - \frac{1}{2}\frac{\delta}{\delta \varphi}\cdot\KoL\cdot \frac{\delta \Sigma_0}{\delta\varphi}.
\end{eqnarray}
Expanding in a similar fashion to the desired loop level, one can extract the $\beta$-function and anomalous dimension up to that loop order by means of taking the flow equation for two choices of $n$-point level and solving for the two unknowns.

\section{Additional regularization}\label{MGIERGSUSY}

Although the effective propagator has been regularized with an ultraviolet momentum cutoff, this is insufficient to remove the ultraviolet divergences from the theory \cite{Slavnov:1972sq,Lee:1972fj}. A particularly elegant way to provide the additional regularization that is required to cancel the divergences at all loop orders is by introducing a supersymmetry between bosonic gauge fields and fermionic Pauli-Villars fields. This is analogous to how Parisi-Sourlas supersymmetry cancels degrees of freedom in spin glasses, as discussed in the Introduction's Subsection \ref{IntroPSSUSY}. Since this thesis is mostly concerned with the ERG for gravity at the classical level, the description of this additional regularization will be kept brief.

Pauli-Villars regularization works by introducing a cancellation term to the propagator such that the propagator tends towards zero as the momentum tends towards infinity, but remains largely unaffected when the momentum is much less than the cutoff scale. In its simplest form, applied to a scalar of mass $m$ and a cutoff $\Lambda$, the propagator in Euclidean signature carrying momentum $p$, which we will denote by $\Delta(p, m,\Lambda)$ is regularized as:
\begin{equation}
 \Delta(p, m,\Lambda) = \frac{1}{p^2-m^2} - \frac{1}{p^2- \Lambda^2}.
\end{equation}
In the continuum limit where $\Lambda\to\infty$, we return to the unregulated propagator, given purely by the first term on the right hand side. Conversely, in the high-momentum limit where $p\to \infty$, the propagator tends towards zero, and thus the extent of ultraviolet divergence is reduced.

The implementation of the manifestly gauge invariant regulator involves adding not only additional fermionic Pauli-Villars fields, but also an extra duplicate gauge field to complete the balancing of bosonic and fermionic modes. To ensure that the regularization only cancels the divergence down to the cutoff scale, a Higgs mechanism is implemented with an additional super-scalar that places the mass scale of the Pauli-Villars fields (and their own mass scale) at the cutoff scale $\Lambda$, thus breaking the supersymmetry at that scale and decoupling the duplicate bosonic field from the field of physical interest at that scale. In particular, the implementation involves promoting the gauge field to a super-matrix:
\begin{equation}\label{supergaugefield}
 \mathcal{A}_{\mu} = \left( \begin{array}{cc}
A^{1}_{\mu} & B_{\mu} \\
\bar{B}_{\mu} & A^{2}_{\mu} \end{array} \right) + \mathcal{A}^{0}_{\mu}\mathcal{I},
\end{equation}
where $A^{1}_{\mu}$ is the original bosonic gauge field of physical interest, $A^2_{\mu}$ is a duplicate bosonic gauge field, $B_{\mu}$ and $\bar{B}_\mu$ are the fermionic Pauli-Villars ghost fields, $A^0_\mu$ is an additional field that can be added freely without impacting on the physics, and $\mathcal{I}$ is the identity matrix. The super-trace of a super-matrix of this form is the trace of the first bosonic half minus the trace of the second bosonic half, \ie ${\rm tr}(A^1_\mu-A^2_\mu)$ in the case of (\ref{supergaugefield}) (the identity is super-traceless). After promoting the gauge field to this new super-matrix, the Yang-Mills contribution to the action is then
\begin{equation}\label{SYMaction}
 S_{\rm YM} = \frac{1}{4g^2}{\rm str}\int \mathcal{F}_{\mu\nu}c^{-1}\left(-\frac{D^2}{\Lambda^2}\right)\mathcal{F}_{\mu\nu},
\end{equation}
where $\mathcal{F}_{\mu\nu}$ has been defined in the same spirit as $F_{\mu\nu}$, \ie $\mathcal{F}_{\mu\nu} := i[D_{\mu}, \mathcal{A}_{\nu}]$. Since the identity is super-traceless and commutes with any matrix, it does not contribute to (\ref{SYMaction}).

To extract real physics, one needs to break the supersymmetry and decouple these extra fields at the cutoff scale using a Higgs mechanism. This is achieved with an extra superscalar matrix $\mathcal{C}$, given by
\begin{equation}\label{superscalar}
  \mathcal{C} = \left( \begin{array}{cc}
C^1 & D \\
\bar{D} & C^2 \end{array} \right),
\end{equation}
where $C^1$ and $C^2$ are bosonic scalar fields and $D$ and $\bar{D}$ are fermionic scalars. Now it remains to write the action for the scalar super-field, whose propagator is also regularized by a cutoff function $\tilde c\pcutarg$ (not necessarily the same cutoff function as for the gauge field):
\begin{equation}
 S_{\mathcal{C}} = {\rm str}\int d^D x\ D_\mu \mathcal{C}(x)\left\{\tilde{c}^{-1}\right\}D_\mu \mathcal{C}(x) + \frac{\lambda}{4}{\rm str}\int d^D x\ \left(\mathcal{C}^2-\Lambda^2\right)^2.
\end{equation}
The first term on the right hand side is the regularized kinetic term for the super-scalar and the second term is the Higgs-like potential for the super-scalar where a spontaneous symmetry breaking places the vacuum expectation value of $C^1$ and $C^2$ at the scale set by $\Lambda$. The super-trace prescription then gives the fermionic field $B_\mu$ a mass while keeping the bosonic fields $A^1_\mu$ and $A^2_\mu$ massless, with interactions also set at the cutoff scale $\Lambda$. This provides the additional regularization required to calculate $\beta$-functions for Yang-Mills theories in the manifestly gauge-invariant formalism at 1-loop and beyond. A generalization of this method for gravity using ``Parisi-Sourlas'' supermanifolds would seem to be an attractive choice of regularization when we construct the manifestly diffeomorphism-invariant ERG in the next chapter \cite{Morris:2016nda}.

\chapter{Manifestly diffeomorphism-invariant ERG}\label{ChapterMDIERG}

\section{Background-independent gravity flow equation}\label{MDIERGFlow}

This chapter reports on original research published in \cite{Morris:2016nda}, which began the construction of the manifestly diffeomorphism-invariant Exact Renormalization Group. Many studies of RG flow in gravity already exist, a few examples of which are \cite{Stelle:1976gc,Adler:1982ri,Reuter:1996cp,Litim:2003vp,Litim:2008tt,Reuter:2012id,Dietz:2012ic,Christiansen:2012rx,Falls:2013bv,Dietz:2016zzu}, but this approach has the advantage that there is no need to fix a gauge, which also offers the opportunity for a genuinely background-independent description. The demand for such a background-independent description is strong, since it avoids a great deal of complication and questions over the validity of results \cite{Reuter:2009kq,Becker:2014qya,Dietz:2015owa,Labus:2016lkh}. We begin with the construction of a suitable flow equation, inspired by the Polchinski flow equation (\ref{scalarSdot}) and its adaptation to gauge theories (\ref{gaugeflow}). As discussed in Section \ref{MGIERGKadanoff}, Kadanoff blocking requires a notion of locality that is not compatible with a metric that has a Lorentzian signature. This is resolved in gravity, as it is in scalar or gauge theories, by performing a Wick rotation into Euclidean signature. Thus the metric signature is completely positive after the transformation. The two remaining sign convention choices for GR (relating to the definitions of the Riemann and Ricci tensors) are as described in and below (\ref{RiemannTensorDef}). Studies of quantum gravity usually fix a background metric (typically a flat background) and study gravity via perturbations to that background, as outlined in Subsection \ref{IntroMetricPert}. We will return to the fixed-background formalism in Section \ref{MDIERGfixed}, but for now we will proceed to construct the diffeomorphism-invariant ERG background-independently. Background-independence is a huge advantage not only because of its elegance and generality, but also because it makes the geometrical meaning of the gravity theory manifest in a philosophically satisfying manner. From now on, we shall be specializing to four dimensional spacetime, although it will be convenient on occasion to comment on general dimensions.

In analogy to (\ref{Kblocking}), we begin by writing down the effective Boltzmann factor as a means to define how microscopic degrees of freedom are averaged to give the macroscopic degrees of freedom via a blocking procedure:
\begin{equation}\label{block}
e^{-S[g]} = \int\mathcal{D}g_{0}\ \delta\left[g-b\left[g_{0}\right]\right]e^{-S_{{\rm bare}}[g_{0}]},
\end{equation}
where Lorentz indices in functional arguments have been suppressed for notational convenience. The microscopic bare action $S_{\rm bare}$ is a functional of the microscopic bare metric $g_{0\mu\nu}$ and macroscopic effective action $S$ is a functional of the macroscopic renormalized metric $g_{\mu\nu}$. The renormalized metric is constructed from the blocked metric $b_{\mu\nu}$, which is a functional of the bare metric and a function of position, via integrating over the delta function in (\ref{block}) with respect to the bare metric. Following the reasoning used to derive (\ref{partinv}), we can note that the partition function is invariant under RG transformations, since the functional integral of (\ref{block}) with respect to the renormalized metric trivially integrates out the delta function on the right hand side of (\ref{block}), leaving behind the usual form of the partition function in terms of the bare metric and the bare action.

Following the step in (\ref{BoltzFlow}), we derive the flow equation by differentiating (\ref{block}) with respect to $\Lambda$ to get
\begin{equation}
 \Lambda\frac{\partial}{\partial\Lambda}e^{-S[g]}=-\int_{x} \frac{\delta}{\delta g_{\mu\nu}(x)}\int\mathcal{D}g_{0}\ \delta\left[g-b\left[g_{0}\right]\right]\Lambda\frac{\partial b_{\mu\nu}(x)}{\partial\Lambda}e^{-S_{{\rm bare}}[g_{0}]}.
\end{equation}
We then integrate out the bare metric and rewrite in terms of the ``rate of change of blocking functional'' $\Psi_{\mu\nu}(x)$, in analogy with (\ref{generalERG}), to get the general ERG flow equation for gravity:
\begin{equation}\label{Psimunu-flow}
 \Lambda\frac{\partial}{\partial\Lambda}e^{-S[g]}=\int_{x} \frac{\delta}{\delta g_{\mu\nu}(x)}\left(\Psi_{\mu\nu}(x)e^{-S[g]}\right).
\end{equation}
Applying the blocking functional to lower the cutoff scale in the exact RG flow is simply an exact reparametrization of the field in the effective action, as is the case in scalar and gauge theories. Because of this, there is a physical equivalence between effective actions at different scales. The correspondence is especially straightforward at the classical level, where the flow equation takes the form  
\begin{equation}
\dot{S} = \int_{x}\Psi_{\mu\nu}(x) \frac{\delta S}{\delta g_{\mu\nu}(x)}\,.
\end{equation}
Expressed explicitly it terms of the action's functional argument, we have
\begin{equation} 
\label{classical_equivalence}
S_{\Lambda-\delta\Lambda}[g_{\mu\nu}] = S_{\Lambda}[g_{\mu\nu}-\Psi_{\mu\nu}\,\delta\Lambda/\Lambda]\,.
\end{equation}
The full quantum flow equation is
\begin{equation}
\dot{S} = \int_{x}\Psi_{\mu\nu}(x) \frac{\delta S}{\delta g_{\mu\nu}(x)} - \int_x\frac{\delta\Psi_{\mu\nu}(x)}{\delta g_{\mu\nu}}\,,
\end{equation}
where the latter term comes from a reparametrization of the measure in (\ref{block}).

To obtain the form of the flow equation that is analogous to the Polchinski flow equation, we define $\Psi_{\mu\nu}(x)$ in a fashion similar to (\ref{Polchange}):
\begin{equation}
\label{Psimunu}
 \Psi_{\mu\nu}(x) = \frac{1}{2} \int_{y} K_{\mu\nu\rho\sigma}(x,y)\frac{\delta\Sigma}{\delta g_{\rho\sigma}(y)},
\end{equation}
where $K_{\mu\nu\rho\sigma}(x,y)$ is a covariant bitensor that we will refer to as the ``kernel'' of the flow equation from now on. The factor of $1/2$ in (\ref{Psimunu}), as in the scalar and gauge cases, is used to ensure that the graviton kinetic term is canonically normalized. The kernel is the object that inspired the comment at the end of Subsection \ref{IntroDiffSym} about the need for a generalized notion of the Lie derivative for two position arguments $x$ and $y$. This point will be returned to in Section \ref{MDIERGWard}. As in the scalar and gauge cases, the kernel performs the r\^{o}le of an effective propagator in the flow equation, although its form is more complicated, as we will see shortly. It is convenient to choose the index notation of $K_{\mu\nu\rho\sigma}(x,y)$ such that it is symmetric under the index label exchange $\mu\leftrightarrow\nu$ and also under $\rho\leftrightarrow\sigma$. The first pair of Lorentz indices are associated with the first position argument and the second pair with the second argument. Neglecting coupling rescaling for now, we can once again set $\Sigma = S - 2\hat{S}$, where $S$ and $\hat{S}$ are set to be the same at the 2-point level. The Polchinski-inspired flow equation for gravity is then 
\begin{equation}
\label{full-flow}
 \dot{S}=\frac{1}{2} \int_{x}\frac{\delta S}{\delta g_{\mu\nu}(x)}\int_{y}K_{\mu\nu\rho\sigma}(x,y)\frac{\delta\Sigma}{\delta g_{\rho\sigma}(y)}-\frac{1}{2}\int_{x}\frac{\delta}{\delta g_{\mu\nu}(x)}\int_{y}K_{\mu\nu\rho\sigma}(x,y)\frac{\delta\Sigma}{\delta g_{\rho\sigma}(y)}\,.
\end{equation}
As already noted for the scalar field equation (\ref{scalarSdot}) and the gauge flow equation (\ref{gaugeflow}), the first term on the right hand side contributes at all loop levels including the classical level, whereas the second term begins at the 1-loop level and contributes at all higher loop levels. 

Diffeomorphism covariance requires that the kernel carries a factor of $1/\sqrt{g}$ (note the lack of a minus sign in Euclidean signature). An easy way to see this is that both $S$ and $\Sigma$ have a factor of $\sqrt{g}$ in their measure, so the kernel must carry a $1/\sqrt{g}$ to balance. The factor of $1\sqrt{g}$ will also be seen to be necessary to satisfy the Ward identities for diffeomorphism covariance in fixed-background formalism, as seen in Section \ref{MDIERGWard}. The covariantized cutoff function itself is constructed in an analogous way to the gauge case, with $\dot{\Delta}$ being a function of $-\nabla^2$ and $-\nabla^2/\Lambda^2$. When we develop the fixed-background formalism in Section \ref{MDIERGfixed}, $\Delta$ will be seen to be the ``effective propagator'' in a similar way to the gauge theory case. However, the strictly background-independent description does not have a notion of gravitons to propagate, so there is no notion of momentum, $n$-point expansions, or a propagator in this sense. In the background-independent description, there is no notion of $\Delta$, even though $\KoL$ exists as part of the kernel. More immediately obvious is the need to incorporate index structure into the kernel. This is done by introducing two instances of the metric tensor to carry the four indices, symmetrized into two pairs, thus completing the bitensor construction. 

Unlike in the gauge theory case, there are two possible independent index structures for the flow equation.
Firstly, we have the ``cross-contracted'' index structure. In the classical part of the flow equation, this takes the form
\begin{equation}\label{c.c.}
\dot{S}|_{c.c.} = \frac{1}{2}\int_{x}\frac{\delta S}{\delta g_{\mu\nu}}\frac{g_{\mu(\rho}g_{\sigma)\nu}}{\sqrt{g}}\dot{\Delta}\cutarg \frac{\delta \Sigma}{\delta g_{\rho\sigma}}.
\end{equation} 
Next, we have the ``two-traces'' part of the classical flow equation:
\begin{equation}\label{t.t.}
\dot{S}|_{t.t.} = \frac{1}{2}\int_{x}\frac{\delta S}{\delta g_{\mu\nu}}\frac{g_{\mu\nu}g_{\rho\sigma}}{\sqrt{g}}\dot{\Delta}\cutarg \frac{\delta \Sigma}{\delta g_{\rho\sigma}}.
\end{equation} 
The flow equation is then a linear combination of these two contributions:
\begin{equation}\label{flowratio}
\dot{S} = \dot{S}|_{c.c.} + j\dot{S}|_{t.t.},
\end{equation}
where $j$ is a dimensionless parameter which, we will see, determines the balance of modes propagating in the flow equation. Put another way, it determines how the different metric modes are integrated out. Although it is found independently here, this parameter has also appeared in the DeWitt supermetric \cite{DeWitt:1967yk}, fulfilling the same r\^{o}le. Since an overall factor can be absorbed into the kernel, there is no need to place a coefficient on the cross-contracted term in (\ref{flowratio}). Putting all of the ingredients discussed together, we can finally write the kernel as the following covariant bitensor:
\begin{equation}
\label{kernel-grav}
 K_{\mu\nu\rho\sigma}(x,y) = \frac{1}{\sqrt{g}}\delta(x-y)\left(g_{\mu(\rho}g_{\sigma)\nu}+jg_{\mu\nu}g_{\rho\sigma}\right)\dot{\Delta}\cutarg.
\end{equation}
To see how the parameter $j$ affects the balance of tensor modes in the flow equation, consider extracting the conformal mode explicitly as an overall rescaling factor outside the metric:
\begin{equation}\label{conformalmetric}
 g_{\mu\nu} = \tilde{g}_{\mu\nu}e^{\sigma},
\end{equation}
where $\tilde{g}_{\mu\nu}$ is a metric with a fixed determinant and $\sigma$ is the conformal rescaling mode. We can then write the derivative of $S$ with respect to $\sigma$ as
\begin{equation}
\label{d-sigma}
 \frac{\delta S}{\delta \sigma} = g_{\mu\nu}\frac{\delta S}{\delta g_{\mu\nu}}.
\end{equation}
This is especially powerful, as we see that the ``two-traces'' structure corresponds to the case where the actions are only differentiated with respect to the conformal factor in the flow equation. Put another way, the ``two-traces'' structure, corresponding to the limit where $j\to\infty$ is actually the limit in which the conformal mode is the only mode propagating in the flow equation. Effective actions for gravity constructed from RG flows using only the conformal mode of the metric are referred to as ``conformally reduced'' gravity models, corresponding to a ``conformal truncation'' of the more complete effective theory for gravity \cite{Dietz:2015owa,Machado:2009ph,Reuter:2008qx,Reuter:2008wj,Bonanno:2012dg}.

There also exists a choice for $j$ that eliminates the conformal mode from the RG flow. Consider how a 2-component tensor $T^{\rho\sigma}$ in $D$ spacetime dimensions can be separated into pure trace and traceless parts:
\begin{equation} 
T^{\rho\sigma} = \underbrace{\frac{1}{D}g^{\rho\sigma} T}_{\rm pure\ trace} + \underbrace{T^{\rho\sigma} - \frac{1}{D}g^{\rho\sigma}T}_{\rm traceless}\,.
\end{equation}
Now let us take a trace over the index structure used in the kernel:
\begin{equation} 
\left(g_{\mu(\rho}g_{\sigma)\nu}+jg_{\mu\nu}g_{\sigma\rho}\right)g^{\rho\sigma} = g_{\mu\nu} (1+jD)\,.
\end{equation}
For the choice that $j=-1/D$, this trace goes to zero. This tells us that the pure trace mode of the variation of an action with respect to the metric does not propagate in the flow equation in the case where $j=-1/D$. This can be easily verified by inserting (\ref{conformalmetric}) into the flow equation for this choice and seeing the result goes to zero if $\sigma$ is the only mode being varied. For this reason, this choice of $j$ decouples the cosmological constant from the flow equation at the classical level, \ie any cosmological constant term in $S$ or $\hat{S}$ disappears entirely from $\dot{S}$, so it can only appear in the effective action as an integration constant with its own $\Lambda$-independent scale: this means that a cosmological constant cannot be introduced into a fixed-point action for this choice of $j$, which is related to unimodular gravity \cite{Unruh:1988in,Saltas:2014cta,Eichhorn:2015bna}. Though these special cases are interesting, they will not be considered further in this thesis. They do, however, illustrate how $j$ determines the balance of modes propagating in the flow equation.

The calculations that follow are sufficiently complicated to merit some condensing of the notation. The flow equation (\ref{full-flow}) will be written at the classical level as a symmetric bilinear mapping $a_0$ of the actions $S$ and $\Sigma$ onto the new differentiated action $\dot{S}$:
\begin{equation} 
\label{a0}
\dot{S} = -a_0[S,\Sigma]\,,
\end{equation}
where $a_0$ is a shorthand for the first term on the right hand side of the flow equation (\ref{full-flow}). To extend to quantum calculations, a linear map $a_1[\Sigma]$ can be defined as shorthand for the second term on the right hand side of (\ref{full-flow}). Sometimes it is also convenient to use similarly condensed notation at the level of the Lagrangian densities:
\begin{equation} 
\label{ca0}
\dot{\cL} = -\ca_0[\cL,\cL-2\hat{\cL}]\,,
\end{equation}
where $\ca_0$ performs a similar r\^{o}le to $a_0$, but formally must be defined separately as the bilinear mapping of two Langrangian densities onto one differentiated Lagrangian density. 

\section{Background-independent calculations}\label{MDIERGBIGeneral}

To begin exploring the flow equation (\ref{full-flow}) for gravity, one can begin by inserting a simple ansatz for $S$ and $\hat{S}$ into the flow equation and seeing what the resulting form of $\dot{S}$ is. In anticipation of constructing the $n$-point expansion in fixed-background formalism in a similar manner to the scalar and gauge cases, it would be wise to set $S=\hat{S}$ up to the quadratic order in the Riemann tensor, since all action terms at quadratic order or lower in the Riemann tensor give contributions to the 2-point function on a flat background, while the cubic order in the Riemann tensor only gives contributions at the 3-point level and above. In this section, we will be mainly concerned with building a fixed-point action, which has all couplings as dimensionless quantities multiplied by powers of $\Lambda$. However, it is also of physical interest to perturb away from the fixed-point action by introducing operators whose coefficients contain dimensionful parameters other than $\Lambda$ in Subsection \ref{MDIERGRunning}.

For a simple ansatz, such as $\cL\sim R\Lambda^2$, one finds that $\dot{S}$ contains new higher order contributions, implying that $S$ needs to be appended with extra higher-derivative operators to satisfy the flow equation. In fact, this na\"{i}ve choice of ansatz leads us into difficulties. First of all, one immediately finds that $\dot{S}$, using the right hand side of (\ref{a0}), lacks an Einstein-Hilbert term, meaning that the original Lagrangian term cannot be reproduced after integrating $\dot{S}$ with respect to RG time. The immediate remedy to this is to append $S$ to include a cosmological constant-like term $\sim\Lambda^4$, such that the original Einstein-Hilbert term reappears from $\ca_0\left[\Lambda^2 R, \Lambda^4\right]$. The new cosmological constant-like term reappears from $\ca_0\left[\Lambda^4,\Lambda^4\right]$. To gain the benefit of this, the $\dot{\Delta}$ derivative expansion must have a zero-derivative term (\ie a term simply proportional to $1/\Lambda^4$), else any $\ca_0\left[\Lambda^4,\cdots\right]$ would trivially vanish. This then raises a problem at the quadratic order in the Riemann tensor where $R^2$ and $R_{\mu\nu}R^{\mu\nu}$ terms appear in $\dot{\cL}$, which give rise to terms proportional to ${\rm ln}(\Lambda)$ when integrated with respect to RG time. A term proportional ln$(\Lambda)$ in the action would imply that the complete expansion contains logarithms of operators constructed out of derivative operators (\eg ln$(R)$ or ln$(\nabla^2)$) to balance, which would violate the locality required for Kadanoff blocking. This problem is resolved in the ``Einstein scheme'' discussed in Subsection \ref{MDIERGEinsteinBI} via a coupling rescaling of the action similar to that already discussed in the context of gauge theories. Before we get to that, we need to develop a systematic way to discuss the operator expansion in the effective action.

Let us denote the Lagrangian as an operator expansion using the following generalized notation:
\begin{equation} 
\label{lagrangian-expansion}
\cL = \sum_{i=0}^\infty \sum_{\alpha_i} g_{2i,\alpha_i}\, \cO_{2i,\alpha_i}\,,
\end{equation}
where $\cO_{2i,\alpha_i}$ is an operator constructed exclusively of the metric and its spacetime derivatives with mass dimension $2i$ and a unique identifying label $\alpha_i$ that distinguishes it from other operators with the same mass dimension, and $g_{2i,\alpha_i}$ is the corresponding coupling coefficient for that operator. Because the metric itself is dimensionless, the label $i$ simply counts the number of pairs of covariant derivative operators and Riemann tensors that make up the operator $\cO_{2i,\alpha_i}$. The implicit assertion that $2i$ is an even number is enforced by Lorentz invariance. For a Lagrangian density of mass dimension $l$, the couplings $g_{2i,\alpha_i}$ are of mass dimension $l-2i$.

In order not to fall foul of locality problems, we must insist that the lowest-order operator has mass dimension $2i=0$, for which there is only one candidate: $\cO_0=1$. The corresponding Lagrangian term is a cosmological constant-like term, however caution is required in its physical interpretation. First of all, if this operator is constructed in the most na\"{i}ve way, \ie $\sim\Lambda^4$, then the complete integrating out of all energy modes $\Lambda\to 0$ takes it to zero, illustrating that it is, in fact, not a physical cosmological constant but rather an unphysical feature of the blocking scheme. Even if one sets the coefficient to carry is own independent mass dimension, one should be wary that further quantum corrections will contribute to the physically observable cosmological constant, which in general would be expected to differ from the value initially written down at the classical level. Thus the coupling $g_0(\Lambda)$ should be viewed as a running parameter, with the actual cosmological constant $\lambda_C$ becoming apparent in the infrared limit $\Lambda\to0$ as $\lambda_C = 8\pi G g_0(0)$, where $G$ is the physically observed value of Newton's constant.

Newton's constant itself is related to the coefficient of the next lowest-order operator, the Ricci scalar, which has $2i=2$. The Ricci scalar is the only diffeomorphism-invariant dimension 2 Lagrangian operator in a pure gravity theory. The Einstein-Hilbert term $\cL\sim R$ is the only curvature term in the action that we have direct experimental evidence of, featuring in the usual action for Einstein gravity given in (\ref{EinsteinGravAct}). It is troublesome then that this term is especially problematic for RG on account of its the positive mass dimension of its coefficient, as already touched on at the beginning of this section. This thesis will discuss two renormalization schemes for gravity at the classical level. These are the ``Weyl scheme'' introduced in Subsection \ref{MDIERGWeylBI} and the ``Einstein scheme'' in Subsection \ref{MDIERGEinsteinBI}. In the Weyl scheme, as we will see, the Einstein-Hilbert term does not appear at the classical level, but rather it would emerge from quantum corrections to the action in the complete quantum gravity description. In the Einstein scheme, it is introduced into the classical action with its coupling rescaled so as to canonically normalize graviton kinetic term, rendering its classical Lagrangian density a dimension 2 operator in any $D$-dimensional space. In any scheme, the physically observed Newton's constant comes from evaluating the running coefficient of the Einstein-Hilbert term in the $\Lambda\to0$ limit of the full quantum theory.

At $2i=4$, we obtain two new independent diffeomorphism-invariant operators that are both quadratic in the Riemann tensor, they are $R^2$ and $R_{\mu\nu}R^{\mu\nu}$. Both of these operators are of interest to cosmology as candidates for inducing Starobinsky inflation. The $R^2$ term has already been mentioned in Subsection \ref{IntroInflation} but, in the specific case of the FLRW metric in (\ref{FLRWmetric}), the field equations for $R_{\mu\nu}R^{\mu\nu}$ match those of $R^2/3$. There also exists an operator of the form $R_{\mu\nu\rho\sigma}R^{\mu\nu\rho\sigma}$, however this is related to the other two in four dimensions via the Gauss-Bonnet topological invariant:
\begin{equation}\label{Gauss-Bonnet}
 E = \frac{1}{32\pi^2}\int d^4x \sqrt{-g}\left(R_{\alpha\beta\gamma\delta}R^{\alpha\beta\gamma\delta} -4 R_{\alpha\beta}R^{\alpha\beta} + R^2\right).
\end{equation}
In four dimensional spacetime, the functional derivative with respect to the metric of the Gauss-Bonnet term is zero. It is a topological invariant in the sense that it simply counts the Euler characteristic of the manifold. Since we are concerned with spacetime without boundaries, this term offers no contributions and can be safely treated as zero. Thus we can choose to eliminate one of the three curvature squared terms in (\ref{Gauss-Bonnet}) from our action, and it is convenient to choose to remove the $R_{\mu\nu\rho\sigma}R^{\mu\nu\rho\sigma}$ term. If we progress to $2i=6$, not only do we encounter terms that are cubic in the Riemann tensor, but we also find more terms that are quadratic in the Riemann tensor but possess explicit covariant derivatives in the Lagrangian. An example of this is the $R\nabla^2 R$ term. Explicit covariant derivatives did not appear at lower orders because they could always be removed from the action via integration by parts, again supposing that we are not concerned with a spacetime boundary.

In keeping with our locality requirement, we define $\dot{\Delta}$ to be a Taylor series expansion in squared covariant derivative operators, starting with a constant term:
\begin{equation} 
\label{kernel-taylor}
\dot{\Delta}(-\nabla^2) = \sum^\infty_{k=0} \frac{1}{k!}\, \dot{\Delta}^{(k)}(0)\left(-{\nabla^2}\right)^k\,,
\end{equation}
where factors of $\Lambda^2$ are hidden in $\dot{\Delta}^{(k)}(0)$. The dimension of $\dot{\Delta}$ is always minus the dimension of the (rescaled) classical Lagrangian density, as implied by the structure of the flow equation, which can be easily seen in (\ref{c.c.}) and (\ref{t.t.}). With this in mind, we can see that inserting any two Lagrangian operators into the bilinear $\ca_0$ results in a series expansion in higher-order operators, starting at a dimension that is the sum of the dimensions of the two input operators: 
\begin{equation} 
\label{ca0k}
\ca_0[\cO_{d_1},\cO_{d_2}] = \sum^\infty_{k=0} 
 \ca^k_0[\cO_{d_1},\cO_{d_2}]\,,
\end{equation}
where $\ca^k_0[\cO_{d_1},\cO_{d_2}]$ is the operator obtained by taking the $k$th order part of the kernel expansion in (\ref{kernel-taylor}), which then has dimension $d = d_1 + d_2 + 2k$. Since we require the kernel to be local, $k$ is always either positive or zero. Since we require that our action and seed action are also local, all $d_{j}$ are also either positive or zero, and $d$ is greater than or equal to $d_1$, $d_2$ and $2k$. Thus, starting from a given ansatz (other than a pure cosmological constant) for the effective action and seed action, the flow equation requires us to include higher-order contributions to give a self-similar flow, but it does not require us to include any lower-order corrections, \ie a coupling $g_{2i}$ corresponding to an operator of dimension $2i$ can appear in the flow of another coupling $g_{2(i+j)}$, where $j>0$, but not {\it vice versa}. This allows us to solve the flow equation for lower dimension operators and iteratively progress to solving up to higher orders. In practice, we can solve the flow equation up to a given power of the Riemann tensor, which is even more useful \eg for use in Chapter \ref{ChapterBack2}. If the effective action has a cosmological constant-like term, couplings can appear in their own flow (\ie their own RG time derivative), but since that is just a first order differential equation, this presents no problems.

To illustrate this point, let us consider the classical flow of the cosmological constant-like piece:
\begin{equation} 
\label{g0-flow}
\dot{g}_0=g_0(2\hat{g}_0-g_0)\, \ca^0_0[1,1]\,,
\end{equation}
where $g_0$ is the coefficient that appears in the effective action and $\hat{g}_0$ is the coefficient that appears in the seed action. If we insist that the 2-point functions of $S$ and $\hat{S}$ should match in the fixed-background formalism, then $g_0$ should equal $\hat{g}_0$ also, supposing that the background is fixed to be flat (although this is arguably not the most natural choice for $\Lambda\neq 0$). This is straightforwardly solved since $\ca_0^0[1,1]$ is simply a number times a power in $\Lambda$ (for a fixed-point action, or a polynomial more generally). Proceeding iteratively, $g_2(\Lambda)$ can be solved at the classical level using the form of $g_0(\Lambda)$ found by solving (\ref{g0-flow}). The flow equation for $g_2$ is  
\begin{equation} 
\label{g2-flow}
\dot{g}_2 = 2 (g_0 \hat{g}_2 +\hat{g}_0 g_2 -g_0 g_2)\, \frac{\ca^0_0[\cO_2,1]}{\cO_2} \,,
\end{equation}
where $\cO_2$ is the operator for the Einstein-Hilbert Lagrangian. Since $\ca^0_0[\cO_2,1]$ is proportional to $\cO_2$, this is also straightforward to solve. There is no dimension 2 contribution from $\ca^1_0[1,1]$ because $\nabla_\alpha g_{\beta\gamma}=0$ and $\nabla_\alpha g_0 = 0$. Indeed, if the cosmological constant-like term is used in either argument of $\ca_0$, then we can see, using integration by parts, that only the zeroth order of its expansion can be non-zero. More explicitly,
\begin{equation} 
\label{k-1}
\ca^k_0[\cO_d,1] =0 \qquad\forall k>0\,.
\end{equation}
In fact, flow equation terms involving a cosmological constant-like term can be very simply evaluated at the classical level in four dimensions via
\begin{equation} 
\label{0-1}
\ca_0[\cO_d,1]=\ca^0_0[\cO_d,1] = \frac{1}{8}(d-4)(1+4j)\dot{\Delta}(0)\,\cO_d\,,
\end{equation}
where we can see explicitly how $j=-1/D$ (where $D=4$ here) decouples the cosmological constant from the flow equation. Alternatively, the structure of the flow equation (\ref{full-flow}) and its kernel (\ref{kernel-grav}), expressed in the language of the bilinear (\ref{a0}), gives us the classical flow for cross-terms with the cosmological constant:
\begin{equation}\label{cosconstcrossterm}
a_0\left[S,\int_x\!\! \sqrt{g}\,\right] = -\frac{1}{4} (1+4j)\dot{\Delta}(0)\int_x\!\! g_{\mu\nu} \frac{\delta S}{\delta g_{\mu\nu}}\,.
\end{equation}
The final factor in this expression, as seen in (\ref{d-sigma}), counts the power in the conformal mode defined in (\ref{conformalmetric}), \ie it counts the power in $g_{\mu\nu}$ of the action term. This metric power counting starts at two because of the $\sqrt{g}$ factor in the measure and decreases by one for each pair of covariant derivative operators or for each Riemann tensor. Thus the metric power count drops by one for every increase by two of the operator dimension. This is because the corresponding Lorentz indices are contracted to form a scalar by an inverse metric, carrying a conformal factor power of minus one. This power counting is the origin of the $-(d-4)/2$ factor in (\ref{0-1}).

\subsection{Effective action in Weyl scheme}\label{MDIERGWeylBI}

As discussed at the beginning of Section \ref{MDIERGBIGeneral}, starting the expansion of the effective action at the classical level with an Einstein-Hilbert term, or even an Einstein-Hilbert plus cosmological constant term, raises problems with locality. The first of two approaches discussed in this thesis that resolve this will be referred to as the ``Weyl scheme''. In the Weyl scheme, the Einstein-Hilbert term is not included in the classical action, instead it is deferred to the quantum corrections. The effective action in Weyl scheme begins its expansion in powers of the Riemann tensor with the dimension four operators, $R^2$ and $R_{\mu\nu}R^{\mu\nu}$. 
The flow equation at this level can be written as
\begin{equation} 
\label{4-flow}
\dot{g}_{4,1} R^2 + \dot{g}_{4,2} R^{\mu\nu}R_{\mu\nu} = 4g^2_2 \ca^0_0[R,R]  +2g_0g_{4,1}\ca^0_0[R^2,1] +2g_0g_{4,2}\ca^0_0[R^2_{\mu\nu},1]\,,
\end{equation}
where $g_{4,1}$ is the coupling for the $R^2$ term, $g_{4,2}$ is the coupling for the $R_{\mu\nu}R^{\mu\nu}$ and $g_0$ is the coupling for a cosmological constant-like term, as before.

Na\"{i}vely, one might imagine that introducing a cosmological constant-like term together with curvature-squared terms would introduce locality problems similar to those discussed for the Einstein-Hilbert term. One might think this because a cross-term like $\ca_0[R^2,1]$ would na\"{i}vely be expected to contribute dimension four operators to $\dot{S}$, which would be integrated with respect to RG time to appear as corrections to $S$ that are proportional to ln$(\Lambda)$, implying that ln$(R)$ operators or similar exist elsewhere in the complete construction. To see why this is not true, let us first consult (\ref{cosconstcrossterm}) to see that $\ca_0[S,1]$ is proportional to the trace of the field equation for $S$. The field equations for $R_{\mu\nu}R^{\mu\nu}$ and $R^2$ are
\begin{eqnarray}\label{derivRmnRmn}
 \frac{\delta}{\delta g_{\mu\nu}}\int_{x}\sqrt{g}R_{\alpha\beta}R^{\alpha\beta} & = & \sqrt{g}\left(\frac{1}{2}g^{\mu\nu}R_{\alpha\beta}R^{\alpha\beta}-2R^{\mu}_{\ \alpha}R^{\nu\alpha} \right.\nonumber \\ 
 & &\left. -\nabla^2 R^{\mu\nu}-\frac{1}{2}g^{\mu\nu}\nabla^2 R+2\nabla_\alpha\nabla^{(\mu} R^{\nu)\alpha}\right),
\end{eqnarray}
and
\begin{equation}\label{derivR2}
 \frac{\delta}{\delta g_{\mu\nu}}\int_{x}\sqrt{g}R^2 = \sqrt{g}\left(\frac{1}{2}g^{\mu\nu}R^2 - 2RR^{\mu\nu}+2\nabla^{\mu}\nabla^{\nu}R - 2g^{\mu\nu}\nabla^2 R\right).
\end{equation}
Note that these field equations have been derived by differentiating with respect to the metric rather than the inverse metric. They have a minus sign factor different to the na\"{i}ve form expected by differentiating with respect to the inverse metric and raising the indices, as one can see from (\ref{raiselowerminus}).
Both of these field equations have traces proportional to $\nabla^2 R$. This operator cannot be fed back into the Lagrangian density to produce a new action term because it is a total covariant derivative, so it vanishes under integration by parts. Thus we can safely include both a cosmological constant-like term and curvature-squared terms in the effective action without introducing any dimension four operators into $\dot{S}$ that would cause non-locality in $S$. 

When constructing the classical fixed-point action in Weyl scheme, we can choose whether or not to include a cosmological constant-like term $\sim\Lambda^4$. Since, as already discussed, the physical cosmological constant is taken from the $\Lambda\to0$ limit, there is no loss of physics by not including the $\Lambda^4$ term. A cosmological constant can still reappear in the quantum corrections or be inserted into the classical effective action as a relevant perturbation away from the fixed-point. As its appearance in this context is an unnecessary complicating feature, there is insufficient motivation to include it in this construction, so we will proceed without it.

Following the discussion of scalar and gauge theories, for example in Section \ref{MGIERGloopwise} or equations (\ref{rescaleactseries}) and (\ref{rescalebeta}), the quantum construction of the Weyl scheme would also follow a loopwise expansion in powers of an overall (asymptotically free) coupling that is rescaled outside the action. Since we are, for now, developing the classical fixed-point action, this is not an immediate concern, but it would follow the same reasoning as the scalar and gauge cases. As with the scalar and gauge cases, we wish to canonically normalize the kinetic term (in fixed-background formalism), with the result that $g_{4,2}=2$.

Let us begin our iterative exploration of the classical fixed-point action by writing the seed action at its lowest order to be
\begin{equation} \label{startaction}
\hat{S} = 2\!\int_{x} \sqrt{g}\left(R_{\mu\nu}R^{\mu\nu}+sR^2 +\cdots\right)\,,
\end{equation}
where $s$ is a dimensionless parameter that we leave general. In previous literature, it has been seen to take a fixed-point value close to, but not equal, to $-1/3$ in the (asymptotically free) UV limit $\Lambda\to\infty$ \cite{Avramidi:1985ki,deBerredoPeixoto:2004if,Codello:2006in,Codello:2008vh}. More precisely, the literature gives $s=-\left(1+\omega_{*}\right)/3$ where $\omega_{*}\approx-0.0228$. In the scalar and gauge cases, we saw that it was useful to set a form for $\hat{S}$ up to the 2-point level to be equal to the form of $S$ at the 2-point level. In the gravity case, this corresponds to setting a form for $\hat{S}$ up to quadratic order in the Riemann tensor, which will set the form of $S$ up to quadratic order and allow us to proceed iteratively from there to determine the form of $S$ at higher orders in the Riemann tensor. Just as 3-point and higher contributions to $\hat{S}$ in the scalar and gauge cases led to reparametrizations of the field with no new physics content, adding higher-curvature contributions to $\hat{S}$ in the gravity case does similar. We set the seed Lagrangian to its simplest form:
\begin{equation} 
\label{Shat-4}
\hat{\cL} = 2R_{\alpha\beta}\,c^{-1}\!(-\nabla^2/\Lambda^2)\, R^{\alpha\beta} + 2sR\, c^{-1}\!(-\nabla^2/\Lambda^2)\, R\,,
\end{equation}
where $c$ is the smooth ultraviolet cutoff function that will be seen in the fixed-background formalism to be the momentum cutoff $c(p^2/\Lambda^2)$ that regulates the propagator to $c/p^4$. Thus, in the momentum representation of the fixed-background formalism,
\begin{equation}
 \label{dotD-Weyl}
\dot{\Delta}(p^2) = -\frac{2}{\Lambda^2 p^2} \,c'(p^2/\Lambda^2)\,.
\end{equation}
Relating this back to the position representation, and to the background-independent formalism in particular, note that locality requires $\KoL$ to follow a covariant derivative operator expansion that starts at the zeroth order, thus $c'(0)=0$. When we have the form of $\dot{S}$, we would integrate this with respect to RG time to get $S$. Since we require that locality is preserved, we choose the integration constant (trivially) such that
\begin{equation}
 \int \frac{d\Lambda}{\Lambda} \dot{\Delta}(p^2) = \frac{c(p^2/\Lambda^2)-1}{p^4} = \frac{1}{p^4} \sum^\infty_{k=2} \frac{c^{(k)}(0)}{k!} \pcutarg^k,
\end{equation}
where $c(0)=1$, ensuring that we have no troublesome $1/p^4$ factors in $S$. Background independently, $\KoL$ can be expanded as
\begin{equation}\label{BIWSKE}
 \dot{\Delta}\left(-\nabla^2\right) = -\frac{2}{\Lambda^4}\,c''(0) + \frac{1}{\Lambda^6}\,c'''(0) \nabla^2+O(\nabla^4)\,.
\end{equation}
Integrating this with respect to RG time, we get
\begin{equation}
 \label{dDelta-replace}
\int \frac{d\Lambda}{\Lambda} \dot{\Delta} = \frac{1}{2\Lambda^4}\,c''(0) -\frac{1}{6\Lambda^6}\,c'''(0) \nabla^2+O(\nabla^4)\,.
\end{equation}

Although we keep the seed Lagrangian to the simple form in (\ref{Shat-4}), the effective Lagrangian $\cL$ will match this at the quadratic order and then continue to include an infinite series of higher-curvature operators.
The flow equation for the effective action can now be expressed as
\begin{equation} 
\label{R4-flow}
\dot{\mathcal{L}} = 4\,\ca_0[R_{\mu\nu}R^{\mu\nu},R_{\alpha\beta}R^{\alpha\beta}] +8s\,\ca_0[R_{\mu\nu}R^{\mu\nu},R^2]+4s^2\ca_0[R^2,R^2]+\cdots\,,
\end{equation}
where the ellipsis stands for contributions to $\dot{\mathcal{L}}$ originating from operators of dimension eight and beyond. 

There are no dimension six operators in $\dot{\mathcal{L}}$ because only operators of dimension $\ge4$ exist in $\mathcal{L}$ to appear as arguments in $\ca$. Following the reasoning discussed in and below (\ref{ca0k}), the series continues with operators of all even-numbered dimensions from eight onwards. Furthermore, (\ref{R4-flow}) already allows us uniquely determine the couplings of all operators at dimension 10. A dimension 12 operator can be constructed by pairing a dimension 8 operator with a dimension 4 operator in $\ca_0$, therefore the dimension 8 operator couplings contribute to the dimension 12 operator coupling flow and so must be determined first. We will not discuss the dimension 12 couplings further, but rather focus on the dimension 8 and 10 operator couplings.
More explicitly, using the (second) Bianchi identity in (\ref{SpecialSecondBianchi}), we find that
\begin{eqnarray}
\label{RR-RR}
2\ca_0[R_{\mu\nu}R^{\mu\nu},R_{\alpha\beta}R^{\alpha\beta}] & = & 
R_{\alpha\beta}R^{\alpha\beta}\KoL R_{\gamma\delta}R^{\gamma\delta}
-4R_{\alpha\beta}R^{\alpha}_{\ \gamma}\KoL R^{\gamma\delta}R^{\beta}_{\ \delta}
-4R_{\alpha\beta}R^{\alpha}_{\ \gamma}\KoL\nabla^2 R^{\beta\gamma}
\nonumber \\ & & 
+8R_{\alpha\beta}R^{\alpha}_{\ \gamma}\KoL\nabla_{\delta}\nabla^{\beta}R^{\gamma\delta}
-\nabla^2 R_{\alpha\beta}\KoL\nabla^2 R^{\alpha\beta} -\nabla^2 R\KoL\nabla^2 R  
\nonumber \\ & &
+4\nabla^2 R_{\alpha\beta}\KoL\nabla_{\gamma}\nabla^{\alpha}R^{\beta\gamma} - 4\nabla_{\alpha}\nabla_{(\beta}R_{\gamma)}^{\ \ \alpha}\KoL\nabla_{\delta}\nabla^{\beta}R^{\gamma\delta} \nonumber \\ &&
-4j\nabla^2 R\KoL\nabla^2 R\,, \\
\label{R2-R2}
2\ca_0[R^2,R^2]
& = & R^2\KoL R^2 - 2R^2\KoL \nabla^2 R - 4RR_{\alpha\beta}\KoL R^{\alpha\beta}R + 8RR^{\alpha\beta}\KoL \nabla_{\alpha}\nabla_{\beta}R 
\nonumber \\ & &
- 4\nabla_{\alpha}\nabla_{\beta}R\KoL\nabla^{\alpha}\nabla^{\beta}R - 8\nabla^2 R\KoL \nabla^{2}R \nonumber \\ &&
-36j\nabla^2 R \KoL \nabla^2 R\,,
\end{eqnarray}
and the cross-term is
\begin{eqnarray}
\label{RR-R2}
2\ca_0[R_{\mu\nu}R^{\mu\nu},R^2]
& = & 
R^2\KoL R_{\alpha\beta}R^{\alpha\beta}-4RR_{\alpha\beta}\KoL R^{\alpha}_{\ \gamma}R^{\beta\gamma} - 2RR_{\alpha\beta}\KoL \nabla^2 R^{\alpha\beta}
\nonumber \\ & &
+4RR^{\alpha\beta}\KoL\nabla_\gamma\nabla_\alpha R^{\gamma}_{\ \beta} -\nabla^2 R\KoL R_{\alpha\beta}R^{\alpha\beta}+4\nabla_{\alpha}\nabla_{\beta}R\KoL R^{\alpha\gamma}R^{\beta}_{\ \gamma}\nonumber \\ & &
+2\nabla_{\alpha}\nabla_{\beta}R\KoL\nabla^{2}R^{\alpha\beta}
-4\nabla^\alpha \nabla^\beta R\KoL\nabla_\gamma \nabla_\alpha R^{\gamma}_{\ \beta} - 3\nabla^2 R\KoL \nabla^2 R
\nonumber \\ & &
-12j
\nabla^2 R\KoL\nabla^2 R\,. 
\end{eqnarray}

Applying (\ref{dDelta-replace}) to these equations, we find that $\mathcal{L}$ has the following dimension 8 and 10 contributions at the quadratic level in the Riemann tensor:
\begin{eqnarray}
 \label{8-10-ops}
&&-\left\{ 1+4j+4s(2+3s)(1+3j) \right\} \left[ 
\frac{1}{\Lambda^4}\,c''(0) R \left(-\nabla^2\right)^2 R 
+\frac{1}{3\Lambda^6}\,c'''(0) R \left(-\nabla^2\right)^3 R \right]\nonumber\\
&&\qquad-\frac{1}{\Lambda^4}\,c''(0) R_{\mu\nu} \left(-\nabla^2\right)^2 R^{\mu\nu}
-\frac{1}{3\Lambda^6}\,c'''(0) R_{\mu\nu} \left(-\nabla^2\right)^3 R^{\mu\nu}\,.
\end{eqnarray}
The contributions on the first line are taken from all of the equations (\ref{RR-RR}), (\ref{R2-R2}) and (\ref{RR-R2}), whereas the second line has exclusively been supplied from (\ref{RR-RR}). Recall that canonical normalization of the graviton kinetic term in fixed-background formalism requires $c(0)=1$ and that locality demands $c'(0)=0$. To ensure that the $\mathcal{L}$ terms in (\ref{8-10-ops}) match the form in (\ref{Shat-4}), the $R_{\mu\nu}c^{-1}R^{\mu\nu}$ part immediately requires that
\begin{equation}
 c^{-1}\left(-\nabla^2/\Lambda^2\right) = 1 -\frac{1}{2\Lambda^4}\,c''(0) \left(-\nabla^2\right)^2 -\frac{1}{6\Lambda^6}\,c'''(0)\left(-\nabla^2\right)^3+O(\nabla^8)\,,
\end{equation}
which matches the form implied by the expansion in (\ref{dDelta-replace}). To make the $Rc^{-1}R$ part of (\ref{8-10-ops}) match (\ref{Shat-4}), we need also to set the condition that
\begin{equation} \label{jconstraint}
1+4j+4s(2+3s)(1+3j) = s\,.
\end{equation}
which is a constraint on the value of $j$ in terms of $s$. Leaving $s$ general, $j$ is determined to be
\begin{equation}
 \label{j-Weyl}
j = -\frac{1}{4}\, \frac{1+4s}{1+3s}\,.
\end{equation}
Alternatively, $s=-1/3$ is a solution (at the classical level) which allows $j$ to be set to any arbitrary value. The remaining dimension 8 and 10 operators that we have not discussed are of cubic or quartic order in the Riemann tensor. With the working performed in this section, we have solved the form of the effective action up to the quadratic order in the Riemann tensor, which is especially useful for comparing to the 2-point level of the fixed-background formalism, given in Subsection \ref{MDIERG2ptW}. The 2-point level of the fixed-background formalism also determines the form of the propagator. Iteratively solving to the $n$-th order in the Riemann tensor corresponds in the fixed-background formalism to iteratively solving the action to the $n$-point level.

\subsection{Running away from the fixed point with dimensionful couplings}\label{MDIERGRunning}

A fixed-point action has $\Lambda$ as its only scale. If we include additional operators with coefficients carrying their own length scale, these operators will perturb the action away from the fixed point, even at the classical level. Note that $\Lambda$-independent terms are at liberty to appear in the effective action as integration constants in $\Lambda$. At the classical level, operators that have $\Lambda$-independent coefficients with positive mass dimension are relevant, since the ratio of the coefficient to the corresponding power of $\Lambda$ grows as $\Lambda$ is decreased (as high-energy modes are integrated out). There are two such diffeomorphism-invariant operators, they are the cosmological constant-like part $\mathcal{O}_0=1$ and the Einstein-Hilbert term $\mathcal{O}_2=-2R$, where the minus sign is because of the rotation to Euclidean signature and, together with the two, results in a canonically normalized kinetic term in the fixed-background formalism. The general flow equations for these operators were already noted in (\ref{g0-flow}) and (\ref{g2-flow}). Specializing to the Weyl scheme, in which the fixed-point and seed action couplings for these operators are zero, we simply have
\begin{equation}
\label{relevant-dirs}
\dot{g}_0 = \alpha\,{g_0^2}/{\Lambda^4}\,,\qquad 
\dot{g}_2 = \alpha\,{g_0g_2}/{\Lambda^4},
\end{equation}
where $\alpha$ is a constant determined by the form of the kernel. Using (\ref{0-1}), knowing the value of $j$ in Weyl scheme (\ref{j-Weyl}) and the value of $\dot{\Delta}(0)$ from (\ref{BIWSKE}), we find that $\alpha=sc''(0)/(1+3s)$. To see that these parameters are relevant, note that the respective dimensionless couplings $\tilde{g}_0 = g_0/\Lambda^4$ and $\tilde{g}_2 = g_2/\Lambda^2$ become large as $\Lambda\to 0$, which is the limit from which we find the physical values of the cosmological constant and Newton's constant.

A nice feature of setting $S=\hat{S}$ for the fixed-point parts of the action, up to quadratic order in the Riemann tensor at least, is that we have $\Sigma=-S$ at this level. Thus, when we include these relevant perturbations with their own length scale, we do not get cross-terms in the classical flow equation, \ie terms in $\dot{S}$ originating from a bilinear of both fixed-point and perturbation terms, at least not up to quadratic order in the Riemann tensor. Noting that the perturbation terms do not appear in $\hat{S}$, the result of this is that the cross-terms terms cancel at all orders where $S=\hat{S}$ is true at the fixed point. To see this explicitly, let $\cL_{\rm total}$ be the complete Lagrangian, $\cL$ be the fixed-point part and $\Delta\cL$ be the perturbation. If $\cL=\hat{\cL}$, then
\begin{eqnarray}\label{nocrossterms}
 \dot{\cL}_{\rm total} = \Lambda\partial_{\Lambda}\left(\cL + \Delta\cL\right) &=& -\ca_0[\cL+\Delta\cL,\cL+\Delta\cL-2\hat{\cL}] \nonumber\\
&=& \ca_0[\cL +\Delta\cL, -\cL + \Delta\cL] \nonumber \\
&=& \ca_0[\cL,\cL]-\ca_0[\Delta\cL,\Delta\cL]\,,
\end{eqnarray}
If $\cL\neq \hat{\cL}$ at some order in the Riemann tensor, \eg cubic, then, in the presence of an Einstein-Hilbert or cosmological constant-like term in $\Delta\cL$, cross-terms appear at that order and higher. At the fixed point, all dimensionless couplings $g_{4,\alpha}$ are constant in $\Lambda$ at the classical level, so $\dot{g}_{4,\alpha}=0$ at the fixed-point. Perturbing away from the fixed point, we obtain an RG flow for $g_{4,\alpha}$ at the classical level from the perturbation terms, and not from any cross-terms, as already discussed. Using (\ref{4-flow}), we find this perturbation-induced flow to be
\begin{equation}
 \label{D4-flow}
\dot{g}_{4,1} R^2 + \dot{g}_{4,2} R^{\mu\nu}R_{\mu\nu} = -4g^2_2 \ca^0_0[R,R] = 2\dot{\Delta}(0)\, g_2^2\left(R_{\mu\nu}R^{\mu\nu}+jR^2\right)\,,
\end{equation}
noting again from (\ref{BIWSKE}) that $\KoL(0) = -2c''(0)/\Lambda^4$. As with the fixed-point expansion, the coupling flow from the perturbation terms can be solved for iteratively. Starting from (\ref{relevant-dirs}), it is easy to solve for $g_0$, and use that solution to get $g_2$. Using (\ref{D4-flow}), the solutions to $g_0$ and $g_2$ can be used to find solutions to $g_{4,\alpha}$. One can continue to follow this logic to obtain the couplings for higher-dimension operators, proceeding iteratively.

When building the fixed-point action, we placed a constraint on the coefficient of the $R_{\mu\nu}R^{\mu\nu}$ to achieve a canonically normalized kinetic term. However, the relevant perturbations from $\mathcal{O}_0$ and $\mathcal{O}_2$ provide us with a running in $g_{4,\alpha}$. As already discussed, we would rescale an overall coupling out of the action in the manner discussed in Section \ref{MGIERGloopwise}, so the classical running of $g_{4,\alpha}$ would also be accounted for outside of the rescaled classical action. The inclusion of the Einstein-Hilbert perturbation would also modify the graviton propagator by introducing a $p^2$ contribution, although we will not develop the complete description here.

\subsection{Effective action in Einstein scheme}\label{MDIERGEinsteinBI}

At the beginning of Section \ref{MDIERGBIGeneral}, we discussed how introducing an Einstein-Hilbert term to the fixed-point action in the na\"{i}ve way runs into locality problems at the curvature-squared level. The Weyl scheme, developed in Subsection \ref{MDIERGWeylBI}, resolved this matter by banishing the Einstein-Hilbert term from the fixed-point action, recovering it at the classical level as a perturbation term in Subsection \ref{MDIERGRunning}. In this subsection, we will instead develop the Einstein scheme, in which the Einstein-Hilbert term appears in the fixed-point action, but the locality issues are resolved by rescaling out the Newton's constant-like coefficient, including its mass dimension. This leaves the effective Lagrangian density, at the classical level, as a dimension 2 operator, given a 4-dimensional spacetime. Rescaling out the Newton's constant-like parameter from the classical action also ensures that it does not appear in the kernel (or the effective propagator in fixed-background formalism). The complete action then has a loopwise expansion, similar to that which appears in gauge theories (\ref{rescaleactseries}) or scalar theories (\ref{scalarloopaction}):
\begin{equation}
\label{mass-expansion}
 S = \frac{1}{\tilde\kappa} S_{0} + S_{1} + \tilde\kappa S_{2} + \tilde\kappa^2 S_{3}+\cdots,
\end{equation}
where $\tilde{\kappa}=32\pi G$, remembering that $\mathcal{O}_2=-2R$. Thus the effective action is seen as a loopwise expansion where higher powers of $\kappa$ correspond to higher loop orders, as is also seen in \cite{Donoghue:1994dn}. The physically measured value of Newton's constant would emerge in the $\Lambda\to0$ limit, whereas the Newton's constant used in this expansion is still a function of $\Lambda$. The expansion in $\tilde\kappa$ is also an expansion in powers of $\hbar$, although this is left implicit by the choice of natural units. Expressed as an expansion in the reduced Planck mass, we have $\tilde\kappa = 4/M_{\rm Planck}^2$. Since $\tilde\kappa$ is a mass dimension $-2$ parameter, the rescaled actions $S_{i}$ at each $i$-loop level have a mass dimension of $2i-2$, thus ensuring that $S$ itself is dimensionless. As we saw with gauge theories in (\ref{Sigma-g}), we also need to rescale $\Sigma$ for use in the flow equation to $\Sigma = \tilde\kappa - 2\hat{S}$. Therefore both $\Sigma$ and $\hat{S}$ are mass dimension $-2$ actions. To consider the complete flow equation in the spirit of Section \ref{MGIERGloopwise} and equations (\ref{scalarloop0}) and (\ref{scalarloop1}) in particular, let us note that the complete flow equation, given by (\ref{full-flow}) with (\ref{kernel-grav}), can be expressed as
\begin{equation}
 \dot{S} = -a_0[S,\Sigma] +a_1[\Sigma]\,,
\end{equation}
where $a_1[\Sigma]$ is the second term on the right hand side of (\ref{full-flow}), which begins to contribute at the 1-loop level. Now the expansion of the flow equation can be elegantly given by
\begin{eqnarray}\label{gravloopexp}
 &&\frac{1}{\tilde\kappa} \dot{S}_{0} + \dot{S}_{1} + \tilde\kappa \dot{S}_{2} 
 + \tilde\kappa^2 \dot{S}_{3}+\cdots +\beta \left(
-\frac{1}{\tilde{\kappa}^2} S_0 + S_2 +2\tilde{\kappa} S_3 +\cdots \right) = \nonumber \\ && -\frac{1}{\tilde\kappa}a_0[S_0,S_0-2\hat{S}]
  -2a_0[S_0-\hat{S},S_1]+a_1[S_0-2\hat{S}] \nonumber\\ 
&& +\tilde{\kappa}\left(-2a_0[S_0-\hat{S},S_2]-a_0[S_1,S_1]+a_1[S_1]\right)+\cdots\,.
\end{eqnarray}
This thesis is mainly concerned with the classical level of this flow equation. The classical limit is given by $\tilde\kappa\to0$ (equivalent to $M_{\rm Planck}\to\infty$), alternatively by $\hbar\to0$. This limit is of physical interest because, although not infinite, the reduced Planck mass is very large ($\sim 10^{18}$ GeV), and so the classical limit is an excellent approximation for currently accessible physics. In the classical limit, we do not expect the dimensionless part of $\tilde{\kappa}$ to run. The classical flow is, as always,
\begin{equation}
 \label{classical-Einstein-flow}
\dot{S}_{0} = -a_0[S_0,S_0-2\hat{S}]\,.
\end{equation}
One can use (\ref{gravloopexp}) to separate the flow loop by loop by solving the classical part first and then iteratively solving higher loop orders until one reaches the desired loop level. The $\beta$-function itself expands out loopwise, such that each loop order in (\ref{gravloopexp}) is at the fixed order in $\tilde\kappa$ given in (\ref{mass-expansion}): 
\begin{equation}
\label{beta}
 \beta := \Lambda\partial_{\Lambda}\tilde\kappa = \beta_{1}\Lambda^2\tilde\kappa^2 + \beta_{2}\Lambda^4\tilde\kappa^3 + \cdots\,.
\end{equation}
Powers of $\Lambda$ have been introduced as factors here to set all $\beta_i$ to be dimensionless numbers. If we instead rescaled $\tilde\kappa$ to be a dimensionless coupling $\upsilon$ times $\Lambda^{-2}$, we would have a $\beta$-function of the form 
\begin{equation}
 \beta(\upsilon) =  \Lambda\partial_{\Lambda}\upsilon = 2\upsilon +\beta_1 \upsilon^2 +\beta_2 \upsilon^3 +\cdots\,.
\end{equation}
Note that this form now possesses a classical part of the $\beta$-function, reflecting the classical flow of Newton's constant-like part of the action in Einstein scheme that follows simply from its mass dimension. Since the Einstein-Hilbert action is a dimension 2 operator rather than a dimension 4 operator, the kernel in Einstein scheme is dimension $-2$ rather than dimension $-4$ (corresponding to a $\Delta = c\left(p^2/\Lambda^2\right)/p^2$ effective propagator rather than a $c\left(p^2/\Lambda^2\right)/p^4$ effective propagator in fixed-background formalism). This balances the dimension 2 Lagrangian density from $\Sigma$ in Einstein scheme rather than the dimension 4 form in Weyl scheme. Thus the factor of $\KoL$ in the kernel becomes
\begin{equation}
 \label{dotD-Einstein}
\dot{\Delta}\left(-\nabla^2\right) = -\frac{2}{\Lambda^2} \,c'(-\nabla^2/\Lambda^2)\,.
\end{equation}
This forms a local expansion that creates no problems for Kadanoff blocking:
\begin{equation}\label{BIESKE}
 \dot{\Delta}\left(-\nabla^2\right) = -\frac{2}{\Lambda^2}\,c'(0) + \frac{2}{\Lambda^4}\,c''(0) \nabla^2+O(\nabla^4)\,,
\end{equation}
consistent with the form expected for creating the effective propagator in fixed-background formalism.

As with the Weyl scheme, we require that the effective action at the fixed point matches the seed action up to the 2-point level by setting both fixed-point actions to be equal up to the quadratic order in the Riemann tensor. The condition for a canonically normalized kinetic term is $g_2=1$, so we also have $\hat{g}_2=1$. Unlike in the Weyl scheme, we did not have to fix a value for $c'(0)$, or indeed any term in $c'(-\nabla^2/\Lambda^2)$. However, also unlike in Weyl scheme, we do have to impose that $g_0=0$, \ie no cosmological constant-like part of the classical fixed-point action. As can be seen in (\ref{g2-flow}), a $g_0$ coupling would introduce a non-zero value for $\dot{g}_2$ at the classical level, which is dimensionless and therefore would integrate back to a ln$(\Lambda)$ term in $S$, resurrecting the locality problem we are trying to avoid by creating the Einstein scheme. As already noted, this sacrifice is not problematic, since the fixed-point $g_0$ is not the physical cosmological constant and it vanishes in the $\Lambda\to0$ limit anyway. Using the unique form for the fixed-point action up to $d=2$, we can determine the coefficients of the $d=4$ operators at the classical level. Using (\ref{D4-flow}) with the Einstein scheme kernel from (\ref{dotD-Einstein}), we find the couplings for the quadratic order in the Riemann tensor to be
\begin{equation}
 \label{R2-Einstein}
g_{4,1} R^2 + g_{4,2} R^{\mu\nu}R_{\mu\nu} = 4\int \frac{d\Lambda}{\Lambda}\,\ca^0_0[R,R] = -2\frac{c'(0)}{\Lambda^2}\, \left(R_{\mu\nu}R^{\mu\nu}+jR^2\right)\,,
\end{equation}
where an integration of (\ref{dotD-Einstein}) has been used such that
\begin{equation}
 \int \frac{d\Lambda}{\Lambda}\, \dot{\Delta}(0) =  \frac{c'(0)}{\Lambda^2}.
\end{equation}
The original Einstein-Hilbert term reappears in $S$ as an integration constant. Since the Einstein-Hilbert term is the only diffeomorphism-invariant dimension 2 operator, it is the only one that can appear as an integration constant of the flow at the fixed point. Any other integration constants would appear as operators that perturb the Lagrangian density away from the fixed point, \eg a cosmological constant.

Thus begins the expansion in higher-curvature terms that make up the fixed-point action. We have defined the kernel in terms of an ultraviolet cutoff function $c$, which is to regulate the propagator in the fixed-background formalism. To ensure that this happens, we need the kinetic term (\ie the expansion up to quadratic order in the Riemann tensor) to be regulated by $c^{-1}$. Two immediate obstacles appear. Firstly, we cannot construct a diffeomorphism-invariant cutoff-regulated term at the linear order in curvature because total covariant derivatives of the Einstein-Hilbert term can be removed via integration by parts and we cannot split the Einstein-Hilbert action in such a way as to sandwich any covariant derivatives inside. Secondly, we have two independent curvature-squared parts which need not remain in fixed proportion as we expand in covariant derivatives inserted between the two Riemann tensor instances.

Since this is a problem concerning the propagator itself and the regularization of that propagator, this issue and its solution become much clearer in the fixed-background formalism, which will be investigated in Subsection \ref{MDIERG2ptE}, where the effective propagator will be solved exactly. The solution can be described background-independently as follows. The Einstein-Hilbert term is kept in its usual form, but a function of $\nabla^2$, related to the inverse cutoff function, is placed between the two Riemann tensor instances in the curvature-squared terms. This function of $\nabla^2$ is chosen to be the same for both structures, thus keeping them in fixed proportion at all orders in $\nabla^2$. This places a constraint on the value of $j$, which turns out to be $j=-1/2$ or $j=-1/3$. The true inverse cutoff function $c^{-1}$ appears in the complete 2-point function in the usual way, but that 2-point function is built out of both the curvature-squared terms and the Einstein-Hilbert term together. Thus the function of covariant derivatives $d\left(-\nabla^2/\Lambda^2\right)$ that appears in the curvature squared terms is related to the cutoff function via \eg $d(0)=-c'(0)$, but it applies only to the 4-derivative and higher parts of the kinetic term, the 2-derivative part coming from the Einstein-Hilbert term itself. The form of the seed Lagrangian is then
\begin{equation}\label{biEHfull}
 \hat{\cL} = -2R + \frac{2}{\Lambda^2}R_{\mu\nu}\,d(-\nabla^2/\Lambda^2) R^{\mu\nu} + \frac{2}{\Lambda^2}jR\, d(-\nabla^2/\Lambda^2) R\,.
\end{equation}
In fact, to construct this regulated 2-point function in fixed-background formalism, we will require that the Einstein-Hilbert action has the same Lorentz index structure as the curvature-squared part, which eliminates the $j=-1/3$ solution and leaves us with $j=-1/2$ in the Einstein scheme. As always, we are at liberty to extend (\ref{biEHfull}) by adding curvature-cubed and higher operators, which again corresponds to reparametrizing the field.

In a pure gravity theory, the only relevant perturbation from this fixed-point is a cosmological constant term $\mathcal{O}_0=1$. Using the classical flow equation for $\mathcal{O}_0$ (\ref{g0-flow}), the result of the bilinear in (\ref{0-1}) and the Einstein-scheme kernel in (\ref{dotD-Einstein}), we find that the flow of the cosmological constant is 
\begin{equation}
 \label{cosmoFlowE}
\dot{g}_0 = -(1+4j)\frac{c'(0)}{\Lambda^2} g_0^2\,.
\end{equation}
This does not introduce a flow for $g_2$ because, as already discussed around (\ref{nocrossterms}), the fixed-point and perturbation actions do not have cross-terms in $\ca_0$ until cubic order in the Riemann tensor.

\section{Fixed-background formalism}\label{MDIERGfixed}

We turn now to the fixed-background formalism in which we fix a background metric and describe gravity via perturbations to that background metric, as outlined in Subsection \ref{IntroMetricPert}. The metric perturbation $h_{\mu\nu}(x)$, which can be thought of as the graviton field, is defined as the difference between full and background metrics, as given in (\ref{MetricPertDef}). Lorentz indices will be raised and lowered using the background metric and the inverse background metric. The full inverse metric, however, expands out as an infinite series in the perturbation, as given in (\ref{InverseMetricExp}). For most of this section, we will choose to focus on the momentum representation, constructed via Fourier transformations (\ref{PertFourier}). In addition to the shorthand notation $\dbar p$ for momentum integration given in (\ref{dbarpshort}), it is also convenient to introduce shorthand notation for the momentum delta function: 
\begin{equation}
 \delbar(p) := (2\pi)^D \delta(p).
\end{equation}
The metric we will be specializing to is the Euclidean metric, given by a Kronecker delta $\delta_{\mu\nu}$. The Euclidean signature is necessary for locality and the flatness is naturally convenient for a momentum representation. The momentum representation for the perturbation field can be physically interpreted classically as corresponding to the frequency spectrum of gravitational waves propagating on the flat background or quantum mechanically as the 4-momentum modes of gravitons propagating on the background.

As with scalar (\ref{Scalarnpointexp}) or gauge theories (\ref{Gaugenpointexp}), but unlike in the background-independent description, we can construct the action as an $n$-point expansion in the graviton field:
\begin{eqnarray}\label{ActionSeries}
& S & = \int \dbar p \ \delbar(p)\mathcal{S}^{\mu\nu}(p)h_{\mu\nu}(p)
            + \frac{1}{2}\int \dbar p \ \dbar q \ \delbar(p+q)\mathcal{S}^{\mu\nu\rho\sigma}(p,q)h_{\mu\nu}(p)h_{\rho\sigma}(q) \nonumber \\ & &
            + \frac{1}{3!}\int \dbar p\ \dbar q\ \dbar r \ \delbar(p+q+r)\mathcal{S}^{\mu\nu\rho\sigma\alpha\beta}(p,q,r)h_{\mu\nu}(p)h_{\rho\sigma}(q)h_{\alpha\beta}(r) + \cdots
\end{eqnarray}
where $\mathcal{S}^{\alpha_1\beta_1\cdots\alpha_n\beta_n}(p_1,\cdots,p_n)$ are the $n$-point functions.
Although it is trivially easy to have action terms at the 0-point level, they have no physical significance and are thus not included in the expansion. Momentum conservation requires that all momentum arguments for a given term sum to zero, as enforced by the delta functions. There is only one kind of 1-point function, which carries no momentum and is simply a constant $\mathcal{S}$ times the inverse flat metric:
\begin{equation} \label{one-point-S}
\mathcal{S}^{\mu\nu}(0) = \mathcal{S}\delta^{\mu\nu}\,.
\end{equation}
It simplifies the notation somewhat to use $\mathcal{S}\delta^{\mu\nu}$ instead of $\mathcal{S}^{\mu\nu}(0)$, so it will be convenient to do so frequently from now on. Given an action $S$, we can extract the $n$-point functions by performing functional differentiation $n$ times with respect to the perturbation field and evaluating the perturbation at zero: 
\begin{equation}\label{npointgrav}
\mathcal{S}^{\alpha_1\beta_1\cdots\alpha_n\beta_n}(p_1,\cdots,p_n) = \frac{\delta}{\delta h_{\alpha_1\beta_1}(p_1)}\cdots\frac{\delta}{\delta h_{\alpha_n\beta_n}(p_n)}S\Big|_{h=0}\,.
\end{equation}
The $n$-point functions are constructed such that the Lorentz indices come in pairs. Each pair corresponds to one of the momentum arguments such that, in the action, the pair is contracted with the indices from a factor of the metric perturbation that shares this corresponding momentum argument, as seen explicitly in (\ref{ActionSeries}). This gives the $n$-point functions some index symmetries: the $n$-point function is symmetric under exchange of index labels within a pair and it is also symmetric under an exchange of two pairs and the corresponding momentum arguments together.

To write the flow equation (\ref{full-flow}) and its kernel (\ref{kernel-grav}) in fixed-background formalism, it is helpful to note that there is a very simple correspondence between functional derivatives with respect to the metric and with respect to the perturbation:
\begin{equation}
\frac{\delta}{\delta g_{\mu\nu}(x)} = \frac{\delta}{\delta h_{\mu\nu}(x)}\,.
\end{equation}
After Fourier-transforming into momentum representation, the fixed-background version of the flow equation at the classical level is
\begin{equation}\label{fixedsdot}
\dot{S} = \frac{1}{2}\int \dbar q\ \dbar r\ \frac{\delta S}{\delta h_{\mu\nu}(-q)}K_{\mu\nu\rho\sigma}(q,r)\frac{\delta \Sigma}{\delta h_{\rho\sigma}(-r)},
\end{equation}
where we choose a sign convention in which momentum arguments for the kernel and $n$-point functions are defined as propagating outwards. We require that $S$, $\Sigma$ and $K$ are all individually momentum-conserving, ensuring that momentum is conserved at every vertex as it propagates through the flow equation. Thus we can employ diagrams much like Figure \ref{fig:Anpoint} to visualise the propagation of momentum through the flow equation and its $n$-point expansions in the fixed-background formalism. As with scalar and gauge theories, the $n$-point expansion of the flow equation is found by taking the $n$th functional derivative of (\ref{fixedsdot}) and evaluating $h_{\alpha\beta}$ at zero, as seen in (\ref{npointgrav}). 

Similarly to how gauge theories have an $n$-point expansion in the kernel due to its covariantization, so too does the gravity kernel, in a manner similar to (\ref{npointgrav}). In fact, the gravity kernel gets its $n$-point expansion not only from the expansion in covariant derivatives in (\ref{kernel-grav}), but also from the metric factors required for index structure and the factor of $1/\sqrt{g}$ required for bitensor covariance. Although the kernel expansion has similarities to (\ref{ActionSeries}), there are also differences. Unlike the action, the kernel begins with a 0-point term that is meaningful. The 0-point kernel term carries momentum from $S$ to $\Sigma$ without diverting any of that momentum elsewhere, whereas $n$-point kernel terms also distribute momentum between $n$ external legs. The kernel expansion begins at the 0-point level with two pairs of Lorentz indices and two momentum arguments that are used to link $S$ to $\Sigma$, after that, it expands such that each $n$-point function has $n+2$ pairs of Lorentz indices and $n+2$ momentum arguments. 

Just as the background-independent flow equation could be iteratively solved to higher orders in the Riemann tensor, we will see that the fixed-background flow equation can be iteratively solved for higher $n$-point functions, starting from the 2-point level in Weyl scheme and Einstein scheme, such that the equivalence of the fixed-background and background-independent formalisms is clear. Section \ref{MDIERGWard} will explore the relations between $n$-point functions imposed by the diffeomorphism invariance of the actions and the diffeomorphism covariance of the kernel.

\section{Ward identities}\label{MDIERGWard}

The diffeomorphism invariance of the action imposes constraints on its $n$-point expansion that are akin to the Ward identities that gauge-invariance imposes on Yang-Mills theories. In this section, we will derive these gravitational Ward identities, which relate $(n+1)$-functions to their corresponding $n$-point functions. The kernel, which is a diffeomorphism-covariant bitensor, also has Ward identities. Ward identities for the kernel will be derived in Subsection \ref{MDIERGKernelWard}. Since the $n$-point expansion of $S$ is constrained by Ward identities, so too is the $n$-point expansion of $\dot{S}$: they obey Ward identities of the same form. Thus, via the flow equation (\ref{fixedsdot}), the Ward identities of the kernel become related to the Ward identities of the actions in a way that also satisfyingly relates to the momentum-conservation properties of the flow equation. An important consistency check will be walked through in Subsection \ref{MDIERGWardflow}, demonstrating by example how the Ward identities for $\dot{S}$, $S$, $K$ and $\Sigma$ relate to each other.

\subsection{Ward identities for an action}\label{MDIERGWardact}
We begin by noting again the diffeomorphism transformation of a metric perturbation given in (\ref{dpertintro}). More specifically, we will wish to specialise to a particular choice of form for this diffeomorphism transformation that is based on the partial derivative operator:
\begin{equation}\label{hLie}
\delta h_{\alpha\beta}= \mathsterling_{\xi}\left(\delta+h\right)_{\alpha\beta} = 2(\delta+h)_{\lambda(\alpha}\partial_{\beta)} \xi^\lambda + \xi\cdot\partial h_{\alpha\beta}\,.
\end{equation}
Choosing the partial derivative rather than the metric compatible-covariant derivative will ensure that the resulting Ward identities that we derive are expressed in closed form, \ie they relate $n$-point functions to their $(n-1)$-point functions and not to the complete series from $(n-1)$ to zero or one. The reason that a metric-compatible covariant derivative would result in more complication is that, although it would have eliminated the second term on the right hand side of (\ref{dpertintro}), the covariant derivative acting on $\xi^\lambda$ in the first term would introduce a connection coefficient that carries an inverse metric, which expands out as an infinite series expansion in the metric perturbation, which is an unnecessary obstacle to constructing a closed-form Ward identity for a general $n$-point level. 

To obtain the Ward identity, take the expansion in $h$ of the action, as given in (\ref{ActionSeries}), and perform the diffeomorphism transformation (\ref{hLie}) in momentum representation. Then perform functional differentiation with respect to $h$ $(n-1)$ times to get the $n$-point level, evaluate at $h=0$ as usual and also differentiate out the $\xi$ factor. Because we must have $\delta S=0$ overall, the transformation is constrained at every $n$-point level to give
\begin{eqnarray}\label{WardIds}
-2p_{1\mu_1} \Sv^{\mu_1\nu_1\cdots\mu_n\nu_n}(p_1,\cdots,p_n) & = &
 \sum_{i=2}^n \pi_{2i} \Big\{\, p_2^{\nu_1}\Sv^{\mu_2\nu_2\cdots\mu_n\nu_n}(p_1+p_2,p_3,\cdots,p_n) \\ &&
+2p_{1\alpha}\delta^{\nu_1(\nu_2}\Sv^{\mu_2)\alpha\mu_3\nu_3\cdots\mu_n\nu_n}(p_1+p_2,p_3,\cdots,p_n)\,\Big\}\,,\nonumber
\end{eqnarray}
where $\pi_{2i}$ is a transposition operator that substitutes $p_2,\mu_2,\nu_2\leftrightarrow p_i\mu_i\nu_i$. For these Ward identities, and indeed all Ward identities derived in this thesis, momentum conservation is satisfied such that the sum of all $n$ momentum arguments appearing here is zero.
The reason that we have a relation between the $n$-point and the $(n-1)$-point levels is that $\delta h$ in (\ref{hLie}) has both an $\mathcal{O}(\xi)$ term and an $\mathcal{O}(h\xi)$ term, so an $\mathcal{O}(h^n)$ term (with a corresponding $n$-point function) transforms to give both an $\mathcal{O}(h^{n-1}\xi)$ term and an $\mathcal{O}(h^n\xi)$ term. Thus a given order in $h$ has contributions from both an $n$-point function and its corresponding $(n-1)$-point function. An important special case is the 2-point Ward identity:
\begin{equation}\label{Ward1to2}
2p_{\mu}\mathcal{S}^{\mu\nu\rho\sigma}(p,-p) = \mathcal{S}p_{\mu}\delta^{\mu\nu}\delta^{\rho\sigma}-2\mathcal{S}p_{\mu}\delta^{\mu(\rho}\delta^{\sigma)\nu}\,.
\end{equation}
If the 1-point function is zero, then the 2-point function is transverse. The 1-point piece, which can only come from a cosmological constant-like term, is momentum-independent. A momentum-independent 1-point function uniquely determines the corresponding momentum-independent 2-point function via (\ref{Ward1to2}). A similar argument applies between the 2-point and 3-point functions and so on {\it ad infinitum}, which can be more readily seen via a momentum-independent specialization of (\ref{WardIds}). 

Indeed, all of these momentum-independent $n$-point functions are in fixed proportion, following the expansion of $\sqrt{g}$, as seen in a cosmological constant-like term. Thus the 2-point function can be split into a momentum-independent piece related to the cosmological constant and a transverse momentum-dependent piece corresponding to the graviton kinetic term. The momentum-dependent part of the 2-point function is transverse because the graviton is massless, as protected by the diffeomorphism symmetry. In fact, we will find in Section \ref{MDIERGTran2} that there are two independent momentum-dependent 2-point structures relating to the two independent curvature-squared structures in the background-independent formalism.

To derive the Ward identities for momentum-independent $n$-point functions, an intermediate step that is useful in its own right is to derive ``differential'' Ward identities, in which only one of the momenta is tended towards zero. This is of particular interest for studying flow equations in which either $S$ or $\Sigma$ on the right hand side of (\ref{fixedsdot}) is taken at the 1-point level. To construct the differential Ward identity, let us begin by relabelling (\ref{WardIds}) such that the left hand side has an $(n+1)$-point function with momentum arguments $\epsilon, p_1-\epsilon, p_2,\cdots,p_n$, where $\epsilon\to0$. Although both sides of (\ref{WardIds}) vanish in this limit, we can still extract the $\mathcal{O}(\epsilon)$ piece by differentiating with respect to $\epsilon$. Na\"{i}vely, one might expect that, given this setup, $p_1$ becomes a specially distinguished momentum in the Ward identity but momentum conservation ultimately forces the complete expression to have the appropriate symmetries under exchanges of indices and momenta. The differential Ward identity is
\begin{eqnarray}
 \label{diffWard}
-2\Sv^{\alpha\beta\mu_1\nu_1\cdots\mu_n\nu_n}(0,p_1,\cdots,p_n) &=& \left(\sum_{i=1}^n p_i^\beta\partial^\alpha_i-\delta^{\alpha\beta}\right) \Sv^{\mu_1\nu_1\cdots\mu_n\nu_n}(p_1,\cdots,p_n)\nonumber\\
&& +2\sum_{i=1}^n\pi_{1i}\,\delta^{\beta(\nu_1}\Sv^{\mu_1)\alpha\mu_2\nu_2\cdots\mu_n\nu_n}(p_1,\cdots,p_n)\,, \nonumber \\ &&
\end{eqnarray}
where, in this context and only in this context, $\partial^\alpha_i$ is an operator that differentiates with respect to $p_{i\alpha}$. This immediately tells us that any $n$-point function for an action with one momentum argument set to zero is uniquely determined by the corresponding $(n-1)$-point function. This knowledge becomes useful in (\ref{fixedsdot}) where, for example, $S$ is expanded out to some $n$-point level, but the kernel is at the 0-point level and $\Sigma$ is at the 1-point level. In this context, $\dot{S}$ would be at the $(n-1)$-point level, so one might na\"{i}vely worry that we might be unable to solve the flow equation iteratively from the 1-point level upwards. However, since $\Sigma$ would be only at the 1-point level, it would receive zero momentum from the kernel, so the kernel would receive zero momentum from $S$ and therefore the $n$-point function used from $S$ would have one of its momentum arguments fixed at zero, making it uniquely determined from its $(n-1)$-point function. This makes the problem of iteratively solving the flow equation by $n$-point functions tractable. This scenario we have just discussed corresponds in background-independent formalism to the case where $\Sigma$ is a cosmological constant-like term and therefore this argument is strongly related to (\ref{0-1}) and (\ref{cosconstcrossterm}).

To see this relationship, first note that such a ``tadpole'' diagram would not only set one of the momenta to zero, but also contract its corresponding indices together. A simple way to see this is that the 1-point function (\ref{one-point-S}) contracts two of the kernel indices in (\ref{kernel-grav}), leaving the zero-point part of the zero-momentum kernel as simply a constant times the background metric, which contracts together the two indices associated with the zero-momentum argument in the $n$-point function of interest. Contracting the indices in (\ref{diffWard}), we have
\begin{equation}
 \label{diffWardcontracted}
\Sv^{\ \alpha\mu_1\nu_1\cdots\mu_n\nu_n}_\alpha(0,p_1,\cdots,p_n) = \left(2-n-\frac{1}{2}\sum_{i=1}^n p_i\cdot\partial_i\right)\Sv^{\mu_1\nu_1\cdots\mu_n\nu_n}(p_1,\cdots,p_n)\,.
\end{equation}
The $p_i\cdot\partial_i$ term simply counts the order in momentum of the $n$-point function to its right. In background-independent terms, this counts the mass dimension of the Lagrangian operator that gives us this term. For an operator of mass dimension $d$, the factor in brackets then becomes $-(d-4+2n)/2$. This factor resembles a factor appearing in (\ref{0-1}) and the discussion below it, except that the additional $2n$ piece appears because the complete cross term in (\ref{0-1}) and (\ref{cosconstcrossterm}) is not constructed exclusively out of tadpoles, but in fact also includes diagrams where the kernel and the cosmological constant-like part are expanded to higher points as well. The complete result for the cross-term of two operators is then the sum over all $n$-point diagrams, which is independent of $n$.

As promised, we now turn to the specialization of the Ward identities to momentum-independent $n$-point functions, which we will find to correspond to cosmological constant-like terms. To do this, we simply take (\ref{diffWard}) in the limit of all momenta going to zero. Since the actions can be expressed as local expansions in powers of momentum (alternatively as local expansions in higher-dimension operators), taking all momenta to zero simply returns the zeroth order, which is the momentum-independent part. The Ward identities for momentum-independent $n$-point functions are
\begin{equation}\label{Wardmomind}
2\Sv^{\mu_1\nu_1\cdots\mu_n\nu_n}(\underline{0}) = \delta^{\mu_1\nu_1}\Sv^{\mu_2\nu_2\cdots\mu_n\nu_n}(\underline{0})
-2\sum_{i=2}^n\pi_{2i} \,\delta^{\nu_1(\nu_2}\Sv^{\mu_2)\mu_1\mu_3\nu_3\cdots\mu_n\nu_n}(\underline{0})\,.
\end{equation}
Equation (\ref{Wardmomind}) exactly and uniquely gives us the form of any momentum-independent $n$-point function provided we already know its corresponding $(n-1)$-point function. The 0-point function of an action has no physical importance, so we can begin with a 1-point function of the form (\ref{one-point-S}) and use (\ref{Wardmomind}) to find the 2-point function:
\begin{equation}\label{Ward2ptCC}
2\mathcal{S}^{\mu\nu\rho\sigma}(0,0) = \mathcal{S}\delta^{\mu\nu}\delta^{\rho\sigma}-2\mathcal{S}\delta^{\mu(\rho}\delta^{\sigma)\nu}.
\end{equation}
We can then substitute the above 2-point function into (\ref{Wardmomind}) to get the 3-point function:
\begin{eqnarray}
 \label{Ward3ptCC}
2\mathcal{S}^{\mu\nu\rho\sigma\alpha\beta}(0,0,0)  & = &  2\mathcal{S}\delta^{(\alpha|(\mu}\delta^{\nu)(\rho}\delta^{\sigma)|\beta)}
- \mathcal{S}\delta^{\mu\nu}\delta^{\rho(\alpha}\delta^{\beta)\sigma} \nonumber \\ &&
 -2\mathcal{S}^{(\mu|\alpha\rho\sigma}(0,0)\delta^{\beta|\nu)}
 +\mathcal{S}^{\mu\nu\rho\sigma}(0,0)\delta^{\alpha\beta}\,.
\end{eqnarray}
Using this technique, we can iterate to any $n$-point level we chose, using the Ward identities to exactly and uniquely determine each momentum-independent $n$-point function. These $n$-point structures correspond to the expansion in metric perturbations of $\sqrt{g}$, which is the source of the $n$-point expansion of a cosmological constant-like term. This will be demonstrated explicitly in Section \ref{MDIERGsqrtg}.

\subsection{Ward identities for the kernel}\label{MDIERGKernelWard}

In the previous subsection, we derived Ward identities for the action based on its invariance under diffeomorphisms, \ie we performed a diffeomorphism transformation on $S$ under the constraint that $\delta S=0$. Deriving Ward identities for the kernel is a similar procedure, except that it has a non-zero diffeomorphism transformation. As hinted at in Subsection \ref{IntroDiffSym}, the kernel transformation is related to (\ref{LieGeneralTens}), except that the kernel is a bitensor, \ie it takes two position arguments. thus we have two copies of the transformation, one for each position argument. The kernel transformation in position representation is given by its Lie derivative:
\begin{eqnarray}\label{proptrans}
 \mathsterling_{\xi}K_{\mu\nu\rho\sigma}(x,y) & = & \xi(x)\cdot\partial_{x}K_{\mu\nu\rho\sigma}(x,y) + \xi(y)\cdot\partial_{y}K_{\mu\nu\rho\sigma}(x,y) \nonumber \\ && + 2K_{\lambda(\mu|\rho\sigma}(x,y)\partial_{x|\nu)}\xi^{\lambda}(x)+ 2K_{\mu\nu\lambda(\rho|}(x,y)\partial_{y|\sigma)}\xi^{\lambda}(y)\,.
\end{eqnarray}
The momentum representation is related to the position representation via a Fourier transformation of both position arguments:
\begin{equation}
 K_{\mu\nu\rho\sigma}(x,y) = \int \dbar q\ \dbar r\ e^{-iq\cdot x-ir\cdot y}K_{\mu\nu\rho\sigma}(q,r).
\end{equation}
Let us adopt notation for the $n$-point expansion of the kernel in the following spirit:
\begin{equation}\label{kernelnpoint}
 K_{\mu\nu\rho\sigma}(q,r) = \mathcal{K}_{\mu\nu\rho\sigma}(q,r) + \int \dbar p_1\ \delbar(p_1+q+r)\mathcal{K}^{\alpha_1\beta_1}_{\ \ \ \ \mu\nu\rho\sigma}(p_1,q,r)h_{\alpha_1\beta_1}(p_1) +\cdots.
\end{equation}
Finding the Ward identities for the kernel follows the same procedure as for the action, except that the overall diffeomorphism transformation of (\ref{kernelnpoint}) is the Fourier transformation of (\ref{proptrans}). Thus we obtain a closed form for the kernel Ward identities:
\begin{eqnarray}\label{kernWard}
&& 2p_{\gamma}'\mathcal{K}^{\gamma\delta\alpha_1\beta_1\cdots\alpha_n\beta_n}_{\ \ \ \ \ \ \ \ \ \ \ \ \ \ \mu\nu\rho\sigma}(p',p_1,\cdots,p_n,q,r) = \nonumber \\
&& -(p'+q)^\delta \mathcal{K}^{\alpha_1\beta_1\cdots\alpha_n\beta_n}_{\ \ \ \ \ \ \ \ \ \ \ \ \mu\nu\rho\sigma}(p_1,\cdots,p_n,q+p',r) \nonumber \\
&& -(p'+r)^\delta \mathcal{K}^{\alpha_1\beta_1\cdots\alpha_n\beta_n}_{\ \ \ \ \ \ \ \ \ \ \ \ \mu\nu\rho\sigma}(p_1,\cdots,p_n,q,r+p') \nonumber \\
&& +2\delta^{\lambda\delta}p'_{(\mu|}\mathcal{K}^{\alpha_1\beta_1\cdots\alpha_n\beta_n}_{\ \ \ \ \ \ \ \ \ \ \ \ \ |\nu)\lambda\rho\sigma}(p_1,\cdots,p_n,q+p',r) \nonumber \\
&& +2\delta^{\lambda\delta}p'_{(\rho|}\mathcal{K}^{\alpha_1\beta_1\cdots\alpha_n\beta_n}_{\ \ \ \ \ \ \ \ \ \ \ \ \ \mu\nu|\sigma)\lambda}(p_1,\cdots,p_n,q,r+p') \nonumber \\
&& -\sum_{i=1}^{n}\pi_{i1}\left\{ p^{\delta}_{1}\mathcal{K}^{\alpha_1\beta_1\cdots\alpha_n\beta_n}_{\ \ \ \ \ \ \ \ \ \ \ \ \mu\nu\rho\sigma}(p'+p_1,p_2,\cdots,p_n,q,r)\right. \nonumber \\
&& \left. +2p'_{\lambda}\delta^{\delta(\alpha_1}\mathcal{K}^{\beta_1)\lambda\alpha_2\beta_2\cdots\alpha_n\beta_n}_{\ \ \ \ \ \ \ \ \ \ \ \ \ \ \ \ \, \mu\nu\rho\sigma}(p'+p_1,p_2,\cdots,p_n,q,r)\right\}.
\end{eqnarray}
This complicated form is related to the action Ward identity (\ref{WardIds}), except that it has additional terms whose r\^{o}le in the flow equation is to protect momentum conservation. The final two terms of (\ref{kernWard}) strongly resemble the right hand side of (\ref{WardIds}) because they are derived in the same way and they fulfill the same function. Were it not for the first four terms on the right hand side, (\ref{kernWard}) would essentially be the same as (\ref{WardIds}), but with some extra index and momentum labels not taking part in the structure. These four new terms originate with the four terms on the right hand side of (\ref{proptrans}), relating directly to the bitensor structure. Looking to the momentum arguments, we can see that $p'$ in these terms has been added to either $q$ or $r$, depending on whether the term comes from the transformation of $x$ or $y$. These extra terms in the kernel Ward identity are required to ensure that the consistent application of Ward identities to the flow equation as a whole is momentum-conserving, both overall and at the level of individual vertices. More precisely, these extra terms cancel momentum-violating contributions that would appear as a result of applying the Ward identities for $S$ and $\Sigma$ on the right hand side of (\ref{fixedsdot}).

To understand how this works in terms of the flow equation diagrams, \eg Figure \ref{fig:Anpoint}, consider applying Ward identities to $\dot{S}$. If we take (\ref{fixedsdot}) at a given $n$-point level and contract one of the indices with its corresponding momentum, the Ward identities (\ref{WardIds}) tell us that we now have an expression in terms of the corresponding $(n-1)$-point function, \ie we have removed one of the external legs. Momentum conservation then requires us to add the momentum from this removed leg another leg. This reallocation of momentum can be seen explicitly by examining the momentum arguments on the right hand side of (\ref{WardIds}) or (\ref{kernWard}). Because we have symmetries under exchanges of index pairs and their corresponding momenta, the Ward identity does not give preference to any particular leg to add this momentum to, but rather sums over all the remaining legs. Again, to see this explicitly, notice the summations on the right hand side of (\ref{WardIds}) or (\ref{kernWard}). When we apply this to the right hand side of (\ref{fixedsdot}), we have to use the kernel Ward identities for $S$, $\Sigma$ and $K$ individually, as the leg we remove can appear in any of these three. The individual Ward identities for $S$ and $\Sigma$ do not distinguish between external legs and the leg that connects to the kernel. As a result, the individual Ward identities for $S$ and $\Sigma$ include terms in which the momentum from the leg we remove is transferred to the leg that connects to the kernel. Taken in the context of the flow equation, such terms violate momentum conservation because the action is transferring a momentum to the kernel that is different to the momentum that the kernel is receiving. This violates momentum not just at the level of that vertex, but also in the external legs of the diagram overall. The first four terms in the right hand side of the kernel Ward identity (\ref{kernWard}) cancel all such terms exactly in every context. Thus we see the pleasing relationship between the diffeomorphism covariance of the kernel and the overall diffeomorphism invariance of the flow equation via momentum conservation in the flow equation Ward identities.

Just as the action has momentum-independent Ward identities, so too does the kernel, which are 
\begin{eqnarray}\label{KernWardMomInd}
 \mathcal{K}^{\gamma\delta\alpha_1\beta_1\cdots\alpha_n\beta_n}_{\ \ \ \ \ \ \ \ \ \ \ \ \ \ \ \mu\nu\rho\sigma}(\underline{0}) & = & -\frac{1}{2}\delta^{\gamma\delta}\mathcal{K}^{\alpha_1\beta_1\cdots\alpha_n\beta_n}_{\ \ \ \ \ \ \ \ \ \ \ \ \mu\nu\rho\sigma}(\underline{0}) 
 + \delta^{\lambda(\gamma}\delta^{\delta)}_{\ \ (\mu|}\mathcal{K}^{\alpha_1\beta_1\cdots\alpha_n\beta_n}_{\ \ \ \ \ \ \ \ \ \ \ \ |\nu)\lambda\rho\sigma}(\underline{0})  \nonumber \\
&& + \delta^{\lambda(\gamma}\delta^{\delta)}_{\ \ (\rho|}\mathcal{K}^{\alpha_1\beta_1\cdots\alpha_n\beta_n}_{\ \ \ \ \ \ \ \ \ \ \ \ \mu\nu|\sigma)\lambda}(\underline{0}) \nonumber \\&&
-\sum_{i=1}^{n}\pi_{i1}\left\{\delta^{(\gamma|(\alpha_1}\mathcal{K}^{\beta_1)|\delta)\cdots\alpha_n\beta_n}_{\ \ \ \ \ \ \ \ \ \ \ \ \ \mu\nu\rho\sigma}(\underline{0})\right\}\,.
\end{eqnarray}
These Ward identities apply to the $n$-point expansion, from zeroth order, of the no-derivative part of the kernel (\ref{kernel-grav}). This no-derivative part of the kernel simply consists of a quadratic structure in the metric, as required to produce the necessary index structure, and a factor of $1/\sqrt{g}$.

\subsection{Ward identity consistency in the flow equation}\label{MDIERGWardflow}

We now turn to a consistency check to see how the Ward identities for $S$, $K$ and $\Sigma$ fit together to give the expected Ward identity for $\dot{S}$. The simplest test that exhibits all the important features is the 2-point Ward identity for $\dot{S}$ where $S$ and $\Sigma$ are allowed to have 1-point functions, \ie possess a cosmological constant-like term. Following from (\ref{Ward1to2}), we have
\begin{equation}\label{Ward2pSdot}
 2p_{\alpha_{1}}\dot{\Sv}^{\alpha_{1}\beta_{1}\alpha_{2}\beta_{2}}(p,-p)=p^{\beta_{1}}\dot{\Sv}^{\alpha_{2}\beta_{2}}(0)-2p_{\lambda}\delta^{\beta_{1}(\alpha_{2}}\dot{\Sv}^{\beta_{2})\lambda}(0).
\end{equation}
The flow equation (\ref{fixedsdot}) is easily written out at the 1-point level to give the expression for $\dot{S}^{\alpha\beta}(0)$:
\begin{equation}\label{1pSdot}
 \dot{\Sv}^{\alpha\beta}(0) = \left(\Sv\left|^{\alpha\beta\mu\nu}(0,0)\mathcal{K}_{\mu\nu\rho\sigma}(0,0)\right|\Sigma\right)^{\rho\sigma}(0) + \Sv^{\mu\nu}(0)\mathcal{K}^{\alpha\beta}_{\ \ \mu\nu\rho\sigma}(0,0,0)\Sigma^{\rho\sigma}(0),
\end{equation}
where the notation has been condensed such that the large round brackets indicate an anticommutation of the actions:
\begin{eqnarray}
&& \left(\Sv\left|^{\alpha\beta\cdots\mu\nu}(p,\cdots,-q)\mathcal{K}_{\mu\nu\rho\sigma}(q,r)\right|\Sigma\right)^{\gamma\delta\cdots\rho\sigma}(p',\cdots,-r)  := \nonumber \\
&& \Sv^{\alpha\beta\cdots\mu\nu}(p, \cdots,-q)\mathcal{K}_{\mu\nu\rho\sigma}(q,r)\Sigma^{\gamma\delta\cdots\rho\sigma}(p',\cdots,-r) + \nonumber \\ 
&& \Sigma^{\alpha\beta\cdots\mu\nu}(p,\cdots,-q)\mathcal{K}_{\mu\nu\rho\sigma}(q,r)\Sv^{\gamma\delta\cdots\rho\sigma}(p',\cdots,-r).
\end{eqnarray}
Substituting (\ref{1pSdot}) into (\ref{Ward2pSdot}), the 2-point Ward identity for $\dot{S}$ becomes
\begin{eqnarray}\label{Ward2pSdotflow}
 2p_{\alpha_{1}}\dot{\Sv}^{\alpha_{1}\beta_{1}\alpha_{2}\beta_{2}}(p,-p) & = & p^{\beta_{1}}\left(\Sv\left|^{\alpha_2 \beta_2 \mu\nu}(0,0)\mathcal{K}_{\mu\nu\rho\sigma}(0,0)\right|\Sigma\right)^{\rho\sigma}(0)\nonumber \\ 
&& + p^{\beta_1}\Sv^{\mu\nu}(0)\mathcal{K}^{\alpha_2 \beta_2}_{\ \ \ \ \ \mu\nu\rho\sigma}(0,0,0)\Sigma^{\rho\sigma}(0) \nonumber \\ 
&& - 2p_{\lambda}\delta^{\beta_1 (\alpha_2}\left(\Sv\left|^{\beta_2)\lambda}(0,0)\mathcal{K}_{\mu\nu\rho\sigma}(0,0)\right|\Sigma\right)^{\rho\sigma}(0) \nonumber \\ 
&& - 2p_{\lambda}\delta^{\beta_1 (\alpha_2|}\Sv^{\mu\nu}(0)\mathcal{K}^{|\beta_2)\lambda}_{\ \ \ \ \ \mu\nu\rho\sigma}(0,0,0)\Sigma^{\rho\sigma}(0).
\end{eqnarray}
To test for consistency, let us substitute the 2-point flow equation into the left hand side of (\ref{Ward2pSdotflow}) and demonstrate that both sides match. The 2-point flow equation is given generally by
\begin{eqnarray}\label{general2ptflow}
 \dot{\Sv}^{\alpha_1 \beta_1 \alpha_2 \beta_2}(p,-p) & = & \left(\Sv\left|^{\alpha_1 \beta_1 \mu\nu}(p,-p)\mathcal{K}_{\mu\nu\rho\sigma}(p,-p)\right|\Sigma\right)^{\alpha_2\beta_2\rho\sigma}(p,-p) + \nonumber \\ 
&& \left(\Sv\left|^{\alpha_1\beta_1\mu\nu}(p,-p)\mathcal{K}^{\alpha_2\beta_2}_{\ \ \ \ \mu\nu\rho\sigma}(-p,p,0)\right|\Sigma\right)^{\rho\sigma}(0) + \nonumber \\
&& \left(\Sv\left|^{\alpha_2\beta_2\mu\nu}(p,-p)\mathcal{K}^{\alpha_1\beta_1}_{\ \ \ \ \mu\nu\rho\sigma}(-p,p,0)\right|\Sigma\right)^{\rho\sigma}(0) + \nonumber \\
&& \Sv^{\mu\nu}(0)\mathcal{K}^{\alpha_1\beta_1\alpha_2\beta_2}_{\ \ \ \ \ \ \ \ \ \ \mu\nu\rho\sigma}(p,-p,0,0)\Sigma^{\rho\sigma}(0) + \nonumber \\
&& \left(\Sv\left|^{\alpha_1\beta_1\alpha_2\beta_2\mu\nu}(p,-p,0)\mathcal{K}_{\mu\nu\rho\sigma}(0,0)\right|\Sigma\right)^{\rho\sigma}(0).
\end{eqnarray}
Now we contract \ref{general2ptflow} with $2p_{\alpha_1}$ to form the left hand side of (\ref{Ward2pSdotflow}). The action and kernel Ward identities can be used to simplify expressions as appropriate. Applying the 2-point Ward identities for $S$ and $\Sigma$ to the result of contracting the first two terms in (\ref{general2ptflow}) with $2p_{\alpha_1}$ gives
\begin{eqnarray}\label{2pmomvio}
&& p^{\beta_1}\left(\Sv\left|^{\alpha_2\beta_2\mu\nu}(-p,p)\mathcal{K}_{\mu\nu\rho\sigma}(p,-p)\right|\Sigma\right)^{\rho\sigma}(0)\nonumber \\ 
&& -2p_{\lambda}\delta^{\beta_1 (\mu}\left(\Sv\left|^{\nu)\lambda}(0)\mathcal{K}_{\mu\nu\rho\sigma}(p,-p)\right|\Sigma\right)^{\alpha_2\beta_2\rho\sigma}(p,-p) \nonumber \\
&& +p^{\beta_1}\left(\Sv\left|^{\mu\nu}(0)\mathcal{K}^{\alpha_2\beta_2}_{\ \ \ \ \mu\nu\rho\sigma}(-p,p,0)\right|\Sigma\right)^{\rho\sigma}(0) \nonumber \\ 
&& -2p_{\lambda}\delta^{\beta_1(\mu}\left(\Sv\left|^{\nu)\lambda}(0)\mathcal{K}^{\alpha_2\beta_2}_{\ \ \ \ \mu\nu\rho\sigma}(-p,p,0)\right|\Sigma\right)^{\rho\sigma}(0),
\end{eqnarray}
which are all momentum-violating terms that must be cancelled. The next two terms in (\ref{general2ptflow}) contract with $2p_{\alpha_1}$ such that we can apply either the 1 or 2-point kernel Ward identities. The momentum-violating (\ie cancelling) parts of those contributions are
\begin{eqnarray}\label{splitcancel}
&& -p^{\beta_1}\left(\Sv\left|^{\alpha_2\beta_2\mu\nu}(-p,p)\mathcal{K}_{\mu\nu\rho\sigma}(p,-p)\right|\Sigma\right)^{\rho\sigma}(0) \nonumber \\
&& +2p_{\lambda}\delta^{\beta_1 (\mu}\left(\Sv\left|^{\nu)\lambda}(0)\mathcal{K}_{\mu\nu\rho\sigma}(p,-p)\right|\Sigma\right)^{\alpha_2\beta_2\rho\sigma}(p,-p) \nonumber \\
&& +2p_{\lambda}\delta^{\beta_1(\mu}\left(\Sv\left|^{\nu)\lambda\alpha_2\beta_2}(-p,p)\mathcal{K}_{\mu\nu\rho\sigma}(0,0)\right|\Sigma\right)^{\rho\sigma}(0) \nonumber \\
&& -p^{\beta_1}\left(\Sv\left|^{\mu\nu}(0)\mathcal{K}^{\alpha_2\beta_2}_{\ \ \ \ \mu\nu\rho\sigma}(-p,p,0)\right|\Sigma\right)^{\rho\sigma}(0) \nonumber \\
&& +2p_{\lambda}\delta^{\beta_1(\mu}\left(\Sv\left|^{\nu)\lambda}(0)\mathcal{K}^{\alpha_2\beta_2}_{\ \ \ \ \mu\nu\rho\sigma}(-p,p,0)\right|\Sigma\right)^{\rho\sigma}(0).
\end{eqnarray}
These exactly cancel the momentum-violating terms from the 2-point Ward identities in $S$ and $\Sigma$ shown in (\ref{2pmomvio}) with the exception of the third term in (\ref{splitcancel}), which we will need shortly.
The 2-point kernel Ward identity, as applied to the fourth term in (\ref{general2ptflow}), gives us also some momentum-conserving non-cancelling pieces, which are
\begin{equation}\label{noncancel}
 p^{\beta_1}S^{\mu\nu}(0)\mathcal{K}^{\alpha_2\beta_2}_{\ \ \ \ \mu\nu\rho\sigma}(0,0,0)\Sigma^{\rho\sigma}(0) - 2p_{\lambda}\delta^{\beta_1(\alpha_2|}\Sv^{\mu\nu}\mathcal{K}^{|\beta_2)\lambda}_{\ \ \ \ \ \mu\nu\rho\sigma}(0,0,0)\Sigma^{\rho\sigma}(0).
\end{equation}
Finally, the last term in (\ref{general2ptflow}) contracts with $2p_{\alpha_1}$ such that we can use the 3-point action Ward identity to get
\begin{eqnarray}\label{finalcancel}
&& p^{\beta_1}\left(\Sv\left|^{\alpha_2\beta_2\mu\nu}(0,0)\mathcal{K}_{\mu\nu\rho\sigma}(0,0)\right|\Sigma\right)^{\rho\sigma}(0)\nonumber \\
&& -2p_{\lambda}\delta^{\beta_1(\alpha_2}\left(\Sv\left|^{\beta_2)\lambda\mu\nu}(0,0)\mathcal{K}_{\mu\nu\rho\sigma}(0,0)\right|\Sigma\right)^{\rho\sigma}(0)\nonumber \\
&& -2p_{\lambda}\delta^{\beta_1(\mu}\left(\Sv\left|^{\nu)\lambda\alpha_2\beta_2}(p,-p)\mathcal{K}_{\mu\nu\rho\sigma}(0,0)\right|\Sigma\right)^{\rho\sigma}(0).
\end{eqnarray}
The last term in (\ref{finalcancel}) is a momentum-violating term, which cancels with the third term in (\ref{splitcancel}), which was left over after the cancellations with (\ref{2pmomvio}). All of the momentum-violating contributions from kernel Ward identities that are listed in (\ref{splitcancel}) come from the first four terms in the general expression for the kernel Ward identities (\ref{kernWard}), which came from the non-zero diffeomorphism transformation of the kernel in (\ref{proptrans}). Putting together the non-cancelling pieces from (\ref{noncancel}) and (\ref{finalcancel}), we reproduce the right hand side of (\ref{Ward2pSdotflow}), as required.

\section{Metric determinant expansion in the metric perturbation}\label{MDIERGsqrtg}

Since the action carries a factor of $\sqrt{g}$ in its measure and the kernel carries a factor of $1/\sqrt{g}$, it is useful to establish the $n$-point expansion of $\sqrt{g}$ and its powers. The determinant of the metric can be expressed more conveniently as 
\begin{equation}
{\det{}^{l/2}(g_{\mu\nu})} = e^{\frac{l}{2}\tr\left(\ln\left(\delta_{\mu\nu}+h_{\mu\nu}\right)\right)}\,.
\end{equation}
For small $h_{\mu\nu}$, the logarithm can be expressed as a Taylor expansion. Taking the trace, we get
$h - \frac{1}{2}h_{\mu\nu}h^{\mu\nu} + \frac{1}{3}h_{\mu\nu}h^{\mu\rho}h^{\nu}_{\ \rho} - \cdots$.
Performing a Taylor expansion of the exponential, we get
\begin{equation}
\sqrt{g}^{\,l} = 1 + l\frac{h}{2}-l\frac{h_{\mu\nu}h^{\mu\nu}}{4}+l^2\frac{h^2}{8}+l\frac{h_{\mu\nu}h^{\mu\rho}h^{\nu}_{\ \rho}}{6}-l^2\frac{h_{\mu\nu}h^{\mu\nu}h}{8}+l^3\frac{h^3}{48}+\cdots
\end{equation}
The $n$-point expansion is then obtained via differentiation with respect to the metric perturbation in the usual way:
\begin{equation}
\label{one-point-cc}
\mathcal{S}_{c}^{\mu\nu} = \frac{l}{2}\delta^{\mu\nu},
\end{equation}
\begin{equation}
\label{two-point-cc}
\mathcal{S}_{c}^{\mu\nu\rho\sigma} = \frac{l^2}{4}\delta^{\mu\nu}\delta^{\rho\sigma} - \frac{l}{2}\delta^{\mu(\rho}\delta^{\sigma)\nu},
\end{equation}
\begin{eqnarray}
\label{three-point-cc}
{\mathcal{S}}_{c}^{\mu\nu\rho\sigma\alpha\beta} & = & \frac{l^3}{8}\delta^{\mu\nu}\delta^{\rho\sigma}\delta^{\alpha\beta}
+l\delta^{(\mu|(\rho}\delta^{\sigma)(\alpha}\delta^{\beta)|\nu)} \nonumber \\ &&
-\frac{l^2}{4}\left(\delta^{\mu\nu}\delta^{\rho(\alpha}\delta^{\beta)\sigma}
+\delta^{\rho\sigma}\delta^{\mu(\alpha}\delta^{\beta)\nu}
+\delta^{\alpha\beta}\delta^{\mu(\rho}\delta^{\sigma)\nu}\right).
\end{eqnarray}
Of particular interest is $l=1$, which gives us the $n$-point expansion of $\sqrt{g}$, \ie the $n$-point expansion of cosmological constant-like action terms. Relating this to the Ward identities for momentum-independent $n$-point functions (\ref{Wardmomind}), we see that (\ref{one-point-cc}), (\ref{two-point-cc}) and (\ref{three-point-cc}) are the forms that uniquely satisfy the Ward identities (\ref{Ward2ptCC}) and (\ref{Ward3ptCC}).

\section{Functional derivatives of the covariantized kernel}\label{MDIERGfuncder}

When discussing the kernel for gauge theories in Figure \ref{fig:Anpoint} and the discussion below it, we noted that covariantization introduced to the kernel an expansion in powers of the field. Similarly with gravity, the kernel expands out in $n$-point functions. To be able claim that we can exactly solve the flow equation iteratively from the 2-point level upwards, it is important to demonstrate that we can calculate the expansion in metric perturbations of a general form of the kernel. Consider the background-independent expression for the kernel in (\ref{kernel-grav}). It is easy to use the working in Section \ref{MDIERGsqrtg} or alternatively the momentum-independent Ward identities in (\ref{KernWardMomInd}) to compute the momentum-independent part of the kernel. However, it is more challenging to compute the expansion in $h$ induced by the covariant derivatives, in part because it comes from the metric connections induced by the tensor structure that the covariant derivatives act on, and in part because of the potentially infinite series of higher covariant derivatives that exist in $\KoL$. To illustrate the method, let us compute the $\mathcal{O}(h)$ part of the general form of $\dot{\Delta}$, which gives us the 1-point function.

The first step is to consider a single $-\nabla^2$ operator acting on a 2-component contravariant tensor $T^{\rho\sigma}$. In the flow equation, this contravariant tensor could be a ``field equation'' $\delta \Sigma / \delta g_{\mu\nu}$, or perhaps a derivative of it, \ie $\left(-\nabla^{2}\right)^m\delta \Sigma / \delta g_{\rho\sigma}$. We will consider the general case here. We can express this term at the 1-point level in momentum representation as
\begin{equation}\label{dalembertian}
(-\nabla^2)(p,r) T^{\rho\sigma}(-r) = H^{\alpha\beta \ \ \rho\sigma}_{\ \ \gamma\delta}(p,r)T^{\gamma\delta}(-r) h_{\alpha\beta}(p)\,,
\end{equation}
where $H^{\alpha\beta \ \ \rho\sigma}_{\ \ \gamma\delta}(p,r)$ is explicitly computed to give 
\begin{multline}\label{dalembert1pt}
H^{\alpha\beta \ \ \rho\sigma}_{\ \ \gamma\delta}(p,r)T^{\gamma\delta}(-r) h_{\alpha\beta}(p)  =  -\left(h^{\alpha\beta}(p)r_{\alpha}r_{\beta} - p_{(\alpha}r_{\beta)}h^{\alpha\beta}(p)+\frac{1}{2}p\cdot r h(p)\right)T^{\rho\sigma}(-r)  \\ 
 +\left((p^2 -2p\cdot r)h_{\lambda}^{\ (\rho|}(p) + p_{\lambda}(p_{\alpha}-2r_{\alpha})h^{\alpha (\rho|}(p)
  -p^{(\rho|}(p_{\alpha}-2r_{\alpha})h^{\alpha}_{\ \lambda}(p)\right)T^{|\sigma)\lambda}(-r). 
\end{multline}
As already hinted, $(-\nabla^2)^m T^{\rho\sigma}$ is also a contravariant tensor. This means that a series of $n$ such operators acting on $T^{\rho\sigma}$ can be evaluated at the $\mathcal{O}(h)$ level by using (\ref{dalembertian}) where the contravariant tensor used already contains $m$ $-\nabla^2$ operators and a further $n-m-1$ of these operators act as total derivatives on the whole of (\ref{dalembertian}). Moving to momentum representation, these extra $-\nabla^2$ operators are transformed simply to squared 4-momenta. More explicitly, we have
\begin{equation}
(-\nabla^2)^{n}(p,r) T^{\rho\sigma}(-r) = \sum_{m=0}^{n-1} |p-r|^{2(n-1-m)} H^{\alpha\beta \ \ \rho\sigma}_{\ \ \gamma\delta}(p,r)\, |r|^{2m}\, T^{\gamma\delta}(-r) h_{\alpha\beta}(p)\,.
\end{equation}
The new factor is simply the summation of a finite geometric progression:
\begin{equation}
 \sum_{m=0}^{n-1} |p-r|^{2(n-1-m)}|r|^{2m} = \frac{(p-r)^{2n}-r^{2n}}{(p-r)^2 -r^2}\,.
\end{equation}
Now all that remains is to generalize this to a function of squared covariant derivatives, which is straightforwardly
\begin{equation}
\dot{\Delta}(-\nabla^2)(p,r) T^{\rho\sigma}(-r) = \frac{\dot{\Delta}\left(|p-r|^2\right) - \dot{\Delta}(r^2)}{|p-r|^2-r^2}H^{\alpha\beta \ \ \rho\sigma}_{\ \ \gamma\delta}(p,r)T^{\gamma\delta}(-r)h_{\alpha\beta}(p)\,.
\end{equation}
This illustrates the principle by which the kernel can be expanded out to any $n$-point level, although the working and solutions will become increasingly complicated as one expands to higher order in $h$. As always, $n$-point functions are obtained from $\mathcal{O}(h^n)$ terms by means of functional differentiation. For the rest of this thesis, only the 0-point part of the kernel is of interest.
\section{Transverse 2-point functions}\label{MDIERGTran2}

We know from the Ward identities of momentum-independent $n$-point functions (\ref{KernWardMomInd}) that there only exists one momentum-independent 2-point function, which corresponds to a cosmological constant-like term. We also know from the full 2-point Ward identity (\ref{Ward1to2}) that all momentum-dependent 2-point functions are transverse, which relates to the masslessness of the graviton, as imposed by the diffeomorphism symmetry. Unlike the momentum-independent $n$-point functions, momentum-dependent $n$-point functions are not uniquely determined by the Ward identities. This section concerns the identification of allowed transverse 2-point functions for a diffeomorphism-invariant action. We will see that there exist two independent transverse 2-point structures with the required symmetry. Demanding a local expansion in powers of momentum, we actually find that only one allowed structure appears at quadratic order in momentum (corresponding to the Einstein-Hilbert action) but there exist two independent structures from quartic order onwards. The structure allowed at quadratic order is a linear combination of the two structures allowed at quartic order. Only even powers of momentum exist in the action because of Lorentz invariance. Momentum conservation allows us to write the 2-point function in terms of only a single momentum argument\footnote{The momentum propagating into a 2-point function at one end is equal to the momentum propagating out of it at the other.}. At quadratic order in momentum, there are four Lorentz invariant $\mathcal{O}(h^2)$ structures, which we can write as a general linear combination:
 \begin{equation}\label{linearquadratic}
 a_{1}h p^2 h + a_{2}h_{\alpha\beta}p^2 h^{\alpha\beta} + a_{3}hp_{\alpha}p_{\beta}h^{\alpha\beta} + a_{4}h^{\alpha\beta}p_{\alpha}p_{\gamma}h_{\beta}^{\ \gamma}\,,
\end{equation}
where $a_i$ are coefficients with a trivially zero diffeomorphism transformation: $\delta a_i=0$. To establish which linear combinations can feature in diffeomorphism-invariant actions, we perform the linearized diffeomorphism $\delta h_{\alpha\beta}\to -2ip_{(\alpha}\xi_{\beta)}$ and solve for those linear combinations that transform to zero:
\begin{eqnarray}\label{linearquadratictrans}
0 & = & 4a_{1}hp^2 p\cdot\xi + 4a_{2}h^{\alpha\beta}p^2 p_{\alpha}\xi_{\beta} + 2a_{3}hp^2 p\cdot\xi \nonumber \\ & & + 2a_{3}h^{\alpha\beta}p_{\alpha}p_{\beta}p\cdot\xi + 2a_{4}h^{\alpha\beta}p^2 p_{\alpha}\xi_{\beta} + 2a_{4}h^{\alpha\beta}p_{\alpha}p_{\beta}p\cdot\xi.
\end{eqnarray}
We only require the linearized transformation to vanish. This is because the the non-linear terms in the diffeomorphism transformation of the $\mathcal{O}(h^2)$ part of the action are involved in the cancellation of terms from the diffeomorphism transformation of the $\mathcal{O}(h^3)$ and higher parts of the action. In (\ref{linearquadratic}), the only set of coefficients (up to an overall factor) that solves (\ref{linearquadratictrans}) is $a_1 = -a_2 = -a_3/2 = a_4/2$. This is the $\mathcal{O}(h^2)$ part of the Einstein-Hilbert Lagrangian:
\begin{equation}
\label{EH2}
\cL^{(2)}_{\rm EH} =
  \frac{1}{2}\left(h_{\mu\nu}p^2 h^{\mu\nu}-hp^2 h+2h^{\mu\nu}p_{\mu}p_{\nu}h-2h^{\mu\nu}p_{\mu}p_{\rho}h_{\nu}^{\ \rho}\right) \,.
\end{equation}
More precisely, we can relate this Lagrangian term to its corresponding background-independent operator:
\begin{equation}
 \int_{x}\!\!\sqrt{g}\,\cO_2= -2\int_{x}\!\!\sqrt{g}R =  \int\! \dbar p \ 
\cL^{(2)}_{EH} +O(h^3)\,.
\end{equation}
This makes good sense because the Einstein-Hilbert action is the only diffeomorphism-invariant gravity operator at the 2-derivative level. An additional Lorentz-invariant 2-point structure emerges when we generalize to quartic or higher order in momentum, giving us a more general linear combination:
\begin{equation}
 b_{1}h^{\alpha\beta}p^4 h_{\alpha\beta} + b_{2}hp^4 h + b_{3}h^{\alpha\beta}p^2 p_{\alpha}p_{\beta}h 
 + b_{4}h^{\alpha\beta}p^2 p_{\alpha}p_{\gamma}h_{\beta}^{\ \gamma} + b_{5}h^{\alpha\beta}p_{\alpha}p_{\beta}p_{\gamma}p_{\delta}h^{\gamma\delta} \,.
\end{equation}
Setting the linearized diffeomorphism transformation to zero gives
\begin{eqnarray}
0 & = & 4b_{1}h^{\alpha\beta}p^4 p_{\alpha}\xi_{\beta} + 4b_{2}hp^4 p\cdot\xi + 2b_{3}h^{\alpha\beta}p^2 p_{\alpha}p_{\beta}p\cdot\xi \nonumber \\ & & 
2b_{3}hp^4 p\cdot\xi + 2b_{4}h^{\alpha\beta}p^4 p_{\alpha}\xi_{\beta} + 2b_{4}h^{\alpha\beta}p^2 p_{\alpha}p_{\beta}p\cdot\xi \nonumber \\ & & 
+ 4b_{5}h^{\alpha\beta}p^2p_{\alpha}p_{\beta}p\cdot\xi,
\end{eqnarray}
which is solved by $b_{5}= b_{1} + b_{2}$, $b_{4}=-2b_{1}$, $b_{3}=-2b_{2}$. This gives us two linearly independent forms for the $\mathcal{O}(h^2)$ Lagrangian at quartic order in momentum:
\begin{eqnarray}\label{transverseAB}
\cL^{(2)}_{a} &=& \frac{1}{2} 
\left(
h^{\mu\nu}p^4 h_{\mu\nu} - 2h^{\mu\nu}p^2 p_{\mu}p_{\rho}h_{\nu}^{\ \rho} + h^{\mu\nu}p_{\mu}p_{\nu}p_{\rho}p_{\sigma}h^{\rho\sigma}\right)\,,\\
\cL^{(2)}_{b} &=& \frac{1}{2} 
\left(
hp^4 h -2h^{\mu\nu}p^2 p_{\mu}p_{\nu}h +h^{\mu\nu}p_{\mu}p_{\nu}p_{\rho}p_{\sigma}h^{\rho\sigma}\right)\,.
\end{eqnarray}
These are the most general structures at $\mathcal{O}(h^2)$ of a momentum-dependent Lagrangian. To see this note that the only way to extend to $\mathcal{O}(p^6)$ and higher is to add factors of $p^2$, which is a scalar with trivial diffeomorphism transformation. Also, the $\mathcal{O}(h^2)$ Einstein-Hilbert Lagrangian is expressible as a linear combination of the Lagrangians in (\ref{transverseAB}). Let us denote this linear combination as $a\cL^{(2)}_a + b\cL^{(2)}_b$, where $a(p^2/\Lambda^2)$ and $b(p^2/\Lambda^2)$ are functions with zero diffeomorphism transformation that locality requires to be Taylor-expandable in $p^2/\Lambda^2$. The Einstein-Hilbert Lagrangian has $a=-b=1/p^2$, whereas the 2-point part of the $R_{\mu\nu\rho\sigma}R^{\mu\nu\rho\sigma}$ action term has $a=2, b=0$, the term from $R^2$ has $b=2,a=0$, and the action term from $R_{\mu\nu}R^{\mu\nu}$ has $a=b=1/2$. 

This information will prove very useful when designing the regularized propagator in the Einstein scheme. We can confirm from these coefficients that the three action terms that are quadratic in the Riemann tensor are related via the Gauss-Bonnet topological invariant (\ref{Gauss-Bonnet}) at the 2-point level. Indeed, these terms are related in the same proportions at every $n$-point level. This remains true at the 2-point level if one inserts extra $-\nabla^2$ operators into these structures: \eg $R_{\mu\nu\rho\sigma}\nabla^2 R^{\mu\nu\rho\sigma} + 4R_{\mu\nu}\nabla^2 R^{\mu\nu}+ R\nabla^2 R$, but not at 3-point or higher levels.
Using the usual method, we can express these $\mathcal{O}(h^2)$ functions as 2-point functions:
\begin{eqnarray}
 \label{EH2pt}
\mathcal{S}^{\mu\nu\rho\sigma}_{\rm EH}(-p,p) &=& p^2(\delta^{\mu(\rho}\delta^{\sigma)\nu}-\delta^{\mu\nu}\delta^{\rho\sigma})+p^{\mu}p^{\nu }\delta^{\rho\sigma}+\nonumber \\ && p^{\rho}p^{\sigma }\delta^{\mu\nu}-2p^{(\mu |}p^{(\rho}\delta^{\sigma)|\nu)}\,,\\
\label{a2pt}
\mathcal{S}^{\mu\nu\rho\sigma}_{a}(-p,p) &=& 
p^4\delta^{\mu(\rho}\delta^{\sigma)\nu}-2p^2 p^{(\mu |}p^{(\rho}\delta^{\sigma) |\nu)} +p^{\mu}p^{\nu}p^{\rho}p^{\sigma} \nonumber \\
&=& \left(p^2 \delta^{(\mu |(\rho}-p^{(\mu |}p^{(\rho}\right)\left(p^2 \delta^{\sigma)|\nu)}-p^{\sigma)}p^{|\nu)}\right)\,,\\
\label{b2pt}
\mathcal{S}^{\mu\nu\rho\sigma}_{b}(-p,p) &=& 
p^4\delta^{\mu\nu}\delta^{\rho\sigma}-p^2p^{\mu}p^{\nu}\delta^{\rho\sigma}-p^2p^{\rho}p^{\sigma}\delta^{\mu\nu}+p^{\mu}p^{\nu}p^{\rho}p^{\sigma} \nonumber
\\
&=& \left(p^2\delta^{\mu\nu}-p^{\mu}p^{\nu}\right)\left(p^2 \delta^{\rho\sigma}-p^{\rho}p^{\sigma}\right)\,.
\end{eqnarray}
All of these momentum-dependent 2-point functions are transverse, as implied by the 2-point Ward identity for a diffeomorphism-invariant action (\ref{Ward1to2}).

\section{Classical 2-point functions at fixed-points}\label{MDIERG2pta}

Let us now turn to the construction of the fixed-background formulations of the two schemes that were discussed background-independently in Subsections \ref{MDIERGWeylBI} and \ref{MDIERGEinsteinBI}. The Einstein scheme in Subsection \ref{MDIERGEinsteinBI} forbade a cosmological constant-like term in the fixed-point action by locality. The Weyl scheme in Subsection \ref{MDIERGWeylBI}, however gave us a choice of whether to include a cosmological constant-like term in the fixed-point action or not. Since this cosmological constant-like term is unphysical and indeed vanishes in the infrared limit, we chose not to include it even in Weyl scheme. For this reason, the fixed-background construction of both fixed-point actions begins at the 2-point level, where they exclusively use the transverse 2-point functions derived in Section \ref{MDIERGTran2}. As we will see, the construction of the 2-point function is also related to the construction of the kernel in each scheme.

\subsection{Classical 2-point function in Weyl scheme}\label{MDIERG2ptW}

We begin by setting a form for the seed Lagrangian. As with the background-independent form of the Weyl scheme, we will choose (\ref{Shat-4}). Unlike in the scalar or gauge cases, this (quite minimal) choice for the seed Lagrangian is already an infinite expansion in $n$-point functions. We cannot truncate this down to a purely 2-point seed Lagrangian without breaking diffeomorphism invariance. However, we are at liberty to keep it at quadratic order in the Riemann tensor. Provided that the classical fixed-point action matches the seed action up to quadratic order in the Riemann tensor, they also match at the 2-point level. Given (\ref{Shat-4}), we can express the 2-point functions as
\begin{equation}
\label{fullaction2}
 \mathcal{S}^{\alpha\beta\gamma\delta} = c^{-1}\mathcal{S}^{\alpha\beta\gamma\delta}_{a}+\left(1+ 4s\right)c^{-1}\mathcal{S}_{b}^{\alpha\beta\gamma\delta}\,,
\end{equation}
where the 2-point functions $\mathcal{S}^{\alpha\beta\gamma\delta}_{a}$ and $\mathcal{S}^{\alpha\beta\gamma\delta}_{b}$ are as defined in (\ref{a2pt}) and (\ref{b2pt}). Momentum arguments have been suppressed for convenient presentation. To make the working more manageable, let us split the flow equation into its cross-contracted and two-traces forms, as defined in (\ref{c.c.}), (\ref{t.t.}) and (\ref{flowratio}). The two-traces part of the 2-point flow is relatively simple:
\begin{equation}\label{f2ptt}
 \dot{\mathcal{S}}^{\alpha\beta\gamma\delta}|_{t.t.} = -16(1+3s)^2c^{-2}p^{4}\KoL\,\mathcal{S}^{\alpha\beta\gamma\delta}_{b}(p,-p)\,.
\end{equation}
However, the cross-contracted part contains both structures:
\begin{equation}
 \label{f2pcc}
\dot{\mathcal{S}}^{\alpha\beta\gamma\delta}|_{c.c.} =
   -4(1+2s)(1+6s)\,c^{-2}p^{4}\KoL\,\mathcal{S}^{\alpha\beta\gamma\delta}_{b}   - c^{-2} p^4 \KoL\left(\mathcal{S}^{\alpha\beta\gamma\delta}_{a}+\mathcal{S}^{\alpha\beta\gamma\delta}_{b}\right)\,.
\end{equation}
Putting these together in the proportion given by (\ref{flowratio}), we can then compare with the original 2-point function in (\ref{fullaction2}) to split the flow equation into flows for $\mathcal{S}^{\alpha\beta\gamma\delta}_{a}+\mathcal{S}^{\alpha\beta\gamma\delta}_{b}$ and $\mathcal{S}^{\alpha\beta\gamma\delta}_{b}$ individually. The flow for $\mathcal{S}^{\alpha\beta\gamma\delta}_{a}+\mathcal{S}^{\alpha\beta\gamma\delta}_{b}$ can be written as
\begin{equation}
 \label{diff1}
 \dot{\left(c^{-1}\right)} =-p^4 c^{-2} \KoL\,,
\end{equation}
and the flow for $\mathcal{S}^{\alpha\beta\gamma\delta}_{b}$ is 
\begin{equation}
\label{diff2}
 s\dot{\left(c^{-1}\right)} = -p^4 c^{-2} \KoL \left[4j(1+3s)^2+(1+2s)(1+6s)\right]\,.
\end{equation}
We can easily solve (\ref{diff1}) by setting
\begin{equation}
 \Delta(p^2) = \frac{c(p^2/\Lambda^2)}{p^4},
\end{equation}
which is the form of a regularized propagator for a theory where the bare kinetic term is of fourth order in derivatives, as is the case here. Turning to (\ref{diff2}), we can make it match (\ref{diff1}) exactly by setting $j= - (1+4s)/4(1+3s)$, as discussed in the background-independent description. Alternatively, we can set $s=-1/3$ for arbitrary $j$, although this is disfavoured since it requires a fixed value for $s$ that is slightly different to the fixed-point value implied in other studies \cite{Avramidi:1985ki,deBerredoPeixoto:2004if,Codello:2006in,Codello:2008vh}. 
 
\subsection{Classical 2-point function in Einstein scheme}\label{MDIERG2ptE}

Once again, we begin our construction by setting a value for the seed action. As with the background-independent formalism, we set the seed Lagrangian to be (\ref{biEHfull}). Setting the classical fixed-point action to match the seed action up to the quadratic order in the Riemann tensor, equivalently up to 2-point level, we have
\begin{equation}
\label{fullEaction2}
 \mathcal{S}^{\alpha\beta\gamma\delta} = \left(\frac{1}{p^2}+\frac{d}{\Lambda^2}\right)\mathcal{S}^{\alpha\beta\gamma\delta}_{a}+\left(-\frac{1}{p^2} + \left(1 + 4j\right)\frac{d}{\Lambda^2}\right)\mathcal{S}_{b}^{\alpha\beta\gamma\delta}\,.
\end{equation}
We will again find it convenient to split the flow equation into cross-contracted and two-traces forms. The two-traces part is
\begin{equation}\label{fEHtt}
 \dot{\mathcal{S}}^{\mu\nu\rho\sigma}|_{t.t.}(p,-p) = 
 -4p^{4}\left(\frac{1}{p^2}-2(1+3j)\frac{d}{\Lambda^2}\right)^2\!\KoL\,\mathcal{S}^{\alpha\beta\gamma\delta}_{b}\,.
\end{equation}
The cross-contracted part is
\begin{eqnarray}
 \label{fEHpcc}
\dot{\mathcal{S}}^{\alpha\beta\gamma\delta}|_{c.c.} & = &
4(1+2j)p^4\frac{d}{\Lambda^2}\left(\frac{2}{p^2}-(1+6j)\frac{d}{\Lambda^2}\right)
\KoL\,\mathcal{S}^{\alpha\beta\gamma\delta}_{b}  \nonumber \\ && 
- p^4\left(\frac{d}{\Lambda^2} +\frac{1}{p^2}\right)^2
 \!\KoL\left(\mathcal{S}^{\alpha\beta\gamma\delta}_{a}
+\mathcal{S}^{\alpha\beta\gamma\delta}_{b}\right)\,.
\end{eqnarray}
The appearance of $j$ in (\ref{fEHtt}) and (\ref{fEHpcc}) may be confusing at first glance: $j$ appears because of the original form of (\ref{biEHfull}).
Once again, we combine these two parts of the flow equation again in the proportion in (\ref{flowratio}) and then split it via (\ref{fullEaction2}) into an $\mathcal{S}^{\alpha\beta\gamma\delta}_{a}+\mathcal{S}^{\alpha\beta\gamma\delta}_{b}$ flow and an $\mathcal{S}^{\alpha\beta\gamma\delta}_{b}$ flow. The flow for structure $\mathcal{S}^{\alpha\beta\gamma\delta}_{a}+\mathcal{S}^{\alpha\beta\gamma\delta}_{b}$ and structure $\mathcal{S}^{\alpha\beta\gamma\delta}_{b}$ can be written respectively as
\begin{eqnarray}
 \label{diffE1}
\Lambda\partial_{\Lambda}\left(\frac{d}{\Lambda^2}\right) &=& -p^4 \KoL\left(\frac{d}{\Lambda^2} +\frac{1}{p^2}\right)^2\,,\\
\label{diffE2}
 j\Lambda\partial_{\Lambda}\left(\frac{d}{\Lambda^2}\right) &=& -p^4 \KoL\left(\frac{j}{p^4}+(1+12j+36j^2+36j^3)\frac{d^2}{\Lambda^4} \right.\nonumber \\ && \left.-2(1+4j+6j^2)\frac{d}{p^2 \Lambda^2}\right).
\end{eqnarray}
Following the method used for the Weyl scheme, we begin by solving the (\ref{diffE1}) for the effective propagator and then solving (\ref{diffE2}) for $j$. Solving (\ref{diffE1}) to consistently give us the effective propagator $\Delta = c/p^2$, where the cutoff function $c$ is expressed in terms of $d$, we get
\begin{equation}
 c= \frac{1}{1+d\,p^2\!/\Lambda^2}\,.
\end{equation}
This can be expanded at first order to confirm that $c'(0)=-d(0)$, as can be seen by inspection of (\ref{biEHfull}). Solving (\ref{diffE2}) to match (\ref{diffE1}), we find two solutions for $j$ in the fixed-background formalism. Either $j=-1/2$ or $j=-1/3$. The $j=-1/3$ option is unattractive when we compare with the background-independent form. To see why, consult (\ref{EH2pt}) to see that the 2-point structure of the Einstein-Hilbert term is proportional to the 2-point structure of a linear combination of the $R^2$ and $R_{\alpha\beta}R^{\alpha\beta}$ terms. When we construct the regularized 2-point function, we would like the same index structure for the 2-point function to apply at all orders in momenta (\ie we have a single momentum cutoff function regulating a single 2-point structure). This is true for the $j=-1/2$ structure but not for the $j=-1/3$ structure, since the ratio of $R_{\alpha\beta}R^{\alpha\beta}$-like to $R^2$-like 2-point structure is $-1/2$ in the Einstein-Hilbert term and $j$ in the higher-order terms. Choosing $j=-1/2$, the 2-point function for the classical fixed-point action is simply the UV-regulated Einstein-Hilbert 2-point function:
\begin{equation}
 \mathcal{S}^{\alpha\beta\gamma\delta}(p,-p) = c^{-1} \mathcal{S}^{\alpha\beta\gamma\delta}_{\rm EH}\,.
\end{equation}

\section{Further regularization via supermanifolds}\label{MDIERGSUSY}

In this chapter, we have developed the manifestly diffeomorphism-invariant ERG at the classical level and outlined how to introduce quantum corrections in a loopwise expansion. By analogy with the manifestly gauge-invariant ERG, the regularization of the propagator via a smooth UV momentum cutoff function $c\left(p^2/\Lambda^2\right)$ is not expected to be sufficient to remove all UV divergences. In the gauge-invariant ERG discussed in Section \ref{MGIERGSUSY}, the gauge field was promoted to a super-matrix (\ref{supergaugefield}) that featured fermionic fields that effect a Pauli-Villars regularization. This is analogous to how Parisi-Sourlas supersymmetry was used as a formal device in Subsection \ref{IntroPSSUSY} to describe the loss of degrees of freedom in a spin system forced by a Gaussian-distributed random external field with vanishing correlation length. Motivated by the success of SU($N|N$) supersymmetry broken at the scale set by $\Lambda$, as used for UV-regulating Yang-Mills theories, future work will likely benefit from using a similar regulator for gravity. The gravity field is the metric itself, therefore extending the metric (in Euclidean signature) to include fermionic degrees of freedom also involves extending the spacetime coordinates to include fermionic components. This can be written as
\begin{equation}
 x^A = (x^\mu,\theta^a)\,,
\end{equation}
where $x^\mu$ are the familiar set of bosonic (\ie commuting) coordinates, $\theta^a$ are the new fermionic (anticommuting) coordinates and $x^A$ are the complete set of supercoordinates. The Lorentz index $\mu$ ranges from 1 to $D$ in $D$ dimensions. The fermionic index $a$ ranges from $D+1$ to $2D$, giving us a total of $2D$ components in $x^A$. We would then express the invariant interval as
\begin{equation}
 ds^2 = dx^A g_{AB} dx^B\,,
\end{equation}
where the fermionic components require us to be careful about the ordering of coordinates and fields. The super-metric $g_{AB}$ can be understood by splitting it into four pieces. The quadrant carrying two bosonic indices $g_{\mu\nu}$ is the familiar form of the metric as it usually appears in GR. As a symmetric tensor, it carries $D(D+1)/2$ degrees of freedom. The quadrant carrying two fermionic indices $g_{ab}$ is itself a bosonic field which, being antisymmetric, carries $D(D-1)/2$ degrees of freedom. The two off-diagonal quadrants both carry one bosonic index and one fermionic index and they are both fermionic fields. These fermionic fields are constrained by $g_{\mu a}=-g_{a\mu}$, so they only carry $D^2$ degrees of freedom. Putting this all together, we can see that there are $D^2$ bosonic degrees of freedom that are cancelled by $D^2$ fermionic degrees of freedom where there exists an unbroken supersymmetry between the bosonic and fermionic degrees of freedom. As with the gauge case, we would want to break this supersymmetry and decouple the bosonic and fermionic degrees of freedom at the cutoff scale $\Lambda$. This would be done using a Higgs-like mechanism induced by a super-scalar mode akin to (\ref{superscalar}). 

Supermanifolds have already been extensively studied as mathematical objects \eg \cite{Rogers:1979vp,Leites:1980rna,dewitt1992supermanifolds}. This procedure should not be mistaken for a supergravity theory: the fermionic modes are still spin-2 tensor modes, in contradiction with the spin-statistics theorem \cite{Pauli:1940zz}. This is acceptable because none of these fermionic degrees of freedom are physical, neither in the length vector $x^A$, nor in the super-metric $g_{AB}$: they are part of the regularization, which is purely a formal device with no physical meaning in itself. The physical gravity theory is recovered below the cutoff scale $\Lambda$, where the bosonic and fermionic modes decouple and the real physics only concerns the pure bosonic metric $g_{\mu\nu}$. This is similar to the gauge case, except that the two bosonic metric quadrants are not duplicates of each other, rather the fermionic-index part $g_{ab}$ is also unphysical, but is nevertheless required as part of the cancellation with the fermionic modes in the high-energy limit. Finally, it is important to note that, at the time of writing, the supermanifold regularization discussed here has not yet been implemented in a fully-constructed manifestly diffeomorphism-invariant quantum ERG, so the content of this section on supermanifolds is currently a conjecture.

\section{Outlook}

In this chapter, we have constructed the manifestly diffeomorphism-invariant classical ERG in analogy to the manifestly gauge-invariant ERG reviewed in Chapter \ref{AWHPreview1}. This is a continuous Wilsonian RG method that constructs effective actions and $\beta$-functions by means of smoothly integrating out high-energy modes down to a cutoff scale $\Lambda$ without fixing a gauge. The lack of gauge-fixing also gave us the opportunity to construct both fixed-background and background-independent formalisms that are equivalent and complementary to each other. Although we have focussed on the classical construction, we have also outlined the strategy for continuing to a fully quantum gravity construction. We have constructed classical fixed-point actions in two different schemes, which are the Weyl and Einstein schemes. We have also discussed flows away from the fixed-point via the introduction of relevant operators at the classical level and how these can ultimately relate to the physical cosmological constant and Newton's constant. More abstractly, we have discussed how the flow parameter $j$ determines the balance of modes in the flow equation, commenting on the special cases for conformally reduced gravity and unimodular gravity as well as determining the values in the Einstein and Weyl schemes. 

We have also noted that the classical construction corresponds to the limit $\hbar\to0$ or indeed to the limit where gravitational interaction is strongly suppressed, \ie the limit of weak coupling. In all currently performed tests of GR, the limit of weak coupling holds, therefore the classical theory is an excellent approximation to observed reality. Indeed, Einstein's original formulation of GR, expressed in (\ref{EinsteinGravAct}), is consistent with every experiment so far. With that said, higher derivatives are already well-motivated phenomenologically as a mechanism for early-universe inflation, as discussed in Subsection \ref{IntroInflation}. As we have seen in this chapter, these higher-derivative Lagrangian operators, \eg $R^2$ or $R_{\alpha\beta}R^{\alpha\beta}$, already appear in the effective action at the classical level and are thus well-motivated theoretically even before quantum corrections become important. 

Since physics is an experimental science, it is fitting that this thesis will now turn to discuss the cosmological implications of these higher-derivative operators. In particular, we will explore the subject of ``cosmological backreaction'', in which physics at very short scales of length can impact on large-scale behaviour via non-linear effects, even though the original linear (metric) fluctuations average to zero. This interplay of physics at different scales of length presents an interesting opportunity for high-energy physics to become apparent in low-energy phenomenology. This new research direction for the thesis relates to the content of this chapter we now finishing because we will be concerned with effective theories of gravity expressed as local expansions in higher-derivative operators, just as we have constructed in the Weyl and Einstein schemes. 

\chapter{Review of cosmological backreaction}\label{AWHPreview2}

\section{Conceptual overview}
As discussed in Subsection \ref{IntroHomoIso}, the standard cosmological model, also known as the concordance model, assumes that we can approximate the universe as homogeneous and isotropic at large scales. This is supported by astronomical evidence, as gathered by galaxy surveys or temperature mapping of the CMB. However, having homogeneity and isotropy at large scales does not rule out the possibility that large-scale behaviour is significantly affected by inhomogeneity at small scales. As has already been noted at the end of Subsection \ref{IntroHomoIso}, the transition to homogeneity occurs at approximately at the 100$h^{-1}$Mpc scale \cite{Hogg:2004vw,Sarkar:2009iga,Yadav:2005vv,Scrimgeour:2012wt,Pandey:2015htc,Pandey:2015xea,Laurent:2016eqo}, below which the universe is not homogeneous. We know that this inhomogeneity becomes very apparent as we move to smaller scales, with the density of the proton being roughly 2$\times 10^{17}$ kg/m$^3$ compared with a cosmological average density of order $10^{-26}$ kg/m$^3$. 

In this thesis, we have already discussed problems of physics at many scales of length, noting in particular how it is not practically feasible to perform computations using every individual microscopic degree of freedom when there are many degrees of freedom. When developing RG methods, we discussed how such problems are addressed by averaging over microscopic degrees of freedom, allowing us to obtain macroscopic observables, which only depend on a greatly reduced set of macroscopic degrees of freedom. Similarly, cosmologists interested in the overall expansion rate of the universe do not attempt to compute the individual gravitational influence of every star and planet, certainly not of every atom or subatomic particle. Rather, they consider an average over all of these microscopic fluctuations (though not normally using RG methods). Thus they solve the gravitational field equation, usually (\ref{EinsteinGravFieldEq}), using only the averaged form of the stress-energy tensor with all microscopic fluctuations already averaged out. Since the universe is homogeneous and isotropic at scales greater than 100$h^{-1}$Mpc, it follows that the FLRW metric expressed in (\ref{FLRWmetric}) is a good approximation at large scales \cite{Green:2014aga}.

There is a problem with this simple reasoning: the problem is not that the metric does not average to (\ref{FLRWmetric}) at large scales, but rather that we have not established whether or not we can neglect the non-linear contributions to the field equation from metric perturbations, such as those discussed in Subsection \ref{IntroMetricPert}. To illustrate the point, consider a simple function like $f(x)={\rm sin}(x/\lambda)$. The average value of this function over a large domain in $x$ tends to zero, however this is not true of $f^2(x)$. Similarly, if we split the metric into a background plus perturbation, as done in (\ref{MetricPertDef}), we can choose a background such that the average over the linear order in $h$ tends towards zero at large scales, but that does not automatically imply that $\mathcal{O}(h^2)$ and higher contributions to the field equation are also vanishing when averaged over large scales. Cosmological backreaction in this context is to be viewed as the macroscopic effect on the field equations introduced by non-linear terms constructed from microscopic perturbations to the metric induced by inhomogeneity in the stress-energy tensor. These non-linear terms can be collected together and averaged over to create a new ``effective stress-energy tensor'' for backreaction in our chosen gravity theory. This is the notion of backreaction that we will focus on in this thesis. 

Backreaction is sometimes motivated as a possible means to induce accelerating expansion into the late universe without introducing exotic new physics, answering the ``why now?'' problem by noting that the necessary level of inhomogeneity is only achieved at late times. For examples of the many studies that have appeared on the subject, see \cite{Zalaletdinov:1992cg,Zalaletdinov:1996aj,Buchert:1995fz,Buchert:1999er,Rasanen:2003fy,Alnes:2005rw,Coley:2006kp,Schwarz:2010px,Roy:2011za,Clarkson:2011zq,Buchert:2011sx,Clifton:2013vxa,Roukema:2013cya,Buchert:2015wwr,Bentivegna:2015flc,Sanghai:2016ucv}.
We will not be concerned in this thesis with exact inhomogeneous solutions to GR that do not average to FLRW at large scales. Nor will we be concerned with claims that there exist important non-local effects that cannot be captured by an averaging scheme. Following the discussions of locality in Chapters \ref{AWHPreview1} and \ref{ChapterMDIERG}, we will consider that any gravity theory whose Lagrangian is Taylor-expandable in the Riemann tensor and its covariant derivatives is sufficiently local that its behaviour at large scales should be well-described by a carefully-constructed averaging scheme. Moreover, for theories that do not possess such a notion of locality, it is not clear if the calculation of any observables is feasible at all, since computing every microscopic degree of freedom at every scale and arbitrarily large separations is beyond our capability to do.

\section{Notation}

Let us introduce some notation now that we will stick to for the rest of this thesis. The covariant derivative compatible with the full metric will be denoted from now on as $D_\mu$. This covariant derivative acts on a tensor $T_{\alpha_1\cdots\alpha_m}^{\ \ \ \ \ \ \ \beta_1\cdots\beta_n}$ such that
\begin{eqnarray}\label{fullcovderiv}
 D_\mu T_{\alpha_1\cdots\alpha_m}^{\ \ \ \ \ \ \ \beta_1\cdots\beta_n} & = & \partial_\mu T_{\alpha_1\cdots\alpha_m}^{\ \ \ \ \ \ \ \beta_1\cdots\beta_n} - \sum_{i=1}^{m}\Gamma^{\lambda}_{\ \ \mu\alpha_i}T_{\alpha_1\cdots\lambda\cdots\alpha_m}^{\ \ \ \ \ \ \ \ \ \ \ \beta_1\cdots\beta_n} \nonumber \\ && 
+ \sum_{i=1}^{n}\Gamma^{\beta_i}_{\ \ \mu\lambda}T_{\alpha_1\cdots\alpha_m}^{\ \ \ \ \ \ \ \beta_1\cdots\lambda\cdots\beta_n}.
\end{eqnarray}
The covariant derivative compatible with the background metric $g^{(0)}_{\mu\nu}$ will be denoted from now on as $\nabla_\mu$. Using this kind of derivative in place of the partial derivative on the right hand side of (\ref{fullcovderiv}), we have
\begin{eqnarray}
 D_\mu T_{\alpha_1\cdots\alpha_m}^{\ \ \ \ \ \ \ \beta_1\cdots\beta_n} & = & \nabla_\mu T_{\alpha_1\cdots\alpha_m}^{\ \ \ \ \ \ \ \beta_1\cdots\beta_n} - \sum_{i=1}^{m}C^{\lambda}_{\ \ \mu\alpha_i}T_{\alpha_1\cdots\lambda\cdots\alpha_m}^{\ \ \ \ \ \ \ \ \ \ \ \beta_1\cdots\beta_n} \nonumber \\ && 
+ \sum_{i=1}^{n}C^{\beta_i}_{\ \ \mu\lambda}T_{\alpha_1\cdots\alpha_m}^{\ \ \ \ \ \ \ \beta_1\cdots\lambda\cdots\beta_n},
\end{eqnarray}
where $C^{\alpha}_{\ \beta\gamma}$ is the difference between full and background connections, which is given by
\begin{equation}\label{diffconnect}
 C^{\alpha}_{\ \beta\gamma} = \frac{1}{2}g^{\alpha\lambda}\left(\nabla_\beta h_{\gamma\lambda}+\nabla_\gamma h_{\beta\lambda}-\nabla_\lambda h_{\beta\gamma}\right).
\end{equation}
Using (\ref{diffconnect}), we can split the full Ricci tensor into its background and perturbation parts by
\begin{equation}
 R_{\alpha\beta} = R^{(0)}_{\alpha\beta} - 2\nabla_{[\alpha}C^{\gamma}_{\ \gamma]\beta} + 2C^{\gamma}_{\ \beta[\alpha}C^{\delta}_{\ \delta]\gamma}.
\end{equation}
The linearized Riemann tensor is defined by
\begin{equation}\label{LinRie}
 R^{(1)}_{\alpha\beta\gamma\delta} := -2\nabla_{[\alpha|}\nabla_{[\gamma}h_{\delta]|\beta]}. 
\end{equation}
The linearized Ricci tensor can be expressed as
\begin{equation}\label{LinRict}
 R^{(1)}_{\alpha\beta} := \frac{1}{2}\left(2\nabla_\gamma \nabla_{(\alpha}h_{\beta)}^{\ \ \gamma} - \nabla^2 h_{\alpha\beta} - \nabla_\alpha\nabla_\beta h\right).
\end{equation}
contracting the indices of (\ref{LinRict}) using the background metric, we obtain the linearized Ricci scalar:
\begin{equation}\label{LinRics}
 R^{(1)} := \nabla_\alpha \nabla_\beta h^{\alpha\beta} - \nabla^2 h.
\end{equation}
The linearized Riemann tensor and its contractions are all invariant under linearized diffeomorphisms:
\begin{equation}\label{LinDiff}
 \delta h_{\alpha\beta} = 2\nabla_{(\alpha}\xi_{\beta)},
\end{equation}
Indices will be raised and lowered by the full metric until the field equation is split into background and perturbation parts, whereupon indices will be raised and lowered by the background metric. In particular, we will consistently use $D^2:=g^{\mu\nu}D_\mu D_\nu$ and $\nabla^2:=g^{(0)\mu\nu}\nabla_\mu \nabla_\nu$.

\section{Isaacson's shortwave approximation}\label{Isaacson}

The method we will use to study backreaction in this thesis is Green and Wald's weak-limit averaging framework \cite{Green:2010qy,Green:2013yua}, which is an adaptation of formalism originally developed for studying gravitational waves. We should bear this in mind because the result for backreaction in Einstein's gravity is rather unsurprising when this historical point is considered. We will see that the backreaction stress-energy tensor is radiation-like, see (\ref{FluidEqState}) and the comments beneath it, just as gravitational waves in GR are radiation-like: it is the same stress-energy tensor. Also, when we extend the Green and Wald formalism to higher derivatives, our result for the backreaction stress-energy tensor in more general theories will be the same as the stress-energy tensor for gravitational waves in those theories. Thus the remaining chapters of this thesis can be motivated either by interest in backreaction or by interest in gravitational waves in higher-derivative theories, the results are identical for both.

We begin this narrative with Isaacson's study of gravitational radiation in the limit of high frequency published in two parts in 1967 and 1968 \cite{Isaacson:1967zz,Isaacson:1968zza}. Aspects of this construction will be carried over into the backreaction study, while some of its idiosyncracies will be discarded as we progress to Burnett's more mathematically rigorous and precise formulation \cite{Burnett:1989gp}. This will not be a complete review of Isaacson's calculations, but will rather focus on the details of interest to this thesis. As an example, Isaacson specialised to Lorenz gauge, whereas this thesis will avoid gauge-fixing completely. 

Isaacson was motivated by gravitational waves emitted by strong gravitational sources, explicitly listing neutron stars, collapsing supernovae and quasars. In particular, he was interested in the limit where the wavelength is much less than the curvature radius of the wavefront. When discussing gravitational radiation propagating (and possibly detected) a large distance from the source, this approximation loses very little precision.

Isaacson's method begins by separating the metric into a background plus a perturbation, as seen in (\ref{MetricPertDef}). The background need not be specified, but a condition is placed that the leading order part of the perturbation is proportional to a small parameter $\epsilon$. Acting on the metric with a derivative operator extracts the length scale of its spacetime variations. To illustrate the idea, let us briefly use some less precise language and suppress Lorentz indices:
\begin{equation}\label{IsaacScales}
 \partial g^{(0)} \sim g^{(0)}/L, \ \ \partial h \sim h/\lambda,
\end{equation}
where $L$ is the length scale of curvature in the background metric and $\lambda$ is the wavelength of gravitational waves propagating on the background. It is assumed that $L$ is much larger than $\lambda$, \ie the gravitational waves are high-frequency perturbations to a slowly varying background. Isaacson makes a distinction between ``gravitational waves'', which are general metric perturbations of some wavelength $\lambda$ and ``gravitational radiation''. The latter are the gravitational waves that escape the system, asymptotically approaching a flat-spacetime horizon also known as future null infinity. Since the Isaacson formulation is concerned with this latter kind of gravitational wave, the matter stress-energy tensor is set to zero, leaving the stress-energy tensor for gravitational radiation dominant in the field equation. Recalling that the Einstein field equation can be written in Ricci form as (\ref{EinsteinFieldEqRicci}), the Isaacson formulation has
\begin{equation}
 R_{\mu\nu}^{(0)} + R_{\mu\nu}^{(1)} = \kappa\left(t_{\mu\nu} -\frac{1}{2}g^{(0)}_{\mu\nu}t\right),
\end{equation}
where $R^{(0)}_{\mu\nu}$ is the Ricci tensor for the background metric, $R^{(1)}_{\mu\nu}$ is the linearized Ricci tensor and $t_{\mu\nu}$ is the effective stress-energy tensor due to gravitational waves, built out of the quadratic and higher parts of the expansion in $h$, which is taken to be the main source of the background curvature.

For a perturbation with a given Fourier mode, the linear order in $h$, being an odd function, averages to zero on large scales. So the linear order is not expected to contribute to the effective stress energy tensor that induces the background curvature. By this reasoning, the background curvature is sourced by the $\mathcal{O}(h^2)$ part of the field equation, which should then have its leading-order contribution at the zeroth order in $\epsilon$. The linear part, which (as we will see) is divergent as $\epsilon\to0$, is expected to satisfy a wave equation that ensures that all divergent contributions to the field equation cancel:
\begin{equation}\label{IsaacWave}
 R_{\mu\nu}^{(1)} = 0.
\end{equation}
This is a very strong assertion that plays an important r\^{o}le, but we will see that it is not completely rigorous, and a much more nuanced version of it will appear in the eventual backreaction formalism. To see why we will wish to revise this, note that although the leading part of $R_{\mu\nu}^{(1)}$ is divergent, it may also contain non-divergent sub-leading parts that affect the field equation at other orders. The field equation can now be simplified to
\begin{equation}
 R^{(0)}_{\mu\nu} -\frac{1}{2}g_{\mu\nu}^{(0)}R^{(0)} = \kappa t_{\mu\nu}.
\end{equation}
The background metric is $\mathcal{O}(1)$ in sense that its components are dimensionless numbers that are close to unity. From (\ref{IsaacScales}), we know that $R_{\mu\nu}^{(0)}\sim 1/L^2$ and that the $\mathcal{O}(h^2)$ part of $t_{\mu\nu}$, which is the leading order when $\epsilon$ is small, is of order $\epsilon^2 /\lambda^2$. Isaacson's formulation matches these terms by asserting that $\lambda = \epsilon L$, which corresponds to the case where the background spacetime curvature is mainly induced by the stress-energy tensor of the gravitational waves. Let us write the limit of $t_{\mu\nu}$ where $\epsilon\to0$ as $t^{(0)}_{\mu\nu}$. This limit selects the quadratic part of $t_{\mu\nu}$ in $h$. This leading order part of $t_{\mu\nu}$ can be written, subject to the constraint in (\ref{IsaacWave}), as
\begin{equation}\label{IsaacUnrefinedStress}
 \kappa t^{(0)}_{\mu\nu} = R^{(2)}_{\mu\nu} - \frac{1}{2}g_{\mu\nu}^{(0)}R^{(2)},
\end{equation}
where $R^{(2)}_{\mu\nu}$ and $R^{(2)}$ are the expansion to quadratic order in $h$ of the Ricci tensor and Ricci scalar respectively. In common with much of the discussion in this thesis, Isaacson noted that the microscopic fluctuations of the perturbation field are not of interest at large scales. Therefore Isaacson made use of the Brill-Hartle spacetime averaging scheme \cite{Brill:1964zz} to find the macroscopic form of the stress-energy tensor. In the Brill-Hartle scheme, the fluctuating tensor field tensor is averaged via an integral over a spacetime volume. Using integration by parts, we can re-express the stress-energy tensor in simpler forms. Total derivatives of the form $\nabla_\alpha \left(h_{\mu\nu}\nabla_\beta h_{\rho\sigma}\right)$ have their order in $\epsilon$ raised by one after performing a spacetime integration, since the total derivative operator can be moved away from the metric perturbations using integration by parts: in this way it does not contribute a factor of $1/\lambda$ by the reasoning in (\ref{IsaacScales}). Such total derivatives vanish in the $\epsilon\to0$ limit. Additionally, we can rearrange other terms into more convenient forms using integration by parts in the $\epsilon\to0$ limit:
\begin{equation}
 \left<\nabla_\alpha h_{\mu\nu}\nabla_\beta h_{\rho\sigma}\right> = - \left<h_{\mu\nu}\nabla_\alpha \nabla_\beta h_{\rho\sigma}\right>.
\end{equation}
The reasoning here is awkward because the boundary is assumed to be flat: ultimately, we will want to avoid this assumption. Using this averaging behaviour, we can write the spacetime average of $R^{(2)}_{\mu\nu}$ as
\begin{equation}\label{IsaacSecond}
 \left<R^{(2)}_{\mu\nu}\right> = \left<-\frac{1}{2}h^{\alpha\beta}R^{(1)}_{\mu\alpha\nu\beta} - \frac{1}{4}h_{\mu\nu}R^{(1)}+R^{(1)}_{\alpha(\mu}h_{\nu)}^{\ \ \alpha}\right>,
\end{equation}
noting that indices are raised and lowered with the background metric. Note, however, that the quadratic part of the Ricci scalar is not just the trace of (\ref{IsaacSecond}) with the background metric, since the perturbation of the inverse metric that contracts the indices of the Ricci tensor also forms part of the expansion in $h$:
\begin{equation}
 \left<R^{(2)}\right> = - \left<\frac{1}{4}\left(hR^{(1)}+2h^{\alpha\beta}R_{\alpha\beta}^{(1)}\right)\right>.
\end{equation}
Using the wave equation in (\ref{IsaacWave}), we can simplify (\ref{IsaacSecond}) down to just its first term. This in turn simplifies (\ref{IsaacUnrefinedStress}) down to its Brill-Hartle averaged form:
\begin{equation}\label{IsaacStress}
 \kappa t^{\rm BH}_{\mu\nu} = \left<\frac{1}{2}h^{\alpha\beta}R_{\mu\alpha\nu\beta}^{(1)}\right>.
\end{equation}
The trace of this is $\left<\frac{1}{2}h^{\alpha\beta}R_{\alpha\beta}^{(1)}\right>$, which vanishes because of the wave equation (\ref{IsaacWave}). Thus the averaged effective stress-energy tensor for high-frequency gravitational radiation in Einstein gravity is traceless, as expected for radiation generally, see Subsection \ref{IntroStressEnergy}, especially the comments following equation (\ref{FluidEqState}). The averaged field equation then becomes
\begin{equation}\label{IsaacField}
 R^{(0)}_{\mu\nu} -\frac{1}{2}g_{\mu\nu}^{(0)}R^{(0)} = \left<\frac{1}{2}h^{\alpha\beta}R_{\mu\alpha\nu\beta}^{(1)}\right>.
\end{equation}
In this form, the stress-energy tensor is diffeomorphism-invariant, given the $\epsilon\to0$ limit, the averaging scheme and the wave equation. However, we will need to reformulate this more rigorously to proceed with the adaptation to backreaction.


\section{Burnett's weak-limit averaging formalism}\label{Burnett}

We turn now to Burnett's 1988 weak-limit averaging scheme, which reformulates Isaacson's high-frequency approximation in a more precise manner \cite{Burnett:1989gp}. Once again, the formalism is concerned with low-amplitude, high-frequency gravitational radiation emitted by dynamical astrophysical processes such as collapsing stars, stellar binaries or coalescing black holes. Again, the notation here will differ from Burnett's notation: the purpose of this section is to discuss how Burnett's formulation is an improvement on Isaacson's. The language here will be less mathematically formal than Burnett's, since we ultimately wish to extend to higher derivatives with minimal notational baggage. The Burnett formalism is concerned with a one-parameter family of metrics $g_{\mu\nu}(x,\lambda)$ that solve the field equation, parametrized by a positive definite variable $\lambda$. This $\lambda$ is physically related to the wavelength of perturbations, as seen in the Isaacson formulation, but it is treated more abstractly here. As $\lambda\to0$, $g_{\mu\nu}(x,\lambda)$ converges uniformly to a background metric $g_{\mu\nu}(x,0)$. Thus we can adapt (\ref{MetricPertDef}) to
\begin{equation}
 g_{\mu\nu}(x,\lambda) = g_{\mu\nu}(x,0) + h_{\mu\nu}(x,\lambda).
\end{equation}
In this way, $\lambda$ parametrizes the metric perturbation, which vanishes in the limit $\lambda\to0$. More precisely, $h_{\mu\nu}(x,\lambda)$ is uniformly bounded by a constant times $\lambda$ for sufficiently small $\lambda$, \ie the leading order part of $h_{\mu\nu}(x,\lambda)$ is of $\mathcal{O}(\lambda)$. More generally, it will be a convenient shorthand for any quantity that is uniformly bounded by a constant times $\lambda^n$ in the small $\lambda$ regime to refer to it as being of $\mathcal{O}(\lambda^n)$. Since the background metric is independent of $\lambda$, it is of $\mathcal{O}(1)$ in this language. It is also assumed that $\nabla_\alpha h_{\mu\nu}$ is of $\mathcal{O}(1)$. For constructing the effective stress-energy tensor, which will once again be the main source of background curvature, it is required that terms of the form $h_{\mu\nu}\nabla_\alpha\nabla_\beta h_{\rho\sigma}$ converge uniformly under averaging in the limit $\lambda\to0$ to a smooth averaged tensor field that is $\mathcal{O}(1)$. However, $\nabla_\alpha\nabla_\beta h_{\mu\nu}$ is of $\mathcal{O}(\lambda^{-1})$, so it is divergent in the $\lambda\to0$ limit, just as in the Isaacson formulation. The background Ricci tensor, being constructed only from the background metric, is $\mathcal{O}(1)$.
From this, we can see that $\lambda$ is related to the wavelength of perturbations, as indeed it appears in the Isaacson scheme.

Up to this point, the setup is essentially the same as Isaacson's but Burnett raised some questions about the rigour of Isaason's approach. Firstly, he asked whether it is rigorous to assert that (\ref{IsaacWave}) holds, using an order-by-order analysis in $\lambda$. Asserting that terms of the form $\nabla_\alpha\nabla_\beta h_{\mu\nu}$ are uniformly bounded by a constant times $\lambda^{-1}$ does not prevent these terms from having $\mathcal{O}(1)$ contributions that only cancel with other $\mathcal{O}(1)$ terms in the field equation, therefore it is not clear at all that (\ref{IsaacWave}) is strictly true. In fact, it is very unlikely to remain true at sub-leading orders in general. Secondly, Burnett questioned the averaging procedure itself. He asked whether the right hand side of (\ref{IsaacField}) is well defined, in particular questioning what size and shape the integration region should be. This latter point is not trivial, since the trick of integrating by parts to suppress total derivatives carries the assumption that the boundary of the averaging region is unaffected by $\lambda$.

The formal construction of Burnett's method begins with four conditions. Firstly,
\begin{equation}
 \mathcal{G}_{\mu\nu}\left(g(\lambda)\right) = 0 \ \ {\rm for} \ \lambda>0.
\end{equation}
where $\mathcal{G}_{\mu\nu}$ is the Einstein tensor defined in (\ref{EinsteinTensor}). If we evaluate at $\lambda=0$, this condition is relaxed in order to still accommodate the non-zero background curvature. When we generalize to higher derivative theories in Chapters \ref{ChapterBack1} and \ref{ChapterBack2}, it will be convenient to define a more general form for this tensor to be
\begin{equation}
 \mathcal{G}_{\mu\nu} := \frac{2\kappa}{\sqrt{-g}}\frac{\delta S_{\rm grav}}{\delta g^{\mu\nu}} = \kappa T_{\mu\nu},
\end{equation}
which is the same as (\ref{EinsteinTensor}) for Einstein gravity.

Secondly, $g_{\mu\nu}(x,\lambda)$ converges uniformly to $g_{\mu\nu}(x,0)$ as $\lambda\to0$. The third condition is that $\nabla_\alpha h_{\mu\nu}$ is uniformly bounded, \ie $\mathcal{O}(1)$ in our language: this translates physically as specifying that both the amplitude and wavelength of fluctuations vanish equally quickly as $\lambda\to0$. Finally, it is specified that the average over a term of the form $h_{\mu\nu}\nabla_\alpha\nabla_\beta h_{\rho\sigma}$ must converge uniformly in the ``weak'' limit of $\lambda\to0$ to a smooth averaged tensor field that we can call $B_{\mu\nu\alpha\beta\rho\sigma}$. More precisely, this means that
\begin{equation}\label{BurnettInt}
 \lim_{\lambda\to0}\int d^4 x\sqrt{-g(x,0)}\left(h_{\mu\nu}\nabla_\alpha\nabla_\beta h_{\rho\sigma}-B_{\mu\nu\alpha\beta\rho\sigma}\right)f^{\mu\nu\alpha\beta\rho\sigma} = 0, 
\end{equation}
for any smooth tensor field $f^{\mu\nu\alpha\beta\rho\sigma}$ of compact support, \ie a smooth tensor field that specifies the finite averaging region. Note that $f$ is not a function of $\lambda$, therefore an integration by parts of a total derivative reveals that a total derivative of some tensor $T_{\alpha_1\cdots\alpha_n}(x,\lambda)$ is of the same order in $\lambda$ as $T_{\alpha_1\cdots\alpha_n}(x,\lambda)$ itself (when weak-limit averaging), in agreement with the Brill-Hartle averaging scheme used by Isaacson. More explicitly, we have
\begin{equation}\label{BurnettMove}
 \int d^4 x\sqrt{-g^{(0)}}f^{\lambda\alpha_1\cdots\alpha_n}\nabla_\lambda A_{\alpha_1\cdots\alpha_n}(\lambda) = -\int d^4x\sqrt{-g^{(0)}}\left(\nabla_\lambda f^{\lambda\alpha_1\cdots\alpha_n}\right)A_{\alpha_1\cdots\alpha_n}(\lambda).
\end{equation}
Also note that this requirement specifies very little about the size and shape of the averaging region. More generally, we can say that a tensor $A_{\alpha_1\cdots\alpha_n}(\lambda)$ converges uniformly in the weak limit to some averaged tensor $B_{\alpha_1\cdots\alpha_n}$ if
\begin{equation}\label{genweaklim}
 \lim_{\lambda\to 0}\int d^4 x \sqrt{-g^{(0)}}f^{\alpha_1\cdots\alpha_n}A_{\alpha_1\cdots\alpha_n}(\lambda)=\int d^4 x\sqrt{-g^{(0)}}f^{\alpha_1\cdots\alpha_n}B_{\alpha_1\cdots\alpha_n}
\end{equation}
for any smooth test tensor field $f^{\alpha_1\cdots\alpha_n}$ of compact support. This is quite inconvenient to write out every time, so instead we will denote equality under weak averaging of both sides using $\w$. Thus if some tensor $A_{\alpha_1\cdots\alpha_n}(\lambda)$ and some other tensor $C_{\alpha_1\cdots\alpha_n}(\lambda)$ both converge in the weak limit to an averaged tensor $B_{\alpha_1\cdots\alpha_n}$, we can write
\begin{equation}
 A_{\alpha_1\cdots\alpha_n} \w B_{\alpha_1\cdots\alpha_n} \w C_{\alpha_1\cdots\alpha_n}.
\end{equation}
As with the Brill-Hartle averaging scheme, we can rearrange derivatives at the quadratic order in $h$ using (\ref{BurnettMove}) to get
\begin{equation}
 h_{\rho\sigma}\nabla_{\alpha}\nabla_{\beta}h_{\mu\nu} \w - \nabla_{\alpha}h_{\rho\sigma}\nabla_{\beta}h_{\mu\nu} \sim \mathcal{O}(1).
\end{equation}

Concerning the linear order in $h$, note that, under weak averaging, all these terms in the field equation are $\mathcal{O}(\lambda)$ under weak averaging. Even expressions of the form $\nabla_\alpha\nabla_\beta h_{\mu\nu}$ are $\mathcal{O}(\lambda)$ because the total derivative operators do not change the order in $\lambda$ under weak averaging. Thus the linear order in $h$, which diverges as $\lambda\to0$ prior to averaging, disappears from the field equation when averaged in the weak limit. One might wonder what happens then to the constraint given by the wave equation (\ref{IsaacWave}) in Isaacson's formalism. This information is preserved in a more subtle way under weak limit averaging. We can take the field equation and multiply both sides by the metric perturbation $h_{\alpha\beta}$ such that all terms become $\lambda$-suppressed apart from the initially divergent $\mathcal{O}(h)$ pieces. Starting from the Ricci form of the field equation, we then have
\begin{equation}\label{EinZero}
 R_{\alpha\beta}^{(1)}h_{\gamma\delta} \w 0.
\end{equation}
Although this statement is not as strong as (\ref{IsaacWave}), it is just as powerful for our purposes and it is a much more rigorous statement. It is convenient to give a name to such an important constraint, so we will from now on refer to this constraint and its generalizations to other theories as the ``zero tensor''. Following this prescription, one ultimately derives the same form for the effective stress-energy tensor (under weak limit averaging) as in the Isaacson case, but the derivation is on a much stronger mathematical foundation as it makes no implicit assumptions about subleading parts of $R^{(1)}_{\mu\nu}$ and the treatment of the integral in (\ref{BurnettInt}) justifies more rigorously how total derivative operators do not change the order in $\lambda$.

\section{Green and Wald weak-limit framework}\label{GreennWald}

The Green and Wald framework \cite{Green:2010qy,Green:2013yua} is an adaptation of Burnett's weak limit formalism \cite{Burnett:1989gp} for studying cosmological backreaction. It is credited with providing a strong theoretical backing to the assertion that cosmological backreaction does not produce significant effects on the expansion rate of the universe. In this thesis, we will not contest this claim in Einstein gravity, although it has attracted controversy more generally \cite{Ellis:2011hk,Buchert:2015iva,Green:2015bma,Green:2016cwo}. Of particular interest to this thesis is the derivation of the effective stress-energy tensor for backreaction. Since this effective stress-energy tensor is constructed out of metric perturbations, it will take the same form as the effective stress-energy tensor in the Isaacson and Burnett formalisms. However, unlike these gravitational wave studies, the physical stress-energy tensor $T_{\mu\nu}(x,\lambda)$, corresponding to energy and momentum that is not constructed from gravitational waves, is not set to zero. This is generally required because, rather than studying gravitational waves propagating in a vacuum region far away from their source, we are studying a cosmological spacetime that will include the usual matter and energy content of the universe.

To reproduce the constraint seen in (\ref{EinZero}), the Green and Wald formalism needs to determine the weak limit of $h_{\alpha\beta}(x,\lambda)\kappa T_{\mu\nu}(x,\lambda)$, where $T_{\mu\nu}(x,\lambda)$ is the stress-energy tensor for matter or any cosmological fluids other than gravitational backreaction. To do this, Green and Wald proved a lemma \cite{Green:2010qy}, which we will now discuss. Let $A_{\alpha_1\cdots\alpha_n}(\lambda)$ be a one-parameter family of smooth tensor fields that converge uniformly as $\lambda\to0$ to $A_{\alpha_1\cdots\alpha_n}(0)$ and let $B(\lambda)$ be a one-parameter family of smooth non-negative functions that converge uniformly as $\lambda\to0$ to $B(0)$. The lemma states that
\begin{equation}\label{GreenWaldLem0}
 A_{\alpha_1\cdots\alpha_n}(\lambda)B(\lambda) \to A_{\alpha_1\cdots\alpha_n}(0)B(0)\  \ {\rm as }\ \ \lambda\to0.
\end{equation}
The proof begins by introducing a smooth tensor field of compact support $F^{\alpha_1\cdots\alpha_n}$ in the same spirit as in (\ref{genweaklim}). Also introduced is $f$, which is a smooth non-negative function, also with compact support and the condition that $f=1$ on the support of $F^{\alpha_1\cdots\alpha_n}$. Thus it follows that $F=fF$. Neither $F^{\alpha_1\cdots\alpha_n}$ nor $f$ are functions of $\lambda$. Now, suppressing indices for notational convenience with the understanding that the indices of $A_{\alpha_1\cdots\alpha_n}(\lambda)$ are contracted with the indices of $F^{\alpha_1\cdots\alpha_n}$ in a weak limit integral of the form (\ref{genweaklim}), we can write
\begin{eqnarray}\label{GreenWaldLem1}
 && \lim_{\lambda\to0}\int d^4x \sqrt{g^{(0)}}\Bigg(A(\lambda)B(\lambda) - A(0)B(0)\Bigg)F = \nonumber \\
 && \lim_{\lambda\to0}\int d^4x \sqrt{g^{(0)}}\Bigg(\Big(A(\lambda)-A(0)\Big)B(\lambda) - A(0)\Big(B(\lambda)-B(0)\Big)\Bigg)F.
\end{eqnarray}
Since $A_{\alpha_1\cdots\alpha_n}(0)F^{\alpha_1\cdots\alpha_n}$ is simply a test function of compact support and $B(\lambda)\to B(0)$ in weak limit, the second term on the right hand side of (\ref{GreenWaldLem1}) vanishes. To show that the first term also vanishes, note that
\begin{equation}\label{GreenWaldLem2}
 \left|\int d^4 x\sqrt{g^{(0)}}\Big(A(\lambda)-A(0)\Big)B(\lambda)F\right| \le \int d^4 x\sqrt{g^{(0)}}\left|\Big(A(\lambda)-A(0)\Big)F\right|\Big|B(\lambda)f\Big|.
\end{equation}
Noting that both $B(\lambda)$ and $f$ are non-negative, both sides of (\ref{GreenWaldLem2}) are less than or equal to
\begin{equation}
 {\rm sup}\left|\Big(A(\lambda)-A(0)\Big)F\right|\int d^4 x\sqrt{g^{(0)}}B(\lambda)f,
\end{equation}
where ``sup'' of a function of $x$ is its supremum, \ie the largest value of the function for any $x$. Green and Wald then introduce an arbitrary positive number $\epsilon$ and note that since $A(\lambda)\to A(0)$ as $\lambda\to0$, there exists a positive value of $\lambda$ below which
\begin{equation}
 \left|\int d^4 x\sqrt{g^{(0)}}\Big(A(\lambda)-A(0)\Big)B(\lambda)F\right| \le \epsilon
\end{equation}
holds. Since $B(\lambda)\to B(0)$ for $\lambda\to0$, there also exists a positive value of $\lambda$ below which
\begin{equation}
 \int d^4 x\sqrt{g^{(0)}}\Big(B(\lambda)-B(0)\Big)f < \epsilon
\end{equation}
holds. Therefore, there exists a positive value of $\lambda$ below which
\begin{equation}
 \left|\int d^4 x\sqrt{g^{(0)}}\Big(A(\lambda)-A(0)\Big)B(\lambda)F\right| < \epsilon \left(\epsilon+\int d^4 x\sqrt{g^{(0)}}B(0)f\right)
\end{equation}
is true. Therefore the expression should vanish in the weak limit $\lambda\to0$, leaving (\ref{GreenWaldLem1}) equal to zero. This proves (\ref{GreenWaldLem0}). Now it remains to relate this to the weak limit average of $h_{\alpha\beta}(x,\lambda)T_{\mu\nu}(x,\lambda)$. Suppose now that $T_{\mu\nu}(x,\lambda)$ satisfies the weak energy condition (\ref{WeakEnergy}), which is cosmologically reasonable. It follows that if we define
\begin{equation}
 B(x, \lambda) := T_{\mu\nu}(x,\lambda)t^\mu(x) t^\nu(x),
\end{equation}
where $t^\alpha(x)$ is a vector field that is timelike on the metric $g_{\mu\nu}(x,\lambda)$ for $\lambda$ smaller than some threshold, then $B(x,\lambda)$ is a positive function below that threshold. We know that $\kappa T_{\mu\nu}(x,\lambda)$ converges weakly because the left hand side of the field equation (\ref{EinsteinGravFieldEq}) converges weakly, therefore so should $B(x,\lambda)$. This $B(x,\lambda)$ can then be used in the working above. The starting assumptions of the formalism include that $h_{\alpha\beta}(x,\lambda)$ converges weakly on compact sets to zero. Therefore we can set $A_{\alpha\beta}(x,\lambda):=h_{\alpha\beta}(x,\lambda)$ and $A(x,0)=0$. Using the lemma above, we know that
\begin{equation}\label{GreenWaldDiLem}
 h_{\alpha\beta}(x,\lambda)T_{\mu\nu}(x,\lambda)t^\mu(x)t^\nu(x) \w 0.
\end{equation}
Since $h_{\alpha\beta}(x,\lambda)T_{\mu\nu}(x,\lambda)$ is symmetric under exchange of the index labels $\mu$ and $\nu$, (\ref{GreenWaldDiLem}) is only true for all vectors $t^\mu(x)$ that are timelike on $g_{\mu\nu}(x,\lambda)$ for small $\lambda$ if 
\begin{equation}\label{GreenWaldZero1}
 h_{\alpha\beta}(x,\lambda)T_{\mu\nu}(x,\lambda) \w 0.
\end{equation}
This now allows us to reproduce the familiar constraint from the linear perturbation to the field equation, as seen in Burnett's (\ref{EinZero}) and similar to Isaacson's wave equation (\ref{IsaacWave}). To see how, let us first split the Einstein tensor into its background $\mathcal{G}^{(0)}_{\mu\nu}$ and perturbation $\delta\left[\mathcal{G}_{\mu\nu}\right]$. The field equation is then written as
\begin{equation}\label{fieldpert}
 \mathcal{G}^{(0)}_{\mu\nu} + \delta\left[\mathcal{G}_{\mu\nu}\right] = \kappa T_{\mu\nu}.
\end{equation}
The ``zero tensor'' constraint is then taken by multiplying both sides of (\ref{fieldpert}) by $h_{\rho\sigma}$ and taking a weak limit:
\begin{equation}
 \underbrace{h_{\rho\sigma}\mathcal{G}^{(0)}_{\mu\nu}}_{\text{vanishes in weak limit}} + h_{\rho\sigma}\delta\left[\mathcal{G}_{\mu\nu}\right] = \underbrace{h_{\rho\sigma}\kappa T_{\mu\nu}}_{\text{vanishes in weak limit}}.
\end{equation}
The vanishing of the right hand side using the result from Green and Wald (\ref{GreenWaldDiLem}) allows us to write
\begin{equation}\label{GenZero}
 h_{\rho\sigma}\delta\left[\mathcal{G}_{\mu\nu}\right] \w 0,
\end{equation}
This reproduces the Burnett result (\ref{EinZero}) for Einstein gravity.

In the weak limit, we write the contribution from the perturbation pieces as an effective stress-energy tensor, just as in the Isaacson and Burnett schemes:
\begin{equation}
 \delta\left[\mathcal{G}_{\mu\nu}\right] \w -\kappa t^{(0)}_{\mu\nu}.
\end{equation}
Unlike in the Isaacson and Burnett schemes, the spacetime need not be empty aside from gravitational radiation. Thus we have an averaging scheme that is applicable to cosmological spacetimes. Recall that the original motivation for taking $\lambda$ to be small was that the wavelength of gravitational radiation should be much smaller than the curvature radius of the wavefront. The corresponding picture for cosmological backreaction is that the perturbations induced by cosmological inhomogeneity should be on a much smaller length scale than the cosmological averaging scale. The field equation can now be expressed in the weak limit as
\begin{equation}
 \mathcal{G}^{(0)}_{\mu\nu} \w \kappa T_{\mu\nu}^{(0)} + \kappa t^{(0)}_{\mu\nu}.
\end{equation}
Since we are concerned with more general cosmological spacetimes, it is reasonable to ask what r\^{o}le, if any, the cosmological constant plays in backreaction. We can choose to move the cosmological constant term into $T_{\mu\nu}$ and refer to it as ``vacuum energy'', which is not unusual. Alternatively, we can note that it is simply a constant times the metric and therefore it seems reasonable to simply state that
\begin{equation}\label{noccc}
 \Lambda g_{\mu\nu} \w \Lambda g^{(0)}_{\mu\nu}.
\end{equation}
A positive cosmological constant term would also satisfy the weak energy condition (\ref{WeakEnergy}), so it would play no r\^{o}le in the zero tensor (\ref{GenZero}). This reasoning is straightforwardly adapted for a negative cosmological constant. Using
\begin{equation}\label{Ein1}
 \delta \left[R_{\mu\nu}\right] \w -\frac{1}{2}h^{\alpha\beta}R^{(1)}_{\mu\alpha\nu\beta} - \frac{1}{4}h_{\mu\nu}R^{(1)}+R^{(1)}_{\alpha(\mu}h_{\nu)}^{\ \ \alpha},
\end{equation}
and
\begin{equation}\label{Ein2}
 \delta\left[g_{\mu\nu}R\right] \w h_{\mu\nu}R^{(1)} - \frac{1}{4}g_{\mu\nu}^{(0)}\left(hR^{(1)}+2h^{\alpha\beta}R_{\alpha\beta}^{(1)}\right),
\end{equation}
the effective stress-energy tensor due to backreaction in Einstein gravity is initially calculated to be
\begin{equation}\label{EinsteinStress}
 \kappa t^{E}_{\mu\nu} \w \frac{1}{2}h^{\alpha\beta}R_{\mu\alpha\nu\beta}^{(1)}+\frac{3}{4}h_{\mu\nu}R^{(1)} - R^{(1)}_{\alpha(\mu}h_{\nu)}^{\ \ \alpha} - \frac{1}{8}g_{\mu\nu}^{(0)}\left(hR^{(1)}+2h^{\alpha\beta}R_{\alpha\beta}^{(1)}\right).
\end{equation}
Thus the effective stress-energy tensor due to backreaction simplifies to
\begin{equation}\label{EinAll}
 \kappa t^{E}_{\mu\nu} \w \frac{1}{2}h^{\alpha\beta}R_{\mu\alpha\nu\beta}^{(1)},
\end{equation}
whose trace is once again zero because it takes the form of the zero tensor for Einstein gravity, given in (\ref{EinZero}).

Before we move on to discuss the generalization of this approach to backreaction in the context of higher-derivative effective gravity theories, let us interpret what we see here. One of the main motivations for studying backreaction is the promise that it might offer an alternative to dark energy, which was discussed in Subsection \ref{IntroDarkEnergy}. In such a view, the hope is that the effective stress-energy tensor for backreaction would create an acceleration effect that mimics a cosmological constant. To understand the implications of (\ref{EinAll}) and (\ref{EinZero}), note that we can split the effective stress-energy tensor due to backreaction into two pieces. These two pieces are the traceless piece and the pure trace piece:
\begin{equation}\label{tlesspuret}
 t^{(0)}_{\mu\nu} = \underbrace{t^{(0)}_{\mu\nu} - \frac{1}{4}g_{\mu\nu}^{(0)}t^{(0)}}_{\text{traceless}}+\underbrace{\frac{1}{4}g^{(0)}_{\mu\nu}t^{(0)}}_{\text{pure trace}}.
\end{equation}
It is the pure trace part of the effective stress-energy tensor that would be capable of causing cosmological acceleration, having $w=-1$ in the language of (\ref{FluidEqState}), whereas the traceless part is radiation-like with $w=1/3$. However, the Green and Wald formalism has told us unambiguously that the pure trace part of the backreaction is zero at large scales in Einstein gravity. This proves that backreaction of this kind cannot be responsible for the observed effect of dark energy, given Einstein's GR. Estimates of the magnitude of this effect, \eg via Newtonian approximations, agree that it should be small anyway, with a Newtonian potential parameter of order $10^{-5}$ \cite{Ishibashi:2005sj,Behrend:2007mf}. These results have generated some controversy. In particular, advocates of backreaction listed their crticisms in \cite{Buchert:2015iva}. Green and Wald responded in \cite{Green:2015bma} to these objections. They then released simpler arguments in \cite{Green:2016cwo}.

However, moving forward to Chapter \ref{ChapterBack1}, we will see that the results for a more general gravity theory including an $R^2$ term are more interesting, offering some hope for backreaction as a significant cosmological effect in our universe \cite{Preston:2014tua}. Chapter \ref{ChapterBack2} is the complete generalization of that work, offering an effective stress-energy tensor for backreaction and gravitational waves that applies to all local diffeomorphism-invariant effective gravity theories that include an Einstein-Hilbert term \cite{Preston:2016sip}.

\chapter{Backreaction in modified gravity}\label{ChapterBack1}

Chapter \ref{AWHPreview2} delved into cosmological backreaction in Einstein's GR. Backreaction is interesting because it is a mechanism by which, physics at short distance scales might impact on physics at large distance scales. This is quite unlike most phenomena in particle physics, where short-distance scale effects are suppressed such that high-energy collisions are required to gain experimental access. In the context of backreaction, the short-distance physics takes the form of metric perturbations away from the averaged metric with short wavelengths compared with the averaging scale. In the schemes discussed in Chapter \ref{AWHPreview2}, the effective stress-energy tensor is allowed to be of $\mathcal{O}(1)$ even though the metric perturbation itself is $\mathcal{O}(\lambda)$ because the derivatives are large: the low amplitude of perturbations is balanced by the high frequency. This motivates us to wonder what would happen if we considered higher-derivative operators, such as the $R^2$ term. Would having fourth-order derivatives allow us to create an appreciable backreaction effect from high-frequency perturbations that could mimic a cosmological constant, even if ordinary GR does not allow this? This chapter reports on original research published in \cite{Preston:2014tua}.

There is some overlap with other studies of gravitational waves and backreaction in modified gravity. Backreaction in modified gravity has previously been studied in the context of Buchert's averaging scheme \cite{Vitagliano:2009zy,Jimenez:2013mwa}. However, the affirmative conclusion of that approach is not so surprising because Buchert's approach already claims a cosmologically significant backreaction in unmodified gravity \cite{Buchert:1995fz,Buchert:1999er,Buchert:2001sa,Visser:2015mur}. Since Green and Wald demonstrated using their rigorous formalism that backreaction is not important in unmodified gravity \cite{Green:2010qy}, there is a strong motivation to use adapt the Green and Wald formalism to test that null result in $R+R^2/6M^2$ gravity. 

There already existed two studies of gravitational radiation in the limit of high frequency for higher-derivative gravity theories that used Isaacson's shortwave approximation \cite{Stein:2010pn,Berry:2011pb}. These studies overlap with the research presented in this chapter and Chapter \ref{ChapterBack2}, agreeing in particular with the conclusion in Chapter \ref{ChapterBack2}, reporting on original research in \cite{Preston:2016sip}, that cubic and higher terms in the Riemann tensor do not contribute to the effective stress-energy tensor. However, they differ from the work discussed here in that they both fix a gauge and that \cite{Stein:2010pn} evaluates the effective stress-energy tensor at asymptotically-flat future null infinity and \cite{Berry:2011pb} sets the background to Minkowski, whereas the work presented here maintains strict diffeomorphism invariance at all times and allows for a general cosmological background, as required for a backreaction calculation.

There also existed a study that claimed that adapting the Green and Wald formalism also gave a traceless effective stress-energy tensor in $R+R^2/6M^2$ gravity \cite{Saito:2012xa}. However, that study used an approximation where the physical stress-energy tensor is treated as its purely background (\ie averaged) form from the beginning. In particular, this assumption allowed that study to avoid a crucial step in which we scale $M$ as as $1/\lambda$, which we will see to be mandatory for the physical consistency of the argument. This will be demanded especially strongly in Chapter \ref{ChapterBack2}. Backreaction has also been studied explicitly in the context of coupling gravity to other fields \cite{Szybka:2015zca}.

In Subsection \ref{IntroInflation}, we noted that $R+R^2/6M^2$ gravity is a model that produces inflation in the early universe. Recall also from equation (\ref{ScalTensJordAct}) that this theory, as an $f(R)$ theory, can be re-expressed as a scalar-tensor theory via a Legendre transformation. This scalar-tensor theory resembles ordinary Einstein gravity when the value of the scalar mode settles towards the value that minimizes its potential. In Jordan frame, this potential is a quadratic function of $\phi$ centred around $\phi=1$, in Einstein frame, it takes an exponential form and comes with a canonically normalized kinetic term. Intuitively, we would expect backreaction to be more interesting for this theory, since perturbing the ``scalaron'' $\phi$ away from its minimum raises its potential above zero, regardless of whether the perturbation in $\phi$ is itself positive or negative. We might expect that metric perturbations should introduce perturbations to $\phi$ that always give a positive effective vacuum energy, since pushing $\phi$ in either direction is effectively pushing it up the potential slope from its average position at late times. So it is not unreasonable to ask whether the presence of an $R^2$ term might actually create a small cosmological constant-like effect in the universe when a supply of high-frequency metric perturbations is applied, despite the null result in Einstein gravity.

The gravity action for the Starobinsky model was given in (\ref{Starobaction}). Including the cosmological constant and matter parts, we have
\begin{equation}\label{Staraction}
 S = \int d^4 x\sqrt{-g}\left[\frac{1}{16\pi G}\left(R+\frac{R^2}{6M^2}-2\Lambda\right)+\mathcal{L}_{\rm Matter}\right].
\end{equation}
The field equation is given by
\begin{equation}\label{Starofield}
 R_{\mu\nu} + \Lambda g_{\mu\nu} -\frac{1}{2}g_{\mu\nu}\left(R+\frac{R^2}{6M^2}\right)+\frac{1}{3M^2}\left(RR_{\mu\nu}-D_\mu D_\nu R + g_{\mu\nu}D^2 R\right)\cdots = \kappa T_{\mu\nu}.
\end{equation}
We have already noted in Subsection \ref{IntroInflation} that this inflation model, with $M\sim10^{13}$ GeV, is in good agreement with current observations. It did, however, briefly run into tension with a claimed primordial B-mode signal in the CMB observed by the BICEP2 telescope \cite{Ade:2014xna}. This B-mode signal was later shown to have been caused by interstellar dust rather than primordial gravitational waves \cite{Liu:2014mpa,Flauger:2014qra,Ade:2015tva}. The controversy generated by the BICEP2 result occured in parallel with the research reported in this chapter.

In this study, we considered a scheme in which perturbations of field equation contributions from both the Einstein and Starobinsky parts of the action are significant. This is important because, as we will see, neither the Einstein nor the Starobinsky parts of the action can produce significant backreaction contributions on their own. Importantly, getting a large effect from the Starobinsky part requires that the metric perturbations are of a sufficiently high wavenumber to overcome the suppression of the Starobinsky term by the mass scale $M$. This will lead us to require that the source of cosmological inhomogeneity is of an exotic kind, \ie belonging to new particle physics rather than to an uneven distribution of large objects, such as galaxies. This will be discussed further in Section \ref{StaroUV}.

The structure of this chapter is as follows. In Section \ref{StaroScales}, we will discuss how to adapt the Green and Wald formalism to incorporate the additional scale of length given by $M$. In Section \ref{StaroZero}, we will calculate the zero tensor in $R+R^2/6M^2$ gravity and discuss how it differs from its relative in Einstein gravity. In Section \ref{StaroBackreact}, we will calculate the stress-energy tensor for backreaction in $R+R^2/6M^2$, paying particular attention to its trace, which we will find to be non-zero and of the correct sign to mimic a positive cosmological constant. In Section \ref{StaroScalarTensor}, we will look at the scalar-tensor description of the theory in Jordan frame, which allows us both to gain more intuition for the result and show equivalence between the two descriptions. In Section \ref{StaroDiffeomorphism}, we will see that the effective stress-energy tensor is invariant under diffeomorphisms in the weak limit, which tells us both that the calculations have been done correctly and that the result is independent of coordinates. Finally, in Section \ref{StaroUV}, we will discuss exotic particle physics candidates that might induce a sufficiently large backreaction effect from the Starobinsky terms to be cosmologically interesting.

\section{Incorporating additional scales of length}\label{StaroScales}

Na\"{i}vely, we would continue in the same spirit as in Chapter \ref{AWHPreview2}, scaling the new higher-derivative terms according to
\begin{equation}\label{wavelength}
 \nabla_{\alpha_1}\cdots\nabla_{\alpha_n}h_{\beta\gamma} \sim \lambda^{1-n},
\end{equation}
leaving $M$ as a constant in $\lambda$. We will discuss in this section and also again in Chapter \ref{ChapterBack2} the problems which arise from this na\"{i}ve approach and the reasons why we actually need to scale $M\sim1/\lambda$ as well as using the scaling in (\ref{wavelength}). To illustrate the problems, let us progress at first without scaling $M$ at all. Since total derivatives do not change the order in $\lambda$, as illustrated in (\ref{BurnettMove}), we know that all linear terms in $h$ are of the same order in $\lambda$, therefore they all vanish in the weak limit. However, we run into difficulty applying (\ref{wavelength}) at the quadratic order in $h$ because the $R^2$ term gives us field equation contributions in (\ref{Starofield}) at the fourth order in derivatives such as from $R_{\mu\nu}^{(1)}R^{(1)}$ which, being $\mathcal{O}(\lambda^{-2})$, would diverge in the weak limit $\lambda\to0$.

The obvious resolution to this problem, which we will ultimately disfavour in this section and see as completely hopeless in Chapter \ref{ChapterBack2}, is to scale $h_{\mu\nu}(x,\lambda)$ as $\mathcal{O}(\lambda^2)$ instead. This causes the Einstein (\ie 2-derivative) parts of the field equation (\ref{Starofield}) to converge trivially to their background values in the weak limit, leaving only the perturbations to the Starobinsky (\ie 4-derivative) terms na\"{i}vely able to contribute to the effective stress-energy tensor. We should feel immediately uneasy that this rescaling is {\it a priori} unable to reproduce the Einstein part of the backreaction which, although cosmologically unimportant at late times, is nevertheless non-zero. Carrying on regardless, we can begin by calculating the zero tensor. Let us begin by multiplying (\ref{Starofield}) by $h_{\rho\sigma}$ and taking a weak limit under the assumption that $T_{\mu\nu}$ satisfies the weak energy condition:
\begin{equation}
 h_{\rho\sigma}\nabla_\mu\nabla_\nu R^{(1)} - g_{\mu\nu}^{(0)}h_{\rho\sigma}\nabla^2 R^{(1)} \w 0.
\end{equation}
Taking the trace over $\mu$ and $\nu$, we find that $h_{\rho\sigma}\nabla^2 R^{(1)}$ also vanishes in the weak limit, therefore the zero tensor is simply
\begin{equation}\label{naivezero}
 h_{\rho\sigma}\nabla_\mu\nabla_\nu R^{(1)} \w 0.
\end{equation}
This is extraordinarily powerful as a constraint because it actually tells us that all the perturbation terms in (\ref{Starofield}) vanish in the weak limit, which can be seen by inspection by noting that all the Starobinsky terms possess a factor of $R$ or a derivative of $R$. 

The temptation then is to accept that the quadratic parts vanish in the weak limit and let the cubic parts form the effective stress-energy tensor by scaling $h_{\alpha\beta}(x,\lambda)$ to be $\mathcal{O}(\lambda^{4/3})$. This is a mistake. In the original construction, the linear part of the field equation was $\mathcal{O}(\lambda^{-1})$ for Einstein gravity, but its weak limit average was $\mathcal{O}(\lambda)$, so the linear order was well-defined in the weak limit. The zero tensor was then easily constructed by multiplying by $h_{\rho\sigma}$ such that the information from the linear part of the field equation was preserved as an $\mathcal{O}(1)$ expression that remained $\mathcal{O}(1)$ under weak averaging. The na\"{i}ve scaling described above, where $h\sim\lambda^2$, had a similar structure, except that the linear terms were $\mathcal{O}(\lambda^{-2})$ before averaging. The $h\sim\lambda^{4/3}$ scaling, on the other hand, is poorly defined. The linear level is $\mathcal{O}(\lambda^{-8/3})$ before averaging and $\mathcal{O}(\lambda)$ after, which present no problems so far. A zero tensor can be constructed to preserve the information from the linear level by multiplying both sides by $h_{\alpha\beta}h_{\gamma\delta}$, which is also gives a well-behaved $\mathcal{O}(1)$ limit. We lose control at the quadratic level, however. The quadratic part of the field equation is $\mathcal{O}(\lambda^{-4/3})$ and remains so under weak averaging, \ie it diverges. One might think to reuse the zero tensor (\ref{naivezero}) that retained information from the linear level by simply multiplying by $h_{\rho\sigma}$ and taking the weak limit. Once again, the left hand side diverges in the weak limit; it would also pick up $\mathcal{O}(1)$ contributions from the quadratic order in the field equation, which would not be required to separately cancel. One might try to remedy this situation by solving the field equation order-by-order in $\lambda$, as Isaacson's formulation originally does. However, this is not rigorous, since the lower-order terms in $\lambda$ can also take part in the field equation at higher orders in $\lambda$ (which motivates Burnett's approach), so we cannot simply cancel the quadratic terms in $h$ at all orders in $\lambda$, and we cannot assume that the weak energy condition holds for individual $\mathcal{O}(\lambda)$ contributions to $T_{\mu\nu}$. In summary, this approach is not under good mathematical control.

When we come to make the complete generalization in Chapter \ref{ChapterBack2}, we will see that the entire project of rescaling the order of $h$ according the number of derivatives in the action was hopeless: a better solution is required. That solution is to scale $M$ as a constant divided by $\lambda$. This modifies (\ref{wavelength}) to
\begin{equation}\label{wavelength2}
 \frac{1}{M^m}\nabla_{\alpha_1}\cdots\nabla_{\alpha_n}h_{\beta\gamma} \sim \lambda^{1+m-n}.
\end{equation}
The intuitive understanding of this begins by recalling that $\lambda\to0$ corresponds to the limit where the wavelength of perturbations is much smaller than the cosmological averaging scale. In this respect, the $\lambda\to0$ limit is really the limit in which we take the averaging scale to be large. If we are incorporating the limit of a large averaging scale in this way, it also makes sense to scale $M$ accordingly, since $1/M$ is also much smaller than the averaging scale. Indeed, it would clearly be wrong to tend the wavelength of perturbations towards zero while fixing $M$ to a finite constant, because this would be tending the ratio of $\lambda/M$ to zero, which would imply that we are using $R+R^2/6M^2$ as an effective theory to describe perturbations far below the cutoff length at which the effective theory ceases to be valid. If we chose $M\sim\lambda^{-n}$ where $n>1$, we would have the Starobinsky parts drop out completely and we would be left with the pure Einstein case. Since natural units give us that mass scales like 1/length, let us suppose for this chapter that $M\sim1/\lambda$, since this is the most interesting case that gives us the most general results. We will return to this discussion in Chapter \ref{ChapterBack2}.

\section{Zero tensor forms for modified gravity}\label{StaroZero}

We derive the zero tensor for $R+R^2/6M^2$ gravity in the same manner we did for Einstein gravity: we multiply (\ref{Starofield}) by $h_{\rho\sigma}$ and take a weak limit. Recall that the Ricci form of the field equation (\ref{EinsteinFieldEqRicci}) was convenient for constructing the zero tensor in Einstein gravity (\ref{EinZero}). This is equivalent to using the trace of the zero tensor to find useful re-expressions of terms in the zero tensor that contain a factor of $g_{\mu\nu}^{(0)}$. In the Einstein case, there was only one such term, which took the form $g_{\mu\nu}^{(0)}R^{(1)}h_{\rho\sigma}$: it was therefore zero in the weak limit. In the Starobinsky case, it is joined also by a $g_{\mu\nu}^{(0)}h_{\rho\sigma}\nabla^2 R^{(1)}$ term. We can use the trace of the zero tensor to re-express one of these expressions as a multiple of the other, but we cannot produce an expression to set eliminate both terms individually in the weak limit. We therefore have two forms of the zero tensor, which are
\begin{equation}\label{Szero1}
 h_{\alpha\beta}R^{(1)}_{\gamma\delta} -\frac{g_{\gamma\delta}^{(0)}}{6M^2}R^{(1)}\nabla^2 h_{\alpha\beta}-\frac{1}{3M^2}\left(\nabla_\gamma\nabla_\delta h_{\alpha\beta}\right)R^{(1)} \w 0
\end{equation}
and
\begin{equation}\label{Szero2}
 h_{\alpha\beta}R^{(1)}_{\gamma\delta} - \frac{g^{(0)}_{\gamma\delta}}{6}h_{\alpha\beta}R^{(1)}-\frac{1}{3M^2}\left(\nabla_\gamma \nabla_\delta h_{\alpha\beta}\right)R^{(1)} \w 0.
\end{equation}
Using the trace over $\gamma$ and $\delta$ of either expression, we obtain the relationship between these two forms of the zero tensor:
\begin{equation}
 R^{(1)}\left(1-\frac{\nabla^2}{M^2}\right)h_{\alpha\beta} \w 0.
\end{equation}
Though these two forms are equivalent, they are most convenient for different uses. Using (\ref{Szero1}), we can easily express most 2-derivative terms as 4-derivative expressions in the weak limit. Indeed, we will see that the trace of the effective stress-energy tensor can be completely re-expressed in the weak limit in terms of 4-derivative terms. This is because the trace would usually vanish in pure Einstein gravity. Conversely, (\ref{Szero2}) is most useful for expressing 4-derivative terms as 2-derivative expressions. Indeed, the entire stress-energy tensor can be re-expressed in the weak limit in 2-derivative form. This is because the entire stress-energy tensor vanishes in a pure $R^2$ theory in the weak limit.

In this way, we have established two useful forms of the same zero tensor constraint that applies to $R+R^2/6M^2$ gravity. Unlike in the Einstein gravity case, these zero tensor forms contain both 2-derivative and 4-derivative terms, making them less powerful at setting stress-energy contributions to zero in the weak limit. Instead, this zero tensor constraint is useful for re-expressing the stress-energy tensor into more convenient forms, whether that be rephrasing the whole thing in 2-derivative form or just the trace in 4-derivative form.

\section{Stress-energy tensor in modified gravity}\label{StaroBackreact}

To derive the effective stress-energy tensor for $R+R^2/6M^2$ gravity, we simply take each field equation term in (\ref{Starofield}) and apply the weak limit to its perturbation. In addition to the limits known from pure Einstein gravity (\ref{noccc}), (\ref{Ein1}) and (\ref{Ein2}), we have 
\begin{eqnarray}\label{wRRmn}
 \delta\left[R_{\mu\nu}R\right]/M^2 & \w & R^{(1)}_{\mu\nu}R^{(1)}/M^2, \\
\label{wR2}
 \delta\left[g_{\mu\nu}R^2\right]/M^2 & \w & g_{\mu\nu}^{(0)}R^{(1)2}/M^2, \\
\label{wDDR}
 \delta\left[D_\mu D_\nu R\right]/M^2 & \w & \frac{1}{2}\left(2R_{\mu\nu}^{(1)}+\nabla_\mu \nabla_\nu h\right)R^{(1)}/M^2, \\
\label{wD2R}
 \delta\left[g_{\mu\nu} D^2 R\right]/M^2 & \w & \left(\nabla^2 h_{\mu\nu}-\frac{1}{2}g_{\mu\nu}^{(0)}\nabla^2 h\right)R^{(1)}/M^2.
\end{eqnarray}
Putting these pieces together, we get the effective stress-energy tensor from the Starobinsky part of the field equation to be
\begin{equation}\label{PureStarostress}
 \kappa t_{\mu\nu}^{S} \w \frac{R^{(1)}}{3M^2}\left(\frac{1}{2}g_{\mu\nu}^{(0)}\nabla^2 h- \nabla^2 h_{\mu\nu}+\frac{1}{2}\nabla_\mu \nabla_\nu h + \frac{1}{4}g_{\mu\nu}^{(0)}R^{(1)}\right),
\end{equation}
prior to applying the zero tensor constraint. In this form, it is clear by inspection why (\ref{naivezero}) would have forced all of this to vanish. Fortunately, we instead have (\ref{Szero2}), which allows us to re-express all of this in 2-derivative form:
\begin{equation}
 \kappa t^{S}_{\mu\nu} \w \frac{1}{2}hR_{\mu\nu}^{(1)} + \frac{g^{(0)}_{\mu\nu}}{4}h^{\alpha\beta}R_{\alpha\beta}^{(1)}-\frac{1}{3}h_{\mu\nu}R^{(1)}-\frac{g_{\mu\nu}^{(0)}}{24}hR^{(1)}.
\end{equation}
Bringing this together with the stress-energy tensor due to the Einstein terms (\ref{EinsteinStress}), we obtain the complete effective stress-energy tensor for $R+R^2/6M^2$ gravity in 2-derivative form:
\begin{equation}\label{CombStaroStress}
 \kappa t^{(0)}_{\mu\nu} \w \frac{1}{2}h^{\alpha\beta}R^{(1)}_{\mu\alpha\nu\beta}-R^{(1)}_{\alpha(\mu}h_{\nu)}^{\ \ \alpha}+\frac{5}{12}h_{\mu\nu}R^{(1)}+\frac{1}{2}hR_{\mu\nu}^{(1)}-\frac{1}{6}g_{\mu\nu}^{(0)}hR^{(1)}.
\end{equation}
In the pure Einstein gravity case, the stress-energy tensor was traceless. For $R+R^2/6M^2$ gravity, the 2-derivative form of the trace is
\begin{equation}
 \kappa t^{(0)} \w \frac{1}{4}hR^{(1)}-\frac{1}{2}h^{\alpha\beta}R_{\alpha\beta}^{(1)}.
\end{equation}
This 2-derivative form is not immediately obvious to interpret, so let us use (\ref{Szero1}) to re-express this in 4-derivative form:
\begin{equation}\label{StarStressTrace}
 \kappa t^{(0)} \w -\frac{R^{(1)2}}{6M^2}.
\end{equation}
This is rather exciting. This trace must be negative definite, except in the trivial case of $R^{(1)}=0$. Recall from (\ref{tlesspuret}) that the stress-energy tensor can be split into a traceless part and a pure trace part. Although these two parts are not separately conserved in general, a negative pure trace part is indeed what is required to mimic a positive cosmological constant. There is, of course, still a traceless part. The precise nature of the effective fluid depends on the balance of traceless and pure trace parts and on the time-dependence of that balance as the universe evolves. We cannot make strong statements about this in general, since the balance of traceless and pure trace parts of the backreaction depends on the nature of the inhomogeneity that is sourcing the metric perturbations. A sceptical reader might still wonder if a different way of expressing the theory might give a different result. In Section \ref{StaroScalarTensor} we will see the corresponding result in the scalar-tensor description and in Section \ref{StaroDiffeomorphism}, we will see that the result is diffeomorphism-invariant and therefore not an artifact of some coordinate scheme.

\section{Alternative derivation in scalar-tensor formulation}\label{StaroScalarTensor}

The action (\ref{Staraction}) is related to (\ref{ScalTensJordAct}) via the Legendre transformation (\ref{Legendrefr}), where the potential in the $R+R^2/6M^2$ case is (\ref{ScalTensStarPot}). Alternatively, one could incorporate the cosmological constant into the potential $V(\phi)$ as a vertical shift in the potential, but it is convenient for us to leave it as a separate term such that the minimum of $V(\phi)$ is $V(1)=0$ rather than $V(1)=2\Lambda$. The metric field equation of (\ref{ScalTensJordAct}) is
\begin{equation}
\label{ST-metric-eom}
 \phi \left(R_{\mu\nu} - \frac{1}{2}g_{\mu\nu}R \right)+ g_{\mu\nu}\Lambda  = \kappa T_{\mu\nu} + \left(D_{\mu}D_{\nu}-g_{\mu\nu}D^2\right)\phi-\frac{1}{2}g_{\mu\nu}V(\phi).
\end{equation}
If instead we functionally differentiate (\ref{ScalTensJordAct}) with respect to $\phi$, we obtain $R=V'(\phi)$. The trace of (\ref{ScalTensJordAct}) is
\begin{equation}\label{ST-metrictrace}
 -R\phi + 4\Lambda = \kappa T - 3D^2\phi -2V(\phi).
\end{equation}
Combining (\ref{ST-metrictrace}) with $R=V'(\phi)$, we get a field equation for the scalar mode, which is
\begin{equation}
\label{ST-scalar-eom}
 3D^2 \phi = \kappa T +\phi V'(\phi)-2V(\phi)-4\Lambda.
\end{equation}
We now wish to adapt the Green and Wald weak averaging formalism for use with this scalar mode. Just as with the metric, we split the scalar mode into a background piece and a perturbation piece:
\begin{equation}
\label{scalar-exp}
 \phi (x, \lambda) = \phi_0(x) + \phi_{p}(x, \lambda).
\end{equation}
Just as the metric converges uniformly in the weak limit to its background, so too does the scalar mode:
\begin{equation}
 \phi(x,\lambda)\to\phi_0(x)\ \ {\rm as}\ \lambda\to0.
\end{equation}
The scalar mode is given in the $f(R)=R+R^2/6M^2$ case by
\begin{equation}\label{phidef}
 \phi = f'(R) = 1 + \frac{R}{3M^2}.
\end{equation}
At late times in the universe, \ie after the end of inflation, one would expect that the vacuum expectation value of $\phi$ tends towards $\phi_0=1$. However, it is more complete, initially, to write
\begin{equation}
\label{vev}
 \phi_{0} = 1 + \frac{R^{(0)}}{M^2}.
\end{equation}
The second term here is suppressed by a large mass scale $M$, making it unimportant in the late universe. Applying the scaling $M\sim\lambda^{-1}$, we see that indeed $\phi\to1$ as $\lambda\to0$, therefore it is perfectly adequate for our purposes to use $\phi_0=1$. Following from (\ref{scalar-exp}) and (\ref{vev}), we can write the scalar perturbation as 
\begin{equation}
\label{pert}
 \phi_{p}(x, \lambda) = \frac{\delta R}{3M^2}\,, 
\end{equation}
where $\delta R = R-R^{(0)}$. However, as we have already noted, the background curvature piece $R^{(0)}/3M^2$ is vanishingly small in the weak limit, therefore we can safely drop the $\delta$ from (\ref{pert}), just as we lose the second term from the right hand side of (\ref{vev}) in the weak limit. Since $R^{(1)}$ is $\mathcal{O}(\lambda^{-1})$, we see that $\phi_p$ is $\mathcal{O}(\lambda)$. It follows also from (\ref{pert}) that
\begin{equation}\label{wavelength3}
 \frac{1}{M^m}\nabla_{\alpha_1}\cdots\nabla_{\alpha_n}\phi_p \sim \lambda^{1+m-n},
\end{equation}
analogously to (\ref{wavelength2}). This offers us a new type of $\mathcal{O}(1)$ contribution to the field equation (\ref{ST-metric-eom}) of the form
\begin{equation}
\label{phi-h}
\phi_{p}\nabla_{\mu}\nabla_{\nu}h_{\rho\sigma},
\end{equation}
which converges uniformly in the weak limit to a smooth tensor field that we would then absorb into the effective stress-energy tensor. We can substitute the expression for the scalar field (\ref{phidef}) into the potential (\ref{ScalTensStarPot}) to re-express the scalar potential in $R+R^2/6M^2$ gravity as
\begin{equation}\label{ScalPotCurv}
 V(\phi) = \frac{3}{2}M^2 \phi_p^2 = \frac{R^2}{6M^2} \w \frac{R^{(1)2}}{6M^2}.
\end{equation}
Thus we see that the potential (\ref{ScalPotCurv}), being of $\mathcal{O}(1)$, converges uniformly in the weak limit to a non-negative smooth scalar field. This scalar field obtained from the weak limit of the potential can be called $V_p(x)$ from now on. We can see that $h_{\mu\nu}V_p$ vanishes in the weak limit, as $V_p$ is already $\mathcal{O}(1)$, even before averaging. The physical stress-energy tensor $T_{\mu\nu}$ has remained unchanged from the form it had in the $f(R)$ gravity description. Since everything else in the field equation converges in the weak limit, so too must $\kappa T_{\mu\nu}$. This in turn ensures that $\phi V'(\phi)$ in the scalar field equation (\ref{ST-scalar-eom}) must have a weak limit, which we can also calculate in the Starobinsky case to be 2$V_p$. With that, we have all of the ingredients in place to study the weak limit average of the field equation (\ref{ST-scalar-eom}).

To calculate zero tensors, we can, as always, multiply both sides of the metric field equation (\ref{ST-metric-eom}) by $h_{\rho\sigma}$ and take a weak limit. Once again, we can use the trace of the metric field equation (\ref{ST-metrictrace}) to acquire Ricci forms of the field equation which, in this case, have either $\phi R$ or $\nabla^2\phi$ substituted out:
\begin{equation}
\label{scalar-tensor-zero-1}
 0 \w h_{\rho\sigma}R^{(1)}_{\mu\nu} -\frac{1}{2}g^{(0)}_{\mu\nu}h_{\rho\sigma}\nabla^2\phi_{p}-h_{\rho\sigma}\nabla_{\mu}\nabla_{\nu}\phi_{p},
\end{equation}
\begin{equation}
\label{scalar-tensor-zero-2}
 0 \w h_{\rho\sigma}R^{(1)}_{\mu\nu} - \frac{1}{6}g^{(0)}_{\mu\nu}h_{\rho\sigma}R^{(1)} - h_{\rho\sigma}\nabla_{\mu}\nabla_{\nu}\phi_{p}.
\end{equation}
We can relate these directly to the zero tensors we obtained before: (\ref{scalar-tensor-zero-1}) is the same as (\ref{Szero1}), (\ref{scalar-tensor-zero-2}) is the same as (\ref{Szero2}). This can be seen by using (\ref{pert}) to substitute $\phi_p$ with $R^{(1)}/3M^2$ in (\ref{scalar-tensor-zero-1}) and (\ref{scalar-tensor-zero-2}).
We can also construct a zero tensor from the scalar field equation:
\begin{equation}\label{ST-zero-scal}
 0 \w 3h_{\rho\sigma}\nabla^2\phi_{p} -h_{\rho\sigma}V'(\phi).
\end{equation}
Using the relation that $R=V'(\phi)$, we can see that this is the same as the zero tensor we could obtain from the trace of the metric field equation (\ref{ST-metrictrace}). Equivalently, (\ref{ST-zero-scal}) is the same constraint that one can obtain from the trace over $\mu$ and $\nu$ of either (\ref{scalar-tensor-zero-1}) or (\ref{scalar-tensor-zero-2}). Therefore (\ref{ST-zero-scal}) is not a separate constraint. If we suppose that there also exists a weak limit for $\phi_p\nabla_\mu\nabla_\nu\phi_p$, which would follow from (\ref{wavelength3}), then we could also construct zero tensor forms by multiplying the field equation (\ref{ST-metric-eom}) by $\phi_p$ and then taking a weak limit. This is less useful for us because, unlike our other zero tensors, they would be relating the 4-derivative terms not to 2-derivative terms but to 6-derivative terms. This is because multiplying by $\phi_p$ is the same as multiplying by $\delta R/3M^2$.

Having constructed the zero tensor, we are finally able to go ahead and compute the the effective stress-energy tensor due to backreaction in the scalar-tensor description. To do this, we need to compute the weak limits of all the terms in the metric field equation (\ref{ST-metric-eom}). Recalling that $\phi_0=1$, the weak limits we need in addition to (\ref{Ein1}) and (\ref{Ein2}) are
\begin{eqnarray}
\label{ST RR}
\phi_{p}R_{\mu\nu} &\w& \phi_{p}R^{(1)}_{\mu\nu},\\
\label{ST R2}
\phi_{p}g_{\mu\nu}R &\w& g^{(0)}_{\mu\nu}R^{(1)}\phi_{p},\\
\label{ST V}
\delta\left[g_{\mu\nu}V\right] &\w& g^{(0)}_{\mu\nu}V_{p} \w \frac{3}{2}g^{(0)}_{\mu\nu}M^{2}\phi_{p}^{2},\\
\label{ST DDR}
\delta\left[D_{\mu}D_{\nu}\phi\right]  &\w& \frac{1}{2}\left(R^{(1)}_{\mu\nu}+\nabla_{\mu}\nabla_{\nu}h\right)\phi_{p},\\
\label{ST D2R}
\delta\left[g_{\mu\nu}D^{2}\phi\right] &\w& \left(\nabla^2 h_{\mu\nu} - \frac{1}{2}g_{\mu\nu}^{(0)}\nabla^2 h\right)\phi_{p}.
\end{eqnarray}
We can relate these terms to corresponding terms in the $f(R)$ description the theory using (\ref{pert}). Explicitly, (\ref{ST RR}) is the same as (\ref{wRRmn}), (\ref{ST R2}) is the same as (\ref{wR2}) and is proportional to (\ref{ST V}) in weak limit, (\ref{ST DDR}) is the same as (\ref{wDDR}), and finally (\ref{ST D2R}) is the same as (\ref{wD2R}). Now all that remains is to put the pieces together to get the effective stress-energy tensor in the scalar-tensor description. Using (\ref{scalar-tensor-zero-2}), we can re-express all of the terms, except for the potential term, in 2-derivative form as
\begin{eqnarray}\label{ST-stress}
\kappa t^{(0)}_{\mu\nu} &\w& \frac{1}{2}h^{\alpha\beta}R^{(1)}_{\mu\alpha\nu\beta}-R^{(1)}_{\alpha(\mu}h_{\nu)}^{\ \alpha} + \frac{5}{12}h_{\mu\nu}R^{(1)}  +\frac{1}{2}hR^{(1)}_{\mu\nu}\nonumber\\ && +\frac{1}{4}g^{(0)}_{\mu\nu}h_{\alpha\beta}R^{(1)\alpha\beta}-\frac{7}{24}g_{\mu\nu}^{(0)}hR^{(1)} -\frac{1}{2}g_{\mu\nu}^{(0)}V_{p}.
\end{eqnarray}
Actually, we can also convert the potential term into 2-derivative form using (\ref{ScalTensStarPot}), (\ref{pert}) and (\ref{Szero2}) together. Let us collect all of the terms on the right hand side of (\ref{ST-stress}) together and call them the ``kinetic'' contribution to the stress-energy tensor, to contrast with the ``potential'' contribution from the last term. The trace $\kappa t^{(0)}$ receives a contribution of $-\frac{1}{3M^2}R^{(1)2}$ from the potential term and a contribution of $+\frac{1}{6M^2}R^{(1)2}$ from the kinetic terms. This gives the intriguing relationship that the trace of the effective stress-energy tensor receives a negative contribution from the potential that is twice the magnitude of the positive contribution of the kinetic term. Since the negative term from the potential wins, we have the negative pure trace component to the effective stress-energy tensor that is required to make it a candidate for mimicking a positive cosmological constant. 

\section{Diffeomorphism invariance}\label{StaroDiffeomorphism}

Since we know that the $f(R)$ and scalar-tensor descriptions are equivalent, it suffices to demonstrate diffeomorphism invariance (in the weak limit) of the effective stress-energy tensor in the $f(R)$ description. Indeed, the scalar mode is actually separately invariant under diffeomorphisms, so the diffeomorphism invariance in the scalar-tensor description will also be made clear by the derivation in this section. Consider the diffeomorphism transformation of a metric perturbation given in (\ref{dpertintro}). For this section it will be useful to specialize this to the case where the derivative we use is the covariant derivative that is compatible with the background metric, \ie $\nabla_\mu$. The linearized diffeomorphisms are given by (\ref{LinDiff})
Note that we only need use the linearized level here because higher orders in $h$ are suppressed by $\lambda$, therefore the higher order terms in the diffeomorphism transformation will vanish in the weak limit. The demonstration in this section that the effective stress-energy tensor (\ref{CombStaroStress}) is invariant under diffeomorphisms only applies because the expression is taken in the weak limit. For (\ref{LinDiff}) to have reasonable scaling behaviour, we require that $\xi_\mu$ has an analogous scaling to (\ref{wavelength2}), which is
\begin{equation}\label{wavelength4}
 \frac{1}{M^m}\nabla_{\alpha_1}\cdots\nabla_{\alpha_n}\xi_\beta \sim \lambda^{2+m-n}.
\end{equation}
As already noted, the linearized Riemann tensor (\ref{LinRie}) and its contractions (\ref{LinRics}) and (\ref{LinRict}) are already invariant under these linearized diffeomorphisms. To show that (\ref{CombStaroStress}) is diffeomorphism-invariant, let us begin by taking the zero tensor (\ref{Szero2}) and antisymmetrize in the indices $[\mu,\rho]$ and $[\nu,\sigma]$. Remembering the form of the linearized Riemann tensor (\ref{LinRie}), we can write the result as
\begin{equation}\label{Zerotran}
 \frac{1}{6}g^{(0)}_{[\mu|[\nu}h_{\sigma]|\rho]}R^{(1)}-h_{[\rho|[\sigma}R^{(1)}_{\nu]|\mu]} \w -\frac{R^{(1)}}{3M^2}\nabla_{[\mu|}\nabla_{[\nu}h_{\sigma]|\rho]} = \frac{1}{6M^2} R^{(1)} R^{(1)}_{\mu\rho\nu\sigma}.
\end{equation}
The right hand side of (\ref{Zerotran}) is invariant under diffeomorphisms. This leaves us with a constraint on the diffeomorphism transformations of 2-derivative terms, which is
\begin{multline}
\nabla_{(\rho}\xi_{\sigma)}R^{(1)}_{\mu\nu} + \nabla_{(\mu}\xi_{\nu)}R^{(1)}_{\rho\sigma} - \nabla_{(\mu}\xi_{\sigma)}R^{(1)}_{\nu\rho} - \nabla_{(\nu}\xi_{\rho)}R^{(1)}_{\mu\sigma} \w \\
\frac{g^{(0)}_{\mu\nu}}{6}\nabla_{(\rho}\xi_{\sigma)}R^{(1)}+\frac{g^{(0)}_{\rho\sigma}}{6}\nabla_{(\mu}\xi_{\nu)}R^{(1)}-\frac{g^{(0)}_{\mu\sigma}}{6}\nabla_{(\nu}\xi_{\rho)}R^{(1)}-\frac{g^{(0)}_{\nu\rho}}{6}\nabla_{(\mu}\xi_{\sigma)}R^{(1)}.
\end{multline}
Contracting the indices $\rho$ and $\sigma$ together, we obtain a very useful specialization, which is
\begin{equation}\label{zerogauge}
 \frac{g_{\mu\nu}^{(0)}}{6}\nabla_{\alpha}\xi^{\alpha}R^{(1)} \w \nabla_{\alpha}\xi^{\alpha}R^{(1)}_{\mu\nu}+\frac{2}{3}\nabla_{(\mu}\xi_{\nu)}R^{(1)} -\nabla_{(\mu|}\xi_{\alpha}R_{\ |\nu)}^{(1)\ \alpha}-\nabla_{\alpha}\xi_{(\mu}R_{\nu)}^{(1)\ \alpha}.
\end{equation}
Varying (\ref{CombStaroStress}) under linearized diffeomorphisms, which is the same as the full variation under diffeomorphisms in the weak limit, we get
 \begin{eqnarray}\label{Starostresstran}
\kappa\, \delta t^{(0)}_{\mu\nu} &\w& \frac{g_{\mu\nu}^{(0)}}{3}\xi^{\alpha}\nabla_{\alpha}R^{(1)} -\frac{1}{3}\xi_{(\mu}\nabla_{\nu)}R^{(1)} - \xi_{\alpha}\nabla^2\nabla_{(\mu}h_{\nu)}^{\ \ \alpha} \nonumber \\
&& + \xi_{\alpha}\nabla_{\beta}\nabla_{\mu}\nabla_{\nu}h^{\alpha\beta} - \xi^{\alpha}\nabla_{\alpha}\nabla_{\beta}\nabla_{(\mu}h_{\nu)}^{\ \beta} + \xi_{\alpha}\nabla^{\alpha}\nabla^2 h_{\mu\nu}.
\end{eqnarray}
The right hand side of (\ref{Starostresstran}) vanishes completely under the constraint given in (\ref{zerogauge}). This proves that the effective stress-energy tensor we have obtained in the weak limit for $R+R^2/6M^2$ gravity is invariant under diffeomorphisms. It also follows that the stress-energy tensor for the ordinary Einstein gravity case is invariant under diffeomorphisms (although this was already known), since this is just the special case where $M\to\infty$. This demonstration not only gives us confidence that our calculation of the effective stress-energy tensor is correct, but eliminates any concern that the result might be dependent on coordinates.

\section{Dark energy and new physics in the ultraviolet}\label{StaroUV}

In this chapter we have shown that, unlike in the Einstein gravity case discussed in Chapter \ref{AWHPreview2}, the effective stress-energy tensor for backreaction in $R+R^2/6M^2$ gravity is not traceless. Indeed, it has a negative trace, as required for a candidate to mimic a positive cosmological constant. This is already very interesting, but it is not enough to make this a viable candidate for explaining the observed effect of dark energy in the late universe. The negative trace (\ref{StarStressTrace}) is $\mathcal{O}(1)$ in the sense that it converges in the weak limit (\ie as the averaging scale becomes large) to a finite value, but we have not determined whether this finite value is cosmologically significant, what the overall fluid equation of state (\ref{FluidEqState}) is or indeed how either of these things evolves in time. To put the problem in perspective, let us imagine that the metric inhomogeneity consists of a single Fourier mode with wavelength $L$ equal to the radius of the Earth, \ie $1/L\sim3\times10^{-23}$ GeV in natural units. An operator of the form $\nabla^2/M^2$ would then effectively introduce a suppression factor of order $1/(ML)^2$. Thus far we have been motivated by Starobinsky inflation, for which we would have $M=3\times10^{13}$ GeV. Plugging this value into $1/(ML)^2$, we see that the suppression is actually around 72 orders of magnitude. If we considered metric fluctuations induced by larger structures, such as galaxy clusters, the suppression would be much greater again. Even the length scale of individual nucleons, which is about 1 fm, corresponds to an energy scale of $\sim$1 GeV, which would result in a suppression from the $\nabla^2/M^2$ operator by roughly 27 orders of magnitude. If higher-derivative gravity terms are to have a significant cosmological effect at late times, we need a source of much higher frequency metric perturbations.

Following this reasoning, we would need to look to new physics beyond the Standard Model to source fluctuations of high enough frequency to create a large effect.
With a view to mimicking a positive cosmological constant, three exotic candidates for sourcing the inhomogeneity have been proposed \cite{Preston:2014tua,Evans:2015zwa}. All of these candidates have problems, but they are indicative of the kinds of new physics that cosmological backreaction might be sensitive to. The first is WIMPzilla dark matter \cite{Kolb:1998ki}. WIMPzillas are stable ultra-heavy dark matter candidates that would have been created during the reheating phase immediately after inflation \cite{Chung:2001cb}. WIMPzilla candidates would typically have a similar mass to the inflaton mass scale $\sim10^{13}$ GeV, although they could be created with a suitable relic abundance with masses up to $\sim10^{16}$ GeV. WIMPzillas, having very large masses, would be used to source large spacetime derivatives of the metric close to the particle. When coupled to gravitational radiation, one might speculate that the cosmological spacetime might be filled with suitably high frequency perturbations which average to the required magnitude. Na\"{i}vely, WIMPzillas would seem to have the attraction that, being non-relativistic, one might model the inhomogeneity using a relatively simple Newtonian limit. However, the necessary coupling to gravitational radiation required to make them a viable candidate would be problematic for such an approximation. An immediate problem with using WIMPzillas to mimic a cosmological constant via backreaction is that their number density dilutes with the expansion of the universe as $a^{-3}$, whereas we need a dark energy-like fluid density to remain close to constant, or at least more slowly diluting, so that its relative share of the total energy density of the universe grows as the universe ages, \ie the accelerating expansion rate only occurs at late times. Again, it is possible that coupling to gravitational radiation might help reduce the dilution effect, as the metric fluctuations propagate into the spacetime further away from the particles themselves. On the other hand, radiation dilutes even faster, as $a^{-4}$, because the wavelength stretches as the universe expands, so it seems unlikely that a viable candidate for mimicking the cosmological constant can be built using WIMPzillas. However, backreaction from WIMPzillas in $R+R^2/6M^2$ gravity might still be cosmologically significant as an effect other than dark energy, further investigation of this possibility might even impose astronomical constraints on this possibility.

The second possibility is that quantum fluctuations in the spacetime might average to smooth classical perturbations at a scale where the effective gravity theory is sensitive and valid. This would scale correctly to mimic a cosmological constant as the universe expands, since the inhomogeneity would be sourced from the vacuum. There is a naturalness problem, however, that this could create an effect that is too large to mimic a small cosmological constant. It is also likely that one would require a more UV complete theory of gravity to study this rigorously, since the perturbations would likely have their shortest length scale below the scale of a valid classical effective theory. However, this possibility would be interesting to bear in mind for the future, because backreaction of this kind might impose strong constraints on such exotic spacetime behaviour in a more complete theory. Although this possibility is highly speculative, translational symmetry breaking in the spacetime of quantum gravity has been studied in other contexts, see for example \cite{Bonanno:2013dja}.

The third possibility, suggested more recently \cite{Evans:2015zwa}, is that a suitably constructed Quantum Chromodynamics-like non-Abelian gauge theory might induce a spontaneous translational symmetry breaking in the vacuum, producing ``striped'' phases in chiral condensates. As with the quantum spacetime idea, this inhomogeneity source is tied to the vacuum and therefore does not dilute as the universe expands, which is what we would want in a candidate for mimicking a cosmological constant. Since the development of this model is not the subject of this thesis, the discussion will be relatively brief here. Defining $\Lambda_{\rm stripe}$ as the energy scale that sets the amplitude and wavenumber of the inhomogeneity, the estimate for the magnitude of the backreaction effect proceeds as follows. The leading part of the perturbation to the field equation goes as
\begin{equation}\label{EMSprox}
 R^{(1)}_{\mu\nu} -\frac{1}{2}g_{\mu\nu}R^{(1)} \approx \kappa \delta T_{\mu\nu},
\end{equation}
where $\delta T_{\mu\nu}$ is the perturbation to the stress-energy tensor $T_{\mu\nu}-T^{(0)}_{\mu\nu}$. The relation (\ref{EMSprox}) is an approximation where the higher-derivative part is suppressed by the large mass scale and the terms at higher order in $h$ are considered to be negligible. This anticipates the conclusion that $\Lambda_{\rm stripe}$ is not required to be as large as $M$ in order to get a significant effect in the late universe. For the purpose of an order of magnitude estimate, let us say that
\begin{equation}\label{EMSlin}
 R^{(1)}\sim\nabla^2 h\sim\kappa\delta\rho,
\end{equation}
where $\delta\rho$ is the local overdensity: $\delta\rho = \rho-\rho_0$ where $\rho_0$ is the average density. Applying the approximate relation (\ref{EMSlin}) to the trace of the effective stress-energy tensor (\ref{StarStressTrace}), we get
\begin{equation}
 \kappa t^{(0)} \sim -\frac{\kappa^2}{6M^2}\left<\delta\rho^2\right>.
\end{equation}
We now substitute in the parameters of interest. First let us substitute $t^{(0)}=E_{\rm vac}^4$, where $E_{\rm vac}$ is the effective vacuum energy from backreaction. In the language of particle physics, $\kappa$ is the inverse square of the reduced Planck mass $M_{\rm Planck} = 2\times10^{18}$ GeV. The local overdensity $\delta\rho$ caused by the striped condensation goes like $\Lambda_{\rm stripe}^4$ by dimensions. Finally, this leaves us with an order of magnitude estimate for the effective vacuum energy from backreaction:
\begin{equation}\label{EMSest}
 E_{\rm vac} \sim \frac{\Lambda^2_{\rm stripe}}{\sqrt{MM_{\rm Planck}}}.
\end{equation}
To mimic the observed cosmological constant, let us set $E_{\rm vac}\approx 10^{-12}$ GeV. This leaves us with the estimate that the stripe scale $\Lambda_{\rm stripe}\sim 10^2$ GeV. At first glance, this is very intriguing because 100 GeV is the scale of electroweak symmetry breaking in particle physics \cite{Glashow:1961tr,Weinberg:1967tq,Agashe:2014kda}. It is also the scale of the most massive elementary particles of the Standard Model. However, translational symmetry breaking would also induce Lorentz violation of a kind that is strongly constrained in the Standard Model \cite{Pospelov:2004fj,Pruttivarasin:2014pja}. Therefore, a striped model would need to be disconnected from the Standard Model in its own dark sector in order to be phenomenologically viable. Another point of view is that if $\Lambda_{\rm stripe}$ were larger, the backreaction effect would need to be cancelled in other ways, such as by the usual cosmological constant. In this sense, equation (\ref{EMSest}) could also be seen as providing an upper bound on the inhomogeneity energy scale in a ``typical'' case where we have exotic new physics together with $R+R^2/6M^2$ gravity. However, such excitement should be moderated by the obvious comment that the infamous cosmological constant problem \cite{Weinberg:1988cp} is still unresolved and we have not ruled out that there might be be some fine-tuned cancellation between the backreaction and the actual cosmological constant. 

In Chapter \ref{ChapterBack2}, we will generalize our calculation of the backreaction to a general local diffeomorphism-invariant torsionless effective theory of gravity, \ie where the action is Taylor expandable in the Riemann tensor and its covariant derivatives. This generalization is easily motivated by the construction of effective actions using RG methods, \eg in Chapter \ref{ChapterMDIERG}.

\chapter{Generalized cosmological backreaction}\label{ChapterBack2}

\section{General scaling properties}

This chapter reports on original research in \cite{Preston:2016sip} that generalizes the work in \cite{Preston:2014tua}, which was reported in the previous chapter, to any effective gravity action that is Taylor-expandable in the Riemann tensor and its covariant derivatives.
The discussion in Section \ref{StaroScales} of how to incorporate additional scales of length reflects the thinking in the original paper. Moving on to perform the complete generalization, the reasoning outlined in \ref{StaroScales} is still useful, but it is not complete. First of all, in Section \ref{StaroScales}, we argued against incorporating $n$-th order derivatives by rescaling $h$ to be of $\mathcal{O}(\lambda^{n/2})$ because the backreaction for pure $R^2$ vanishes completely in the weak limit via the zero tensor, leaving us without even the known radiation-like backreaction from Einstein gravity. We further argued that the na\"{i}ve response of rescaling again to extract the $\mathcal{O}(h^3)$ part is also a mistake because we lose control of the quadratic order in $h$. A stronger argument is that this procedure is hopeless in an infinite series expansion in higher derivatives: we simply cannot keep rescaling the order in $\lambda$ of $h$ to incorporate more derivatives because the series keeps going {\it ad infinitum}, therefore we must tame the divergences caused by these higher derivatives instead by scaling $M$ with $\lambda$.

In $R+R^2/6M^2$ gravity, it was convenient to set $M$ to be proportional to $\lambda^{-1}$ because this allowed both the Einstein and the Starobinsky contributions to the field equation to contribute in the weak limit. It also maintained the intuitive relation that mass $\sim$1/length. The physical understanding of this was that the $\lambda\to0$ limit corresponds to taking the averaging scale to be much larger than the perturbation length scale. To prevent the scalaron length scale from being taken to be much larger than the perturbation scale in the weak limit, which would be beyond the validity of the effective theory, we require $M$ to be $\mathcal{O}(\lambda^{-n})$ where $n\ge1$. 

Requiring that the formalism makes good physical sense actually allows us to make a stronger comment than this. We have been interpreting $\lambda$ to be proportional to the wavelength of a monochromatic perturbation because of equation (\ref{wavelength}). Indeed, we could construct the complete perturbation as a Fourier spectrum of monochromatic perturbations, all with wavelength proportional to $\lambda$. An operator of the form $\nabla^2/M^2$, which occurs frequently in our expansion, effectively reads the ratio $\lambda^{-2}/M^2$. More physically, it is reading the square of the ratio of the perturbation wavenumber to the mass scale $M$. This ratio is physical, since both $M$ and the wavenumber of perturbations are, in principle, physically measurable quantities. Therefore the weak limit should not alter this ratio: $M$ must scale as $\lambda^{-1}$ for the averaging procedure to leave the macroscopic properties invariant, \ie to keep physically measurable quantities in fixed proportion as the averaging scale is taken to be large (as $\lambda\to0$).

In equation (\ref{noccc}), we commented that the cosmological constant term in the field equation converges in the weak limit to its background value, since it is just a constant times the metric. We also commented immediately below that equation that the same lemma Green and Wald derived to eliminate $T_{\mu\nu}$ from the zero tensor also applies to the cosmological constant term such that it too does not appear in the zero tensor. With that covered, we already know that it plays no r\^{o}le in the backreaction calculation. Na\"{i}vely, one might think that the dimension of the cosmological constant implies a $\lambda^{-2}$ scaling, but since it does not balance any of the higher-derivative operators, this is unnecessary. Moreover, the convergence of the field equation in the weak limit and (\ref{noccc}) in particular require us to leave it as $\mathcal{O}(1)$. Also note that the scaling of $M$ with $\lambda$ ensures that the higher-derivative background terms are suppressed in the weak limit, such that their only effect on the field equation in the limit of a large averaging scale, \ie much larger than $1/M$, is through their backreaction:
\begin{equation}
 R_{\mu\nu}^{(0)} -\frac{1}{2}g_{\mu\nu}^{(0)}R^{(0)} \w \kappa T_{\mu\nu} + \kappa t^{(0)}_{\mu\nu}.
\end{equation}

\section{Contributing action terms}

In this section, we will learn that not all gravitational action terms in a local diffeomorphism-invariant theory contribute to the effective stress-energy tensor for backreaction in the weak limit. In particular, we will see that cubic and higher orders in the Riemann tensor can be neglected. This section derives the minimal form of the local diffeomorphism-invariant action, which is a closed form, that does not lose any generality in the final calculation for the effective stress-energy tensor. The action is chosen to be local to ensure that the averaging scheme is well-defined and that it converges in the weak limit.

\subsection{Riemann tensor expansions}

The simple $R+R^2/6M^2$ gravity theory considered in Chapter \ref{ChapterBack1} is a truncation of a local $f(R)$ theory, \ie a theory that is Taylor-expandable in the Ricci scalar:
\begin{equation}
 S = \int d^4 x\sqrt{-g}\left(f(R)-2\Lambda\right) + S_{\rm matter},
\end{equation}
where
\begin{equation}
 f(R) = R + \frac{R^2}{6M^2} + {\rm const}\times \frac{R^3}{M^4} + \cdots.
\end{equation}
Note that since we have
\begin{equation}\label{soleform}
 \frac{1}{M^{k-2}}h_{\mu\nu}\nabla_{\alpha_1}\cdots\nabla_{\alpha_k}h_{\rho\sigma} \w \mathcal{O}(1),
\end{equation}
we also have
\begin{equation}\label{vanishing}
 \frac{1}{M^{2(m+n)}}h_{\alpha\beta}\nabla_\gamma\nabla_\delta R^{(1)n}R^{(0)m}\sim \lambda^{2m+n-1}.
\end{equation}
Which implies that the only non-vanishing contributions to the effective stress-energy tensor in the weak limit come from the case where $n=1$ and $m=0$, which implies that $R^3$ and higher terms in the action do not contribute in the weak limit. More explicitly, $R^3$ and higher terms in the action give field equation terms that are either suppressed by being of $\mathcal{O}(h^3)$ or higher, or by the mass scale $M$ balancing derivatives that do not contribute to the order in $\lambda$ (either because they are total derivatives or because they are being used in background Ricci tensors). 

We can relate the introduction of these higher-order terms in the Ricci scalar to the scalar-tensor description as the introduction of terms in the scalar potential (\ref{ScalTensStarPot}) at higher order in $\phi$. We already know that the perturbations to the $\phi$ field, which go as (\ref{pert}), are suppressed at $\mathcal{O}(\lambda)$. Therefore, these higher orders in $\phi$, which represent the higher orders in $R$, do not contribute in the weak limit. From this we have learnt that the stress-energy tensor (\ref{CombStaroStress}), derived for $R+R^2/6M^2$, is already general for any local $f(R)$ theory.

The argument used in (\ref{vanishing}) applies equally well to a general expansion in the Riemann tensor, \ie cubic and higher order terms in the Riemann tensor also vanish under weak limit averaging. At the quadratic order in the Riemann tensor, we have the Lagrangian terms $R^2$, $R_{\mu\nu}R^{\mu\nu}$ and $R_{\mu\nu\rho\sigma}R^{\mu\nu\rho\sigma}$. However, in four dimensions, the Gauss-Bonnet structure (\ref{Gauss-Bonnet}) is a topological invariant, therefore these three terms do not give independent field equation contributions, and so we can choose to express one in terms of the other two. The most general diffeomorphism-invariant action where the gravity part is an expansion in the Riemann tensor can be written up to quadratic order as
\begin{equation}
 S = \int d^4 x \sqrt{-g}\frac{1}{2\kappa}\left(-2\Lambda + R + \frac{a}{M^2}R_{\mu\nu}R^{\mu\nu} + \frac{b}{M^2}R^2 + \cdots\right)+S_{\rm matter}.
\end{equation}

\subsection{Covariant derivative expansions}

We can further generalize this action without compromising diffeomorphism invariance or locality by introducing covariant derivatives of the Riemann tensor into the expansion. As already noted in the comments under equation (\ref{Gauss-Bonnet}), these explicit covariant derivative operators first appear at the quadratic order in the Riemann tensor. Because the effective stress-energy tensor in the weak limit only receives contributions from the quadratic order in the Riemann tensor, the ordering of the covariant derivative operators is unimportant in this study. This is because any non-zero commutators of covariant derivative operators raise the order in the Riemann tensor, as can be seen for example in (\ref{EinsteinFieldStrengthAnalogy}). Rewriting (\ref{EinsteinFieldStrengthAnalogy}) in the notation conventions for this section, we have
\begin{equation}
 \left[D_{\mu},D_{\nu}\right]v_{\alpha} = R^{\lambda}_{\ \alpha\nu\mu}v_{\lambda}.
\end{equation}

Even armed with the knowledge that the order of the covariant derivative operators is unimportant in this work, we might still imagine that there would be a large number of independent terms coming from the different index structures. This is not the case. To see this, recall the second Bianchi identity (\ref{SecondBianchi}), which we can rewrite in the present notation as
\begin{equation}\label{2ndBianchi}
 D_{\lambda}R_{\alpha\beta\gamma\delta} + D_{\gamma}R_{\alpha\beta\delta\lambda} + D_{\delta}R_{\alpha\beta\lambda\gamma} = 0,
\end{equation}
we can rewrite (\ref{SpecialSecondBianchi}) in the current notation as
\begin{equation}
 g^{\alpha\beta}D_{\alpha}R_{\beta\gamma} = \frac{1}{2}D_{\gamma}R.
\end{equation}
The second Bianchi identity (\ref{2ndBianchi}), together with the index (anti-)symmetry properties of the Riemann tensor, can be used to rewrite every index structure for a local diffeomorphism-invariant theory up to quadratic order in the Riemann tensor as
\begin{eqnarray}\label{unsimpact}
 S & = & \int d^4x \sqrt{-g}\frac{1}{2\kappa}\left(R -2\Lambda + \frac{1}{M^2}R_{\alpha\beta}a\derarg R^{\alpha\beta} + \frac{1}{M^2}Rb\derarg R \right. \nonumber \\ && \left.+ \frac{1}{M^2}R_{\alpha\beta\gamma\delta}c\derarg R^{\alpha\beta\gamma\delta}+ \cdots\right) + S_{\rm matter}.
\end{eqnarray}
Thus we now have a closed form for the action terms that contribute to the effective stress-energy tensor in the weak limit. This form is now reasonably straightforward to proceed with calculating results for. However, we can simplify further, since not all of these terms are independent. As already noted, the dimension 4 operators are related via the Gauss-Bonnet topological invariant (\ref{Gauss-Bonnet}). The introduction of the explicit covariant derivative operators breaks the topological invariance. However, the entire $a=-4b=c$ structure disappears from the field equation in the weak limit. To see this, we can look ahead to equations (\ref{derivA}) and (\ref{derivB}), which give the field equation contributions of the first two structures that are non-vanishing in the weak limit. Except for the final terms with $a'$ and $b'$, these expressions have the same structure as if $a$ and $b$ were constants. In the case of an $a=-4b=c$ structure, these final terms would also vanish in the weak limit because
\begin{equation}\label{GB2}
 \frac{1}{M^4}\left(R_{\alpha\beta\gamma\delta}^{(1)}a' \nabla_\mu \nabla_\nu R^{\alpha\beta\gamma\delta(1)} -4R_{\alpha\beta}^{(1)}a' \nabla_\mu \nabla_\nu R^{\alpha\beta(1)} + R^{(1)}a'\nabla_\mu \nabla_\nu R^{(1)}\right) \w 0,
\end{equation}
which is strongly related to the finding in Section \ref{MDIERGTran2} that there only exist two independent transverse 2-point functions in a diffeomorphism-invariant theory. Thus, while the higher-derivative generalization of the Gauss-Bonnet structure is not a topological invariant, it behaves as though it were in the weak limit. More precisely, this means that
\begin{eqnarray}\label{genGauss-Bonnet}
 0 &\w& \frac{\delta}{\delta g^{\mu\nu}}\int d^4x \sqrt{-g}\frac{1}{M^2}\left(R_{\alpha\beta\gamma\delta}a\derarg R^{\alpha\beta\gamma\delta} \right. \nonumber \\ && \left. -4 R_{\alpha\beta}a\derarg R^{\alpha\beta} + Ra\derarg R\right).
\end{eqnarray}
Remember that the scaling of $M$ in the weak limit also makes the background higher-derivative terms vanish. In fact, the reason why the terms that would break topological invariance vanish in the weak limit is that they come from varying the connections from the explicit covariant derivatives with respect to the inverse metric, so they actually appear at the same order as the action terms that are cubic in the Riemann tensor. We can also notice from this result that, for a diffeomorphism-invariant theory, the only non-vanishing field equations terms come from Lagrangian terms that would themselves have a non-vanishing weak limit. Given that the three index structures at quadratic order in (\ref{genGauss-Bonnet}) are not independent, we can choose to eliminate one by re-expressing it in terms of the other two. Finally, we are left with a much simplified form of the action, which now contains the minimum number of terms required to preserved full generality:
\begin{eqnarray}\label{TotalAction}
 S & = & \int d^4x \sqrt{-g}\frac{1}{2\kappa}\left(R -2\Lambda + \frac{1}{M^2}R_{\alpha\beta}a\derarg R^{\alpha\beta} + \frac{1}{M^2}Rb\derarg R + \cdots\right) \nonumber \\ && + S_{\rm matter}.
\end{eqnarray}
To proceed with the calculations, we need to know the field equations. The adaptation and generalization of (\ref{derivRmnRmn}) we require is
\begin{eqnarray}\label{derivA}
  \frac{\delta}{\delta g^{\mu\nu}}\int d^4x \sqrt{-g}R_{\alpha\beta}aR^{\alpha\beta} & = & \sqrt{-g}\left(-\frac{1}{2}g_{\mu\nu}R_{\alpha\beta}aR^{\alpha\beta}+2R_{\mu\alpha}aR_{\nu}^{\ \alpha} \right.\nonumber \\ & &\left. +D^2 aR_{\mu\nu}+\frac{1}{2}g_{\mu\nu}D^2 aR-2D_\alpha D_{(\mu} aR_{\nu)}^{\ \ \alpha} \right.\nonumber \\ & &\left. + \frac{1}{M^2}R_{\alpha\beta}D_{\mu}D_{\nu}a'R^{\alpha\beta}+\cdots\right),
\end{eqnarray}
where $a'$ is the derivative of $a$ with respect to $D^2/M^2$. The ellipsis stands for those terms which vanish in the weak limit (which also do not contribute to the zero tensor). These vanishing terms come from varying with respect to the inverse metric the metric connections appearing on account of the explicit covariant derivative operator expansion $a$. Such terms vanish because they lack terms of the form (\ref{soleform}) for reasons similar to the argument expressed by (\ref{vanishing}). The corresponding equation related to (\ref{derivR2}) is
\begin{eqnarray}\label{derivB}
  \frac{\delta}{\delta g^{\mu\nu}}\int d^4x \sqrt{-g}RbR & = & \sqrt{-g}\left(-\frac{1}{2}g_{\mu\nu}RbR + 2RbR_{\mu\nu}-2D_{\mu}D_{\nu}bR + 2g_{\mu\nu}D^2 bR \right. \nonumber \\ & & \left. + \frac{1}{M^2}R D_{\mu}D_{\nu}b'R+\cdots\right),
\end{eqnarray}
where $b'$ is the derivative of $b$ with respect to $D^2/M^2$. With that, we can now proceed to calculate zero tensors and field equation contributions.

\section{Zero tensors}

As with the pure Einstein and Starobinsky cases, the information from the linearized field equation is preserved in the zero tensor, whose general form is (\ref{GenZero}). Once again, we multiply the field equation by $h_{\rho\sigma}$ and take a weak limit to find
\begin{eqnarray}\label{rawzero}
 0 & \w & h_{\rho\sigma}R^{(1)}_{\mu\nu} - \frac{1}{2}g^{(0)}_{\mu\nu}h_{\rho\sigma}R^{(1)} + \frac{1}{M^2}\left(-2h_{\rho\sigma}\nabla_\mu \nabla_\nu bR^{(1)} + 2g_{\mu\nu}^{(0)}h_{\rho\sigma}\nabla^2 bR^{(1)} \right. \nonumber \\ &&
\left. h_{\rho\sigma}\nabla^2 aR_{\mu\nu}^{(1)}+\frac{1}{2}g_{\mu\nu}^{(0)}h_{\rho\sigma}\nabla^2 aR^{(1)} - h_{\rho\sigma}\nabla_\mu \nabla_\nu aR^{(1)}\right).
\end{eqnarray}
This is a rather complicated form, so once again we wish to find more convenient forms by taking the trace and substituting it back. The trace over $\mu$ and $\nu$ of (\ref{rawzero}) is
\begin{equation}\label{trzero}
 h_{\rho\sigma}R^{(1)} \w \frac{2}{M^2}h_{\rho\sigma}(a+3b)\nabla^2 R^{(1)}.
\end{equation}
The most useful form of the zero tensor for this work comes from using (\ref{trzero}) to eliminate the second term in (\ref{rawzero}) to get
\begin{equation}\label{genzero}
 h_{\rho\sigma}R^{(1)}_{\mu\nu} \w \frac{1}{M^2}\left(h_{\rho\sigma}\nabla_\mu \nabla_\nu (a+2b)R^{(1)} + \frac{1}{2}g_{\mu\nu}^{(0)}h_{\rho\sigma}\nabla^2 (a+2b)R^{(1)}-h_{\rho\sigma}\nabla^2 a R_{\mu\nu}^{(1)}\right).
\end{equation}
In addition to the contraction over $\mu$ and $\nu$ to get (\ref{trzero}), it is also useful to know the cross-contraction of index pairs:
\begin{equation}
 h_{\alpha\beta}R^{(1)\alpha\beta} \w \frac{1}{M^2}\left(h^{\alpha\beta}\nabla_\alpha \nabla_\beta (a+2b)R^{(1)}+\frac{1}{2}h\nabla^2 (a+2b)R^{(1)}-h^{\alpha\beta}\nabla^2 a R^{(1)}_{\alpha\beta}\right).
\end{equation}
The zero tensor for Einstein gravity (\ref{EinZero}) was able to hugely simplify the stress-energy tensor and demonstrate that it is traceless in the weak limit. When generalized to $R+R^2/6M^2$, the zero tensor forms (\ref{Szero1}) and (\ref{Szero2}) were powerful tools for converting between 2-derivative and 4-derivative forms. The zero tensor for the generalized theory (\ref{genzero}) can be used to re-express the eventual stress energy tensor (\ref{totalstress}), but not to such a powerful effect. The zero tensor will be used in Section \ref{Back2Diff} to prove that the stress-energy tensor derived in Section \ref{Back2Stress} is diffeomorphism-invariant.   

\section{Stress-energy tensor}\label{Back2Stress}

To calculate the stress-energy tensor, we need to evaluate the weak limits of the field equations terms that are given by (\ref{derivA}) and (\ref{derivB}). The Einstein gravity terms have already been evaluated in (\ref{noccc}), (\ref{Ein1}) and (\ref{Ein2}). The Starobinsky terms were evaluated in (\ref{wRRmn}), (\ref{wR2}), (\ref{wDDR}) and (\ref{wD2R}). Concerning the functions of covariant derivatives, \eg $a\derarg$, we can treat them as if they converge in the weak limit to their background forms, \ie $a\derbarg$ in the example given. This is because they never appear as total derivatives in the field equation and higher-order corrections in $h$ would be suppressed by $\lambda$. To give an example, we have
\begin{equation}
 \frac{1}{M^2}Rb\derarg R_{\mu\nu} \w \frac{1}{M^2} R^{(1)}b\derbarg R^{(1)}_{\mu\nu}.
\end{equation}
Since this is simple to do, we will not repeat the calculation for terms that are higher-derivative generalizations of the Starobinsky terms. New expressions for weak-averaged field equation terms are
\begin{eqnarray}
 \frac{1}{M^4}\delta\left[R_{\alpha\beta}D_{\mu}D_{\nu}a'\derarg R^{\alpha\beta}\right] & \w & \frac{1}{M^4}R_{\alpha\beta}^{(1)}\nabla_{\mu}\nabla_{\nu}a'\derbarg R^{(1)\alpha\beta}, \\
 \frac{1}{M^4}\delta\left[R D_{\mu}D_{\nu}b'\derarg R\right] & \w & \frac{1}{M^4}R^{(1)} \nabla_{\mu}\nabla_{\nu}b'\derbarg R^{(1)}, \\
 \frac{1}{M^2}\delta\left[g_{\mu\nu}R_{\alpha\beta}a\derarg R^{\alpha\beta}\right] & \w & \frac{g^{(0)}_{\mu\nu}}{M^2} R^{(1)}_{\alpha\beta}a\derbarg R^{(1)\alpha\beta}, \\
 \frac{1}{M^2}\delta\left[R_{\mu\alpha}a\derarg R_{\nu}^{\ \alpha}\right] & \w & \frac{1}{M^2}R^{(1)}_{\mu\alpha}a\derbarg R_{\nu}^{(1) \alpha}, \\
 \frac{1}{M^2}\delta\left[D^2 a\derarg R_{\mu\nu}\right] & \w & \frac{1}{M^2}\left(-\frac{1}{2}ha\derbarg\nabla^2 R^{(1)}_{\mu\nu} \right.\nonumber \\ && 
\left.+ h_{\alpha(\mu}a\derbarg\nabla^2 R_{\nu)}^{(1)\alpha}\right. \nonumber \\ && 
\left.+ h^{\alpha\beta}a\derbarg\nabla_{\alpha}\nabla_{(\mu} R^{(1)}_{\nu)\beta}\right. \nonumber \\ && \left. -\frac{1}{2}h^{\alpha}_{\ (\mu}\nabla_{\nu)}\nabla_{\alpha} a\derbarg R^{(1)}\right), \\
 \frac{1}{M^2}\delta\left[D_{\alpha}D_{(\mu}a\derarg R_{\nu)}^{\ \ \alpha}\right] & \w & \frac{1}{M^2}\left(\frac{1}{4}h_{\ (\mu}^{\alpha}\nabla_{\nu)}\nabla_{\alpha}a\derbarg R^{(1)} \right.\nonumber \\ && 
\left.+ \frac{1}{2}h^{\alpha\beta}a\derbarg \nabla_{\alpha}\nabla_{(\mu}R^{(1)}_{\nu)\beta}\right. \nonumber \\ && 
- \frac{1}{2}h^{\alpha}_{\ (\mu|}a\derbarg\nabla^2 R^{(1)}_{|\nu)\alpha} \nonumber \\ &&
+\frac{1}{2}h^{\alpha\beta}a\derbarg\nabla_{\mu}\nabla_\nu R^{(1)}_{\alpha\beta} \nonumber \\ &&
\left.- \frac{1}{4}ha\derbarg\nabla_\mu\nabla_\nu R^{(1)}\right).
\end{eqnarray}
Finally, putting all of these pieces together, we have the general form of the effective stress-energy tensor derived from an action that is Taylor-expandable in the Riemann tensor and its covariant derivatives: 
\begin{eqnarray}\label{totalstress}
 \kappa t^{(0)}_{\mu\nu} & = & \frac{1}{2}h^{\alpha\beta}R^{(1)}_{\mu\alpha\nu\beta} + \frac{3}{4}h_{\mu\nu}R^{(1)}-R^{(1)}_{\alpha(\mu}h_{\nu)}^{\ \ \alpha}-\frac{1}{8}g^{(0)}_{\mu\nu}\left(hR^{(1)}+2h^{\alpha\beta}R^{(1)}_{\alpha\beta}\right) \nonumber \\ &&
+\frac{1}{M^2}\left(\frac{1}{2}g^{(0)}_{\mu\nu}R^{(1)}bR^{(1)}+h\nabla_{\mu}\nabla_{\nu}bR^{(1)}-2h_{\mu\nu}\nabla^2 bR^{(1)}+g^{(0)}_{\mu\nu}h\nabla^2 bR^{(1)} \right. \nonumber \\ &&
+ \frac{1}{2}g_{\mu\nu}^{(0)}R^{(1)}_{\alpha\beta}aR^{(1)\alpha\beta}-2R_{\mu\alpha}^{(1)}aR_{\nu}^{(0)\alpha}+ \frac{1}{2}h\nabla^2 a R_{\mu\nu}^{(1)}-2h^{\alpha}_{\ (\mu}\nabla^2aR_{\nu)\alpha}^{(1)} \nonumber \\ &&
+h^{\alpha\beta}\nabla_\mu \nabla_\nu aR_{\alpha\beta}^{(1)} + h^\alpha_{\ (\mu}\nabla_{\nu)}\nabla_\alpha a R^{(1)} + \frac{1}{4}g_{\mu\nu}^{(0)}h\nabla^2 a R^{(1)}-\frac{1}{2}h\nabla_\mu \nabla_\nu a R^{(1)}\nonumber \\ &&
\left.  - \frac{1}{2}h_{\mu\nu}\nabla^2 aR^{(1)}-R^{(1)}\frac{\nabla_\mu \nabla_\nu}{M^2}b'R^{(1)} -R^{(1)}_{\alpha\beta}\frac{\nabla_\mu \nabla_\nu}{M^2}a'R^{(1)\alpha\beta}\right).
\end{eqnarray}
This effective stress-energy tensor can be expressed in different ways by applying the zero tensor constraint in (\ref{genzero}).

\section{Effective stress-energy tensor trace}

Backreaction was shown to be radiation-like in Einstein gravity because the trace of the effective stress-energy tensor (\ref{EinAll}) is traceless under the zero tensor constraint (\ref{EinZero}). The $R+R^2/6M^2$ case, however, had a non-vanishing negative trace (\ref{StarStressTrace}), which caused excitement at the possibility that this might mimic a positive cosmological constant, motivating the estimate (\ref{EMSest}) of the effective vacuum energy from backreaction when the fluctuations are associated with the vacuum. Therefore it is once again of interest to evaluate the trace of (\ref{totalstress}) and interpret it physically.

\subsection{General consistency of the trace}

There are two ways to evaluate this trace. Either we can simply take the trace with respect to the background metric of (\ref{totalstress}) or we can take the trace of the field equation first and then perform the weak limit. It is important for the consistency of the formalism that these two methods are equivalent, and therefore it is demonstrated explicitly here. Let us begin by noting that the effective stress-energy tensor can be defined by the weak-limit equation
\begin{equation}\label{generalstress}
 \kappa t^{(0)}_{\mu\nu} \w -\delta\left[\frac{2\kappa}{\sqrt{g}}\frac{\delta S_{\rm grav}}{\delta g^{\mu\nu}}\right].
\end{equation}
The corresponding zero tensor can be defined by
\begin{equation}\label{generalzero}
 0 \w h_{\rho\sigma}\delta\left[\frac{2\kappa}{\sqrt{g}}\frac{\delta S_{\rm grav}}{\delta g^{\mu\nu}}\right].
\end{equation}
To see the equivalence of the two methods, let us initially take the first approach, which is to calculate the effective stress energy tensor first and then to take the trace using the background metric:
\begin{equation}\label{tracecon}
 \kappa t^{(0)} \w -g^{(0)\mu\nu}\delta\left[\frac{2\kappa}{\sqrt{g}}\frac{\delta S_{\rm grav}}{\delta g^{\mu\nu}}\right] \w -\delta\left[g^{\mu\nu}\frac{2\kappa}{\sqrt{g}}\frac{\delta S_{\rm grav}}{\delta g^{\mu\nu}}\right]-\underbrace{h^{\mu\nu}\delta\left[\frac{2\kappa}{\sqrt{g}}\frac{\delta S_{\rm grav}}{\delta g^{\mu\nu}}\right]}_{\text{zero, via the zero tensor}}.
\end{equation}
The far right hand side split the initial expression into a piece that is precisely the form obtained by the second method (\ie taking the field equation trace first and then performing the weak limit) and another term that is clearly vanishing in the weak limit, using the zero tensor (\ref{generalzero}). Thus (\ref{tracecon}) demonstrates that the two approaches are consistent, which bodes well for the consistency of the formalism in general.

\subsection{Evaluation of the generalized trace}

To evaluate the trace of the effective stress-energy tensor (\ref{totalstress}), we will pursue the second approach explicitly here. The trace of the field equation is
\begin{equation}\label{fieldtr}
 -R + \frac{2}{M^2}D^2 (a+3b)R + \frac{1}{M^4}\left(R_{\alpha\beta}D^2 a'R^{\alpha\beta} + RD^2 b'R\right)+\cdots= \kappa T.
\end{equation}
To evaluate the weak limit, it is useful to know that
\begin{eqnarray}
 -R^{(2)} & \w & \frac{1}{4}hR^{(1)}+\frac{1}{2}h^{\alpha\beta}R^{(1)}_{\alpha\beta} \nonumber \\ & \w &
\frac{1}{2M^2}\left(h^{\alpha\beta}\nabla_\alpha \nabla_\beta (a+2b)R^{(1)}+\frac{1}{2}h\nabla^2 (3a+8b)R^{(1)} \right. \nonumber \\ && \left. -h^{\alpha\beta}\nabla^2 a R^{(1)}_{\alpha\beta}\right),
\end{eqnarray}
and
\begin{equation}
 \frac{1}{M^2}\delta\left[D^2 (a+3b)R\right] \w -\frac{1}{2M^2}h\nabla^2 (a+3b)R^{(1)}.
\end{equation}
Using these to take the weak limit of (\ref{fieldtr}), we obtain
\begin{eqnarray}\label{intermtrace}
 \kappa t^{(0)} & \w & -\frac{1}{2M^2}\left(h^{\alpha\beta}\nabla_\alpha \nabla_\beta (a+2b)R^{(1)}-h\nabla^2 (\frac{a}{2}+2b)R^{(1)}-h^{\alpha\beta}\nabla^2 aR^{(1)}_{\alpha\beta}\right. \nonumber \\ && 
\left. +2R^{(1)}_{\alpha\beta}a'\frac{\nabla^2}{M^2}R^{(1)\alpha\beta} + 2R^{(1)}b'\frac{\nabla^2}{M^2}R^{(1)}\right).
\end{eqnarray}
At first glance, this is not very illuminating at all. However, we can use our experience of the transverse 2-point functions in Section (\ref{MDIERGTran2}) to rewrite (\ref{intermtrace}) into a more intuitive form. We will take the first three terms on the right hand side of (\ref{intermtrace}) and separate them into two parts. First of all, we have
\begin{equation}
 \frac{1}{M^2}\left(h^{\alpha\beta}\nabla_{\alpha}\nabla_{\beta}(a+2b)R^{(1)} - h\nabla^2 (a+2b)R^{(1)}\right) \w \frac{1}{M^2}R^{(1)}(a+2b)R^{(1)},
\end{equation}
but we also have the less obvious relation that
\begin{equation}
 \frac{1}{M^2}\left(\frac{1}{2}h\nabla^2 aR^{(1)}-h^{\alpha\beta}\nabla^2 a R^{(1)}_{\alpha\beta}\right) \w \frac{1}{M^2}\left(2R^{(1)}_{\alpha\beta}aR^{(1)\alpha\beta} - R^{(1)}aR^{(1)}\right).
\end{equation}
Using these two expressions, we can simplify (\ref{intermtrace}) to
\begin{equation}\label{totalstresstrace}
 \kappa t^{(0)} \w  -\frac{1}{M^2}\left(R^{(1)}_{\alpha\beta}\left(a+ a'\frac{\nabla^2}{M^2}\right)R^{(1)\alpha\beta} + R^{(1)}\left(b+b'\frac{\nabla^2}{M^2}\right)R^{(1)}\right).
\end{equation}
This form is much more elegant and obviously diffeomorphism-invariant. It also clearly contains (\ref{StarStressTrace}) as the special case of $a=0$, $b=1/6$. The extension to include the generalized index structure and expansion in derivatives is also very elegant and intuitive. Once again, we see that the trace is non-zero in the weak limit and we see that the $R^2$ term is not unique in being able to provide this. It is also obvious at a glance from this form what expression we would have if we chose a form of the action with the $R_{\mu\nu\rho\sigma}c\derarg R^{\mu\nu\rho\sigma}$ term explicitly included.
\section{Diffeomorphism invariance}\label{Back2Diff}

As always, the transformation of the metric perturbation under diffeomorphisms is given by the Lie derivative of the metric, which we can write as
\begin{equation}\label{yetanotherLie}
 \delta h_{\alpha\beta} = \mathsterling_\xi g_{\alpha\beta} = 2g_{\lambda(\alpha}\nabla_{\beta)}\xi^\lambda + g^{\gamma\delta}\xi_\gamma \nabla_\delta h_{\alpha\beta},
\end{equation}
where we choose to use the covariant derivative compatible with the background metric. As in previous cases, contributions to the diffeomorphism transformation of the effective stress-energy tensor coming from the second term in (\ref{yetanotherLie}) trivially vanish in the weak limit, on account of being suppressed by the extra factor of $h$, \ie we only need consider the linearized diffeomorphism transformation. Just as before, the order in $\lambda$ of $\xi$ and its derivatives is given by (\ref{wavelength4}). Applying the diffeomorphism transformation (\ref{yetanotherLie}) directly to the effective stress-energy tensor (\ref{totalstress}) in the weak limit, we initially have
\begin{eqnarray}\label{totaldifftrans}
 \kappa \delta t^{(0)}_{\mu\nu} & \w & -\xi_{(\mu}\nabla_{\nu)}R^{(1)}-\xi\cdot\nabla R^{(1)}_{\mu\nu} + 2\xi_{\alpha}\nabla_{(\mu}R^{(1)\alpha}_{\nu)}+\frac{1}{2}g^{(0)}_{\mu\nu}\xi\cdot\nabla R^{(1)} \nonumber \\ &&
+\frac{2}{M^2}\left(-\xi\cdot\nabla\nabla_\mu \nabla_\nu bR^{(1)} + 2\xi_{(\mu}\nabla_{\nu)}\nabla^2 bR^{(1)} - g^{(0)}_{\mu\nu}\xi\cdot\nabla\nabla^2 bR^{(1)}\right) \nonumber \\ &&
+\frac{1}{M^2}\left(-\xi\cdot\nabla\nabla^2 aR^{(1)}_{\mu\nu}+2\xi^{\alpha}\nabla_{(\mu}\nabla^2 a R^{(1)}_{\nu)\alpha}-\frac{1}{2}g^{(0)}_{\mu\nu}\xi\cdot\nabla\nabla^2 aR^{(1)}\right.\nonumber \\ &&
\left.-\xi\cdot\nabla\nabla_{\mu}\nabla_{\nu}aR^{(1)}+\xi_{(\mu}\nabla_{\nu)}\nabla^2 aR^{(1)}\right).
\end{eqnarray}
To demonstrate that the right hand side of (\ref{totaldifftrans}) vanishes in the weak limit, we need to apply the zero tensor constraint (\ref{genzero}). The diffeomorphism transformation of (\ref{genzero}) gives
\begin{eqnarray}\label{zerotransf}
 \xi_{(\rho}\nabla_{\sigma)}R^{(1)}_{\mu\nu} & \w & \frac{1}{M^2}\left(\xi_{(\rho}\nabla_{\sigma)}\nabla_\mu \nabla_\nu (a+2b)R^{(1)} + \frac{1}{2}g^{(0)}_{\mu\nu}\xi_{(\rho}\nabla_{\sigma)}\nabla^2 (a+2b)R^{(1)} \right.\nonumber \\ && 
\left.-\xi_{(\rho}\nabla_{\sigma)}\nabla^2 aR^{(1)}_{\mu\nu}\right).
\end{eqnarray}
The first four terms on the right hand side of (\ref{totaldifftrans}), which are all third order derivative forms, can be converted into higher-order derivative forms by substituting in the diffeomorphism transformation of the zero tensor given in (\ref{zerotransf}). More specifically, the contracted forms that we need to substitute in are
\begin{eqnarray}
 \xi_{(\mu}\nabla_{\nu)}R^{(1)} & \w & \frac{1}{M^2}\xi_{(\mu}\nabla_{\nu)}\nabla^2 (2a+6b)R^{(1)}, \\
 \xi\cdot\nabla R^{(1)}_{\mu\nu} & \w & \frac{1}{M^2}\left(\xi\cdot\nabla\nabla_\mu \nabla_\nu (a+2b)R^{(1)}+\frac{1}{2}g^{(0)}_{\mu\nu}\xi\cdot\nabla\nabla^2 (a+2b)R^{(1)} \right. \nonumber \\ && \left. -\xi\cdot\nabla\nabla^2aR^{(1)}_{\mu\nu}\right), \\
 -2\xi_{\alpha}\nabla_{(\mu}R^{(1)\alpha}_{\nu)} & \w & \frac{1}{M^2}\left(-2\xi\cdot\nabla \nabla_\mu \nabla_\nu (a+2b)R^{(1)}-\xi_{(\mu}\nabla_{\nu)}\nabla^2 (a+2b)R^{(1)} \right. \nonumber \\ && 
\left.+2\xi_{\alpha}\nabla_{(\mu}\nabla^2 aR_{\nu)}^{\ \ \alpha}\right), \\
 -\frac{1}{2}g^{(0)}_{\mu\nu}\xi\cdot\nabla R^{(1)} & \w & -\frac{1}{M^2}g^{(0)}_{\mu\nu}\xi\cdot\nabla\nabla^2 (a+3b)R^{(1)}.
\end{eqnarray}
After substituting these forms in for the first four terms on the right hand side of (\ref{totaldifftrans}), the entire right hand side of (\ref{totaldifftrans}) cancels exactly to zero. With that finding, we know that the effective stress-energy tensor (\ref{totalstress}) is diffeomorphism-invariant. This important check demonstrates the consistency of that result and shows that it is not dependent on any coordinate scheme. This demonstration applies to all effective theories of gravity whose Lagrangians can be expressed as a local expansion in the Riemann tensor and its covariant derivatives.

\section{Generalized backreaction discussion}

Cosmological backreaction has been proposed as an alternative mechanism to explain the accelerating expansion of the universe at late times. The original motivations of researchers in this field were that a small positive cosmological constant is unnatural (although it is reasonable to suggest that a zero cosmological constant might be infinitely so), that the recentness of the effect poses a ``why now?'' problem (\ie why is the acceleration starting to happen at roughly the same time as there are humans to observe it?) and that it is appealing to provide a mechanism that does not rely on any exotic undiscovered physics. The usual kind of cosmological backreaction, where an inhomogeneous distribution of large structures like galaxies in a cosmos governed by Einstein gravity creates the effect, could be motivated in this way, since the ``why now?'' issue is answered by saying that the universe is only now becoming sufficiently inhomogeneously structured to create a large enough effect. However, estimates of the expected size of the effect in this non-exotic case are far too small \cite{Ishibashi:2005sj,Behrend:2007mf}. More damningly than that, Green and Wald rigorously demonstrated that the effective stress-energy tensor in Einstein gravity is traceless, corresponding to gravitational radiation, which cannot create an accelerating effect \cite{Green:2010qy}.

The motivation here is quite different. The arguments and calculations presented here do not enter into the original backreaction controversy, rather they open up a new front on the question. Extending Einstein's GR to include an $R^2$ term or indeed an infinite expansion in higher derivatives is already introducing new physics of the kind that backreaction theorists are usually trying to avoid. Modified gravity is also popular as a means of explaining dark energy in the context of purely homogeneous cosmology. Such theories usually modify gravity at large length scales with screening mechanisms to avoid solar system-scale constraints, which is hard to motivate theoretically. For examples of the great wealth of literature on such modified gravity theories of cosmology, see \cite{Hu:2007nk,Starobinsky:2007hu,Tsujikawa:2007xu,Appleby:2007vb,Cognola:2007zu,Linder:2009jz,Amendola:2006we,Li:2007xn,DeFelice:2010aj,Tsujikawa:2010sc,Nojiri:2010wj,Abebe:2013zua,Clifton:2011jh,Guo:2013swa,Mukherjee:2014fna}. 

A local expansion in higher-derivative operators suppressed by a large mass scale, on the other hand, is easily motivated by RG flows, as we saw in Chapter \ref{ChapterMDIERG}. An advantage of the manifestly diffeomorphism-invariant ERG that makes its results so easy to implement here is its background-independent formulation, developed in Section \ref{MDIERGBIGeneral}. The classical Einstein scheme action calculated in Subsections \ref{MDIERGEinsteinBI} and \ref{MDIERG2ptE} is easily implemented by setting $a=-2b$ in the language of this chapter. The classical Weyl scheme action calculated in Subsections \ref{MDIERGWeylBI} and \ref{MDIERG2ptW} can also be implemented when we include the Einstein-Hilbert term as a relevant operator with its own length scale, as seen in Subsection \ref{MDIERGRunning}. In this case, we would have $b=sa$, where the parameter $s$ is as defined in the Weyl scheme. Cosmologists studying the late universe are usually less interested in this kind of modified gravity because it is only sensitive to physics at short distance scales, not distances comparable to the size of the present universe. This is where backreaction becomes interesting. Backreaction is a means by which fluctuations in the metric at small length scales can average to create a significant effect at large scales, forming an effective fluid described by the effective stress-energy tensor. Thus, if the metric perturbations are of high enough frequency, the higher-derivative terms can become part of a significant late-universe effect. As we saw in (\ref{StarStressTrace}), the trace of the effective stress-energy tensor for $R+R^2/6M^2$ gravity is negative and non-vanishing under weak-limit averaging. This is the ingredient required for mimicking a positive cosmological constant, although the equation of state and time-evolution would depend on the type of inhomogeneity sourcing the fluctuations.

To achieve a large backreaction effect from the higher-derivative contributions to the effective action, we require that the perturbation contains Fourier modes with very high frequency. The effective theory is obtained by integrating out high-frequency modes down to a given cutoff scale, so it is not appropriate to use such an effective theory to describe physics beyond that cutoff scale. The cutoff scale sets the mass scale of the higher-derivative terms in the expansion. Correspondingly, in the backreaction formalism, the series expansion in higher derivatives would not converge if one introduced Fourier modes of higher frequency than that given by the mass scale of the theory. Thus we require that the length scale of the effective theory should be shorter than the length scale of the perturbations. The largest backreaction effect is created when the length scale of the perturbations is closest to that very short length scale of the effective theory.

As discussed in Section \ref{StaroUV}, candidates for this very short-scale physics must be sub-nuclear to achieve a significant effect: they must come from exotic new BSM particle physics. The candidates suggested were WIMPzilla dark matter, which has problems with dilution, quantum spacetime, which would likely require a more UV complete theory to treat properly, and ``stripy'' chiral condensates \cite{Evans:2015zwa}, which would need to be disconnected from the Standard Model to contain its otherwise highly-constrained Lorentz-violating effects. In the case of $R+R^2/6M^2$ gravity, the effective vacuum energy estimate for the stripy condensates is given by (\ref{EMSest}). It is natural then to ask how this would change in the fully generalized theory developed in this chapter. First of all, in the generalized trace (\ref{totalstresstrace}), even at the level of the dimension 4 operators, we have two independent structures with independent coefficients whose sign has been left undetermined. Thus the comment that the stress-energy tensor trace is negative no longer applies in general. Indeed the two coefficients need not even have the same sign, nor do they in either the Einstein or Weyl schemes of the classical ERG. For a ``typical'' choice of parameters, (\ref{EMSest}) would still seem to be a reasonable estimate for the magnitude of the effect from the dimension 4 operators. However, the generalized trace (\ref{totalstresstrace}) also receives corrections from higher derivatives, since $a$ and $b$ are both functions of $\nabla^2/M^2$. For each additional $\nabla^2/M^2$ operator found in a term, the corresponding additive contribution to $t^{(0)}\sim E_{\rm vac}^4$ would be suppressed by an extra factor of $(\Lambda_{\rm stripe}/M)^2$. Since the value of $\Lambda_{\rm stripe}$ that mimics the observed cosmological constant was estimated to be many orders of magnitude less than $M$, these additional corrections from the higher-derivative expansion are typically small.

The aim of creating a model that realistically mimics the cosmological constant in this way is ambitious. Even if a reasonable model is produced that satisfies all experimental bounds and is shown to produce all of the correct behaviour, the original cosmological constant problem remains unsolved at the present time. It might be that the eventual solution to the cosmological constant problem does set the vacuum energy scale at the observed value, or at least allows it. Would this kill off any need for considering backreaction? Much of the discussion here has been about trying to create a large enough effect to mimic dark energy, but we should also stay alert to any proposed new physics that creates too large an effect. Of course, there could be detailed cancellations with the actual cosmological constant, but this would seem rather unreasonable. Backreaction, possessing the useful property that higher-frequency perturbations create a larger effect, might be used in the future to impose constraints on exotic new UV physics that would otherwise be hard to reach by experiments whose exploratory capability is bounded by an {\it maximum} accessible energy scale, \eg colliders. Putting these speculations aside, the stress energy tensor (\ref{totalstress}), its zero tensor constraint (\ref{genzero}) and its trace (\ref{totalstresstrace}) all apply to gravitational waves in the generalized gravity theory, without having had to fix a gauge or choose a specific background, this in itself is worthy of interest. In this way, we have begun opening the way for the high-energy physics of gravity, as developed in Chapter \ref{ChapterMDIERG} to meet with observable cosmology, as discussed in this chapter and Chapter \ref{ChapterBack1}. 

\chapter{Summary}\label{Summary}

This thesis has developed methods for calculating the physics of gravity at different scales of length. In section \ref{IntroGentle}, we discussed the challenges of computing macroscopic physical observables in systems with many microscopic degrees of freedom. In Section \ref{IntroStatMech}, we saw how this challenge is faced in statistical mechanics, paying particular attention to the Renormalization Group (RG) flow discussed \ref{IntroFlow}, which is underpinned by Kadanoff blocking, discussed in Subsection \ref{IntroKadanoff}. In Section \ref{IntroGR}, we reviewed Einstein's General Relativity theory (GR), starting from its conceptual underpinnings and covering most of the technical details required for this thesis. We finished up the Introduction with an overview of physical cosmology in Section \ref{LCDM}, starting from its foundations and setting the scene for the challenges faced in modern cosmology.

In this thesis, we have developed two diffeomorphism-invariant averaging methods. The first is the manifestly diffeomorphism-invariant Exact Renormalization Group (ERG). This is a method for calculating the continuous RG flow of a gravitational field theory without fixing a gauge. This is strongly analogous to the manifestly gauge-invariant ERG reviewed in Chapter \ref{AWHPreview1}. Section \ref{MGIERGKadanoff} describes how the Kadanoff blocking procedure discussed in Subsection \ref{IntroKadanoff} is adapted to effect a continuous RG flow in scalar field theory via integrating out the high-energy modes of the field down to a cutoff scale, $\Lambda$. Section \ref{ReviewScalar} introduces the Polchinski form of the flow equation for massless scalar fields, and discusses how it is iteratively solved from the 2-point level to higher-point levels at the classical level. Section \ref{YangMills} introduces the generalization of the Polchinski flow equation to Yang-Mills theories, which is the starting point for the manifestly gauge-invariant ERG. Section \ref{MGIERGloopwise} outlines the loopwise expansion. Beyond the classical level, additional regularization besides the cutoff function is required to tame the ultraviolet (UV) divergences. This additional regularization, discussed in \ref{MGIERGSUSY} is provided by promoting the field to a supermatrix containing fermionic Pauli-Villars fields that cancel the bosonic degrees of freedom above the cutoff scale, where the supersymmetry is broken by a super-scalar Higgs mechanism. This is analogous to how Parisi-Sourlas supersymmetry cancels degrees of freedom in a spin system driven by a disordered external field, as discussed in Subsection \ref{IntroPSSUSY}.

Chapter \ref{ChapterMDIERG} is the first chapter concerning original research conducted for this thesis, which developed the manifestly diffeomorphism-invariant ERG itself. Section \ref{MDIERGFlow} developed the flow equation in a background-independent form. Section \ref{MDIERGBIGeneral} demonstrated how the flow equation is solved at the classical level. In particular, the fixed-point action in Weyl scheme was developed in Subsection \ref{MDIERGWeylBI} and the fixed-point action in Einstein scheme was developed in Subsection \ref{MDIERGEinsteinBI}. Flows away from the fixed-point at the classical level are brought about by relevant operators with their own length scales, as discussed in Subsection \ref{MDIERGRunning}. Section \ref{MDIERGfixed} explained the fixed-background form of the flow equation, where the metric is split into a flat (Euclidean) background and a perturbation. Both the action and the kernel are expressed in fixed-background formalism as an infinite expansion in $n$-point functions. Sections \ref{MDIERGsqrtg} and \ref{MDIERGfuncder} provide the necessary tools to compute the $n$-point expansion the kernel, while Section \ref{MDIERGTran2} demonstrated that there only exist two independent momentum-dependent 2-point structures for the action that satisfy the requirement of diffeomorphism invariance. The Ward identities, following from the diffeomorphism symmetry, that relate $n$-point functions to their corresponding $(n-1)$-point functions in the fixed-background formalism were derived in Section \ref{MDIERGWard}. In particular, the Ward identities for a diffeomorphism-invariant action were derived in Subsection \ref{MDIERGWardact}, while the Ward identities for a diffeomorphism-covariant kernel were derived in Subsection \ref{MDIERGKernelWard}. The non-zero diffeomorphism transformation of the kernel was seen to be related via the Ward identity to momentum conservation in the diffeomorphism-invariant kernel, as illustrated in Subsection \ref{MDIERGWardflow}. Section \ref{MDIERG2pta} demonstrated the fixed-point solutions to the classical flow equation at the 2-point level. The fixed-point classical 2-point function was solved for the Weyl scheme in Subsection \ref{MDIERG2ptW} and for the Einstein scheme in Subsection \ref{MDIERG2ptE}. In analogy to the manifestly gauge invariant ERG, a method of introducing additional regularization for the quantum calculations is suggested in Section \ref{MDIERGSUSY} that involves promoting the metric to a supermetric for a supermanifold with both bosonic and fermionic degrees of freedom. The loop expansion for gravity is discussed in Subsection \ref{MDIERGEinsteinBI}.

From Chapter \ref{AWHPreview2} onwards, this thesis changed direction to discuss cosmological backreaction. Backreaction was also studied using a diffeomorphism-invariant averaging scheme. In Section \ref{Isaacson}, we reviewed Isaacson's shortwave approximation for gravitational radiation, focussing in particular on how it is used to derive an effective stress-energy tensor for gravitational radiation in the limit where the wavelength is much shorter than the radius of curvature of the wavefront. In Section \ref{Burnett}, we saw how the rigour of Isaacson's approach was challenged by Burnett. Burnett suggested that it is not clear that one can simply solve the field equation order-by-order in the metric perturbation, since the leading terms might also contain sub-leading contributions that are are only cancelled when other sub-leading field equation contributions are introduced. Burnett also questioned the generality of Isaacson's averaging scheme, questioning in particular whether it depends on the size and shape of the averaging region. As discussed in Section \ref{Burnett}, Burnett introduced a more rigorous approach that answers these issues, but achieves the same final results as Isaacson. Section \ref{GreennWald} reviews how this weak-limit formalism was adapted by Green and Wald to study backreaction, leading up to the revelation that the effective stress-energy tensor for backreaction in Einstein's GR is the same as for gravitational radiation: a traceless stress-energy tensor that cannot be used to induce accelerating expansion in the universe.

Undeterred by this, Chapter \ref{ChapterBack1} reported original research conducted for this thesis in which the Green and Wald formalism for backreaction was generalized to include an additional scale of length that parametrizes a higher-derivative theory of gravity. In particular, a gravity theory with Lagrangian $R+R^2/6M^2$ was considered, which is motivated as a mechanism for cosmological inflation in the early universe in good agreement with current CMB observations, as mentioned in Subsection \ref{IntroInflation}. The crucial finding here was that the effective stress-energy tensor (\ref{CombStaroStress}) had a non-vanishing negative trace (\ref{StarStressTrace}), which raised the possibility that backreaction in this modified gravity theory might be able to mimic a positive cosmological constant. Scenarios for achieving this were discussed in Section \ref{StaroUV}. Chapter \ref{ChapterBack2} concerns the complete generalization of this formalism to an effective theory of gravity whose action is Taylor-expandable in the Riemann tensor and its covariant derivatives. This has the great advantage that we can take an effective action from the background-independent form of the manifestly diffeomorphism-invariant ERG and apply it directly to the generalized effective stress-energy tensor (\ref{totalstress}) and find the trace (\ref{totalstresstrace}).

That concludes this thesis. A manifestly diffeomorphism-invariant ERG has been constructed at the classical level. Two classical fixed-point actions have been developed and RG flow along relevant directions away from the fixed-point has been illustrated at the classical level. The loopwise expansion required for the full quantum gravity construction has been laid out. A Pauli-Villars regularization scheme using supermanifolds has been suggested to tame the UV divergences at the quantum level. The formalism avoids the need for fixing a gauge by never needing to invert the 2-point function. With this manifest diffeomorphism invariance comes the freedom to choose whether or not to fix a background, with results being equivalent either way. The generalized stress-energy tensor for backreaction raises the possibility that higher-derivative theories of gravity, such as those derived using RG methods might become important in late universe cosmology if there exists an exotic source of high-frequency metric perturbations. It now remains to delve into the loopwise expansion of the manifestly diffeomorphism-invariant ERG to complete the quantum gravity construction and to seek out possible sources of high-frequency metric perturbations that might induce a significant backreaction, either to mimic a cosmological constant or to rule out such a combination of gravity theory and exotic source. Thus we have begun a new line of attack on the problem of quantum gravity and a new way that it might be found to impact on realistically observable physics.

\addcontentsline{toc}{chapter}{Bibliography} 
\bibliographystyle{hunsrt}
\bibliography{AWHPthesis}

\end{document}